\documentclass[11pt]{ucsddissertation}
\usepackage[T1]{fontenc}
\usepackage{notomath}
\usepackage[NoDate]{currvita}
\usepackage{array}
\usepackage{tabularx}
\usepackage{booktabs}
\usepackage{ragged2e}
\usepackage{microtype}
\usepackage{xcolor}
\definecolor{myblue}{rgb}{0,0.3,0.6}
\usepackage[hidelinks]{hyperref} %colorlinks=true,linktocpage
\usepackage{graphicx}
\AtBeginDocument{%
    \settowidth\cvlabelwidth{\cvlabelfont 0000--0000}%
}

\title{\textbf{\large Generative AI for Music and Audio}}
\author{Hao-Wen Dong}
\degree{Computer Science}{Doctor of Philosophy}
\cochair{Professor Taylor Berg-Kirkpatrick}
\cochair{Professor Julian McAuley}
\committee{Professor Shlomo Dubnov}
\committee{Professor Lawrence Saul}
\degreeyear{2024}

\titleformat{\paragraph}[runin]{\normalfont\fontseries{bm}\selectfont}{\theparagraph}{1em}{}
\titlespacing{\paragraph}{.5in}{.5ex}{1em}

\usepackage{changepage}
\newenvironment{abstract}{%
    \phantomsection%
    \addcontentsline{toc}{section}{Abstract}%
    \begin{adjustwidth}{.5in}{.5in}%
        \begin{center}\large\bfseries Abstract\end{center}}%
    {\end{adjustwidth}\bigskip\par}

\usepackage{csquotes}
\usepackage[backend=biber,style=authoryear,language=american,date=year,dashed=false,isbn=false,natbib=true,maxbibnames=99,maxcitenames=2,uniquelist=false,backref=true,uniquename=false]{biblatex} %sorting=none
\DefineBibliographyStrings{english}{backrefpage={cited on page},backrefpages={cited on pages}}
\DeclareNameAlias{author}{given-family}
\renewbibmacro{in:}{}
\DeclareCiteCommand{\cite}
  {\usebibmacro{prenote}}
  {\usebibmacro{citeindex}%
   \printtext[bibhyperref]{\usebibmacro{cite}}}
  {\multicitedelim}
  {\usebibmacro{postnote}}

\DeclareCiteCommand*{\cite}
  {\usebibmacro{prenote}}
  {\usebibmacro{citeindex}%
   \printtext[bibhyperref]{\usebibmacro{citeyear}}}
  {\multicitedelim}
  {\usebibmacro{postnote}}

\DeclareCiteCommand{\parencite}[\mkbibparens]
  {\usebibmacro{prenote}}
  {\usebibmacro{citeindex}%
    \printtext[bibhyperref]{\usebibmacro{cite}}}
  {\multicitedelim}
  {\usebibmacro{postnote}}

\DeclareCiteCommand*{\parencite}[\mkbibparens]
  {\usebibmacro{prenote}}
  {\usebibmacro{citeindex}%
    \printtext[bibhyperref]{\usebibmacro{citeyear}}}
  {\multicitedelim}
  {\usebibmacro{postnote}}

\DeclareCiteCommand{\footcite}[\mkbibfootnote]
  {\usebibmacro{prenote}}
  {\usebibmacro{citeindex}%
  \printtext[bibhyperref]{ \usebibmacro{cite}}}
  {\multicitedelim}
  {\usebibmacro{postnote}}

\DeclareCiteCommand{\footcitetext}[\mkbibfootnotetext]
  {\usebibmacro{prenote}}
  {\usebibmacro{citeindex}%
   \printtext[bibhyperref]{\usebibmacro{cite}}}
  {\multicitedelim}
  {\usebibmacro{postnote}}

\DeclareCiteCommand{\textcite}
  {\boolfalse{cbx:parens}}
  {\usebibmacro{citeindex}%
   \printtext[bibhyperref]{\usebibmacro{textcite}}}
  {\ifbool{cbx:parens}
     {\bibcloseparen\global\boolfalse{cbx:parens}}
     {}%
   \multicitedelim}
  {\usebibmacro{textcite:postnote}}

\usepackage{amsmath}
\usepackage{enumitem}
\usepackage[capitalise,noabbrev]{cleveref}
\usepackage{bm}
\usepackage{bbm}
\usepackage{multirow}
\usepackage{tabularx}
\usepackage{url}
\usepackage{fontawesome5}
\usepackage{makecell}
\usepackage{appendix}
\appendixtocoff

\usepackage{etoolbox}
\AtBeginEnvironment{subappendices}{%
    \phantomsection%
    \addcontentsline{toc}{section}{Appendices}%
    \begin{center}\Large\bfseries Appendices\end{center}%
    \vspace{1ex}%
    \renewcommand{\addcontentsline}[3]{}}

\hypersetup{linkcolor=myblue,urlcolor=myblue,citecolor=myblue,anchorcolor=myblue}
\newcommand{\secbreak}{\begin{center}$\ast\ast\ast$\end{center}}

\newcommand{\gcheck}{\checkmark}
\newcommand{\rcross}{}
\newcommand{\ytriangle}{{\footnotesize$\triangle$}}

\DeclareMathOperator*{\argmin}{arg\,min}
\crefformat{footnote}{#2\footnotemark[#1]#3}

\definecolor{tab-blue}{HTML}{1f77b4}
\definecolor{tab-orange}{HTML}{ff7f0e}
\definecolor{tab-green}{HTML}{2ca02c}
\definecolor{tab-red}{HTML}{d62728}
\definecolor{tab-purple}{HTML}{9467bd}

\crefrangeformat{footnote}{#3\footnotemark[#1]#4--#5\footnotemark[#2]#6}
\crefmultiformat{footnote}{#2\footnotemark[#1]#3}{\textsuperscript{,}#2\footnotemark[#1]#3}{\textsuperscript{,}#2\footnotemark[#1]#3}{\textsuperscript{,}#2\footnotemark[#1]#3}
\newcommand{\notesize}{\footnotesize}
\newlength{\figwidth}
\newlength{\figheight}

\newcolumntype{C}{>{\centering\arraybackslash}X}
\definecolor{darkred}{rgb}{0.6,0.0,0.0}
\definecolor{darkblue}{rgb}{0.0,0.0,0.75}

\usepackage{pifont}
\newcommand{\cmark}{\ding{51}}
\newcommand{\xmark}{\ding{55}}

\newcolumntype{L}[1]{>{\raggedright\arraybackslash}m{#1}}

\AtEndPreamble{%
    \hypersetup{%
        pdfauthor = {Hao-Wen Dong},%
        pdftitle = {Generative AI for Music and Audio},%
        pdfsubject = {},%
        pdfkeywords = {}%
    }%
}

\begin{document}

\addtocontents{toc}{\bigskip}
\addtocontents{lof}{\bigskip}
\addtocontents{lot}{\bigskip}

% Begin with frontmatter and so forth
\frontmatter
\maketitle
\makecopyright
\makesignature

% Optional
\begin{dedication}
\vfill
\begin{center}
    \itshape
    To my wife Eltha, my parents Nancy and David, and my pooh bear,\\
    without whom nothing would have been possible.
\end{center}
\vfill
%     \setsinglespacing
%     \raggedright % It would be better to use \RaggedRight from ragged2e
%     \parindent0pt\parskip\baselineskip
%     \todo
%     In recognition of reading this manual before beginning to format the
%     doctoral dissertation or master's thesis; for following the
%     instructions written herein; for consulting with OGS Academic Affairs
%     Advisers; and for not relying on other completed manuscripts, this
%     manual is dedicated to all graduate students about to complete the
%     doctoral dissertation or master's thesis.
    
%     In recognition that this is my one chance to use whichever
%     justification, spacing, writing style, text size, and/or textfont that
%     I want to while still keeping my headings and margins consistent.
\end{dedication}

% Optional
\begin{epigraph}
\vfill
\begin{center}
    \itshape
    Without music, life would be a mistake.\\
    \normalfont
    ---Friedrich Nietzsche
\end{center}
% \begin{center}
%     \itshape
%     Works of art make rules;\\
%     rules do not make works of art.\\
%     \normalfont
%     ---Claude Debussy
% \end{center}

% \begin{adjustwidth}{1.8in}{1.8in}
%     \raggedright\itshape
%     Works of art make rules;\\
%     rules do not make works of art.\\
%     \raggedleft\normalfont
%     ---Claude Debussy
% \end{adjustwidth}
\vfill
% \vskip0pt plus.5fil
% \vfil

% \setsinglespacing
% {\flushright
% True ease in writing comes from art, not chance,\\
% As those move easiest who have learn'd to dance.\\
% 'T is not enough to no harshness gives offence,---\\
% The sound must seem an echo to the sense.

% \vskip\baselineskip
% \textit{Alexander Pope}\par}
% \vfil
% \begin{center}
% You write with ease to show your breeding,\\
% But easy writing's curst hard reading.

% \vskip\baselineskip
% \textit{Richard Brinsley Sheridan}
% \end{center}
% \vfil
% \noindent Writing, at its best, is a lonely life. Organizations for
% writers palliate the writer's loneliness, but I doubt if they improve
% his writing. He grows in public stature as he sheds his loneliness and
% often his work deteriorates. For he does his work alone and if he is a
% good enough writer he must face eternity, or the lack of it, each day.

% \vskip\baselineskip
% \hskip0pt plus1fil\textit{Ernest Hemingway}\hskip0pt plus4fil\null

% \vfil
\end{epigraph}

% Next comes the table of contents, list of figures, list of tables,
% etc. If you have code listings, you can use \listoflistings (or
% \lstlistoflistings) to have it be produced here as well. Same with
% \listofalgorithms.
\tableofcontents
\listoffigures
\listoftables

% Preface
% \begin{preface}
% \todo
% Almost nothing is said in the manual about the preface. There is no
% indication about how it is to be typeset. Given that, one is forced to
% simply typeset it and hope it is accepted. It is, however, optional
% and may be omitted.
% \end{preface}

% Your fancy acks here. Keep in mind you need to ack each paper you
% use. See the examples here. In addition, each chapter ack needs to
% be repeated at the end of the relevant chapter.
\begin{acknowledgements}
% Julian & Taylor
First and foremost I want to express my deepest gratitude to my advisors Julian McAuley and Taylor Berg-Kirkpatrick for their guidance and support throughout my PhD years. While my PhD topic is not in their core fields of expertise, Julian and Taylor have always been providing me insightful and constructive advice on my research. I am fortunate to have such amazing advisors who trust and push me to pursue my own research agenda, and give me much freedom on doing internships elsewhere frequently. For the past few years, Julian and Taylor have greatly shaped my attitude towards research and life, which I will carry on to my future endeavors. In particular, I want to thank Julian for providing me invaluable advice on my career and job search, and I want to thank Taylor for guiding me how to think about and approach my research.

I would also like to thank Shlomo Dubnov for serving on my doctoral committee and for his thoughtful advice and genuine support on my research and career. I also thank Lawrence Saul for serving on my doctoral committee and revamping my writing skills in his amazing Scientific Teaching course. I also want to thank Yi-Hsuan Yang for bringing me into this field and for his continued guidance along my career.

% Family
I want to extend my sincere gratitude to my wife, my parents, my pooh bear and my family, without whom nothing would have been possible. My warmest thanks to my wife Eltha Teng for her love, support and companionship. It is almost like a dream to have done our PhD together at UC San Diego, and I am fortunate to have gone through all the ups and downs with you. It is you that make this journey joyful and memorable. My heartfelt thanks to my parents Nancy Yen and David Dong for their unconditional love and support that have shaped me into the person I am today. My sweetest thanks to my pooh bear for being my best buddy for 26 years and accompanying me through all the deadlines.

% Colleagues and friends
I would also like to thank all my colleagues and friends I met at UC San Diego. Huge thanks to Ke Chen for collaborating on several projects and being my go-to person whenever I want to discuss some new ideas. Big thanks to Zachary Novack for co-organizing the AI Music reading group and collaborating on several projects. Special thanks to Chris Donahue for introducing me to Julian and Taylor when I just started my PhD. Thanks to Weihan Xu, Haven Kim, Sanjayan Sreekala, Chris Francis, Sachinda Edirisooriya, Wanning Lu, Hoang Phan, Diego Reyes, Phillip Long, Junda Wu and Zhouhang Xie for collaborating with me on several projects. Thanks to Zhankui He and Noveen Sachdeva for co-maintaining our lab machines and fixing them in the noisy server room at SDSC once in a while. Thanks to my labmates in McAuley Lab and Berg Lab for their helpful feedback on my work and all the fun we had. Thank you to my friends Yueh-Hua Wu, I-Lin Yeh, Chun-Jhen Lai, Kai-En Lin, Chih-Chun Hsu, I-Da Chiang and Fang-Chi Leong (just to name a few) for enriching my life outside of research.
% I would also like to thank all my colleagues and friends I met at UC San Diego (to name a few): Ke Chen, Zachary Novack, Yueh-Hua Wu, Chun-Jhen Lai, I-Lin Yeh, Chih-Chun Hsu, I-Da Chiang, Fang-Chi Leong, Weihan Xu, Zhankui He, Zexue He, Danlu Chen, Nikita Srivatsan, Noveen Sachdeva, Zhouhang Xie, Junda Wu, Sanjayan Sreekala, Chris Francis, Sachinda Edirisooriya, Wanning Lu, Hoang Phan and Diego Reyes. In particular, big thanks to Chris Donahue for introducing me to Julian and Taylor when I just started my PhD. Huge thanks to Ke Chen for collaborating on several projects and being my go-to person whenever I want to discuss some new ideas. Many thanks to Zachary Novack for co-organizing the AI Music reading group and collaborating on several projects.

% Internships
Part of this dissertation was done during internships at NVIDIA, Adobe, Dolby, Amazon and Sony. I want to express my sincere gratitude to my internship mentors (in chronological order): Cong Zhou, Naoya Takahashi, Wenbo Zhao, Gunnar Sigurdsson, Xiaoyu Liu, Justin Salamon, Oriol Nieto, Siddharth Gururani and Ming-Yu Liu for their guidance and support during my internships as well as the time and effort they devoted in my project. I would also like to thank all my colleagues and friends I met during my internships. Special thanks to Julia Wilkins, Jordi Pons and Chenyang Tao for collaborating and offering useful advice on my projects. Thanks to Junghyun Koo, Tung-Cheng Wu, Vincent Cheung, Yun-Hsuan Chen, Siddharth Nijhawan, Kin Wai Cheuk, Ilaria Manco, Jiawen Huang and Marco Martínez for all the fun we had in Tokyo during my internship at Sony.
% Jiaojiao Fan, Tingle Li, Geneping Yang
% Wei-Hsiang Liao and Yuki Mitsufuji.

% Collaborators & co-authors
% Thank you to all my co-authors who co-authored the publications included in this dissertation (in chronological order): Julian McAuley, Taylor Berg-Kirkpatrick, Ke Chen, Chris Donahue, Cong Zhou, Naoya Takahashi, Yuki Mitsufuji, Shlomo Dubnov, Xiaoyu Liu, Jordi Pons, Gautam Bhattacharya, Santiago Pascual and Joan Serrà.
Thank you to all my collaborators (in alphabetical order): Taylor Berg-Kirkpatrick, Ke Chen, Tagyoung Chung, Chris Donahue, Shlomo Dubnov, Sachinda Edirisooriya, Chris Francis, Satoru Fukayama, Benjamin Genchel, Arpit Gupta, Siddharth Gururani, Jing Huang, Dasaem Jeong, Jiun-Yu Kao, Haven Kim, Tetsuro Kitahara, Yu-Hsiang Lin, Hao-Min Liu, Xiaoyu Liu, Ming-Yu Liu, Phillip Long, Julian McAuley, Yuki Mitsufuji, Anjali Narayan-Chen, Oriol Nieto, Zachary Novack, Nanyun Peng, Jordi Pons, Keijiro Saino, Justin Salamon, Joan Serrà, Gunnar Sigurdsson, Sanjayan Sreekala, Naoya Takahashi, Chih-Pin Tan, Chenyang Tao, Julia Wilkins, Junda Wu, Wen-Yi Xiao, Zhouhang Xie, Weihan Xu, Li-Chia Yang, Yi-Hsuan Yang, Yin-Cheng Yeh, Wenbo Zhao and Cong Zhou. It has been wonderful working with you, and I look forward to collaborating with you again in the future.
% Unsorted list
% Ke Chen, Zachary Novack, Sachinda Edirisooriya, Junda Wu, Zhouhang Xie, Phillip Long, Wen-Yi Xiao, Li-Chia Yang, Hao-Min Liu, Satoru Fukayama, Tetsuro Kitahara, Sanjayan Sreekala, Chris Francis, Haven Kim, Chris Donahue, Weihan Xu, Dasaem Jeong, Yin-Cheng Yeh, Chih-Pin Tan, Benjamin Genchel, Naoya Takahashi, Yuki Mitsufuji, Julian McAuley, Taylor Berg-Kirkpatrick, Shlomo Dubnov, Yi-Hsuan Yang, Cong Zhou, Xiaoyu Liu, Joan Serrà, Justin Salamon, Oriol Nieto, Julia Wilkins, Wenbo Zhao, Gunnar Sigurdsson, Chenyang Tao, Jiun-Yu Kao, Yu-Hsiang Lin, Jordi Pons, Keijiro Saino, Siddharth Gururani, Ming-Yu Liu, Anjali Narayan-Chen, Arpit Gupta, Tagyoung Chung, Jing Huang, Nanyun Peng. 

% Funding
Finally, I want to thank UCSD ECE Department, J. Yang and Family Foundation, and Taiwan Ministry of Education for their generous support for my PhD study.

\secbreak

\itshape

% Publications
This dissertation contains material from the following publications:
\begin{itemize}
    \item \cref{chap:muspy}, in full, is a reprint of the material as it appears in ``MusPy: A Toolkit for Symbolic Music Generation'' by Hao-Wen Dong, Ke Chen, Julian McAuley and Taylor Berg-Kirkpatrick, which was published in the Proceedings of the International Society for Music Information Retrieval Conference (ISMIR) in 2020. The dissertation author was the primary investigator and author of this paper.
    \item \cref{chap:mmt}, in full, is a reprint of the material as it appears in ``Multitrack Music Transformer'' by Hao-Wen Dong, Ke Chen, Shlomo Dubnov, Julian McAuley and Taylor Berg-Kirkpatrick, which was published in the Proceedings of the IEEE International Conference on Acoustics, Speech and Signal Processing (ICASSP) in 2023. The dissertation author was the primary investigator and author of this paper.
    \item \cref{chap:arranger}, in full, is a reprint of the material as it appears in ``Towards Automatic Instrumentation by Learning to Separate Parts in Symbolic Multitrack Music'' by Hao-Wen Dong, Chris Donahue, Taylor Berg-Kirkpatrick and Julian McAuley, which was published in the Proceedings of the International Society for Music Information Retrieval Conference (ISMIR) in 2021. The dissertation author was the primary investigator and author of this paper.
    \item \cref{chap:deepperformer}, in full, is a reprint of the material as it appears in ``Deep Performer: Score-to-Audio Music Performance Synthesis'' by Hao-Wen Dong, Cong Zhou, Taylor Berg-Kirkpatrick and Julian McAuley, which was published in the Proceedings of the IEEE International Conference on Acoustics, Speech and Signal Processing (ICASSP) in 2022. The dissertation author was the primary investigator and author of this paper.
    \item \cref{chap:clipsep}, in full, is a reprint of the material as it appears in ``CLIPSep: Learning Text-queried Sound Separation with Noisy Unlabeled Videos'' by Hao-Wen Dong, Naoya Takahashi, Yuki Mitsufuji, Julian McAuley and Taylor Berg-Kirkpatrick, which was published in the Proceedings of the International Conference on Learning Representations (ICLR) in 2023. The dissertation author was the primary investigator and author of this paper.
    \item \cref{chap:clipsonic}, in full, is a reprint of the material as it appears in ``CLIPSonic: Text-to-Audio Synthesis with Unlabeled Videos and Pretrained Language-Vision Models'' by Hao-Wen Dong, Xiaoyu Liu, Jordi Pons, Gautam Bhattacharya, Santiago Pascual, Joan Serrà, Taylor Berg-Kirkpatrick and Julian McAuley, which was published in the Proceedings of the IEEE Workshop on Applications of Signal Processing to Audio and Acoustics (WASPAA) in 2023. The dissertation author was the primary investigator and author of this paper.
\end{itemize}

\end{acknowledgements}

\def\arraystretch{1.5}

% Stupid vita goes next
\begin{vita}
\bigskip
\noindent
\textbf{\large Education}
\medskip

\noindent
\begin{tabular}{@{}l@{~}c@{~}ll@{}}
    2019 &-- &2024 &\textit{Doctor of Philosophy in Computer Science}, University of California San Diego\\
    2019 &-- &2021 &\textit{Master of Science in Computer Science}, University of California San Diego\\
    2013 &-- &2017 &\textit{Bachelor of Science in Electrical Engineering}, National Taiwan University
\end{tabular}
% \begin{cv}{}
%     \begin{cvlist}{}
%         \item[2013--2017] Bachelor of Science, National Taiwan University
%         \item[2019--2021] Master of Science, University of California San Diego
%         \item[2019--2024] Doctor of Philosophy, University of California San Diego
%     \end{cvlist}
% \end{cv}

\bigskip\bigskip
\noindent
\textbf{\large Professional Experience}
\medskip

\noindent
\begin{tabular}{@{}l@{~}l@{~}c@{~}l@{~}ll@{}}
    Sep &2023 &-- &Dec &2023 &\textit{Research Intern}, NVIDIA\\
    May &2023 &-- &Sep &2023 &\textit{Research Scientist/Engineer Intern}, Adobe\\
    Jan &2023 &-- &Apr &2023 &\textit{Speech/Audio Deep Learning Intern}, Dolby\\
    Sep &2022 &-- &Jan &2023 &\textit{Applied Scientist Intern}, Amazon\\
    May &2022 &-- &Sep &2022 &\textit{Student Intern}, Sony\\
    Jun &2021 &-- &Sep &2021 &\textit{Deep Learning Audio Intern}, Dolby\\
    May &2019 &-- &Aug &2019 &\textit{Research Intern}, Yamaha\\
    Jul &2017 &-- &Apr &2019 &\textit{Research Assistant}, Academia Sinica\\
\end{tabular}

% This puts in the PUBLICATIONS header. Note that it appears inside
% the vita environment. It is optional.
% \publications
% \noindent``Distributions of Control Points in a System for Analysis of Stress
% Distribution'' IRE Transactions of the I.R.E\@. Professional Group on
% Automatic Control, vol. AC-7, pp 272--289, September 2005
\bigskip\bigskip
\noindent\textbf{\large Publications}
\begin{itemize}[label={},leftmargin=*,labelsep=0pt]
    \item Weihan Xu, Julian McAuley, Shlomo Dubnov, and \underline{Hao-Wen Dong}, ``\href{https://arxiv.org/pdf/2311.12257.pdf}{Equipping Pretrained Unconditional Music Transformers with Instrument and Genre Controls},'' \textit{IEEE Big Data Workshop on AI Music Generation (AIMG)}, 2023. 
    \item \underline{Hao-Wen Dong}, Xiaoyu Liu, Jordi Pons, Gautam Bhattacharya, Santiago Pascual, Joan Serrà, Taylor Berg-Kirkpatrick, and Julian McAuley, ``\href{https://arxiv.org/pdf/2306.09635.pdf}{CLIPSonic: Text-to-Audio Synthesis with Unlabeled Videos and Pretrained Language-Vision Models},'' \textit{IEEE Workshop on Applications of Signal Processing to Audio and Acoustics (WASPAA)}, 2023.
    \item \underline{Hao-Wen Dong}, Gunnar A. Sigurdsson, Chenyang Tao, Jiun-Yu Kao, Yu-Hsiang Lin, Anjali Narayan-Chen, Arpit Gupta, Tagyoung Chung, Jing Huang, Nanyun Peng, and Wenbo Zhao, ``\href{https://sightsound.org/papers/2023/Dong_CLIPSynth_Learning_Text-to-audio_Synthesis_from_Videos.pdf}{CLIPSynth: Learning Text-to-audio Synthesis from Videos using CLIP and Diffusion Models},'' \textit{CVPR Workshop on Sight and Sound (WSS)}, 2023.
    \item \underline{Hao-Wen Dong}, Ke Chen, Shlomo Dubnov, Julian McAuley, and Taylor Berg-Kirkpatrick, ``\href{https://arxiv.org/pdf/2207.06983.pdf}{Multitrack Music Transformer},'' \textit{IEEE International Conference on Acoustics, Speech and Signal Processing (ICASSP)}, 2023.
    \item \underline{Hao-Wen Dong}, Naoya Takahashi, Yuki Mitsufuji, Julian McAuley, and Taylor Berg-Kirkpatrick, ``\href{https://arxiv.org/pdf/2212.07065.pdf}{CLIPSep: Learning Text-queried Sound Separation with Noisy Unlabeled Videos},'' \textit{International Conference on Learning Representations (ICLR)}, 2023.
    \item Ke Chen, \underline{Hao-Wen Dong}, Yi Luo, Julian McAuley, Taylor Berg-Kirkpatrick, Miller Puckette, and Shlomo Dubnov, ``\href{https://arxiv.org/pdf/2209.02871.pdf}{Improving Choral Music Separation through Expressive Synthesized Data from Sampled Instruments},'' \textit{International Society for Music Information Retrieval Conference (ISMIR)}, 2022.
    \item \underline{Hao-Wen Dong}, Cong Zhou, Taylor Berg-Kirkpatrick, and Julian McAuley, ``\href{https://arxiv.org/pdf/2202.06034.pdf}{Deep Performer: Scoreto-Audio Music Performance Synthesis},'' \textit{IEEE International Conference on Acoustics, Speech and Signal Processing (ICASSP)}, 2022.
    \item \underline{Hao-Wen Dong}, Chris Donahue, Taylor Berg-Kirkpatrick, and Julian McAuley, ``\href{https://arxiv.org/pdf/2107.05916.pdf}{Towards Automatic Instrumentation by Learning to Separate Parts in Symbolic Multitrack Music},'' \textit{International Society for Music Information Retrieval Conference (ISMIR)}, 2021.
    \item Sachinda Edirisooriya, \underline{Hao-Wen Dong}, Julian McAuley, and Taylor Berg-Kirkpatrick, ``\href{https://arxiv.org/pdf/2108.01769.pdf}{An Empirical Evaluation of End-to-End Polyphonic Optical Music Recognition},'' \textit{International Society for Music Information Retrieval Conference (ISMIR)}, 2021.
    \item Yin-Cheng Yeh, Wen-Yi Hsiao, Satoru Fukayama, Tetsuro Kitahara, Benjamin Genchel, Hao-Min Liu, \underline{Hao-Wen Dong}, Yian Chen, Terence Leong, and Yi-Hsuan Yang, ``\href{https://arxiv.org/pdf/2001.02360.pdf}{Automatic Melody Harmonization with Triad Chords: A Comparative Study},'' \textit{Journal of New Music Research (JNMR)}, 50(1):37–51, 2021.
    \item \underline{Hao-Wen Dong}, Ke Chen, Julian McAuley, and Taylor Berg-Kirkpatrick, ``\href{https://arxiv.org/pdf/2008.01951.pdf}{MusPy: A Toolkit for Symbolic Music Generation},'' \textit{International Society for Music Information Retrieval Conference (ISMIR)}, 2020.
    \item \underline{Hao-Wen Dong} and Yi-Hsuan Yang, ``\href{https://arxiv.org/pdf/1804.09399.pdf}{Convolutional Generative Adversarial Networks with Binary Neurons for Polyphonic Music Generation},'' \textit{International Society for Music Information Retrieval Conference (ISMIR)}, 2018.
    \item \underline{Hao-Wen Dong}, Wen-Yi Hsiao, and Yi-Hsuan Yang, ``\href{https://salu133445.github.io/pypianoroll/pdf/pypianoroll_ismir2018_lbd_paper.pdf}{Pypianoroll: Open Source Python Package for Handling Multitrack Pianorolls},'' \textit{ISMIR Late-Breaking Demos}, 2018.
    \item \underline{Hao-Wen Dong}\textsuperscript{*}, Wen-Yi Hsiao\textsuperscript{*}, Li-Chia Yang, and Yi-Hsuan Yang, ``\href{https://arxiv.org/pdf/1709.06298.pdf}{MuseGAN: Multi-Track Sequential Generative Adversarial Networks for Symbolic Music Generation and Accompaniment},'' \textit{AAAI Conference on Artificial Intelligence (AAAI)}, 2018. {\footnotesize(\textsuperscript{*}equal contribution)}
    \item \underline{Hao-Wen Dong}\textsuperscript{*}, Wen-Yi Hsiao\textsuperscript{*}, Li-Chia Yang, and Yi-Hsuan Yang, ``\href{https://salu133445.github.io/musegan/pdf/musegan-ismir2017-lbd-paper.pdf}{MuseGAN: Demonstration of a Convolutional GAN Based Model for Generating Multi-track Piano-rolls},'' \textit{ISMIR Late-Breaking Demos}, 2017. {\footnotesize(\textsuperscript{*}equal contribution)}
\end{itemize}

% This puts in the FIELDS OF STUDY. Also inside vita and also
% optional.
% \fieldsofstudy
% \noindent Major Field: Engineering (Specialization or Focused Studies)
% \vskip\baselineskip
% Studies in Applied Mathematics\par
% Professors Alpha Beta and Gamma Delta
% \vskip\baselineskip
% Studies in Mechanices\par
% Professors Epsilon Zeta and Eta Theta
% \vskip\baselineskip
% Studies in Electromagnetism\par
% Professors Iota Kappa and Lambda Mu
\end{vita}

\def\arraystretch{1}

% Put your maximum 350 word abstract here.
\begin{dissertationabstract}
Generative AI has been transforming the way we interact with technology and consume content. In the next decade, AI technology will reshape how we create audio content in various media, including music, theater, films, games, podcasts, and short videos. In this dissertation, I introduce the three main directions of my research centered around \textit{generative AI for music and audio}: 1) multitrack music generation, 2) assistive music creation tools, and 3) multimodal learning for audio and music. Through my research, I aim to answer the following two fundamental questions: 1) \textit{How can AI help professionals or amateurs create music and audio content?} 2) \textit{Can AI learn to create music in a way similar to how humans learn music?} My long-term goal is to lower the barrier of entry for music composition and democratize audio content creation.
\end{dissertationabstract}

% This is where the main body of your dissertation goes!
\mainmatter

% Optional Introduction
% \begin{dissertationintroduction}
% \todo
% This optional section is barely described in the OGS manual other than
% saying it is optional and that it appears in the table of contents
% between the Abstract and the first chapter.

% No formatting guidelines appear so presumably, it should be formatted
% like an ordinary chapter. It should appear after the
% \verb!\mainmatter! macro because it should start on page~1.
% \end{dissertationintroduction}

\graphicspath{{chapters/intro/figs/}}
\chapter{Introduction}
\label{chap:intro}

\begin{center}
    \itshape
    AI is the study of how to make computers do things\\at which, at the moment, people are better.\\
    \normalfont
    ---Elaine Rich and Kevin Knight
\end{center}

\bigskip

Generative AI has been transforming the way we interact with technology and consume content. The recent success of large language model-based chatbots (e.g., OpenAI's ChatGPT\footnote{\url{https://chat.openai.com/}} and Google's Gemini\footnote{\url{https://gemini.google.com/}}), AI assistants (e.g., GitHub and Microsoft Copilot) and text-to-image generation systems (e.g., Adobe Firefly,\footnote{\url{https://firefly.adobe.com/}} Midjourney\footnote{\url{https://www.midjourney.com/}} and Stable Diffusion\footnote{\url{https://stability.ai/stable-image}}) showcases how AI-powered technology can be integrated into professional workflows and boost human productivity. In the next decade, generative AI technology will also reshape how we create audio content in the \$2.3 trillion global entertainment industry, including the music, film, TV, podcast and gaming sectors. Take AI-powered music creation for example: On one hand, we have witnessed major progress in automatic music composition \citep{briot2017survey,huang2020aisongcontest}, which has long been considered as a grand challenge of AI. On the other hand, our expectations of \textit{AI Music} today has expanded to cover the whole music creation process---from composition, arrangement, sound production, recording to mixing \citep{deman2019production}. With a growing momentum in both academia and industry, AI-powered audio creation has been gaining attention in the broader AI community.

My research springs from two fundamental questions: 1) \textit{How can AI help professionals or amateurs create music and audio content?} 2) \textit{Can AI learn to create music in a way similar to how humans learn music?} From a musical perspective, technology has always been a driving factor of music evolution. For example, the study of acoustics and musical instrument making fostered the development of classical music; the invention of synthesizers and drum machines helped popularize electronic music. I am thus interested in exploring how the latest AI technology can empower artists to create novel contents. From a technical perspective, music possesses a unique complexity in that music follows rules and patterns while being creative and expressive at the same time. I am thus fascinated about the idea of building intelligent systems that can learn, create and play music like humans do. I envision the future development of AI Music to be a two-way process---\textit{new technology creates new music; new music inspires new technology}.

\begin{figure}
    \centering
    \includegraphics[width=\linewidth]{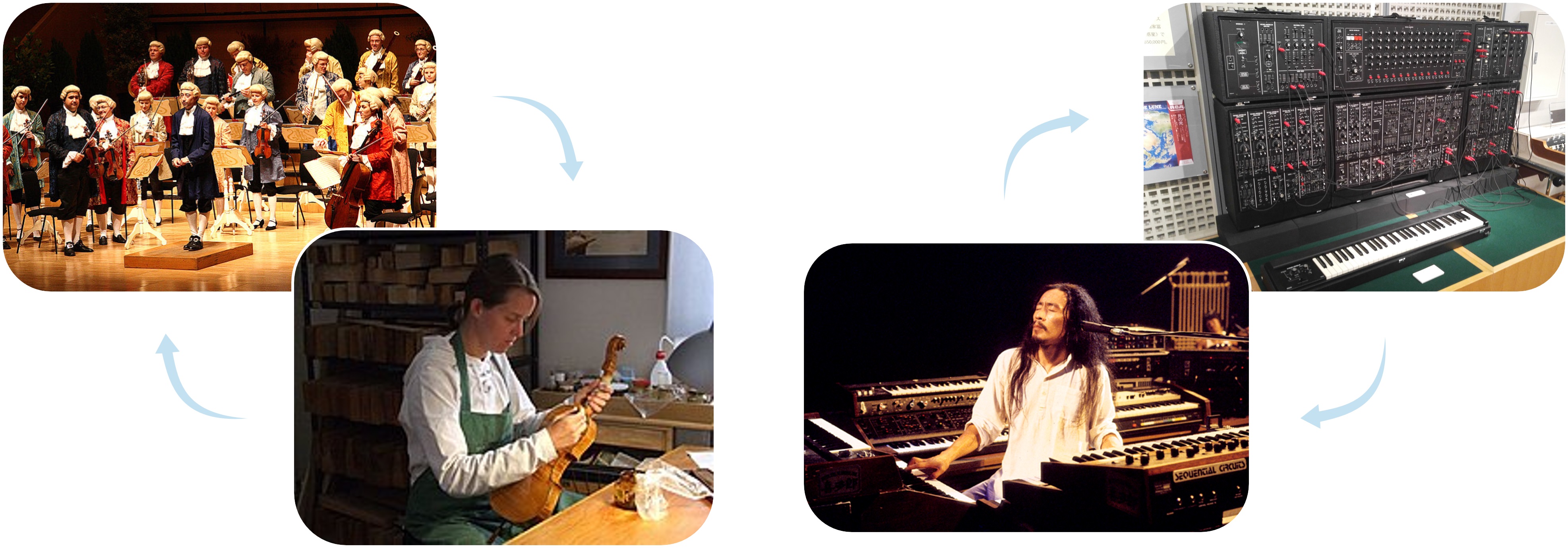}
    \caption[Looking back to the past, how music and technology interacts has always been a two-way process.]{Looking back to the past, how music and technology interacts has always been a two-way process. (Left) the violin-making industry grows with the classical music, and together create the golden age of classical music. (Right) the invention and development of synthesizers and drum machines helped popularize electronic music. Image sources (from left to right): 1) Mozart83, Public domain, via Wikimedia Commons, 2) Hildegard Dodel, Public domain, via Wikimedia Commons, 3) yan, CC BY-SA 4.0, via Wikimedia Commons, and 4) taken at Hamamatsu Museum of Musical Instruments, August 2019.}
    \label{intro:fig:music_tech}
\end{figure}

Motivated by this belief, I study a wide range of topics centered around \textit{Generative AI for Music and Audio}, including multitrack music generation \citep{dong2018musegan,dong2017musegandemo,dong2018binarymusegan,dong2023mmt,xu2023mmt2,liu2018musegangtc}, automatic instrumentation \citep{dong2021arranger}, automatic arrangement \citep{dong2018musegan,liu2018musegangtc}, automatic harmonization \citep{yeh2020harmonization}, music performance synthesis \citep{dong2021deepperformer}, text-queried sound separation \citep{dong2023clipsep}, text-to-audio synthesis \citep{dong2023clipsynth,dong2023clipsonic} and symbolic music processing software \citep{dong2018pypianoroll,dong2020muspy}.

% \begin{figure}
%     \centering
%     \includegraphics[width=.8\linewidth]{ai_music_audio.jpg}
%     \caption{An illustration on how my work aim to empower music and audio creation through machine learning.}
%     \label{intro:fig:overview}
% \end{figure}

\begin{figure}
    \centering
    \includegraphics[width=\linewidth]{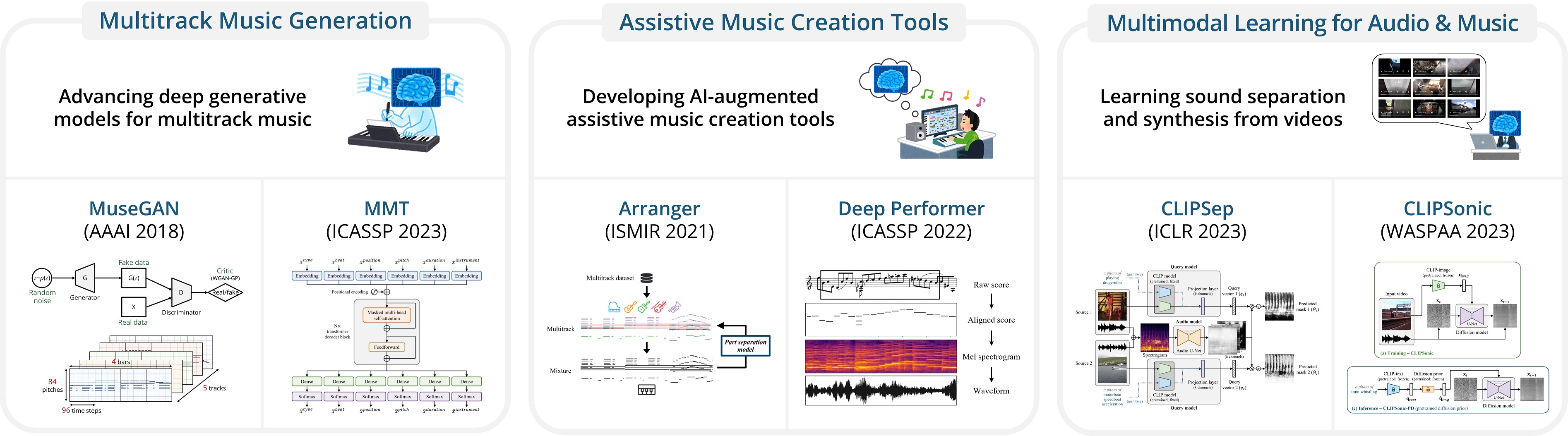}
    \caption{An overview of the three main directions of my research.}
    \label{intro:fig:overview}
\end{figure}

My research can be categorized into three main directions, as shown in \cref{intro:fig:overview}:
% \textit{multitrack music generation}, \textit{assistive music creation tools} and \textit{multimodal learning for audio and music}.
1) \textbf{multitrack music generation}---\textit{advancing deep generative models for multitrack music}, 2) \textbf{assistive music creation tools}---\textit{developing AI tools that can help musicians and amateurs create music}, and 3) \textbf{multimodal learning for audio and music}---\textit{learning sound separation and synthesis from noisy videos}.
% \begin{itemize}
%     \item \textbf{Multitrack music generation}---advancing deep generative models for multitrack music
%     \item \textbf{Assistive music creation tools}---developing AI tools that can help musicians and amateurs create music
%     \item \textbf{Multimodal learning for audio and music}---learning sound separation and synthesis from noisy videos
% \end{itemize}

%====================================
\section{Multitrack Music Generation}
%====================================

Researchers have been working on automatic music composition for decades, and it has long been viewed as a grand challenge of AI. I started this thread of research on multitrack music generation in 2017. Back then, prior work on deep learning-based music generation had focused on generating melodies, lead sheets (i.e., melodies and chords) or four-part chorales \cite{briot2017survey}. However, modern pop music often consists of multiple instruments or tracks. To make deep learning technology applicable in modern music production workflow, it is important to modernize deep learning models for multitrack music generation.

Witnessing the lack of infrastructure for symbolic music generation
when conducting these research projects,
I developed libraries for processing symbolic music to consolidate the infrastructure of music generation research \citep{dong2018pypianoroll,dong2020muspy}. The Python toolkits I developed to process symbolic music for machine learning applications have been widely used in the field. With these libraries, researchers can easily download commonly used datasets programmatically and be liberated from reimplementing tedious data processing routines. The toolkits also allowed me to conduct \textit{the first large-scale experiment that measures the cross-dataset generalizability of deep neural networks for music} \citep{dong2020muspy}. This work will be presented in \cref{chap:muspy}.

Moreover, I study efficient music generation model with an eye to enabling real time improvisation or near real time creative applications. In \cite{dong2023mmt}, I proposed the Multitrack Music Transformer (MMT) model that achieves comparable performance with state-of-the-art systems, landing in between two recently proposed models in a subjective listening test, while achieving substantial speedups and memory reductions over both. Further, I presented \textit{the first systematic analysis of musical self-attention}, where I showed that the trained model learns a relative attention in certain aspects of music. With the great success brought by large-scale generative pretraining in natural language processing, my ongoing work aims to scale this method up using the largest ever symbolic music dataset containing more than one million scores, which I collected and compiled from the MuseScore forum. This work will be presented in \cref{chap:mmt}.

%=======================================
\section{Assistive Music Creation Tools}
%=======================================

Music creation today is still largely limited to professional musicians for it requires a certain level of knowledge in music theory, music notation and music production tools. Apart from generating new music content from scratch, another line of my research focuses on developing AI-augmented tools to assist amateurs to create and perform music. My long-term goal along this research direction is to lower the entry barrier of music composition and make music creation accessible to everyone.

% In this direction, I study automatic instrumentation \citep{dong2021arranger}, automatic arrangement \citep{dong2018musegan,liu2018musegangtc}, automatic harmonization \citep{yeh2020harmonization} and music performance synthesis \citep{dong2021deepperformer}.
For example, in \citep{dong2021arranger}, I developed \textit{the first deep learning model for automatic instrumentation}. Instrumentation refers to the process where a musician arranges a solo piece for a certain ensemble such as a string quartet or a rock band. This can be challenging for amateur composers as it requires domain knowledge of each target instrument. In this work, I proposed a new machine learning model that can produce convincing instrumentation for a solo piece by framing this problem as a sequential multi-class classification problem. Such an automatic instrumentation system can suggest potential instrumentation for amateur composers, especially useful when arranging for an unfamiliar ensemble. Further, the proposed model can empower a musician to play multiple instruments on a single keyboard at the same time. This work will be presented in \cref{chap:arranger}.

Another example is my work on music performance synthesis \citep{dong2021deepperformer}. While synthesizers play a critical role and are intensively used in modern music production, existing synthesizers either requires an input with expressive timing or allows only monophonic inputs. In light of the similarities between text-to-speech (TTS) and score-to-audio synthesis, I showed in this work that we can adapt a state-of-the-art TTS model for music performance synthesis. Moreover, I proposed a novel mechanism to enable polyphonic music synthesis. This work represents \textit{the first deep learning based polyphonic synthesizer that can synthesize a score into a natural, expressive performance}. This work will be presented in \cref{chap:deepperformer}.

%================================================
\section{Multimodal Learning for Audio and Music}
%================================================

The third line of my research focuses on multimodal learning for audio and music. Sound is an integral part of movies, dramas, documentaries, podcasts, games, short videos and audiobooks. In these media, audio and music production tools need to interact with inputs from other modalities such as text and images, and thus multimodal models are critical in enabling controllable creation tools for music and audio in these applications.

Along this direction, I have worked on text-queried sound separation \citep{dong2023clipsep} and text-to-audio synthesis \citep{dong2023clipsynth,dong2023clipsonic}. Unlike existing work that relies on a large amount of paired audio-text data, I explore a new direction of approaching bimodal learning for text and audio through leveraging the visual modality as a bridge. The key idea behind my study is to combine the naturally-occurring audio-visual correspondence in videos and the multimodal representation learned by contrastive language-vision pretraining (CLIP). Based on this idea, I developed \textit{the first text-queried sound separation model that can be trained without any text-audio pairs} \citep{dong2023clipsep}. Text-queried sound separation aims to separate a specific sound out from a mixture of sounds given a text query, which has many downstream applications in audio post-production such as editing and remixing. I showed that the proposed model can successfully learn text-queried sound separation using only noisy unlabeled videos, and it even achieves competitive performance against a supervised model in some settings. Moreover, I built \textit{the first text-to-audio synthesis model that requires no text-audio pairs during training} \citep{dong2023clipsynth,dong2023clipsonic}. The proposed model learns to synthesize audio given text queries, which can find applications in video and audio editing software. One of the key benefits of the approach studied in my work lies in its scalability to large video datasets in the wild as we only need unlabeled videos for training. This work will be presented in \cref{chap:clipsep,chap:clipsonic}.

%==================================
\section{Dissertation Organization}
%==================================

The rest of this dissertation is organized as follows:
\cref{chap:muspy,chap:arranger,chap:deepperformer,chap:clipsep,chap:clipsonic,chap:mmt} are reprints of six conference papers published during my PhD. Specifically, \cref{chap:muspy} is a reprint of \textit{``\href{https://arxiv.org/pdf/2008.01951.pdf}{MusPy: A Toolkit for Symbolic Music Generation}''} \citep{dong2020muspy} published in ISMIR 2020. \cref{chap:mmt} is a reprint of \textit{``\href{https://arxiv.org/pdf/2207.06983.pdf}{Multitrack Music Transformer}''} \citep{dong2023mmt} published in ICASSP 2023. \cref{chap:arranger} is a reprint of \textit{``\href{https://arxiv.org/pdf/2107.05916.pdf}{Towards Automatic Instrumentation by Learning to Separate Parts in Symbolic Multitrack Music}''} \citep{dong2021arranger} published in ISMIR 2021. \cref{chap:deepperformer} is a reprint of \textit{``\href{https://arxiv.org/pdf/2202.06034.pdf}{Deep Performer: Score-to-Audio Music Performance Synthesis}''} \citep{dong2021deepperformer} published in ICASSP 2022. \cref{chap:clipsep} is a reprint of \textit{``\href{https://arxiv.org/pdf/2212.07065.pdf}{CLIPSep: Learning Text-queried Sound Separation with Noisy Unlabeled Videos}''} \citep{dong2023clipsep} published in ICLR 2023. \cref{chap:clipsonic} is a reprint of \textit{``\href{https://arxiv.org/pdf/2306.09635.pdf}{CLIPSonic: Text-to-Audio Synthesis with Unlabeled Videos and Pretrained Language-Vision Models}''} \citep{dong2023clipsonic} published in WASPAA 2023. Finally, \cref{chap:conclusion} concludes this dissertation and discusses my future research directions.

\vfill

\begin{center}
    \itshape
    Think, Think, Think\\
    \normalfont
    ---Winnie the Pooh
\end{center}

\vfill

\graphicspath{{chapters/muspy/figs/}}
\chapter{MusPy: A Toolkit for Symbolic Music Generation}
\label{chap:muspy}

\begin{abstract}
In this paper, we present MusPy, an open source Python library for symbolic music generation. MusPy provides easy-to-use tools for essential components in a music generation system, including dataset management, data I/O, data preprocessing and model evaluation. In order to showcase its potential, we present statistical analysis of the eleven datasets currently supported by MusPy. Moreover, we conduct a cross-dataset generalizability experiment by training an autoregressive model on each dataset and measuring held-out likelihood on the others---a process which is made easier by MusPy's dataset management system. The results provide a map of domain overlap between various commonly used datasets and show that some datasets contain more representative cross-genre samples than others. Along with the dataset analysis, these results might serve as a guide for choosing datasets in future research. Source code and documentation are available at \url{https://github.com/salu133445/muspy}.
\end{abstract}

%=====================
\section{Introduction}
%=====================
\label{muspy:sec:introduction}

\begin{figure}
    \small
    \centering
    \includegraphics[width=.75\linewidth]{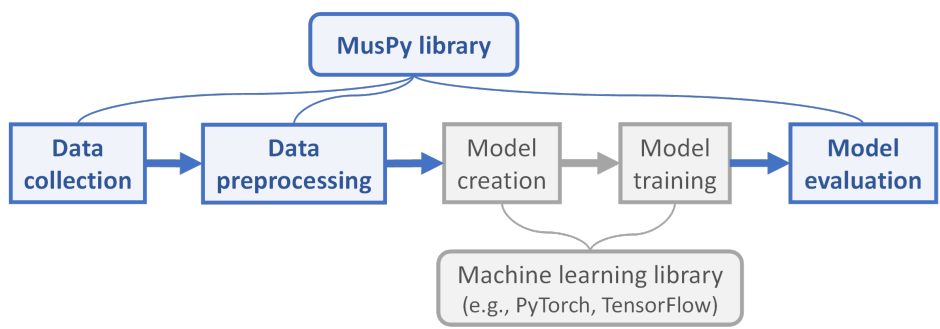}
    \caption[An example of a learning-based music generation system.]{An example of a learning-based music generation system. MusPy provides basic routines specific to music as well as interfaces to machine learning frameworks.}
    \label{muspy:fig:pipeline}
\end{figure}

% Motivation
Recent years have seen progress on music generation, thanks largely to advances in machine learning \citep{briot2017survey}. A music generation pipeline usually consists of several steps---data collection, data preprocessing, model creation, model training and model evaluation, as illustrated in \cref{muspy:fig:pipeline}. While some components need to be customized for each model, others can be shared across systems. For symbolic music generation in particular, a number of datasets, representations and metrics have been proposed in the literature \citep{briot2017survey}. As a result, an easy-to-use toolkit that implements standard versions of such routines could save a great deal of time and effort and might lead to increased reproducibility. However, such tools are challenging to develop for a variety of reasons.

% Datasets
First, though there are a number of publicly-available symbolic music datasets, the diverse organization of these collections and the various formats used to store them presents a challenge. 
These formats are usually designed for different purposes. Some focus on playback capability (e.g.,~MIDI), some are developed for music notation softwares (e.g.,~MusicXML \citep{good2001musicxml} and LilyPond \citep{lilypond}), some are designed for organizing musical documents (e.g.,~Music Encoding Initiative (MEI) \citep{hankinson2011mei}), and others are research-oriented formats that aim for simplicity and readability (e.g.,~MuseData \citep{hewlett1997musedata} and Humdrum \citep{huron1997humdrum}. Oftentimes researchers have to implement their own preprocessing code for each different format. Moreover, while researchers can implement their own procedures to access and process the data, issues of reproducibility due to the inconsistency of source data have been raised in \citep{bittner2019mirdata} for audio datasets.

% Representations
Second, music has hierarchy and structure, and thus different levels of abstraction can lead to different representations \citep{dannenberg1993survey}. Moreover, a number of music representations designed specially for generative modeling of music have also been proposed in prior art, for example, as a sequence of pitches \citep{mozer1994neural,eck2002lstm,bl2012jsb,roberts2018musicvae}, events \citep{oore2020performancernn,huang2019musictransformer,donahue2019lakhnes,huang2020remi}, notes \citep{mogren2016crnngan} or a time-pitch matrix (i.e.,~a piano roll) \citep{yang2018midinet,dong2018musegan}.

% Evaluations
Finally, efforts have been made toward more robust objective evaluation metrics for music generation systems \citep{yang2018evaluation} as these metrics provide not only an objective way for comparing different models but also indicators for monitoring training progress in machine learning-based systems. Given the success of mir\_eval \citep{raffel2014mireval} in evaluating common MIR tasks, a library providing implementations of commonly used evaluation metrics 
for music generation systems could help improve reproducibility.

To manage the above challenges, we find a toolkit dedicated for music generation a timely contribution to the MIR community. Hence, we present in this paper a new Python library, MusPy, for symbolic music generation. It provides essential tools for developing a music generation system, including dataset management, data I/O, data preprocessing and model evaluation.

With MusPy, we provide a statistical analysis on the eleven datasets currently supported by MusPy, with an eye to unveiling statistical differences between them. Moreover, we conduct three experiments to analyze their relative diversities and cross-dataset domain compatibility of the various datasets. These results, along with the statistical analysis, together provide a guide for choosing proper datasets for future research. Finally, we also show that combining multiple heterogeneous datasets could help improve generalizability of a music generation system.

%=====================
\section{Related Work}
%=====================
\label{muspy:sec:related-work}

% Platform - Magenta
Few attempts, to the best of our knowledge, have been made to develop a dedicated library for music generation. The Magenta project \citep{magenta} represents the most notable example. While MusPy aims to provide fundamental routines in data collection, preprocessing and analysis, Magenta comes with a number of model instances, but is tightly bound with TensorFlow \citep{abadi2016tensorflow}. In MusPy, we leave the model creation and training to dedicated machine learning libraries, and design MusPy to be flexible in working with different machine learning frameworks.

% Libraries - music21 jSymbolic
There are several libraries for working with symbolic music. music21 \citep{cuthbert2010music21} is one of the most representative toolkits and targets studies in computational musicology. While music21 comes with its own corpus, MusPy does not host any dataset. Instead, MusPy provides functions to download datasets from the web, along with tools for managing different collections, which makes it easy to extend support for new datasets in the future. jSymbolic \citep{mckay2006jsymbolic} focuses on extracting statistical information from symbolic music data. While jSymbolic can serve as a powerful feature extractor for training supervised classification models, MusPy focuses on generative modeling of music and supports different commonly used representations in music generation. In addition, MusPy provides several objective metrics for evaluating music generation systems.

% Cross-dataset generalizability - LakhNES
Related cross-dataset generalizability experiments \citep{donahue2019lakhnes} show that pretraining on a cross-domain data can improve music generation results both qualitatively and quantitatively. MusPy's dataset management system makes it easier for us to thoroughly verify this hypothesis by examining pairwise generalizabilities between various datasets.

%==============
\section{MusPy}
%==============
\label{muspy:sec:muspy}

\begin{figure*}
    \small
    \centering
    \includegraphics[width=\linewidth]{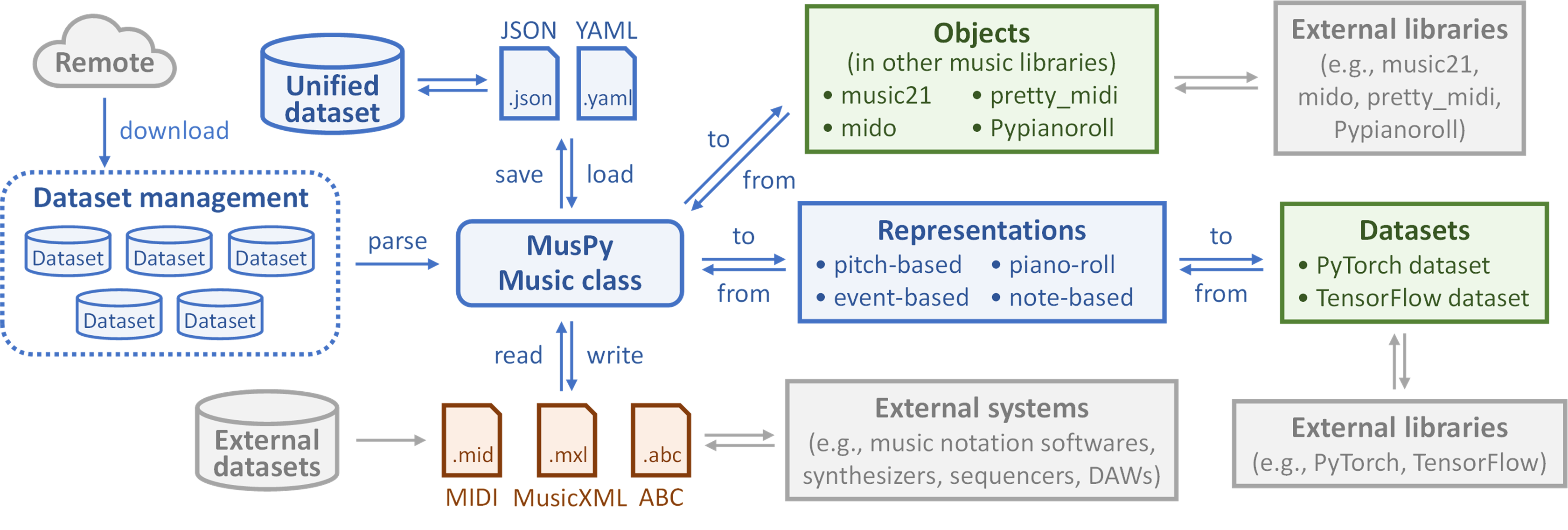}
    \caption[System diagram of MusPy.]{System diagram of MusPy. The MusPy Music object at the center is the core element of MusPy.}
    \label{muspy:fig:system}
\end{figure*}

MusPy is an open source Python library dedicated for symbolic music generation. \cref{muspy:fig:system} presents the system diagram of MusPy. It provides a core class, MusPy Music class, as a universal container for symbolic music. Dataset management system, I/O interfaces and model evaluation tools are then built upon this core container. We provide in \cref{muspy:fig:pipeline_example} examples of data preparation and result writing pipelines using MusPy.

%------------------------------------------------
\subsection{MusPy Music class and I/O interfaces}
%------------------------------------------------
\label{muspy:sec:io}

We aim at finding a middle ground among existing formats for symbolic music and design a unified format dedicated for music generation. MIDI, as a communication protocol between musical devices, uses velocities to indicate dynamics, beats per minute (bpm) for tempo markings, and control messages for articulation, but it lacks the concepts of notes, measures and symbolic musical markings. In contrast, MusicXML, as a sheet music exchanging format, has the concepts of notes, measures and symbolic musical markings and contains visual layout information, but it falls short on playback-related data. For a music generation system, however, both symbolic and playback-specific data are important. Hence, we follow MIDI's standard for playback-related data and MusicXML's standard for symbolic musical markings.

\begin{figure}
    \small
    \centering
    \begin{tabular}{@{}cc@{}}
        \includegraphics[width=.48\linewidth]{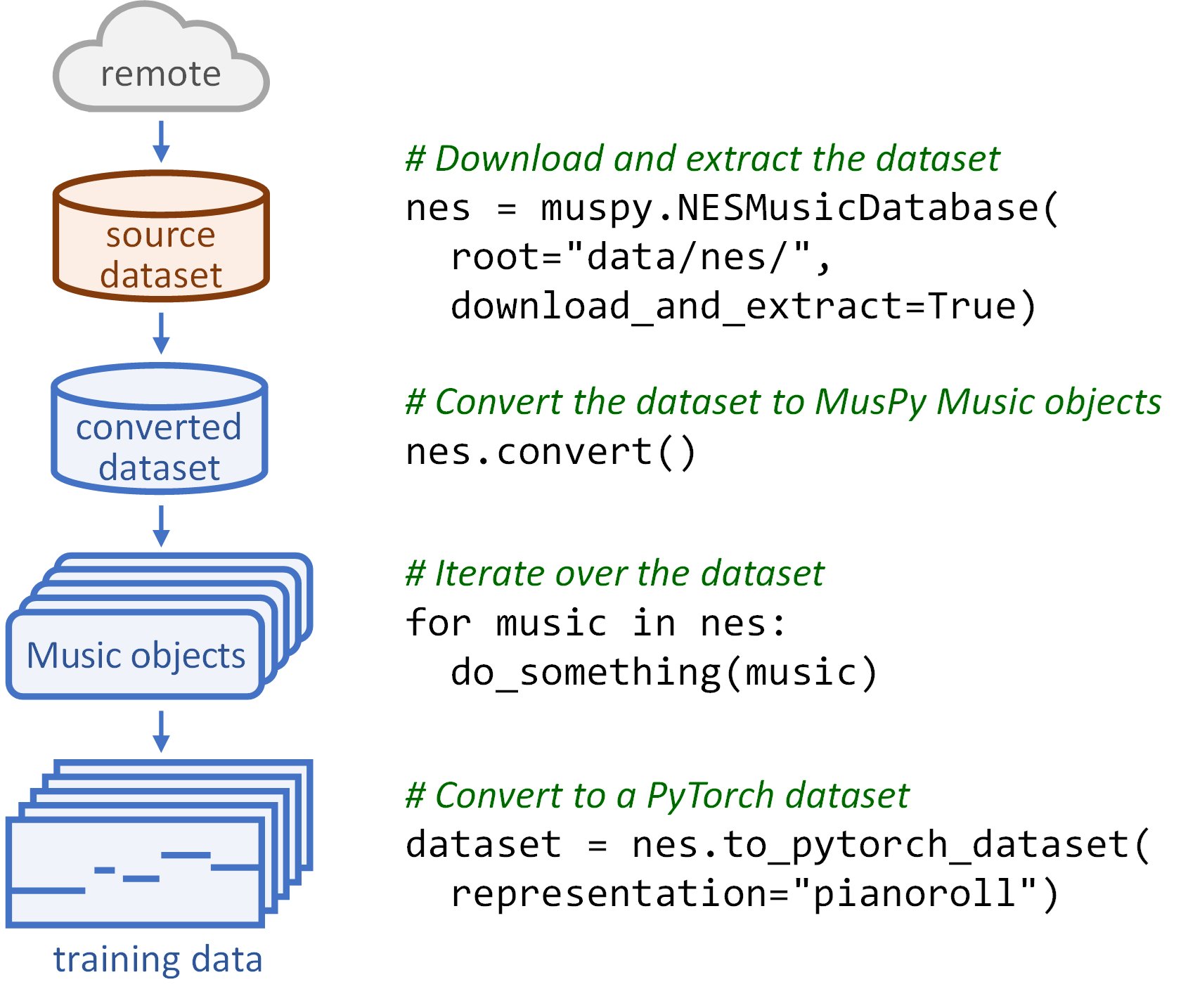} &\includegraphics[width=.48\linewidth]{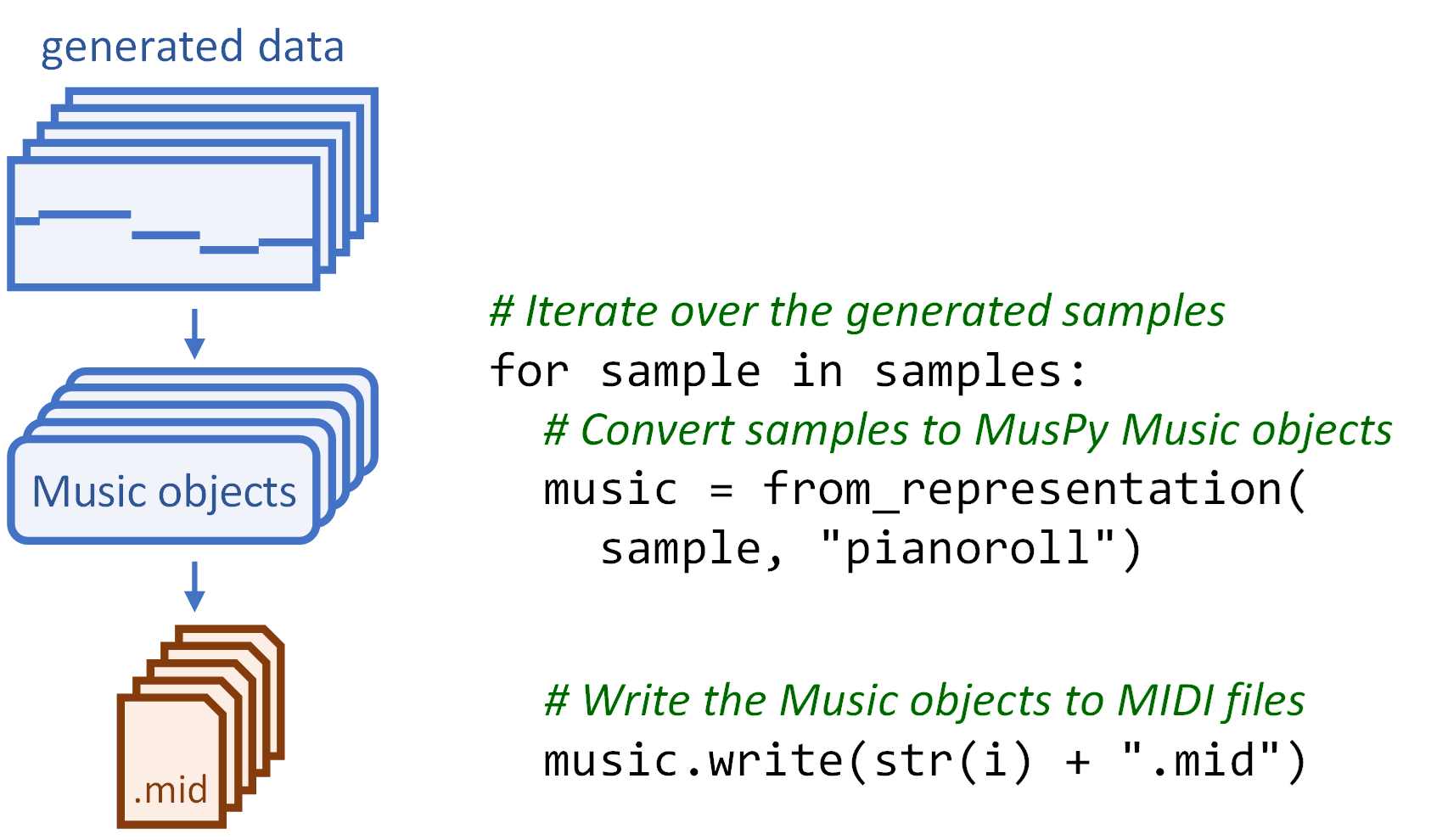} \\
        (a) & (b)
    \end{tabular}
    \caption{Examples of (a) training data preparation and (b) result writing pipelines using MusPy.}
    \label{muspy:fig:pipeline_example}
\end{figure}

\begin{table}
    \small
    \centering
    \caption[Comparisons of MIDI, MusicXML and the proposed MusPy formats.]{Comparisons of MIDI, MusicXML and the proposed MusPy formats. Triangle marks indicate optional or limited support.}
    \label{muspy:tab:comparison}
    \begin{tabular}{lccc}
        \toprule
                                &MIDI       &MusicXML   &MusPy\\
        \midrule
        Sequential timing       &\gcheck    &\rcross    &\gcheck\\
        Playback velocities     &\gcheck    &\ytriangle &\gcheck\\
        Program information     &\gcheck    &\ytriangle &\gcheck\\
        \cmidrule(lr){1-4}
        Layout information      &\rcross    &\gcheck    &\rcross\\
        Note beams and slurs    &\rcross    &\gcheck    &\rcross\\
        Song/source meta data   &\ytriangle &\gcheck    &\gcheck\\
        Track/part information  &\ytriangle &\gcheck    &\gcheck\\
        Dynamic/tempo markings  &\rcross    &\gcheck    &\gcheck\\
        Concept of notes        &\rcross    &\gcheck    &\gcheck\\
        Measure boundaries      &\rcross    &\gcheck    &\gcheck\\
        Human readability       &\rcross    &\ytriangle &\gcheck\\
        \bottomrule
    \end{tabular}
\end{table}

In fact, the MusPy Music class naturally defines a universal format for symbolic music, which we will refer to as the MusPy format, and can be serialized into a human-readable JSON/YAML file. \cref{muspy:tab:comparison} summarizes the key differences among MIDI, MusicXML and the proposed MusPy formats. Using the proposed MusPy Music class as the internal representation for music data, we then provide I/O interfaces for common formats (e.g., MIDI, MusicXML and ABC) and interfaces to other symbolic music libraries (e.g.,~music21 \citep{cuthbert2010music21}, mido \citep{mido}, pretty\_midi \citep{raffel2014prettymidi} and Pypianoroll \citep{dong2018pypianoroll}). \cref{muspy:fig:pipeline_example}(b) provides an example of result writing pipeline using MusPy.

%------------------------------
\subsection{Dataset management}
%------------------------------
\label{muspy:sec:management}

\begin{figure}
    \centering
    \includegraphics[scale=0.25]{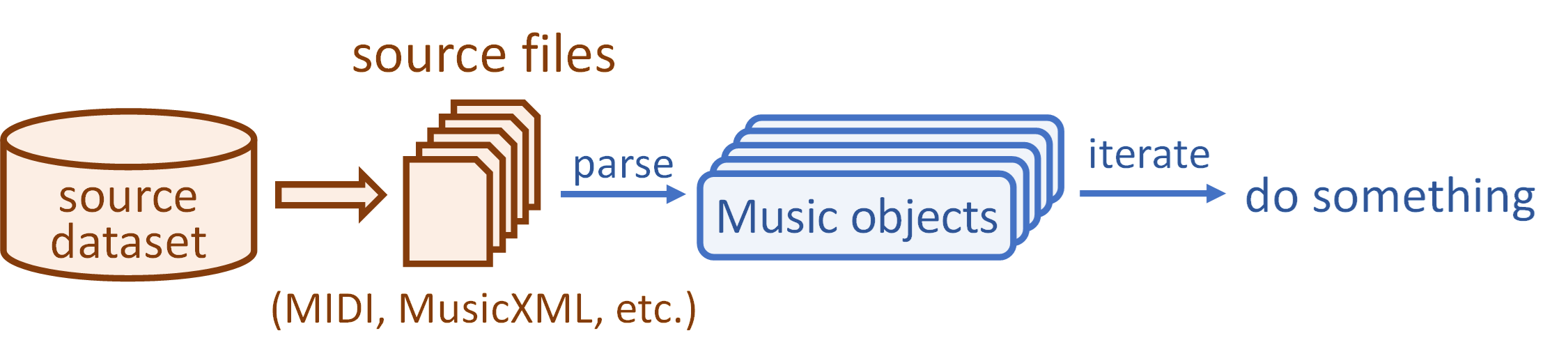}\\
    (a) on-the-fly mode\\[1ex]
    \includegraphics[scale=0.25]{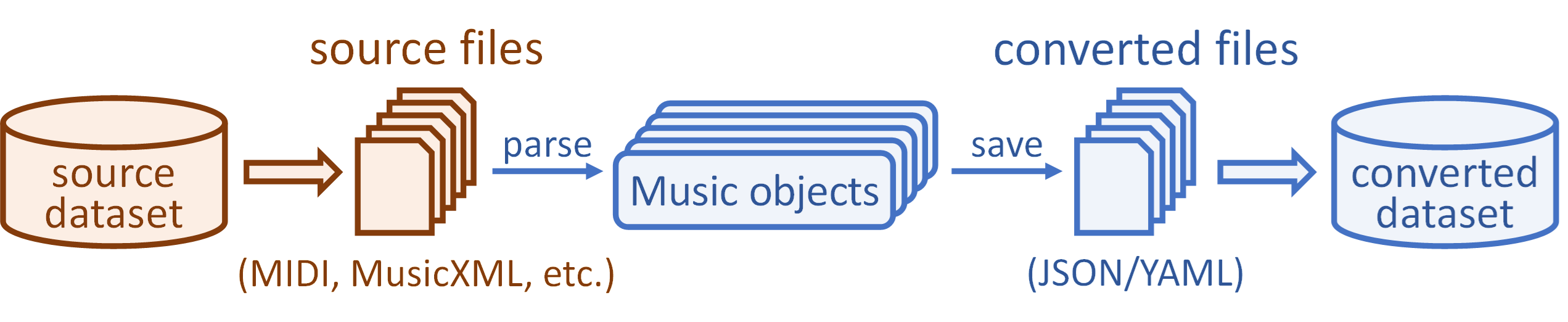}\\
    \includegraphics[scale=0.25]{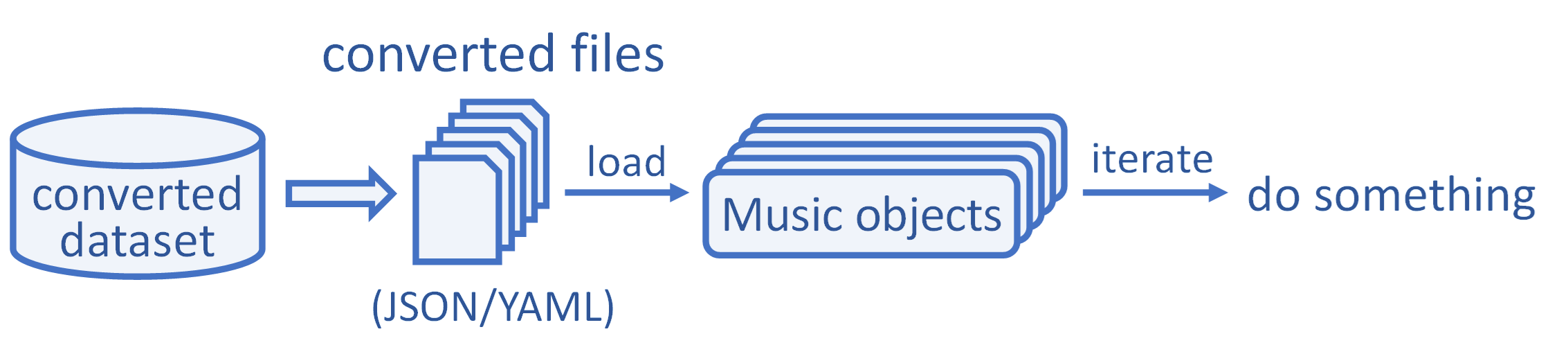}\\
    (b) preconverted mode
    \caption{Two internal processing modes for iterating over a MusPy Dataset object.}
    \label{muspy:fig:dataset_modes}
\end{figure}

MusPy provides an easy-to-use dataset management system similar to torchvision datasets \citep{torchvision} and TensorFlow Dataset \citep{tfds}. \cref{muspy:tab:datasets} presents the list of datasets currently supported by MusPy and their comparisons. Each supported dataset comes with a class inherited from the base MusPy Dataset class. The modularized and flexible design of the dataset management system makes it easy to handle local data collections or extend support for new datasets in the future. \cref{muspy:fig:dataset_modes} illustrates the two internal processing modes when iterating over a MusPy Dataset object. In addition, MusPy provides interfaces to PyTorch \citep{paszke2019pytorch} and TensorFlow \citep{abadi2016tensorflow} for creating input pipelines for machine learning (see \cref{muspy:fig:pipeline_example}(a) for an example).

\begin{table*}
    \small
    \centering
    \caption[Comparisons of datasets currently supported by MusPy.]{Comparisons of datasets currently supported by MusPy. Triangle marks indicate partial support. Note that, in this version, only MusicXML and MIDI files are included for the music21 Corpus.}
    \label{muspy:tab:datasets}
    \footnotesize
    \begin{tabularx}{\linewidth}{Xlrrcc@{~~~~}c@{~~~~}c}
        \toprule
        Dataset                                            &Format   &Hours  &Songs   &Genre     &Melody     &Chords     &Multitrack\\
        \midrule
        \makecell[X]{Lakh MIDI Dataset (LMD)\\ \citep{raffel16lmd}}         &MIDI     &>9000  &174,533 &misc      &\ytriangle &\ytriangle &\ytriangle\\
        \cmidrule(lr){1-8}
        \makecell[X]{MAESTRO Dataset\\\citep{hawthorne2019maestro}}        &MIDI     &201.21 &1,282   &classical &\rcross    &\rcross    &\rcross\\
        \cmidrule(lr){1-8}
        \makecell[X]{Wikifonia Lead Sheet Dataset\\\citep{wikifonia}}      &MusicXML &198.40 &6,405   &misc      &\gcheck    &\gcheck    &\rcross\\
        \cmidrule(lr){1-8}
        \makecell[X]{Essen Folk Song Database\\\citep{essen}}              &ABC      &56.62  &9,034   &folk      &\gcheck    &\gcheck    &\rcross\\
        \cmidrule(lr){1-8}
        \makecell[X]{NES Music Database\\\citep{donahue2018nesmdb}}         &MIDI     &46.11  &5,278   &game      &\gcheck    &\rcross    &\gcheck\\
        \cmidrule(lr){1-8}
        \makecell[X]{Hymnal Tune Dataset\\\citep{hymnal}}                  &MIDI     &18.74  &1,756  &hymn      &\gcheck    &\rcross    &\rcross\\
        \cmidrule(lr){1-8}
        \makecell[X]{Hymnal Dataset\\\citep{hymnal}}                       &MIDI     &17.50  &1,723   &hymn      &\rcross    &\rcross    &\rcross\\
        \cmidrule(lr){1-8}
        \makecell[X]{music21 Corpus\\\citep{cuthbert2010music21}}         &misc     &16.86  &613     &misc      &\ytriangle &\rcross    &\ytriangle\\
        \cmidrule(lr){1-8}
        \makecell[X]{Nottingham Database (NMD)\\\citep{nmd}}               &ABC      &10.54  &1,036   &folk      &\gcheck    &\gcheck    &\rcross\\
        \cmidrule(lr){1-8}
        \makecell[X]{music21 JSBach Corpus\\\citep{cuthbert2010music21}} &MusicXML &3.46   &410     &classical &\rcross    &\rcross    &\gcheck\\
        \cmidrule(lr){1-8}
        \makecell[X]{JSBach Chorale Dataset\\\citep{bl2012jsb}}            &MIDI     &3.21   &382     &classical &\rcross    &\rcross    &\gcheck\\
        \bottomrule
    \end{tabularx}
\end{table*}

%---------------------------
\subsection{Representations}
%---------------------------

Music has multiple levels of abstraction, and thus can be expressed in various representations. For music generation in particular, several representations designed for generative modeling of symbolic music have been proposed and used in the literature \citep{briot2017survey}. These representations can be broadly categorized into four types---the pitch-based \citep{mozer1994neural,eck2002lstm,bl2012jsb,roberts2018musicvae}, the event-based \citep{oore2020performancernn,huang2019musictransformer,donahue2019lakhnes,huang2020remi}, the note-based \citep{mogren2016crnngan} and the piano-roll \citep{yang2018midinet,dong2018musegan} representations. \cref{muspy:tab:representations} presents a comparison of them. We provide in MusPy implementations of these representations and integration to the dataset management system. \cref{muspy:fig:pipeline_example}(a) provides an example of preparing training data in the piano-roll representation from the NES Music Database using MusPy.

\begin{table*}
    \small
    \centering
    \caption[Comparisons of representations supported by MusPy.]{Comparisons of representations supported by MusPy. $T$ and $N$ denote the numbers of time steps and notes, respectively. Note that the configurations can be modified to meet specific requirements and use cases.}
    \label{muspy:tab:representations}
    \begin{tabularx}{\linewidth}{lllX}
        \toprule
        Representation &Shape &Values &Default configurations\\
        \midrule
        Pitch-based &$T \times 1$   &$\{0, 1,\dots, 129\}$          &$128$ note-ons, $1$ hold, $1$ rest (\textit{support only monophonic music})\\
        \cmidrule(lr){1-4}
        Event-based &$T \times 1$   &$\{0, 1,\dots, 387\}$          &$128$ note-ons, $128$ note-offs, $100$ time shifts, $32$ velocities\\
        \cmidrule(lr){1-4}
        Piano-roll   &$T \times 128$ &$\{0, 1\}$ or $\mathbb{R}^+$     &$\{0, 1\}$ for binary piano rolls; $\mathbb{R}^+$ for piano rolls with velocities\\
        \cmidrule(lr){1-4}
        Note-based  &$N \times 4$   &$\mathbb{N}$ or $\mathbb{R}^+$ &List of $(time, pitch, duration, velocity)$ tuples\\
        \bottomrule
    \end{tabularx}
\end{table*}

%----------------------------------
\subsection{Model evaluation tools}
%----------------------------------
\label{muspy:sec:rendering_visualization}

Model evaluation is another critical component in developing music generation systems. Hence, we also integrate into MusPy tools for audio rendering as well as score and piano-roll visualizations. These tools could also be useful for monitoring the training progress or demonstrating the final results. Moreover, MusPy provides implementations of several objective metrics proposed in the literature \citep{mogren2016crnngan,dong2018musegan,wu2020jazz}. These objective metrics, as listed below, could be used to evaluate a music generation system by comparing the statistical difference between the training data and the generated samples, as discussed in \citep{yang2018evaluation}. 
\begin{itemize}
    \item \textit{Pitch-related metrics}---polyphony, polyphony rate, pitch-in-scale rate, scale consistency, pitch entropy and pitch class entropy.
    \item \textit{Rhythm-related metrics}---empty-beat rate, drum-in-pattern rate, drum pattern consistency and groove consistency.
\end{itemize}

%-------------------
\subsection{Summary}
%-------------------
\label{muspy:sec:summary}

To summarize, MusPy features the following:
\begin{itemize}
    \item Dataset management system for commonly used datasets with interfaces to PyTorch and TensorFlow.
    \item Data I/O for common symbolic music formats (e.g., MIDI, MusicXML and ABC) and interfaces to other symbolic music libraries (e.g., music21, mido, pretty\_midi and Pypianoroll).
    \item Implementations of common music representations for music generation, including the pitch-based, the event-based, the piano-roll and the note-based representations.
    \item Model evaluation tools for music generation systems, including audio rendering, score and piano-roll visualizations and objective metrics.
\end{itemize}
All source code and documentation can be found at \url{https://github.com/salu133445/muspy}.

%=========================
\section{Dataset Analysis}
%=========================
\label{muspy:sec:analysis}

\begin{figure}
    \centering
    \includegraphics[width=.8\linewidth]{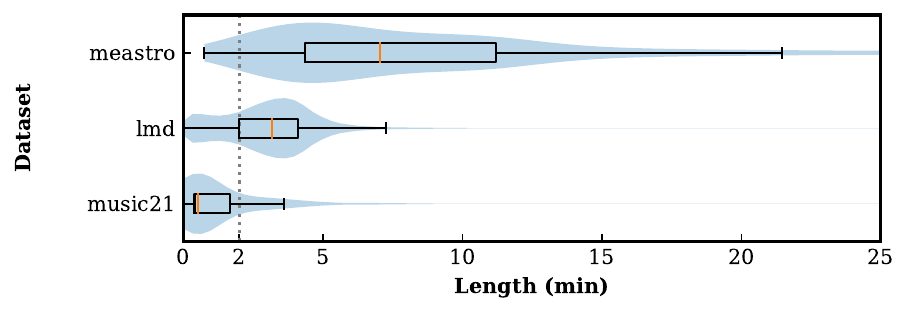}\\
    \includegraphics[width=.8\linewidth]{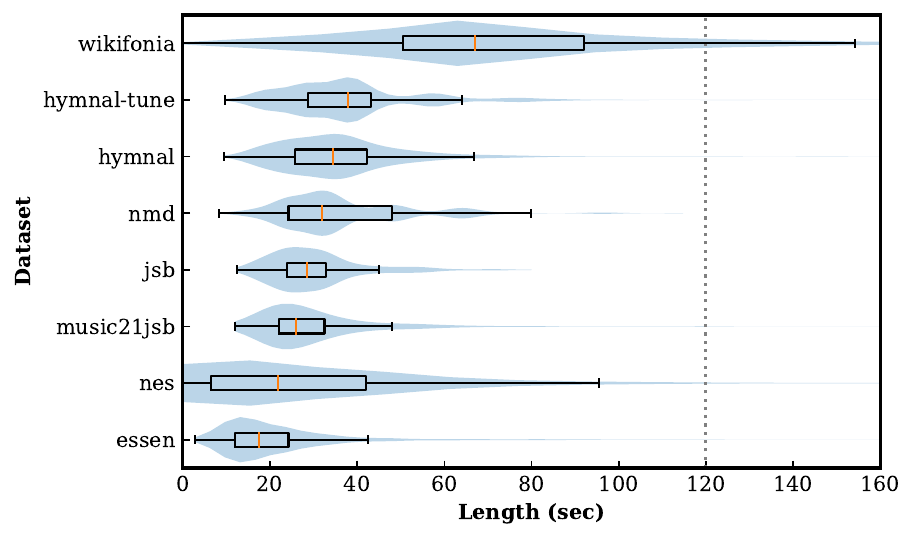}
    \caption{Length distributions for different datasets.}
    \label{muspy:fig:length_dist}
\end{figure}

\begin{figure}
  \centering
  \includegraphics[width=.8\linewidth]{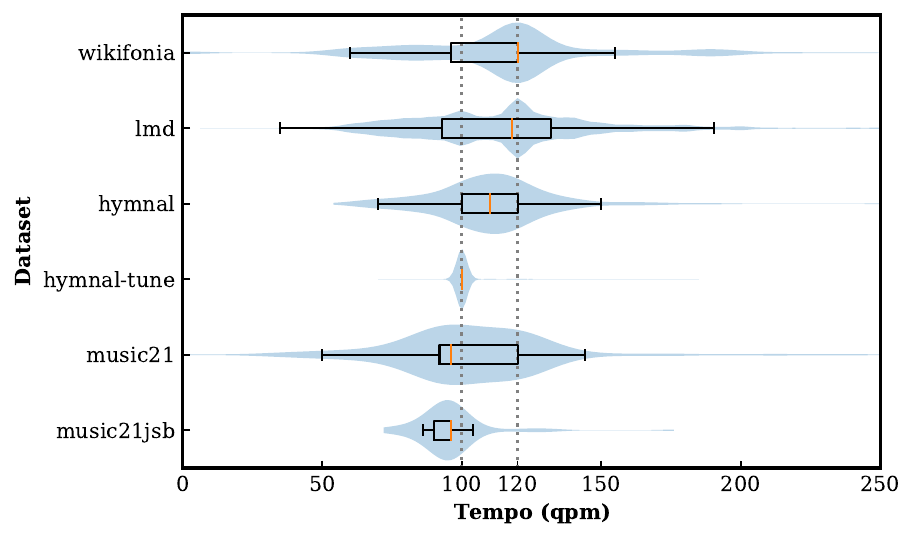}
  \caption{Initial-tempo distributions for different datasets (those without tempo information are not presented).}
  \label{muspy:fig:tempo_dist}
\end{figure}

Analyzing datasets is critical in developing music generation systems. With MusPy's dataset management system, we can easily work with different music datasets. Below we compute the statistics of three key elements of a song---length, tempo and key using MusPy, with an eye to unveiling statistical differences among these datasets. First, \cref{muspy:fig:length_dist} shows the distributions of song lengths for different datasets. We can see that they differ greatly in their ranges, medians and variances.

Second, we present in \cref{muspy:fig:tempo_dist} the distributions of initial tempo for datasets that come with tempo information. We can see that all of them are generally bell-shaped but with different ranges and variances. We also note that there are two peaks, $100$ and $120$ quarter notes per minute (qpm), in Lakh MIDI Dataset (LMD), which is possibly because these two values are often set as the default tempo values in music notation programs and MIDI editors/sequencers. Moreover, in Hymnal Tune Dataset, only around ten percent of songs have an initial tempo other than $100$ qpm.

\begin{figure}
    \centering
    \includegraphics[width=.7\linewidth]{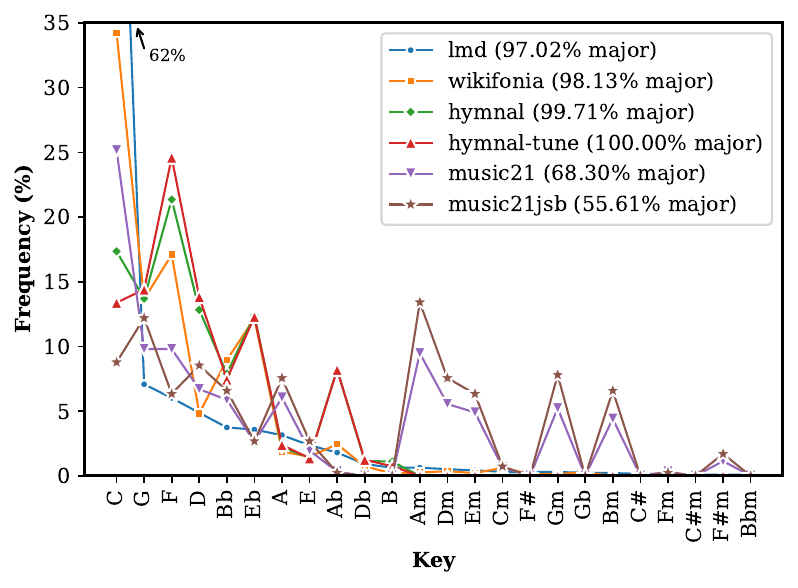}
    \caption[Key distributions for different datasets.]{Key distributions for different datasets. The keys are sorted w.r.t.~their frequencies in Lakh MIDI Dataset.}
    \label{muspy:fig:key_hist}
\end{figure}

Finally, \cref{muspy:fig:key_hist} shows the histograms of keys for different datasets. We can see that the key distributions are rather imbalanced. Moreover, only less than 3\% of songs are in minor keys for most datasets except the music21 Corpus. In particular, LMD has the most imbalanced key distributions, which might be due to the fact that C major is often set as the default key in music notation programs and MIDI editors/sequencers.\footnote{Note that key information is considered as a meta message in a MIDI file. It does not affect the playback and thus can be unreliable sometimes.} These statistics could provide a guide for choosing proper datasets in future research.

%================================
\section{Experiments and Results}
%================================
\label{muspy:sec:exp}

In this section, we conduct three experiments to analyze the relative complexities and the cross-dataset generalizabilities of the eleven datasets currently supported by MusPy (see \cref{muspy:tab:datasets}). We implement four autoregressive models---a recurrent neural network (RNN), a long short-term memory (LSTM) network \citep{hochreiter1997lstm}, a gated recurrent unit (GRU) network \citep{cho2014gru} and a Transformer network \citep{vaswani2017transformer}.

%-------------------------------
\subsection{Experiment settings}
%-------------------------------
\label{muspy:sec:exp_settings}

For the data, we use the event representation as specified in \cref{muspy:tab:representations} and discard velocity events as some datasets have no velocity information (e.g.,~datasets using ABC format). Moreover, we also include an end-of-sequence event, leading to in total $357$ possible events. For simplicity, we downsample each song into four time steps per quarter note and fix the sequence length to $64$, which is equivalent to four measures in $4/4$ time. In addition, we discard repeat information in MusicXML data and use only melodies in Wikifonia dataset. We split each dataset into train--test--validation sets with a ratio of $8:1:1$. For the training, the models are trained to predict the next event given the previous events. We use the cross entropy loss and the Adam optimizer \citep{kingma2015adam}. For evaluation, we randomly sample $1000$ sequences of length $64$ from the test split, and compute the perplexity of these sequences. We implement the models in Python using PyTorch. For reproducibility, source code and hyperparmeters are available at {\small\url{https://github.com/salu133445/muspy-exp}}.

%-------------------------------------------------------
\subsection{Autoregressive models on different datasets}
%-------------------------------------------------------
\label{muspy:sec:exp_datasets}

\begin{figure}
    \centering
    \includegraphics[width=.7\linewidth]{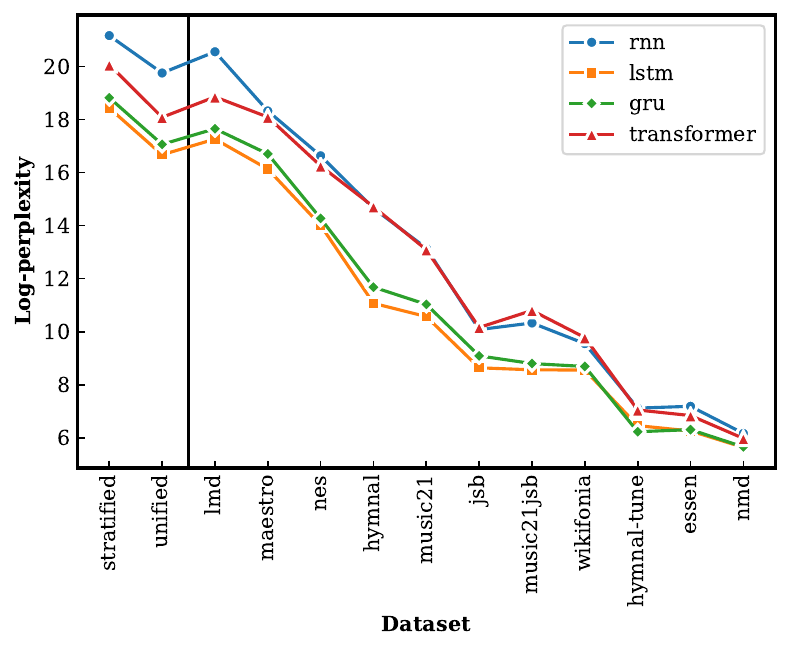}
    \caption{Log-perplexities for different models on different datasets, sorted by the values for the LSTM model.}
    \label{muspy:fig:exp_perplexity}
\end{figure}

\begin{figure}
    \centering
    \includegraphics[width=.7\linewidth]{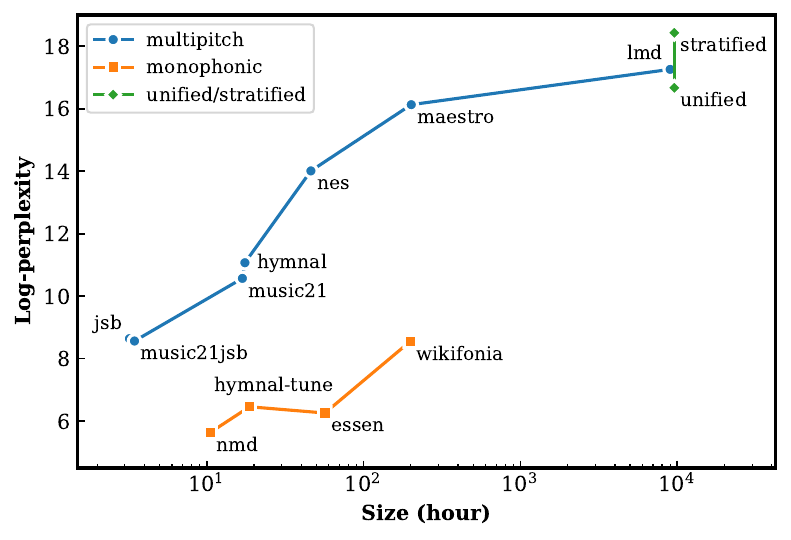}
    \caption[Log-perplexities for the LSTM model versus dataset size in hours.]{Log-perplexities for the LSTM model versus dataset size in hours. Each point corresponds to a dataset.}
    \label{muspy:fig:exp_hour_perplexity}
\end{figure}

In this experiment, we train the model on some dataset $\mathcal{D}$ and test it on the same dataset $\mathcal{D}$. We present in \cref{muspy:fig:exp_perplexity} the perplexities for different models on different datasets. We can see that all models have similar tendencies. In general, they achieve smaller perplexities for smaller, homogeneous datasets, but result in larger perplexities for larger, more diverse datasets. That is, the test perplexity could serve as an indicator for the diversity of a dataset. Moreover, \cref{muspy:fig:exp_hour_perplexity} shows perplexities versus dataset sizes (in hours). By categorizing datasets into multi-pitch (i.e.,~accepting any number of concurrent notes) and monophonic datasets, we can see that the perplexity is positively correlated to the dataset size within each group.

%------------------------------------------
\subsection{Cross-dataset generalizability}
%------------------------------------------
\label{muspy:sec:exp_cross_datasets}

\begin{figure}
    \centering
    \includegraphics[width=.6\linewidth]{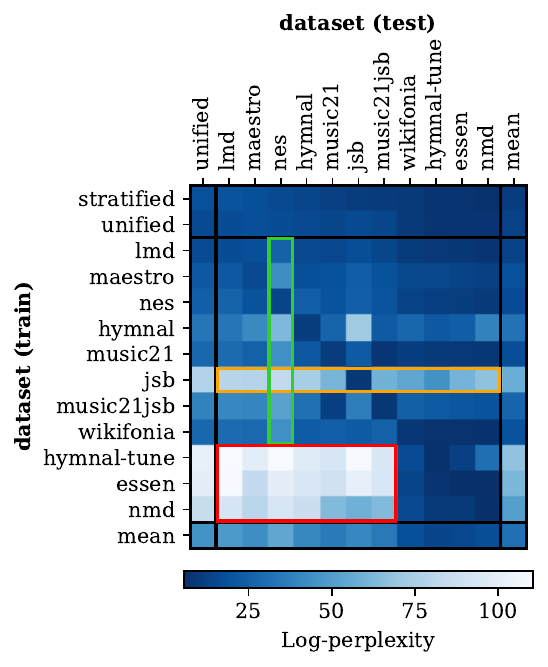}\\
    \includegraphics[width=.6\linewidth]{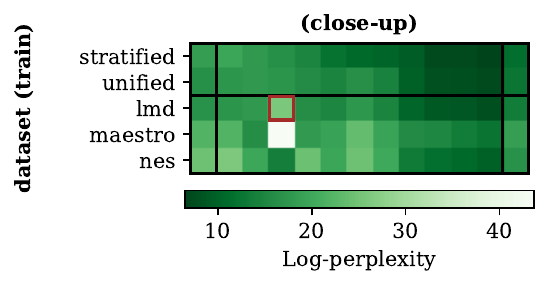}
    \caption[Cross-dataset generalizability results.]{Cross-dataset generalizability results. The values and colors represent the log-perplexities of a LSTM model trained on a specific dataset (row) and tested on another dataset (column). The datasets are sorted by the diagonal values, i.e., trained and tested on the same dataset.}
    \label{muspy:fig:exp_cross_datasets}
\end{figure}

In this experiment, we train a model on some dataset $\mathcal{D}$, while in addition to testing it on the same dataset $\mathcal{D}$, we also test it on each other dataset $\mathcal{D'}$. We present in \cref{muspy:fig:exp_cross_datasets} the perplexities for each train--test dataset pair. Here are some observations:
\begin{itemize}
    \item Cross dataset generalizability is not symmetric in general. For example, a model trained on LMD generalizes well to all other datasets, while not all models trained on other datasets generalize to LMD, which is possibly due to the fact that LMD is a large, cross-genre dataset.
    \item Models trained on multi-pitch datasets generalize well to monophonic datasets, while models trained on monophonic datasets do not generalize to multi-pitch datasets (see the red block in \cref{muspy:fig:exp_cross_datasets}).
    \item The model trained on JSBach Chorale Dataset does not generalize to any of the other datasets (see the orange block in \cref{muspy:fig:exp_cross_datasets}). This is possibly because its samples are downsampled to a resolution of quarter note, which leads to a distinct note duration distribution.
    \item Most datasets generalize worse to NES Music Database compared to other datasets (see the green block in \cref{muspy:fig:exp_cross_datasets}). This is possibly due to the fact that NES Music Database contains only game soundtracks.
\end{itemize}

%-------------------------------------------------------
\subsection{Effects of combining heterogeneous datasets}
%-------------------------------------------------------
\label{muspy:sec:exp_large_dataset}

From \cref{muspy:fig:exp_cross_datasets} we can see that LMD has the best generalizability, possibly because it is large, diverse and cross-genre. However, a model trained on LMD does not generalize well to NES Music Database (see the brown block in the close-up of \cref{muspy:fig:exp_cross_datasets}). We are thus interested in whether combing multiple heterogeneous datasets could help improve generalizability.

We combine all eleven datasets listed in \cref{muspy:tab:datasets} into one large \textit{unified} dataset. Since these datasets differ greatly in their sizes, simply concatenating the datasets might lead to severe imbalance problem and bias toward the largest dataset. Hence, we also consider a version that adopts stratified sampling during training. Specifically, to acquire a data sample in the \textit{stratified} dataset, we uniformly choose one dataset out of the eleven datasets, and then randomly pick one sample from that dataset. Note that stratified sampling is disabled at test time.

We also include in \cref{muspy:fig:exp_perplexity,muspy:fig:exp_hour_perplexity,muspy:fig:exp_cross_datasets} the results for these two datasets. We can see from \cref{muspy:fig:exp_cross_datasets} that combining datasets from different sources improves the generalizability of the model. This is consistent with the finding in \citep{donahue2019lakhnes} that models trained on certain cross-domain datasets generalize better to other unseen datasets. Moreover, stratified sampling alleviates the source imbalance problem by reducing perplexities in most datasets with a sacrifice of an increased perplexity on LMD.

%===================
\section{Conclusion}
%===================
\label{muspy:sec:conclusion}

We have presented MusPy, a new toolkit that provides essential tools for developing music generation systems. We discussed the designs and features of the library, along with data pipeline examples. With MusPy's dataset management system, we conducted a statistical analysis and experiments on the eleven currently supported datasets to analyze their relative diversities and cross-dataset generalizabilities. These results could help researchers choose appropriate datasets in future research. Finally, we showed that combining heterogeneous datasets could help improve generalizability of a machine learning model.

\secbreak

\begin{adjustwidth}{.5in}{.5in}
    \itshape
    \hspace{\parindent}
    This chapter, in full, is a reprint of the material as it appears in ``MusPy: A Toolkit for Symbolic Music Generation'' by Hao-Wen Dong, Ke Chen, Julian McAuley and Taylor Berg-Kirkpatrick, which was published in the Proceedings of the International Society for Music Information Retrieval Conference (ISMIR) in 2020. The dissertation author was the primary investigator and author of this paper.
\end{adjustwidth}

\graphicspath{{chapters/mmt/figs/}}
\chapter{Multitrack Music Transformer}
\label{chap:mmt}

\begin{abstract}
Existing approaches for generating multitrack music with transformer models have been limited in terms of the number of instruments, the length of the music segments and slow inference. This is partly due to the memory requirements of the lengthy input sequences necessitated by existing representations. In this work, we propose a new multitrack music representation that allows a diverse set of instruments while keeping a short sequence length. Our proposed Multitrack Music Transformer (MMT) achieves comparable performance with state-of-the-art systems, landing in between two recently proposed models in a subjective listening test, while achieving substantial speedups and memory reductions over both, making the method attractive for real time improvisation or near real time creative applications. Further, we propose a new measure for analyzing musical self-attention and show that the trained model attends more to notes that form a consonant interval with the current note and to notes that are 4N beats away from the current step.
\end{abstract}

%=====================
\section{Introduction}
%=====================
\label{mmt:sec:introduction}

Prior work has investigated various approaches for symbolic music generation \citep{briot2017survey,ji2020survey}, among which, the transformer model \citep{vaswani2017transformer} has become popular given its recent successes in piano music generation \citep{huang2019musictransformer,huang2020remi,hsiao2021transformer,muhamed2021transformergan}. At the core of a transformer model is the self-attention mechanism that allows the model to dynamically attend to different parts of the input sequence and aggregate information from the whole sequence. Such capabilities make it suitable for modeling the complex structures and textures in music. However, while prior work has also explored applying transformer models to generate multitrack music \citep{musenet,donahue2019lakhnes,ens2020mmm,vonrutte2022figaro}, successful implementations have only been reported either on a limited set of instruments \citep{musenet,donahue2019lakhnes} or short music segments \citep{ens2020mmm,vonrutte2022figaro}. This is partly due to the long sequence length in existing multitrack music representations, which results in a large memory requirement in training. For example, a GPU with 11GB of memory can only generate 29 seconds of music on average using the REMI+ representation \citep{vonrutte2022figaro} on an orchestral music dataset. Moreover, it can only generate less than four notes per second. These limitations together pose a challenge in scaling transformer models to longer music with many instruments, e.g., orchestral music, and for real-time use cases, e.g., automatic improvisation and human-AI music co-creation.

In this paper, we propose a new multitrack music representation to address the long sequence issue in existing multitrack music representations. Using the proposed representation, we present the Multitrack Music Transformer (MMT) for multitrack music generation. Unlike a standard transformer model, the proposed model uses a decoder-only transformer with multi-dimensional inputs and outputs to reduce its memory complexity. On an orchestral dataset, we show that our proposed model can generate longer music in a faster inference speed than two existing approaches. Through a subjective listening test, we show that the proposed model achieves reasonably good performance in terms of coherence, richness and arrangement as well as the overall quality. Moreover, our proposed representation allows a trained autoregressive model to generate music for a specific set of instruments, a task that has not been well studied in prior work.

Further, while the transformer model has been widely used on symbolic music, it remains unclear how self-attention work for symbolic music. Understanding musical self-attention could reveal future research directions in improving transformer models for music. To the best of our knowledge, existing analysis \citep{huang2019musictransformer,huang2018visualizing,chen2021transformer,ziyu2021musebert} provides only case studies on few selected samples, lacking a systematic analysis on self-attention for music. Hence, we propose a new quantity to measure the average attention weights that a transformer model assigns to a certain key of a certain difference from the query. Our analysis shows that the proposed model learns a relative self-attention for certain aspects of music, specifically, beat, position and pitch.

Our proposed model provides a novel foundation for future work exploring longer-form and real-time capable multitrack music generation. The systematic analysis also provide insights into improving the self-attention mechanism for music. Audio samples can be found on our demo website: \url{https://salu133445.github.io/mmt/}. For reproducibility, all source code and hyperparameters are made publicly available at \url{https://github.com/salu133445/mmt}.

\begin{table}
    \small
    \centering
    \caption{Comparisons of related transformer-based music models.}
    \label{mmt:tab:comparison}
    \begin{tabularx}{\linewidth}{lCCCC}
        \toprule
        Model &Multitrack &Instrument control &Compound tokens &Generative modeling\\
        \midrule
        REMI \citep{huang2020remi} &&&&\checkmark\\
        MMM \citep{ens2020mmm} &\checkmark &&&\checkmark\\
        CP \citep{hsiao2021transformer} &&&\checkmark &\checkmark\\
        MusicBERT \citep{zeng2021musicbert} &\checkmark &&\checkmark\\
        FIGARO \citep{vonrutte2022figaro} &\checkmark &&&\checkmark\\
        \cmidrule(lr){1-5}
        MMT (ours) &\checkmark &\checkmark &\checkmark &\checkmark\\
        \bottomrule
    \end{tabularx}
\end{table}

%=====================
\section{Related Work}
%=====================

\paragraph{Multitrack music generation.} Prior work has explored various approaches for symbolic music generation \citep{briot2017survey,ji2020survey}, among which generating multitrack music is considered more challenging for its complex interdependency between voices and instruments. In \citep{dong2018musegan,dong2018binarymusegan}, the authors used a convolutional generative adversarial network to generate short, five-track pop music segments. In \citep{simon2018multitrackmusicvae}, the authors used a variational autoencoder with recurrent neural networks to learn a latent space for multitrack measures. In \citep{donahue2019lakhnes,musenet}, the authors used decoder-only transformer models to generate four-track game music and multi-instrument classical music, respectively. In \citep{vonrutte2022figaro}, the authors used a transformer model to generate multitrack music given a fine-grained description of the characteristics of the desired music. Unlike these systems, our proposed model is built upon a more compact representation that allows it to accommodate longer sequences under the same GPU memory constraint.

\paragraph{Transformers for symbolic music.} Another relevant line of research is on modeling symbolic music with transformer models \citep{vaswani2017transformer}. Some prior work focused on unconditioned generation, including generating piano music \citep{huang2020remi,hsiao2021transformer}, lead sheets \citep{wu2020jazz} and guitar tabs \citep{chen2020guitar} from scratch. Others studied controllable music generation \citep{shih2022themetransformer,vonrutte2022figaro}, polyphonic music score infilling \citep{chang2021infiliing} and general-purpose pre-training for symbolic music understanding \citep{chou2021midibert,zeng2021musicbert,ziyu2021musebert}. In this work, we focus on unconditioned generation for evaluation purposes. However, our proposed model can also generate music for a specific set of instruments.

%========================
\section{Proposed Method}
%========================
\label{mmt:sec:method}

%-------------------------------
\subsection{Data Representation}
%-------------------------------
\label{mmt:sec:representation}

\begin{figure*}
    \small
    \centering
    \includegraphics[width=\linewidth]{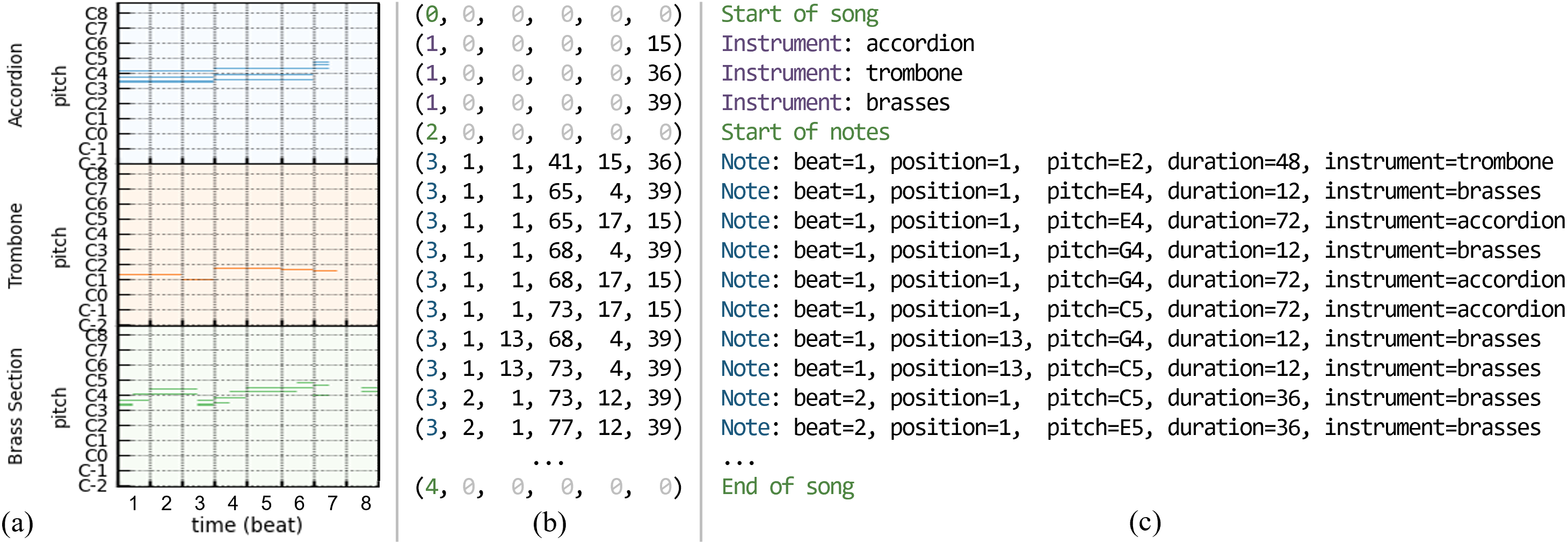}
    \caption[An example of the proposed representation.]{An example of the proposed representation---(a) an example of the first eight beats of a song in the orchestra dataset, shown as a multitrack piano roll, (b) the same song encoded by our proposed representation, where the grayed out zeros denote undefined values and (c) a human-readable translation of the codes shown in (b).}
    \label{mmt:fig:representation}
\end{figure*}

We represent a music piece as a sequence of events $\bm{x} = (\bm{x}_1, \dots, \bm{x}_n)$, where each event $\bm{x}_i$ is encoded as a tuple of six variables: $$(x_i^\mathit{type}, x_i^\mathit{beat}, x_i^\mathit{position}, x_i^\mathit{pitch}, x_i^\mathit{duration}, x_i^\mathit{instrument})\,.$$ The first variable $x^\mathit{type}$ determines the type of the event, among the following five event types:
\begin{itemize}
    \item \textit{Start-of-song}: Indicates the beginning of the song.
    \item \textit{Instrument}: Specifies an instrument used in the song.
    \item \textit{Start-of-notes}: Indicates the end of the instrument list and the beginning of the note list. (This event splits the sequence into two parts: a list of instrument events followed by a list of note events, making a trained autoregressive model readily applicable to instrument-informed generation task; see \cref{mmt:sec:model}.)
    \item \textit{Note}: Specifies a note, whose onset, pitch, duration and instrument are defined by the other five variables: $x^\mathit{beat}$, $x^\mathit{position}$, $x^\mathit{pitch}$, $x^\mathit{duration}$ and $x^\mathit{instrument}$.
    \item \textit{End-of-song}: Indicates the end of the song.
\end{itemize}
For any non-note-type event, the variables $x^\mathit{beat}$, $x^\mathit{position}$, $x^\mathit{pitch}$, $x^\mathit{duration}$, $x^\mathit{instrument}$ are set to zero, which is reserved for undefined values. \cref{mmt:fig:representation} shows an example of our proposed representation.

% \begin{figure*}
%     \small
%     \centering
%     \begin{minipage}{0.75\linewidth}
%     \includegraphics[width=\linewidth]{representation-alt.png}
%     \end{minipage}
%     \hfill
%     \begin{minipage}{0.21\linewidth}\vspace{-1ex}
%     \caption{An example of the proposed representation---(a) an example of the first eight beats of a song in the orchestra dataset, shown as a multitrack piano roll, (b) the same song encoded by our proposed representation, where the grayed out zeros denote undefined values and (c) a human-readable translation of the codes shown in (b).}
%     \label{mmt:fig:representation}
%     \end{minipage}
% \end{figure*}

Following \citep{huang2020remi}, we decompose the note onset information into  beat and position information, where $x^\mathit{beat}$ denotes the index of the beat that the note lies in, and $x^\mathit{position}$ the position of the note within that beat. To be specific, the actual onset of the note is equivalent to $r\cdot x^\mathit{beat} + x^\mathit{position}$, where $r$ is the temporal resolution of a beat. For simplicity, we assume that the beats are always a quarter note apart in this work. This decomposition reduces the size of the vocabulary and helps the model learn the music meter system, as evidenced by \citep{huang2020remi}. For the duration field, following \citep{vonrutte2022figaro}, we only allow a carefully-chosen set of common note duration values and replace any duration outside of this set with the closest known duration. For the instrument field, we map similar MIDI programs to the same instrument to reduce the total number of instruments, resulting in 64 unique instruments from the 128 MIDI programs. For example, both `acoustic grand piano' and `bright acoustic piano' are mapped to the same `piano' instrument.

We note that the proposed representation leads to a significantly shorter sequence length as compared to two existing representations \citep{ens2020mmm,vonrutte2022figaro} for multitrack music generation. On an orchestral dataset \citep{crestel2017lop}, an encoded sequence of length 1,024 using our propsoed representation can represent 2.6 and 3.5 times longer music samples compared to \citep{ens2020mmm} and \citep{vonrutte2022figaro}, respectively. Further, because the timing information is embedded into each note event, the proposed representation is invariant to permutation, i.e., reordering the note events do not affect the decoded music. For the sake of autoregressive training for the transformer model, we sort the notes with respect to the beat field, and subsequently the position, pitch, duration, instrument fields. This allows a trained autoregressive model to be readily applicable to the song continuation task.

%-----------------
\subsection{Model}
%-----------------
\label{mmt:sec:model}

\begin{figure}
    \small
    \centering
    \includegraphics[width=.75\linewidth]{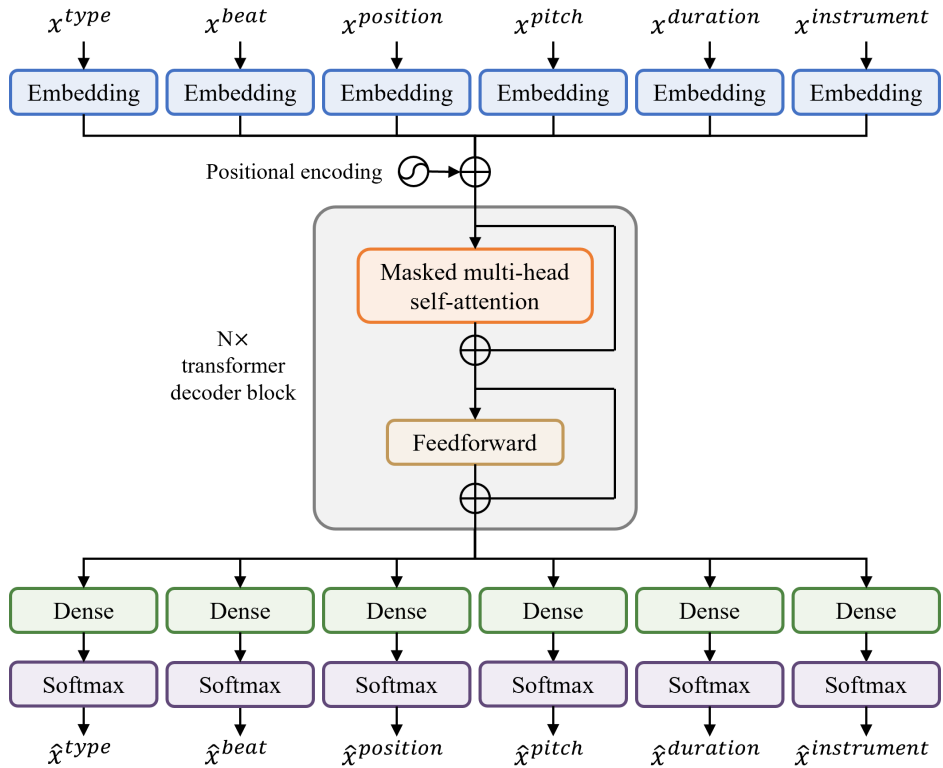}
    \caption{Illustration of the proposed MMT model.}
    \label{mmt:fig:model}
\end{figure}

We present the Multitrack Music Transformer (MMT) for generating multitrack music using the representation proposed in \cref{mmt:sec:representation}. We base the proposed model on a decoder-only transformer model \citep{liu2018transformerdecoder,brown2020gpt3}. Unlike a standard transformer model, whose inputs and outputs are one-dimensional, the proposed model has multi-dimensional input and output spaces similar to \citep{hsiao2021transformer}, as illustrated in \cref{mmt:fig:model}. The model is trained to minimize the sum of the cross entropy losses of different fields under an autoregressive setting. We adopt a learnable absolute positional embedding \citep{vaswani2017transformer}. Once the training is done, the trained transformer model can be used in three different modes, depending on the inputs given to the model to start the generation:
\begin{itemize}
    \item \textbf{Unconditioned generation}: Only a `start-of-song' event is provided to the model. The model generates the instrument list and subsequently the note sequence.
    \item \textbf{Instrument-informed generation}: The model is given a `start-of-song' event followed by a sequence of instrument codes and a `start-of-notes' event to start with. The model then generates the note sequence. Note that we need the `start-of-notes' event as it marks the end of the instrument list, otherwise the model may continue to generate instrument events.
    \item \textbf{$N$-beat continuation}: All instrument and note events in the first $N$ beats are provided to the model. The model then generates subsequent note events that continue the input music.
\end{itemize}

During inference, the sampling process is stopped when an `end-of-song' event is generated or the maximum sequence length is reached. We adopt the top-$k$ sampling strategy on each field and set $k$ to 10\% of the number of possible outcomes per field. Moreover, since the type and beat fields in our representation are always sorted, we further enforce a monotonic constraint during decoding. For example, when sampling for $x_{i + 1}^\mathit{type}$, we set the probability of getting a value smaller than $x_{i}^\mathit{type}$ to zero. This prohibits the model from generating events in certain invalid order, e.g., an `note' event before an `instrument' event.

Finally, while existing multitrack music generation systems \citep{ens2020mmm,vonrutte2022figaro} need to combine several generated tokens to form a note, the proposed MMT model generates a note at each inference step, i.e., a line in \cref{mmt:fig:representation}(b) and (c). This offers MMT a significantly faster inference speed and smaller memory footprint thanks to the reduced size of the self-attention matrix. However, since MMT predicts the six output fields nonautoregressively (i.e., independently), it cannot model the interdependencies between these fields of the same note. We will discuss this trade-off between time/memory complexity and modeling capacity in \cref{mmt:sec:subjective}.

%================
\section{Results}
%================
\label{mmt:sec:result}

%----------------------------
\subsection{Experiment Setup}
%----------------------------

In this work, we consider the Symbolic Orchestral Database (SOD) \citep{crestel2017lop}. We set the temporal resolution to 12 time steps per quarter note. We discard tempo and velocity information as not all data contains such information. Further, we discard all drum tracks. We end up with 5,743 songs (357 hours). We reserve 10\% of the data for validation and 10\% for testing. We use MusPy \citep{dong2020muspy} to process the data. For the proposed MMT model, we use 6 transformer decoder blocks, with a model dimension of 512 and 8 self-attention heads. All input embeddings have 512 dimensions. We trim the code sequences to a maximum length of 1,024 and a maximum beat of 256. During training, we augment the data by randomly shifting all the pitches by $s \sim U(-5, 6)\ (s \in \mathbb{Z})$ semitones and randomly selecting a starting beat. We validate the model every 1K steps and stop the training at 200K steps or when there was no improvements for 20 validation rounds. We render all audio samples using FluidSynth with the MuseScore General SoundFont. We encourage the readers to listen to the sample generated music for the unconditional generation, instrument-informed generation and $N$-beat continuation tasks on our demo website: \url{https://salu133445.github.io/mmt/}.

%-------------------------------------
\subsection{Subjective Listening Test}
%-------------------------------------
\label{mmt:sec:subjective}

To assess the quality of music samples generated by our proposed model, we conducted a listening test with 9 music amateurs recruited from our social networks, where all survey participants can play at least one musical instrument. In the questionnaire, each participant was asked to listen to 10 audio samples generated by each model and rate each audio sample according to three criteria---\textit{coherence}, \textit{richness} and \textit{arrangement}.\footnote{To be specific, we ask the following questions: \textit{coherence}---``Is it temporally coherent? Is the rhythm steady? Are there many out-of-context notes?''; \textit{richness}---``Is it rich and diverse in musical textures? Are there any repetitions and variations? Is it too boring?''; \textit{arrangement}---``Are the instruments used reasonably? Are the instruments arranged properly?''} We compared the MMT model against two baseline models based on the standard decoder-only transformer model. The first baseline model used the MultiTrack representation proposed in the MMM model \citep{ens2020mmm}, where we replaced the bar tokens with beat tokens. The other used a simplified version of the REMI+ representation used in the FIGARO model \citep{vonrutte2022figaro}, where we removed the time signature, tempo and chord tokens as such information is not generally available in our datasets. We will refer to the two baseline models as the MMM and REMI+ models. For a fair comparison, we trimmed all generated samples to a maximum of 64 beats. Moreover, as discussed in \cref{mmt:sec:introduction}, the long sequence length of existing multitrack music representations restricts the model from learning long-term dependencies. Hence, we also computed the mean length of the generated samples and the inference speed in this experiment.

\begin{table*}
    \footnotesize
    \centering
    \caption[Performance comparison of our proposed model against the baseline models.]{Performance comparison of our proposed model against the baseline models. Mean values and 95\% confidence intervals are reported.}
    \label{mmt:tab:performance}
    \begin{tabularx}{\linewidth}{p{7.6em}@{~~~}m{4.4em}@{~~~}m{5.5em}@{~~~}m{5.8em}@{~~~}c@{~~~}c@{~~~}c@{~~~}c}
        \toprule
        &\multirow{2}{4.4em}{Number of parameters} &\multirow{2}{5.5em}{Average sample length (sec)} &\multirow{2}{5.8em}{Inference speed (notes per second)} &\multicolumn{4}{c}{Subjective listening test results}\vspace{.3ex}\\
        \cmidrule(r){5-8}
        &&& &Coherence &Richness &Arrangement &Overall\vspace{.3ex}\\
        \midrule
        MMM \citep{ens2020mmm} &19.81 M &\underline{38.69} &\underline{5.66} &3.48 $\pm$ 0.35 &3.05 $\pm$ 0.38 &3.28 $\pm$ 0.37 &3.17 $\pm$ 0.43\\
        \cmidrule(lr){1-8}
        REMI+ \citep{vonrutte2022figaro} &20.72 M &28.69 &3.58 &\textbf{3.90 $\pm$ 0.52} &\textbf{3.74 $\pm$ 0.21} &\textbf{3.74 $\pm$ 0.44} &\textbf{3.77 $\pm$ 0.41}\\
        \cmidrule(lr){1-8}
        MMT (ours) &19.94 M &\textbf{100.42} &\textbf{11.79} &\underline{3.55 $\pm$ 0.46} &\underline{3.53 $\pm$ 0.35} &\underline{3.40 $\pm$ 0.44} &\underline{3.33 $\pm$ 0.47}\\
        \bottomrule
    \end{tabularx}
\end{table*}

We summarize in \cref{mmt:tab:performance} the evaluation results. Compared to the MMM model, our proposed MMT model achieves a higher score across all criteria. Further, MMT generates 2.6 times longer samples and is twice faster in inference speed. As compared to the REMI+ model, our proposed model achieves a mean opinion score (MOS) of 3.33, while the REMI+ model achieves an MOS of 3.77. However, MMT can generate 3.5 times longer samples and is 3.3 times faster in inference speed. This is because the baseline models need multiple inference passes to combine several generated tokens and form a note, whereas the MMT model generate a note in a single inference pass. Finally, we note that while offering a faster inference speed and longer generated sample length, our proposed model cannot model the interdependencies between the six output heads as it predicts each field independently. For example, the REMI+ model first generates an instrument token and then generates the pitch token given the instrument token, which allows the model to rule out unsuitable pitches for that particular instrument. In contrast, the MMT model samples from each output head independently. We can clearly observe this trade-off between quality and between time/memory complexity can be clearly observed from \cref{mmt:tab:performance}.

%--------------------------------
\subsection{Objective Evaluation}
%--------------------------------
\label{mmt:sec:objective}

In addition the subjective listening test, we follow \citep{mogren2016crnngan,wu2020jazz} and measure the pitch class entropy, scale consistency and groove consistency for evaluating the performance of the proposed model on the unconditioned generation task. For these metrics, we consider a closer value to that of the ground truth better. \cref{mmt:tab:objective} shows the evaluation results. We can see that the REMI+ model achieves closest values to those of the ground truth. We also notice that while the MMM model result in closer values of pitch class entropy and scale consistency to those of the ground truth, it achieves a lower score in the subjective listening test presented in \cref{mmt:sec:subjective} than our proposed MMT model.

\begin{table}
    \small
    \centering
    \caption[Objective evaluation results.]{Objective evaluation results. Mean values and 95\% confidence intervals are reported. A closer value to that of the ground truth is considered better.}
    \label{mmt:tab:objective}
    \begin{tabularx}{\linewidth}{lCCC}
        \toprule
        &Pitch class entropy &Scale consistency (\%) &Groove consistency (\%)\\
        \midrule
        Ground truth &2.974 $\pm$ 0.018 &92.26 $\pm$ 1.25 &93.05 $\pm$ 1.00\\
        \cmidrule(lr){1-4}
        MMM \citep{ens2020mmm} &\underline{2.884 $\pm$ 0.023} &\underline{93.13 $\pm$ 0.49} &91.90 $\pm$ 0.64\\
        REMI+ \citep{vonrutte2022figaro} &\textbf{2.897 $\pm$ 0.019} &\textbf{93.12 $\pm$ 0.51} &\textbf{92.90 $\pm$ 0.49}\\
        MMT (ours) &2.802 $\pm$ 0.025 &94.74 $\pm$ 0.42 &\underline{92.09 $\pm$ 0.49}\\
        \bottomrule
    \end{tabularx}
\end{table}

%----------------------------------
\subsection{Musical Self-attention}
%----------------------------------
\label{mmt:sec:attention}

Despite the growing interests in applying transformer models to music, little effort has been made to understand how self-attention works for symbolic music---existing analyses \citep{huang2019musictransformer,huang2018visualizing,chen2021transformer,ziyu2021musebert} provide only case studies on few selected samples. In this section, we aim to investigate musical self-attention in a systematic way. To this end, we propose two new quantities to measure the average relative attention. Mathematically, given a test set $\mathcal{D}$, we define the \textit{mean relative attention} for a field $d$ (e.g., pitch or beat) as:
\begin{equation}
    \gamma_k^{(d)} = \frac{\sum_{\bm{x} \in \mathcal{D}} \sum_{s > t} a_{s, t}(\bm{x})\;\mathbbm{1}_{\left(\bm{x}^{(d)}_t - \bm{x}^{(d)}_s\right) = k}}{\sum_{k'} \sum_{\bm{x} \in \mathcal{D}} \sum_{s > t} a_{s, t}(\bm{x})\;\mathbbm{1}_{\left(\bm{x}^{(d)}_t - \bm{x}^{(d)}_s\right) = k'}}\,,
\end{equation}
where $\mathbbm{1}[\cdot]$ is the indicator function and $a_{s, t}(\bm{x}) \in [0, 1]$ denotes the attention weight assigned by $\bm{x}_s$ to $\bm{x}_t$. Intuitively, $\gamma_k^{(d)}$ measures the average attention weight that model assigns to a certain key of a certain difference from the query. Note that each attention head has its own attention weight $a_{s, t}$ and thus its own $\gamma_k$. Moreover, we notice that $\gamma_k^{(d)}$ is biased towards differences that occur more frequently. Thus we further propose the \textit{mean relative attention gain}:
\begin{equation}
    \tilde{\gamma}_k^{(d)} = \gamma_k^{(d)} - \frac{\sum_{\bm{x} \in \mathcal{D}} \sum_{s > t}\;\mathbbm{1}_{\left(\bm{x}^{(d)}_t - \bm{x}^{(d)}_s\right) = k}}{\sum_{k'} \sum_{\bm{x} \in \mathcal{D}} \sum_{s > t}\;\mathbbm{1}_{\left(\bm{x}^{(d)}_t - \bm{x}^{(d)}_s\right) = k'}}\,,
\end{equation}
which measures the difference between $\gamma_k^{(d)}$ and the same quantity obtained by assuming a uniform attention matrix.

\begin{figure}
    \newlength{\temp}
    \setlength{\temp}{0.795cm}
    \small
    \raggedright
    \hspace*{4.1cm}$\star$\hspace{\temp}$\star$\hspace{\temp}$\star$\hspace{\temp}$\star$\hspace{\temp}$\star$\hspace{\temp}$\star$\hspace{\temp}$\star$\hspace{\temp}$\star$\hspace{\temp}$\star$\\
    \centering
    \includegraphics[width=.8\linewidth,trim={0 0 0 2.5mm},clip]{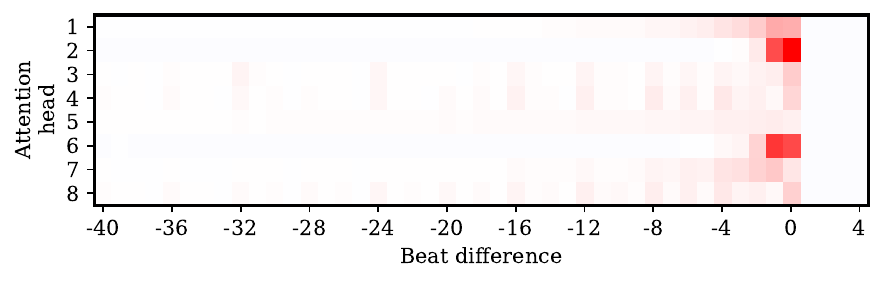}\\[-1ex]
    (a)\\[2ex]
    \raggedright
    \hspace*{5.75cm}{\footnotesize$_\nearrow$~\small 8th note behind}{\footnotesize$_\nearrow$~\small same position}\hspace{.05cm}{\footnotesize$_\nearrow$~\small 8th note ahead}\\[-.5ex]
    \centering
    \includegraphics[width=.8\linewidth,trim={0 0 0 2.5mm},clip]{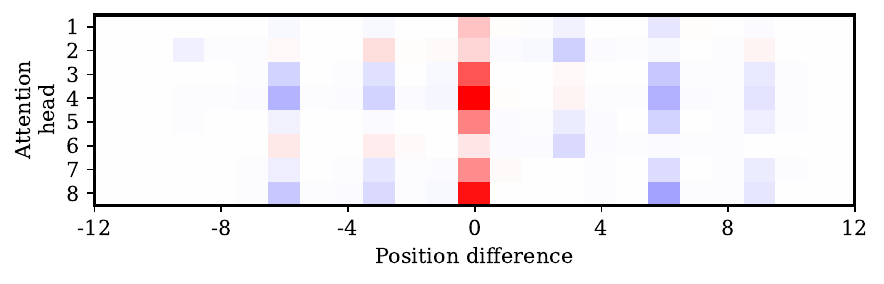}\\[-1ex]
    (b)\\[2ex]
    \raggedright
    \hspace*{6.8cm}{\small same pitch\footnotesize$_\nwarrow$}\hspace{.25cm}{\small 4th\footnotesize$_\nwarrow$}\hspace{.35cm}{\footnotesize$_\nearrow$\small 5th}\hspace{.35cm}{\footnotesize$_\nearrow$\small 8th (octave)}\\[-.5ex]
    \centering
    \includegraphics[width=.8\linewidth,trim={0 0 0 2.5mm},clip]{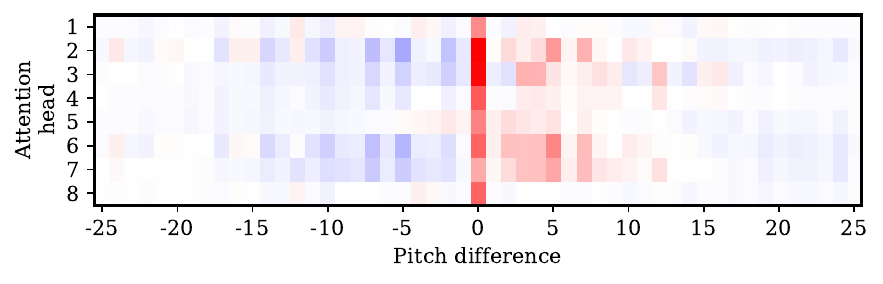}\\[-1ex]
    (c)
    \caption[Mean relative attention gains (a) $\tilde{\gamma}_k^\mathit{beat}$, (b) $\tilde{\gamma}_k^\mathit{position}$ and (c) $\tilde{\gamma}_k^\mathit{pitch}$ (see \cref{mmt:sec:attention} for definitions) of a trained MMT model.]{Mean relative attention gains (a) $\tilde{\gamma}_k^\mathit{beat}$, (b) $\tilde{\gamma}_k^\mathit{position}$ and (c) $\tilde{\gamma}_k^\mathit{pitch}$ (see \cref{mmt:sec:attention} for definitions) of a trained MMT model. Red and blue colors indicate positive and negative values, respectively.} % \textcolor{darkred}{positive} and \textcolor{darkblue}{negative}
    \label{mmt:fig:attention}
\end{figure}

In this experiment, we compute $\tilde{\gamma}_k^\mathit{beat}$, $\tilde{\gamma}_k^\mathit{position}$ and $\tilde{\gamma}_k^\mathit{pitch}$ on 100 test samples for the last attention layer of a trained MMT model. As shown in \cref{mmt:fig:attention}(a), we can see that the 2nd and 6th attention heads attend more to nearby beats, while the other attention heads attend to beats in further past. In addition, several attention heads assign relatively larger weights to the beats that are $4N$ (i.e., 4, 8, 12, 16, etc.) beats away from the current one, as highlighted by the `$\star$' symbols. From \cref{mmt:fig:attention}(b) we observe that the model pays most attention to notes that have the same position as the current note. That is, a note on beat attends more to the last note on beat, and a note off beat attends more to the last note off beat. \cref{mmt:fig:attention}(c) shows that the model attends more to pitches within one octave above, and it pays more attention to pitches that form a consonant interval with the current note, e.g., a 4th, a 5th and an octave. We note that the learned self-attention generally comply with music theory principles.

While recent advances in symbolic music generation has borrowed various techniques from natural language modeling, music is fundamentally different from text in that music has a underlying temporal axis embedded and contains strong recurrence patterns in many aspects. Our analysis here shows that our proposed model learns a relative self-attention for certain aspects of music, specifically, beat, position and pitch. We hope our analysis can shed light on further improvements in optimizing the self-attention mechanism for symbolic music modeling.

%===================
\section{Conclusion}
%===================

We have presented the Multitrack Music Transformer for multitrack music generation. Built upon a new multitrack representation, our proposed model can generate longer multitrack music in a faster inference speed than two existing approaches. We showed in a subjective listening test that the proposed model perform reasonably well against the two baseline models in terms of the quality of the generated music. Through a systematic analysis, we showed that our proposed model learns relative self-attention in certain aspects of music such as beats, positions and pitches. Our findings provide a novel foundation for future work exploring longer-form, real-time capable multitrack music generation and improving the self-attention mechanism for music.

\secbreak

\begin{adjustwidth}{.5in}{.5in}
    \itshape
    \hspace{\parindent}
    This chapter, in full, is a reprint of the material as it appears in ``Multitrack Music Transformer'' by Hao-Wen Dong, Ke Chen, Shlomo Dubnov, Julian McAuley and Taylor Berg-Kirkpatrick, which was published in the Proceedings of the IEEE International Conference on Acoustics, Speech and Signal Processing (ICASSP) in 2023. The dissertation author was the primary investigator and author of this paper.
\end{adjustwidth}

\graphicspath{{chapters/arranger/figs/}}
\chapter{Towards Automatic Instrumentation by Learning to Separate Parts in Symbolic Multitrack Music}
\label{chap:arranger}

\begin{abstract}
Modern keyboards allow a musician to play multiple instruments at the same time by assigning zones---fixed pitch ranges of the keyboard---to different instruments. In this paper, we aim to further extend this idea and examine the feasibility of automatic instrumentation---dynamically assigning instruments to notes in solo music during performance. In addition to the online, real-time-capable setting for performative use cases, automatic instrumentation can also find applications in assistive composing tools in an offline setting. Due to the lack of paired data of original solo music and their full arrangements, we approach automatic instrumentation by learning to separate parts (e.g.,~voices, instruments and tracks) from their mixture in symbolic multitrack music, assuming that the mixture is to be played on a keyboard. We frame the task of part separation as a sequential multi-class classification problem and adopt machine learning to map sequences of notes into sequences of part labels. To examine the effectiveness of our proposed models, we conduct a comprehensive empirical evaluation over four diverse datasets of different genres and ensembles---Bach chorales, string quartets, game music and pop music. Our experiments show that the proposed models outperform various baselines. We also demonstrate the potential for our proposed models to produce alternative convincing instrumentations for an existing arrangement by separating its mixture into parts. All source code and audio samples can be found at \url{https://salu133445.github.io/arranger/}.
\end{abstract}

%=====================
\section{Introduction}
%=====================

% Motivation
Music is an art of time and sound. It often contains complex textures and possibly \emph{parts} for multiple voices, instruments and tracks. While jointly following the global style and flow of the song, each part possesses its own characteristics and can develop different musical ideas independently. For example, in pop music, guitar and piano tend to play chords and might span across a large pitch range, while bass is usually monophonic and stays in a lower range. While playing multiple instruments usually requires multiple performers, keyboardists potentially have the ability to control many instruments at once. Modern keyboards often offer the functionality of \emph{zoning}, which allows a player to divide the pitch range into zones and assign each zone to a certain instrument. However, zoning is not ideal given its low flexibility that requires careful configuration and sometimes rearrangement of the music, and incapability for handling certain genres of music that have close and possibly overlapping harmony.

% Applications
In this paper, we aim for the more ambitious goal of automatic instrumentation---a process that we define as dynamically assigning instruments to notes in solo music. A real-time, online automatic instrumentation model could allow a musician to have their keyboard performance instantaneously and seamlessly performed by a different ensemble. In addition to performative use cases, an offline automatic instrumentation model can also be useful to assist composers in suggesting proper instrumentation or providing a starting point for arranging a solo piece, especially for composers who have less experience arranging for a particular ensemble.

\begin{figure}
    \small
    \centering
    \includegraphics[width=.8\linewidth]{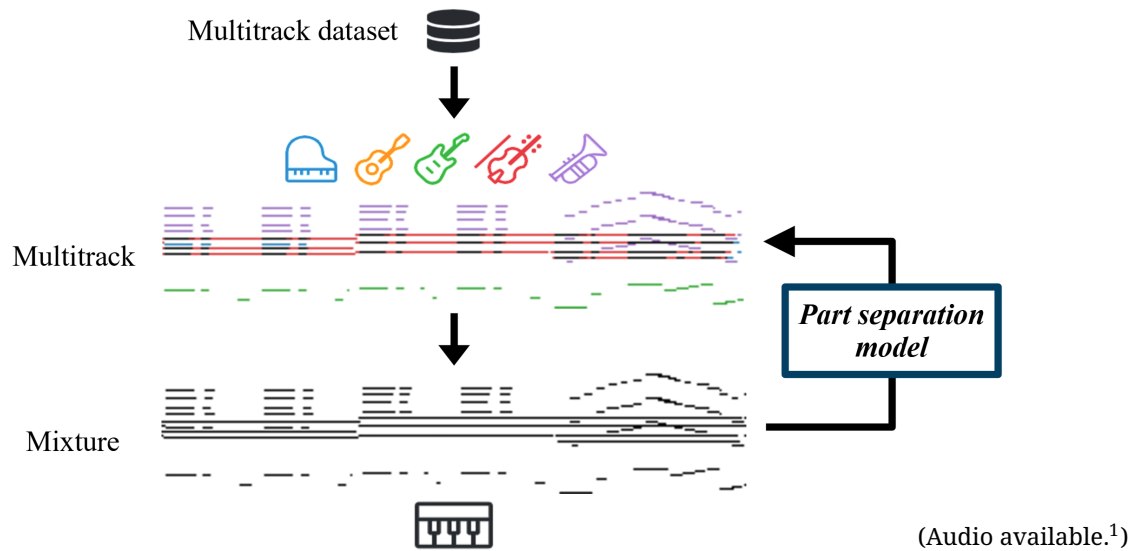}\\[-3ex]
    \raggedleft{\small(Audio available.\cref{arranger:fn:demo})}
    \caption[Proposed pipeline.]{Proposed pipeline. By downmixing a symbolic multitrack into a single-track mixture, we acquire paired data of solo music and its instrumentation. We then use these paired data to train a \textit{part separation} model that aims to infer the part label (e.g., one out of the five instruments in this example) for each single note in a mixture. Automatic instrumentation can subsequently be accomplished by treating input from a keyboard player as a downmixed mixture (bottom) and separating out the relevant parts (top). The music is visualized in the piano roll representation, where the x- and y-axes represent time and pitch, respectively. Colors indicate the instruments.}
    \label{arranger:fig:pipeline}
\end{figure}

% Challenges & Formulation
Automatic instrumentation is challenging as it requires domain knowledge of each target instrument, i.e., which pitches, rhythms, chords, and sequences thereof are playable, and it is hard to specify such knowledge by some fixed set of rules. In view of recent advances in machine learning, we propose to adopt a data-driven approach to this task. However, it can be laborious to acquire paired data of original solo music and their full arrangements. Given the abundance of multitrack music data, we approach automatic instrumentation by learning to separate parts from their mixture in multitrack music, a task we call \emph{part separation} (see \cref{arranger:fig:pipeline}). Assuming that the mixture is to be played on a keyboard and the multitrack is the target arrangement that we want to generate by an automatic instrumentation model, we thereby have paired data for solo music and full arrangements.

% Methods & Results
We frame the new task of part separation as a sequential multi-class classification problem that aims to map sequences of notes into sequences of part labels. We adopt long short-term memory~(LSTM) \citep{hochreiter1997lstm} and Transformer \citep{hsiao2021transformer} models for the task. We conduct an extensive empirical evaluation showing the superiority of our proposed models to baselines for the related task of voice separation as well as strategies found in commodity keyboards. To showcase the potential of our proposed models, we also demonstrate their ability to produce alternative convincing instrumentations for existing arrangements. Audio for all examples and more samples are available on the demo website.\footnote{\url{https://salu133445.github.io/arranger/}\label{arranger:fn:demo}} All source code can be found in the project repository.\footnote{\url{https://github.com/salu133445/arranger}}

%===================
\section{Prior Work}
%===================

Voice separation is a related task to part separation which involves separating blended scores into individual monophonic voices. While useful, voice separation is agnostic to constraints imposed by specific instruments---a composer using a voice separation algorithm would have to manually align voices to appropriate instruments. Some prior work investigates voice separation in small, carefully-annotated pop music datasets \citep{gray2016voiceseparation,guiomardkagan2015voice}. Some prior work on voice separation allows synchronous or overlapping notes in a voice \citep{kilian2002voiceseparation,cambouropoulos2006voice,karydis2007visa,cambouropoulos2008voice}. However, their results are only reported on small test sets in certain genres. Others have adopted multilayer perceptrons \citep{gray2016voiceseparation,devalk2019deepseparation} and convolutional neural networks \citep{gray2020voiceseparation} with hand-crafted input features for voice separation. Another relevant work on hand detection in piano music used LSTMs to separate notes played by right and left hands in piano MIDI data \citep{hadjakos2019hands}. To the best of our knowledge, no past work has examined the task of part separation in a general setting for multiple music genres.

In addition to voice separation, prior work has explored automatic music arrangement. 
The primary focus of prior work for automatic music arrangement has been on reduction---mapping musical scores for large ensembles to parts playable by a single specific instrument such as the piano \citep{chiu2009automatic,onuma2010piano,huang2012towards,nakamura2015automatic,takamori2017automatic,nakamura2018statistical}, guitar \citep{tuohy2005genetic,hori2012automatic,hori2013input} or bass \citep{abe2012automatic}. This past work focuses on identifying the least important notes to delete so that the resultant score is playable on a single instrument, whereas our work seeks to preserve the original score in its entirety and satisfy playability for multiple instruments simultaneously. As an exception, Crestel and Esling \citep{crestel2016live} explore strategies for arranging orchestral music from piano, though their approach does not guarantee that all notes in the input piano map to parts in the output.

Music generation is another body of work that has used neural network sequential models for processing symbolic music \citep{briot2017survey}. 
Simon and Oore \citep{simon2017performance} proposed a convenient approach for music generation which involved training recurrent neural network language models on a language-like ``event-based'' representation of music. 
Subsequently, recent work has explored event-based representations using Transformers \citep{huang2019musictransformer,musenet,donahue2019lakhnes,huang2020remi,ens2020mmm,hsiao2021transformer,muhamed2021transformergan}. 
In this work, we explore a more compact input representation of music that passes all of the information about a note into the model at once, rather than spreading it out across several events.
We also note that Payne \citep{musenet} generate music which contains parts for several instruments, but their model cannot be directly used to perform part separation of existing musical material.

%============================
\section{Problem Formulation}
%============================

Mathematically, we consider a piece of music $x$ as a sequence of notes $(x_1,\dots,x_N)$, where $N$ is the number of notes. Each note is represented by a tuple of time $t_i$ and pitch $p_i$, i.e., $x_i = (t_i, p_i)$. Alternatively, we could also include duration $d_i$ as an input and have $x_i = (t_i, p_i, d_i)$. Each note is associated with a label $y_i \in \{1,\dots,K\}$ that represents the part it is in, where $K$ is the number of parts. The goal of part separation is to learn the mapping between notes and part labels. This formulation is rather flexible and has no assumptions on whether a part is monophonic or not---it can be a voice, an instrument, a track, etc.

In terms of the context given for predicting the label of each note, we can categorize part separation models into three classes: An \emph{independent model} predicts the label for each note independently, without any context. An \emph{online model} predicts the label of the current note $x_i$ given only past information, i.e., notes $(x_1,\dots,x_{i-1})$, as context. An \emph{offline model} predicts the label of the current note $x_i$ given past and future information, i.e.,~the full sequence of $(x_1,\dots,x_N)$, as context. While independent and online models are preferable for use cases that require real-time outputs,~e.g., live performance. Moreover, the inability to look into the future makes the real-time setting more challenging than the offline setting. On the other hand, offline models can find applications in assisstive composing tools.

%===============
\section{Models}
%===============
\label{arranger:sec:model}

We consider the following input features for our models---(1) \emph{time}: onset time, in time step,\footnote{Assuming that the music is in metrical timing, a time step is a factor of some musically-meaningful unit (e.g., a quarter note) and can be adjusted to match the desired temporal resolution.} (2) \emph{pitch}: pitch as a MIDI note number, (3) \emph{duration}: note length, in time step, and (4) \emph{frequency}: fundamental frequency of the pitch, in Hz, computed by the formula $f = 440 \cdot 2^{(p - 69)/12}$. In addition, we also consider features that encode the metric time grid of music similar to the \textsc{Bar} and \textsc{Position} events proposed in \citep{huang2020remi}---(5) \emph{beat}: onset time, in beat, and (6) \emph{position}: position within a beat, in time step.

Moreover, to help the models better disambiguate parts, we also include two simple hints---(7) \emph{entry hints}: onset position for each instrument, encoded as a unit step function centered at its onset time and all zero if the instrument is not used, and (8) \emph{pitch hints}: average pitch of each track. These hints allow the musician to use interactively to make the instrumentation process more controllable. For example, entry hints can be used to control the instruments available as they serve as switches for the instruments.

For the machine learning models, we consider the LSTM \citep{hochreiter1997lstm} and its bidirectional version (BiLSTM) \citep{schuster1997bilstm}. We use a three-layer stacked LSTM with 128 hidden units in each layer (64 hidden units per layer for BiLSTM). We also consider two variants of Transformer \citep{vaswani2017transformer}---one based on the encoder (Transformer-Enc) and one based on the decoder (Transformer-Dec). They share the same architecture that is composed of three Transformer blocks, each of which has 128 hidden units and 8 heads in self-attention computation and 256 hidden units in the internal feedforward network. However, they have different attention masks: Transformer-Enc uses only the padding mask, while Transformer-Dec uses both the padding mask and the lookahead mask, which blocks its access to future information and makes it a online model. In this paper, the LSTM and Transformer-Dec models are made online models, and the BiLSTM and Transformer-Enc models are made offline models that take durations as inputs.

%========================
\section{Baseline Models}
%========================

In order to gain an insight into how the proposed models perform, we include two heuristic algorithms and a voice separation model from the literature in our empirical study.

%--------------------------------
\subsection{Zone-based algorithm}
%--------------------------------

This algorithm simulates a common feature in modern keyboards where a player can preassign a pitch range (i.e., the `zone') for each instrument and notes will automatically be assigned to the corresponding instrument as the player performs. This algorithm finds the optimal zones for the whole training data and uses these optimal zones at test time. For the oracle case, the optimal zones for each sample are computed and used at test time. We note that the oracle case might not be easily achievable as it can be hard for a musician to set the zones optimally beforehand, especially for improvisation.

%-----------------------------------
\subsection{Closest-pitch algorithm}
%-----------------------------------

The closest-pitch algorithm keeps track of the last active pitches $p'_i$ for each track $i$. For each incoming pitch $p$, it finds the pitch among the last active pitches that has the closest pitch to $p$ and assigns the upcoming note with the same label as the chosen pitch. This is a casual model and it also relies on the onset hints. We can formulate this algorithm as follows. For $i = 1,\dots,N$, we have $$\hat{y}_i = \begin{cases}y_i, &\text{if $x_i$ is an onset}\\\displaystyle\argmin_{j\in\{1,\dots,K\}} (p_i - p'_j)^2 + M a_i, &\text{otherwise}\end{cases}\,,$$ where $p'_i$ is the last active pitch of track $i$ before time $t$ and $a_i$ indicates whether track $i$ is active, i.e., a concurrent note has not yet been released. We set $M$ to a large positive number when we assume each part is monophonic, which we will refer to as the `mono' version of this algorithm, otherwise set $M=0$.

%---------------------------------------
\subsection{Multilayer perceptron (MLP)}
%---------------------------------------

We adapt the voice separation model proposed in \citep{devalk2019deepseparation} to the task of part separation. This model uses multilayer perceptron (MLP) to predict the label for the current note based on hand-crafted features that encodes its nearby context. We use entry hints rather than predicting them by the proposed voice entry estimation heuristics. We remove the `interval' feature as there is no upper bound for the number of concurrent notes and change the proximity function to L1 distance. The oracle case of this model replaces error-prone prior predictions with ground truth history labels. In our implementation, we use three fully-connected layers with 128 hidden units each.

\begin{table*}
    \small
    \centering
    \caption{Statistics of the four datasets considered in this paper.}
    \label{arranger:tab:data}
    \small
    \begin{tabularx}{\linewidth}{X@{~~~}llllll}
        \toprule
        Dataset &Hours &Files &Notes &Parts &Ensemble &Most common label\\
        \midrule
        \makecell[Xt]{Bach chorales\\\citep{cuthbert2010music21}}     &3.23  &409   &96.6K &4 &\makecell[tl]{soprano,\\alto,\\tenor,\\bass}                 &\makecell[tl]{bass\\(27.05\%)}\\
        \cmidrule(lr){1-7}
        \makecell[Xt]{String quartets\\\citep{thickstun2017musicnet}} &6.31  &57    &226K  &4 &\makecell[tl]{first violin,\\second violin,\\viola,\\cello}  &\makecell[tl]{first violin\\(38.72\%)}\\
        \cmidrule(lr){1-7}
        \makecell[Xt]{Game music\\\citep{donahue2018nesmdb}}          &45.05 &4.61K &2.46M &3 &\makecell[tl]{pulse wave I,\\pulse wave II,\\triangle wave} &\makecell[tl]{pulse wave II\\(39.35\%)}\\
        \cmidrule(lr){1-7}
        \makecell[Xt]{Pop music\\\citep{raffel16lmd}}              &1.02K &16.2K &63.6M &5 &\makecell[tl]{piano,\\guitar,\\bass,\\strings,\\brass}        &\makecell[tl]{guitar\\(42.50\%)}\\
        \bottomrule
    \end{tabularx}
\end{table*}

%=============
\section{Data}
%=============

In order to examine the effectiveness of the proposed models, we consider four datasets---(1) \emph{Bach chorales} in Music21 \citep{cuthbert2010music21}, (2) \emph{string quartets} in MusicNet \citep{thickstun2017musicnet}, (3) \emph{game music} in Nintendo Entertainment System (NES) Music Database \citep{donahue2018nesmdb} and (4) \emph{pop music} in Lakh MIDI Dataset \citep{raffel16lmd}, which are diverse in their genres, sizes and ensembles (see \cref{arranger:tab:data} for a comparison).

As these datasets are noisy in different ways, we need to further clean the data. For the game music dataset, we discard the percussive noise track in the original dataset as they do not follow the standard 128-pitch system used in other tracks. For the pop music dataset, we use a cleaned subset derived in \citep{dong2018musegan}, which contains only pop songs. We mapped the instruments to the five most common instrument families---piano, guitar, bass, strings and brass. We follow the General MIDI 1 specification on the mapping from an instrument to its instrument family. Instruments that fall outside of these five families are discarded. We note that the lead melody track might occasionally be discarded during the mapping process due to the high variance on instruments used for the melody track.

Moreover, we discard songs with only one active track as the task becomes trivial in this case. We note that all Bach chorales, string quartets and most pop songs are in metrical timing, where a time step corresponds to some fraction of a quarter note. Thus, we downsample them into 24 time steps per quarter note, which can cover 32nd notes and triplets. As songs in the game music dataset are in absolute time, we downsample them to a temporal resolution equivalent to 24 time steps per quarter note in a tempo of 125 quarter notes per minute (qpm).

Finally, we split each dataset into train--test--validation sets with a ratio of $8 : 1 : 1$ except the game music dataset, where we use the original splits provided with the NES Music Database. We use MusPy \citep{dong2020muspy} and music21 \citep{cuthbert2010music21} for processing MIDI and MusicXML files.

\begin{figure}
    \small
    \centering
    \def\imgheight{5ex}
    \def\imgwidth{\linewidth}
    \begin{tabularx}{\linewidth}{L{.16\linewidth} X}
        Musical score
          &\includegraphics[width=\imgwidth]{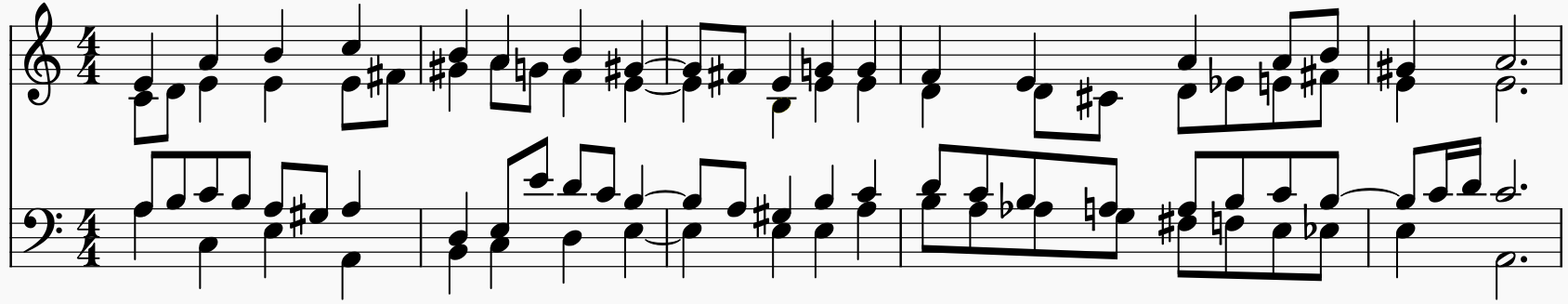}\\
        \cmidrule(lr){1-2}
        Ground truth
          &\includegraphics[width=\imgwidth,height=\imgheight]{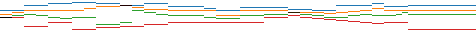}\\
        \cmidrule(lr){1-2}
        Online LSTM prediction
          &\includegraphics[width=\imgwidth,height=\imgheight]{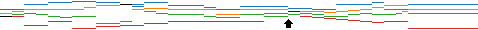}\\
        \cmidrule(lr){1-2}
        Offline BiLSTM prediction
          &\includegraphics[width=\imgwidth,height=\imgheight]{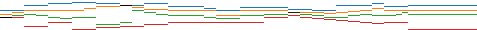}\\
        \multicolumn{2}{r}{\footnotesize (Audio available.\cref{arranger:fn:demo} Colors: \textcolor{tab-blue}{soprano}, \textcolor{tab-orange}{alto}, \textcolor{tab-green}{tenor}, \textcolor{tab-red}{bass}.)}
    \end{tabularx}
    \caption[Example of the Bach chorales dataset---\textit{Wer nur den lieben Gott läßt walten}, BWV 434, measures 1--5.]{Example of the Bach chorales dataset---\textit{Wer nur den lieben Gott läßt walten}, BWV 434, measures 1--5. The LSTM model makes two errors for the bass, as indicated by the arrow. The BiLSTM model gives a perfect prediction.}
    \label{arranger:fig:sample_bach}
\end{figure}

%====================
\section{Experiments}
%====================

%----------------------------------
\subsection{Implementation details}
%----------------------------------

We use a batch size of 16, a sequence length of 500 for training and a maximum sequence length of 2000 for validation and testing. We clip the time by 4096 time steps (i.e., roughly 170 quarter notes), the beat by 4096 beats, and durations by 192 time steps (i.e., 8 quarter notes). We randomly transpose the music by -5 to +6 semitones during training for data augmentation. We use the cross entropy loss with the Adam optimizer with $\alpha = 0.001$, $\beta_1 = 0.9$ and $\beta_2 = 0.999$ \citep{kingma2015adam}. We apply dropout \citep{srivastava2014dropout} to prevent overfitting and layer normalization \citep{ba2016layernorm} to speed up the training. All models are implemented in TensorFlow \citep{abadi2016tensorflow} and experiments are run on NVIDIA GeForce RTX 2070s.

%--------------------------------------------------
\subsection{Qualitative results and error analysis}
%--------------------------------------------------

\begin{figure*}
    \small
    \centering
    \def\imgheight{9ex}
    \def\imgwidth{\linewidth}
    \begin{tabularx}{\linewidth}{L{.16\linewidth} X}
        Musical score &\includegraphics[width=\imgwidth]{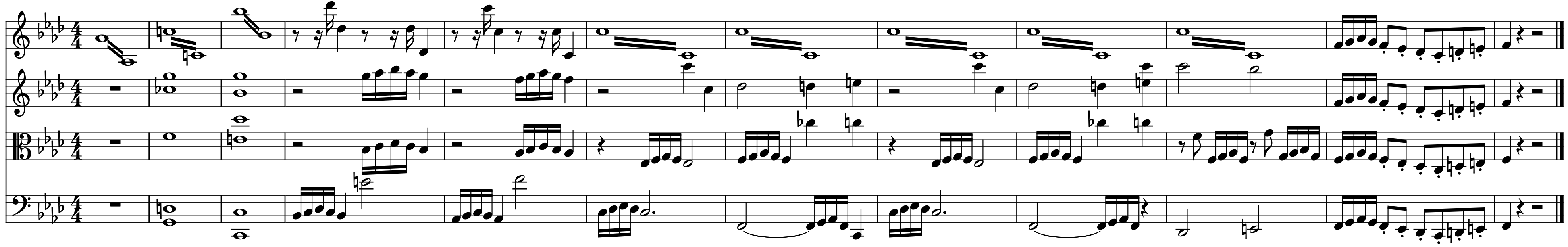}\\
        \cmidrule(lr){1-2}
        Mixture (input) &\includegraphics[width=\imgwidth,height=\imgheight]{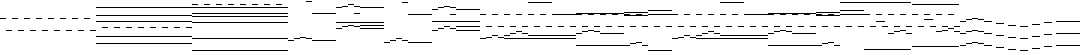}\\
        \cmidrule(lr){1-2}
        Ground truth &\includegraphics[width=\imgwidth,height=\imgheight]{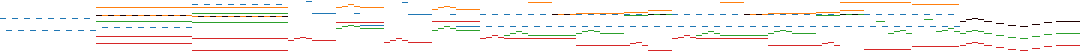}\\
        \cmidrule(lr){1-2}
        Online LSTM\hspace{1ex} prediction &\includegraphics[width=\imgwidth,height=\imgheight]{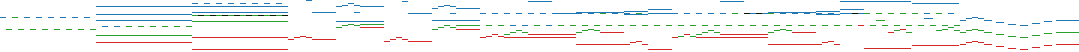}\\
        \cmidrule(lr){1-2}
        Offline BiLSTM prediction &\includegraphics[width=\imgwidth,height=\imgheight]{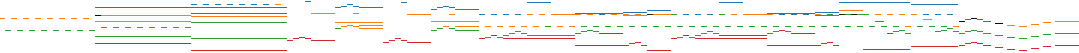}\\
        \multicolumn{2}{r}{\footnotesize (Audio available.\cref{arranger:fn:demo} Colors: \textcolor{tab-blue}{first violin}, \textcolor{tab-orange}{second violin}, \textcolor{tab-green}{viola}, \textcolor{tab-red}{cello}.)}
    \end{tabularx}
    \caption[Hard excerpt in the string quartets dataset---Beethoven's \textit{String Quartet No. 11 in F minor, Op. 95,} movement 1, measures 72--83.]{Hard excerpt in the string quartets dataset---Beethoven's \textit{String Quartet No. 11 in F minor, Op. 95,} movement 1, measures 72--83. The tremolos of the first violin (measures 1--3 and 6--10), the double stops for the second violin, viola and cello (measures 2--3) and the overlapping pitch ranges (measures 2--5) together compose a complex texture. Both models fail to handle the violins and viola properly, especially the second violin.}
    \label{arranger:fig:sample_musicnet}
\end{figure*}

\begin{figure}
    \small
    \centering
    \def\imgheight{7.5ex}
    \def\imgwidth{\linewidth}
    \begin{tabularx}{\linewidth}{L{.16\linewidth} X}
        Ground truth &\includegraphics[width=\imgwidth,height=\imgheight]{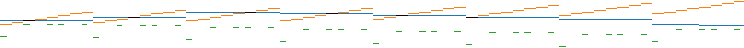}\\
        \cmidrule(lr){1-2}
        Online LSTM\hspace{1ex} prediction &\includegraphics[width=\imgwidth,height=\imgheight]{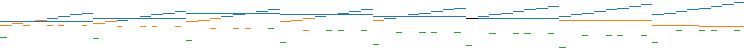}\\
        \cmidrule(lr){1-2}
        Offline BiLSTM prediction &\includegraphics[width=\imgwidth,height=\imgheight]{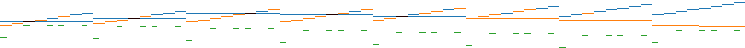}\\
        \multicolumn{2}{r}{\footnotesize (Audio available.\cref{arranger:fn:demo} Colors: \textcolor{tab-blue}{pulse wave I}, \textcolor{tab-orange}{pulse wave II}, \textcolor{tab-green}{triangle wave}.)}
    \end{tabularx}
    \caption[Hard excerpt in the game music dataset---\textit{Theme of Universe} from \textit{Miracle Ropit's Adventure in 2100}.]{Hard excerpt in the game music dataset---\textit{Theme of Universe} from \textit{Miracle Ropit's Adventure in 2100}. Both models perform poorly when there is a sequence of short notes crossing a single long note.}
    \label{arranger:fig:sample_nes}
\end{figure}

\begin{figure}
    \small
    \centering
    \def\imgheight{8ex}
    \def\imgwidth{\linewidth}
    \begin{tabularx}{\linewidth}{L{.16\linewidth} X}
        Ground truth
          &\includegraphics[width=\imgwidth,height=\imgheight]{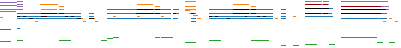}\\
        \cmidrule(lr){1-2}
        Online LSTM\hspace{1ex} prediction
          &\includegraphics[width=\imgwidth,height=\imgheight]{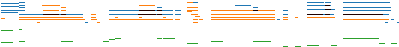}\\
        \cmidrule(lr){1-2}
        Offline BiLSTM prediction
          &\includegraphics[width=\imgwidth,height=\imgheight]{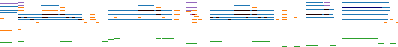}\\
        \multicolumn{2}{r}{\footnotesize (Audio available.\cref{arranger:fn:demo} Colors: \textcolor{tab-blue}{piano}, \textcolor{tab-orange}{guitar}, \textcolor{tab-green}{bass}, \textcolor{tab-red}{strings}, \textcolor{tab-purple}{brass}.)}
    \end{tabularx}
    \caption[Hard excerpt in the pop music dataset---\textit{Blame It On the Boogie} by The Jacksons.]{Hard excerpt in the pop music dataset---\textit{Blame It On the Boogie} by The Jacksons. The BiLSTM model correctly identify and separate the overlapping guitar melody and piano chords, while the LSTM model fails in this case.}
    \label{arranger:fig:sample_lmd}
\end{figure}

We present in \cref{arranger:fig:sample_bach,arranger:fig:sample_musicnet,arranger:fig:sample_nes,arranger:fig:sample_lmd} several examples in the four datasets. Some representative cases include overlapping pitch ranges or chords for two polyphonic instruments (see \cref{arranger:fig:sample_musicnet,arranger:fig:sample_lmd}), overlapping melodies and chords (see \cref{arranger:fig:sample_lmd}) and a sequence of short notes crossing a single long note (see \cref{arranger:fig:sample_nes}).

%--------------------------------
\subsection{Quantitative results}
%--------------------------------

We conduct an extensive empirical evaluation over different dataset and models in different settings. We present the results in \cref{arranger:tab:exp}.\footnote{Due to high computation cost, we report the oracle cases for the zone-based algorithm and MLP model on a subset of $100$ test samples, and omit the oracle case of the zone-based algorithm for the pop music dataset.} First, we notice the improved performance for the oracle cases on the MLP baseline. The large gap of performance is possibly because it predicts each note independently and the errors can propagate over time. This emphasizes the need to incorporate sequential models for this task. Moreover, the BiLSTM model outperforms its LSTM counterpart for most cases. This is reasonable as the BiLSTM model has access to the future information, which could, for example, help identify the direction of an arpeggio. Further, the LSTM and BiLSTM models outperform their Transformer counterparts---Transformer-Dec and Transformer-Enc, respectively---across all settings. However, the Transformer models benefits from faster inference speed at test time as compared to the LSTM models. Finally, we notice that the proposed models perform relatively poorly on the string quartets and game music datasets, possibly because the two violins in the string quartets dataset and the two pulse waves in the game music dataset are sometimes used interchangeably. We examine the use of pitch hints to help the models in the following section.

\begin{table}
    \small
    \centering
    \caption[Comparison of our proposed models and baseline algorithms.]{Comparison of our proposed models and baseline algorithms. Performance is measured in accuracy (\%).}
    \label{arranger:tab:exp}
    \newcommand{\colsep}{\hspace{1.25em}}
    \begin{tabular}{ll@{\colsep}l@{\colsep}l@{\colsep}l}
        \toprule
        Model &Bach &String &Game &Pop\\
        \midrule
        \textbf{Online models}\\
        Zone-based                                   &73.14 &58.85 &43.67 &57.07\\
        MLP \citep{devalk2019deepseparation}          &81.63 &29.85 &43.08$^*$ &33.50$^*$\\
        LSTM                                         &\textbf{93.02} &\textbf{67.43} &\textbf{50.22} &\textbf{74.14}\\
        Transformer-Dec                              &91.51 &57.03 &45.82 &62.14\\
        \cmidrule[.25pt](lr){1-5}
        Zone-based (oracle)                          &78.33 &66.89 &79.54$^*$ &$^\dag$\\
        MLP \citep{devalk2019deepseparation} (oracle) &97.59 &58.16 &65.30 &44.62\\
        \midrule
        \textbf{Offline models}\\
        BiLSTM                                       &\textbf{97.13} &\textbf{74.3}8 &\textbf{52.93} &\textbf{77.23}\\
        Transformer-Enc                              &96.81 &58.86 &49.14 &66.57\\
        \midrule
        \multicolumn{5}{l}{\textbf{Online models (+entry hints)}}\\
        Closest-pitch                                &68.87 &50.69 &57.14 &47.45\\
        Closest-pitch (mono)                         &89.76 &42.82 &49.91 &32.28\\
        LSTM                                         &\textbf{92.70} &\textbf{62.64} &\textbf{62.11} &\textbf{74.19}\\
        Transformer-Dec                              &91.17 &62.12 &56.73 &67.19\\
        \midrule
        \multicolumn{5}{l}{\textbf{Offline models (+entry hints)}}\\
        BiLSTM                                       &\textbf{97.39} &\textbf{71.51} &\textbf{64.79} &\textbf{75.59}\\
        Transformer-Enc                              &93.81 &56.72 &54.67 &67.23\\
        \bottomrule
        \multicolumn{5}{l}{\footnotesize $^*$Reported on a subset of $100$ test samples due to high computation cost.}\\
        \multicolumn{5}{l}{\footnotesize $^\dag$Omitted due to high computation cost.}
    \end{tabular}
    % \\[1ex]
    % \raggedright{\footnotesize $^*$Reported on a subset of $100$ test samples due to high computation cost.\quad$^\dag$Omitted due to high computation cost.}
\end{table}

\begin{table}
    \small
    \centering
    \caption[Effects of input features to the online LSTM model.]{Effects of input features to the online LSTM model. Performance is measured in accuracy (\%). Abbreviations: `Emb'---pitch, beat and position embedding, `Dur'---duration, `EH'---entry hints, `PH'---pitch hints.}
    \label{arranger:tab:exp_features}
    \begin{tabular}{llllllll}
        \toprule
        Emb        &Dur        &EH          &PH          &Bach &String &Game &Pop\\
        \midrule
                   &           &            &            &92.10 &37.29 &43.89 &58.78\\
        \checkmark &           &            &            &93.02 &67.43 &50.22 &74.14\\
        \checkmark &\checkmark &            &            &\textbf{96.17} &66.96 &51.38 &\textbf{78.17}\\
        \checkmark &           &\checkmark  &            &92.70 &62.64 &62.11 &74.19\\
        \checkmark &\checkmark &\checkmark  &            &95.95 &68.17 &63.35 &74.74\\
        \checkmark &           &            &\checkmark  &92.87 &\textbf{70.20} &\textbf{67.45} &75.89\\
        \bottomrule
    \end{tabular}
\end{table}

%------------------------------------
\subsection{Effect of input features}
%------------------------------------

In order to compare the effectiveness of different input features, we also report in \cref{arranger:tab:exp_features} the performance for the LSTM model with different input features. First of all, pitch, note and beat embedding leads to improvements on all datasets, especially significant on the string quartet (30\% gain) and pop music (15\% gain) datasets. Second, entry hints improve the performance by 10\% for the game music dataset, which is possibly because it helps disambiguate the two interchangeable pulse wave tracks. Interestingly, they have negative impacts on the Bach chorales and pop music datasets. Third, duration inputs are always helpful and help achieve the highest accuracy on the Bach chorales and pop music datasets. For example, durations would be critical in distinguishing the overlapping guitar melody and piano chords in the example shown in \cref{arranger:fig:sample_lmd}. Last, pitch hints improve the performance for all datasets but Bach chorales, possibly because the vocal ranges for SATB are strict in chorales. Pitch hints help achieve the highest accuracies for the string quartets and game music datasets as they help disambiguate interchangeable tracks.

%------------------------------------
\subsection{Effects of time encoding}
%------------------------------------

In this experiment, we examine the effects of time encoding. In particular, we consider four variants---(1) raw time as a number, (2) raw beat and position as two numbers, (3) time embedding and (4) beat and position embedding (see \cref{arranger:sec:model} for the definition of beat and position). We report in \cref{arranger:tab:ablation} the results and we can see that using raw time gives the worst performance. Interestingly, the other three encoding strategies achieve comparable performance.

\begin{table}
    \small
    \centering
    \caption[Comparisons of time encoding and data augmentation strategies for the online LSTM model.]{Comparisons of time encoding and data augmentation strategies for the online LSTM model. Performance is measured in accuracy (\%).}
    \label{arranger:tab:ablation}
    \begin{tabular}{lllll}
        \toprule
        Strategy &Bach &String &Game &Pop\\
        \midrule
        \textbf{Time encoding}\\
        Raw time                    &91.97 &37.26 &44.10 &37.92\\
        Raw beat and position       &\textbf{93.13} &66.72 &48.60 &68.42\\
        Time embedding              &92.21 &\textbf{68.31} &49.32 &70.64\\ 
        Beat and position emb.      &93.02 &67.43 &\textbf{50.22} &\textbf{74.14}\\
        \midrule
        \textbf{Data augmentation}\\
        No augmentation             &\textbf{93.03} &\textbf{69.36} &49.03 &70.73\\
        Light augmentation          &92.85 &68.66 &46.38 &71.10\\
        Strong augmentation         &93.02 &67.43 &\textbf{50.22} &\textbf{74.14}\\
        \bottomrule
    \end{tabular}
\end{table}

%----------------------------------------
\subsection{Effects of data augmentation}
%----------------------------------------

In this experiment, we compare the following three strategies of data augmentation---(1) no augmentation, (2) \emph{light augmentation}, where each song is randomly transposed by -1 to +1 semitone during training and (3) \emph{strong augmentation}, where each song is randomly transposed by -5 to +6 semitones during training. We report in \cref{arranger:tab:ablation} the results. We can see that data augmentation is generally harmful for the Bach chorales and string quartets datasets, possibly because classical music has strict rules on the pitch ranges of voices and instruments. However, for game and pop music datasets, where rules on keys and pitch ranges in classical music are loosened, the models yield better performance with proper data augmentation.

\begin{figure*}
    \small
    \centering
    \def\imgheight{8ex}
    \def\imgwidth{\linewidth}
    \begin{tabularx}{\linewidth}{L{.05\linewidth}@{}X}
        (a) &\includegraphics[width=\imgwidth,height=\imgheight]{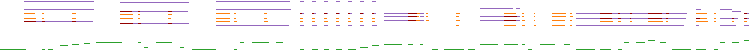}\\
        \cmidrule(lr){1-2}
        (b) &\includegraphics[width=\imgwidth,height=\imgheight]{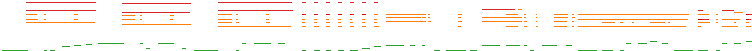}\\
        \cmidrule(lr){1-2}
        (c) &\includegraphics[width=\imgwidth,height=\imgheight]{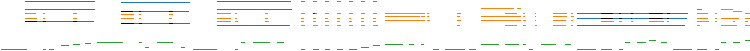}\\
        \multicolumn{2}{r}{\footnotesize (Audio available.\cref{arranger:fn:demo} Colors: \textcolor{tab-blue}{piano}, \textcolor{tab-orange}{guitar}, \textcolor{tab-green}{bass}, \textcolor{tab-red}{strings}, \textcolor{tab-purple}{brass}.)}
    \end{tabularx}
    \caption[\textit{Quando Quando Quando} by Tony Renis---(a) original instrumentation and the versions produced by (b) the online LSTM model without entry hints and (b) the offline BiLSTM model with entry hints.]{\textit{Quando Quando Quando} by Tony Renis---(a) original instrumentation and the versions produced by (b) the online LSTM model without entry hints and (b) the offline BiLSTM model with entry hints. The LSTM model assigns the chords to the guitar, the most common instrument in the pop music dataset except the high pitches, which are assigned to the strings. The BiLSTM model is able to separate the long chords from the short ones and assigns the former to the piano.}
    \label{arranger:fig:sample_instrumentation}
\end{figure*}

%===================
\section{Discussion}
%===================

In \cref{arranger:fig:sample_instrumentation}, we depict the original instrumentation of the song \textit{Quando Quando Quando} alongside the instrumentations generated by our best performing models for both the online and offline settings. While neither model produces an instrumentation identical to that of the original, both produce instrumentations that ``cluster'' notes similarly to the original and are reasonable rearrangements of the song. This indicates a fundamental ambiguity of the task, though we note that such ambiguity is less present in some genres than others---our models are able to achieve high accuracy on the Bach chorales dataset despite its small size. However, for larger and more diverse datasets (e.g., the pop music dataset), accuracy might not be the best metric for measuring the performance of the models, and we plan to include human evaluations in future work.

One limitation of this work lies in the generalizability to real keyboard music since the downmixed music might not be playable on a keyboard, e.g., having more than ten concurrent notes or impossible fingering. Moreover, we did not use the MIDI velocity information in our models, and it could provide an additional signal for separation.

Finally, in addition to its immediate musical applications, we believe that our proposed part separation task may be useful for large-scale pre-training of symbolic music models. Pre-training music generation models on large, heterogeneous music corpora has already been observed to improve performance \citep{donahue2019lakhnes,hung2019improving}. Given that our proposed task represents an additional source of musical knowledge supervision, we speculate that additionally pre-training on this task could improve performance for many downstream tasks, e.g., genre classification and melody extraction.

%===================
\section{Conclusion}
%===================

In this paper, we have proposed a new task of part separation in multitrack music and examined its feasibility under both the online and offline settings. Through a comprehensive empirical evaluation over four diverse datasets, we showed the effectiveness of our proposed models against various baselines. We also presented promising results for applying part separation models to automatic instrumentation. Moreover, we discussed the fundamental ambiguity and limitations of the task and future research directions.

\secbreak

\begin{adjustwidth}{.5in}{.5in}
    \itshape
    \hspace{\parindent}
    This chapter, in full, is a reprint of the material as it appears in ``Towards Automatic Instrumentation by Learning to Separate Parts in Symbolic Multitrack Music'' by Hao-Wen Dong, Chris Donahue, Taylor Berg-Kirkpatrick and Julian McAuley, which was published in the Proceedings of the International Society for Music Information Retrieval Conference (ISMIR) in 2021. The dissertation author was the primary investigator and author of this paper.
\end{adjustwidth}

\graphicspath{{chapters/deepperformer/figs/}}
\chapter{Deep Performer: Score-to-Audio Music Performance Synthesis}
\label{chap:deepperformer}

\begin{abstract}
Music performance synthesis aims to synthesize a musical score into a natural performance. In this paper, we borrow recent advances in text-to-speech synthesis and present the Deep Performer---a novel system for score-to-audio music performance synthesis. Unlike speech, music often contains polyphony and long notes. Hence, we propose two new techniques for handling polyphonic inputs and providing a fine-grained conditioning in a transformer encoder-decoder model. To train our proposed system, we present a new violin dataset consisting of paired recordings and scores along with estimated alignments between them. We show that our proposed model can synthesize music with clear polyphony and harmonic structures. In a listening test, we achieve competitive quality against the baseline model, a conditional generative audio model, in terms of pitch accuracy, timbre and noise level. Moreover, our proposed model significantly outperforms the baseline on an existing piano dataset in overall quality.
\end{abstract}

%=====================
\section{Introduction}
%=====================
\label{deepperformer:sec:intro}

\begin{figure}
    \centering
    \includegraphics[width=.8\linewidth]{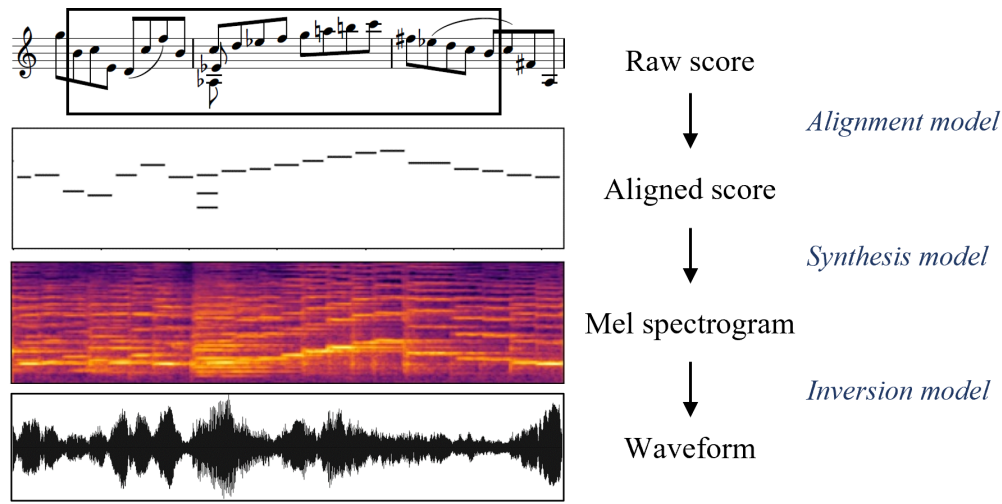}
    \caption{An overview of the proposed three-stage pipeline for score-to-audio music performance synthesis.}
    \label{deepperformer:fig:overview}
\end{figure}

Music synthesis is a complex process that involves both the physics behind a musical instrument and the art of music performance. It remains challenging for a machine to synthesize a natural performance for several reasons. First, it requires a computational model for interpreting and phrasing a musical score. Second, it requires either an explicit or implicit model of the physics and acoustics by which a musical instrument sounds. Third, it requires an understanding of different playing techniques and styles for a musical instrument. While most existing systems only address one of these three challenges at a time, we aim to tackle all these challenges with a data-driven approach using machine learning in this work. We present the Deep Performer---a novel three-stage system for score-to-audio music synthesis, as illustrated in \cref{deepperformer:fig:overview}. 

Prior work has studied music synthesis via various approaches. One line of research focuses on generating realistic samples of musical notes \citep{engel2017nsynth,defossez2018sing,engel2019gansynth}, while in this work we aim to generate the full performance. Some approach music synthesis by conditioning generative audio models with aligned piano rolls \citep{manzelli2018wavenet,hawthorne2019maestro}, which we will include as the baseline model in our experiments. Others study synthesizing audio from the fundamental frequency (F0) contour and loudness curve extracted from a recording \citep{engel2020ddsp,hayes2021waveshaping}, or from lyrics and demo singing audio \citep{ren2020deepsinger}. On the other hand, some use neural networks to generate expressive timing and dynamics from raw scores \citep{oore2020performancernn}. Many have also studied inverting mel spectrograms back to waveforms \citep{shen2018tacotron2,prenger2019waveglow,kumar2019melgan,kong2020hifigan}, including Hifi-GAN \citep{kong2020hifigan}, which we will use as the inversion model in our proposed system. To the best of our knowledge, prior work on deep neural network based music synthesis either requires an input with expressive timing \citep{manzelli2018wavenet,wang2019performancenet,hawthorne2019maestro,schimbinschi2019synthnet,kim2019mel2mel,ren2020deepsinger,engel2020ddsp,hayes2021waveshaping} or allows only monophonic (i.e., one pitch at a time) inputs \citep{ren2020deepsinger,engel2020ddsp,anonymous2022mididdsp}. Our proposed system represents the first that allows unaligned, polyphonic scores as inputs.

In light of the similarity between text-to-speech (TTS) and score-to-audio synthesis, we borrow recent advances from TTS synthesis \citep{tan2021ttssurvey} to music synthesis and propose a three-stage system for score-to-audio music synthesis. Despite the similarity, music synthesis differs from speech synthesis in that music often contains polyphony, and that long notes are common in music. In order to handle polyphonic music, we propose a new \textit{polyphonic mixer} for aligning the encoder and decoder in a transformer encoder-decoder network~ \citep{vaswani2017transformer,wang2017tacotron}. To provide a fine-grained conditioning to the model, we propose a new \textit{note-wise positional encoding} so that the model can learn to behave differently at the beginning, middle and end of a note. Due to the lack of a proper dataset for training a score-to-audio music synthesis model, we collect and release a new dataset of 6.5 hours of high-quality violin recordings along with their scores and estimated alignments. Through our experiments, we show the effectiveness of our proposed system both qualitatively and quantitatively. Finally, we conduct a subjective listening test to compare our proposed model against a baseline model that uses Hifi-GAN \citep{kong2020hifigan} to synthesize the waveform directly from an aligned piano roll. Audio samples can be found on our project website.\footnote{\url{https://salu133445.github.io/deepperformer/}\label{deepperformer:fn:website}}

%================
\section{Methods}
%================
\label{deepperformer:sec:methods}

We illustrate in \cref{deepperformer:fig:overview} the proposed three-stage system for score-to-audio music synthesis, which consists of the following three components: (1) an \textit{alignment model} that predicts the expressive timing for each note from a musical score, (2) a \textit{synthesis model} that synthesizes the mel spectrogram from the aligned score, and (3) an \textit{inversion model} that generates the audio waveform given the synthesized mel spectrogram.

\subsection{Alignment model}
%---------------------------
\label{deepperformer:sec:alignment-model}

The alignment model consists of a transformer encoder that takes as inputs a sequence of notes and the tempo, followed by a fully-connected layer that outputs the onset and duration of each note. The input score uses metric time with a musically-meaningful unit, e.g., quarter notes, while the output alignment is in the unit of frames. Each note is specified by its pitch, onset, duration and (optional) velocity. In addition, we provide the performer IDs so that the model can learn the different playing styles of performers. The alignment model is trained to minimize the mean squared error (MSE) between the ground truth and predicted onsets and durations, in frames.

\subsection{Synthesis model}
%---------------------------
\label{deepperformer:sec:synthesis-model}

\begin{figure}
    \centering
    \includegraphics[width=.8\linewidth]{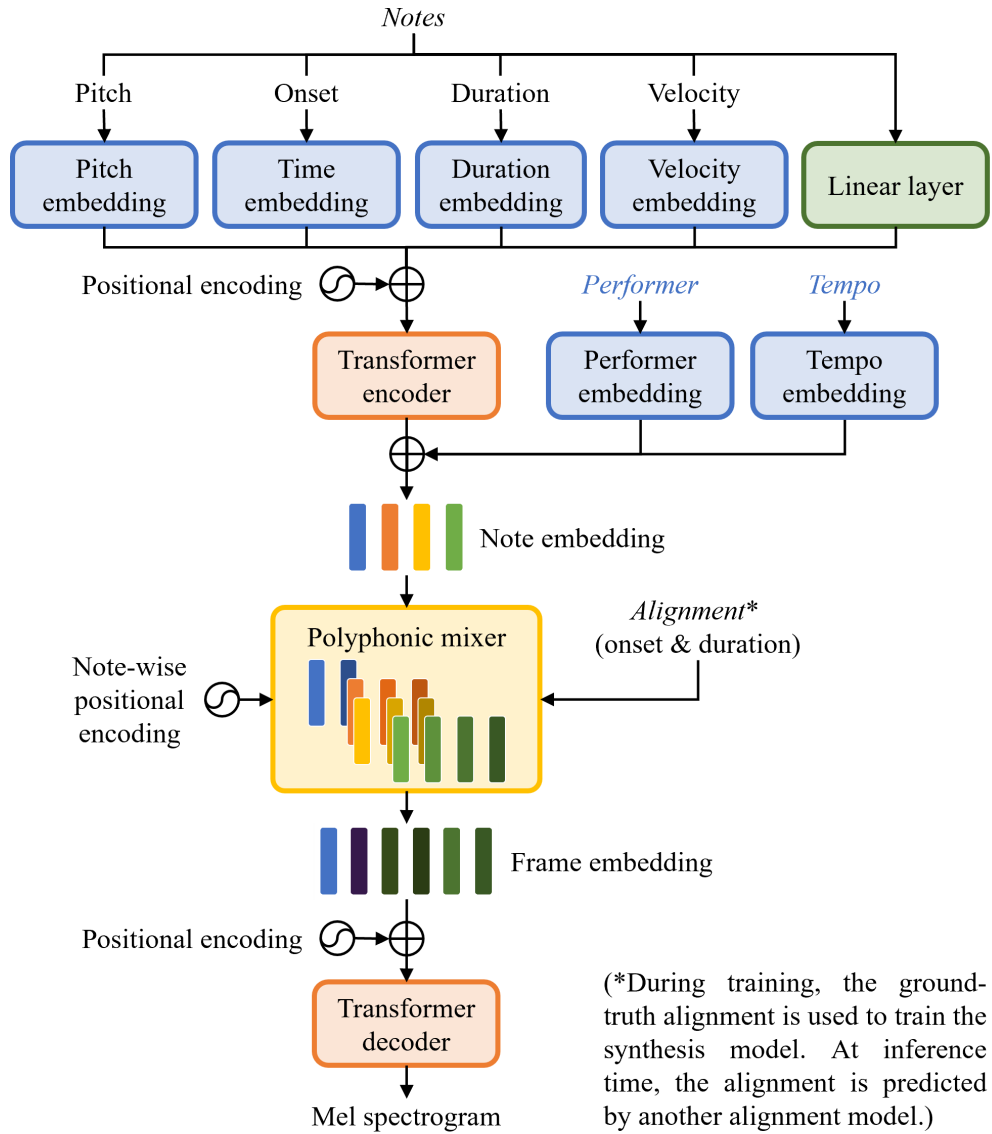}
    \caption{An illustration of the proposed synthesis model.}
    \label{deepperformer:fig:model}
\end{figure}

Given the similarity between TTS and score-to-audio synthesis, we propose a transformer encoder-decoder model for our synthesis model based on \citep{ren2019fastspeech}. In \citep{ren2019fastspeech}, each text embedding produced by the encoder is expanded multiple times according to its duration, and then the expanded text embeddings are concatenated to obtain the frame embeddings to be fed to the decoder. This is called the state expansion mechanism \citep{ren2019fastspeech,yu2020durian}. However, unlike speech, music often contains polyphony. In order to handle polyphonic inputs, we propose the \textit{polyphonic mixer}. As illustrated in \cref{deepperformer:fig:model}, the encoder first encodes the input notes into a sequence of note embeddings. Then, the polyphonic mixer mixes the note embeddings into a sequence of frame embeddings by summing up the note embeddings for the same frame according to their onsets and durations. Finally, the decoder decodes the frame embeddings into a sequence of mel spectrogram frames.

In the state expansion mechanism \citep{ren2019fastspeech,yu2020durian}, the output vectors remain constant for the duration of a note, and the positional information within each note is missing. However, we argue that such note-wise positional information is critical for the model to behave differently at the beginning, middle and end of a note. Hence, we propose the \textit{note-wise positional encoding} to provide a fine-grained conditioning to the decoder. Mathematically, let $p \in [0, 1]$ be the relative position within a note. For a note embedding $\bm{\mathrm{v}}_{\text{note}}$, we have the corresponding frame embedding at position $p$ as $\bm{\mathrm{v}}_{\text{frame}} = (1 + p \bm{\mathrm{w}}) \odot \bm{\mathrm{v}}_{\text{note}}$, where $\bm{\mathrm{w}}$ is a learnable vector initialized to small random numbers so that $\bm{\mathrm{v}}_{\text{frame}} \approx \bm{\mathrm{v}}_{\text{note}}$ initially. The synthesis model is trained to minimize the MSE between the synthesized mel spectrograms and the ground truth, in log scale.

\subsection{Inversion model}
%---------------------------
\label{deepperformer:sec:inversion-model}

Prior work has studied various approaches for synthesizing waveforms from mel spectrograms \citep{shen2018tacotron2,prenger2019waveglow,kumar2019melgan,kong2020hifigan}. In this work, we adopt the state-of-the-art Hifi-GAN model \citep{kong2020hifigan} as our inversion model. We note that the proposed three-stage pipeline allows us to use different datasets for training the models. For example, training the inversion model does not require aligned data and thus it can be trained on a larger dataset as unaligned data are relatively easier to acquire.

%=============
\section{Data}
%=============
\label{deepperformer:sec:data}

\begin{figure}
    \centering
    \includegraphics[width=\linewidth,height=1.5in]{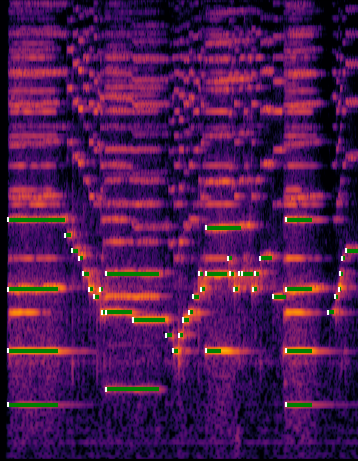}
    \caption{An example of the constant-Q spectrogram of the first 20 seconds of a violin recording and the estimated onsets (white dots) and durations (green lines).}
    \label{deepperformer:fig:alignment}
\end{figure}

Due to the lack of a dataset that provides paired audios and scores with fine alignments for training our proposed system, we compile a new dataset of 6.5 hours of professional violin recordings along with their scores and estimated alignments. For copyright concern, we choose Bach's sonatas and partitas for solo violin (BWV 1001--1006) for the ease to acquire high-quality public recordings from the web. The dataset consists of performances by 17 violinists recorded in various recording setups. To acquire the alignment between a recording and its score, we synthesize the scores using FluidSynth \citep{fluidsynth}, an open-source software synthesizer, with MuseScore General SoundFont \citep{musescoresoundfont} and perform dynamic time warping on the constant-Q spectrogram of the synthesized audio and that of the recording. We present in \cref{deepperformer:fig:alignment} an example of the dataset and its estimated alignment. To facilitate future research on score-to-audio music synthesis, we release the dataset and the source code for the alignment process to the public.\footnote{\url{https://salu133445.github.io/bach-violin-dataset/}} As discussed in \cref{deepperformer:sec:inversion-model}, the inversion model does not require aligned data for training, and thus we also collect an internal dataset of 156 hours of commercial recordings to train the inversion model. Apart from violin, we also consider the MAESTRO dataset \citep{hawthorne2019maestro}, which contains 200 hours of piano recordings with finely-aligned MIDI recordings for 10 competition years of the International Piano-e-Competition \citep{piano-e-competition}. However, since it does not provide the raw scores, we can only train the synthesis and inversion models on this dataset.

%===============================
\section{Experiments \& Results}
%===============================
\label{deepperformer:sec:exp}

\subsection{Implementation details}
%----------------------------------

We use 3 transformer layers in the encoder for the alignment model. The synthesis model shares the same encoder architecture as the alignment model and has 6 transformer layers in the decoder. We use 128 dimensions for all embeddings. For the inversion model, we use the same network architecture as the Hifi-GAN v2 model in \citep{kong2020hifigan}. We use velocity information only for the piano dataset as it is only available in this dataset. Since performer information is unavailable for the piano dataset, we use the competition years as the performer IDs. We use a temporal resolution of 24 time steps per quarter note for the scores. We downsample the audios to 16 kHz mono and use a hop size of 256 in spectrogram computation, i.e., a temporal resolution of 16 ms. The audios are sliced into 5-second clips for training, where 10\% of them are reserved for validation purpose. We use the Adam optimizer \citep{kingma2015adam} with a batch size of 16. Unlike \citep{ren2019fastspeech}, we train the alignment and synthesis models separately as we find that joint training hinders convergence. We train the alignment model for 10K steps and all the synthesis models for 100K (violin) and 250K (piano) steps. For each dataset, the inversion model is trained for 1M steps and shared by different synthesis models. We base our implementation on the code kindly released in \citep{chien2021tts,kong2020hifigan}. We use pretty\_midi \citep{raffel2014prettymidi} and MusPy \citep{dong2020muspy} to process the scores.

\begin{figure}
    \centering
    \small
    \begin{tabularx}{\linewidth}{m{.02\linewidth} X}
        (a) &\includegraphics[width=\linewidth,height=.5in]{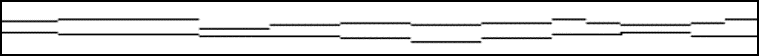}\\
        (b) &\includegraphics[width=\linewidth,height=.5in]{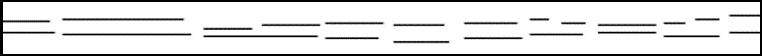}\\
        (c) &\includegraphics[width=\linewidth]{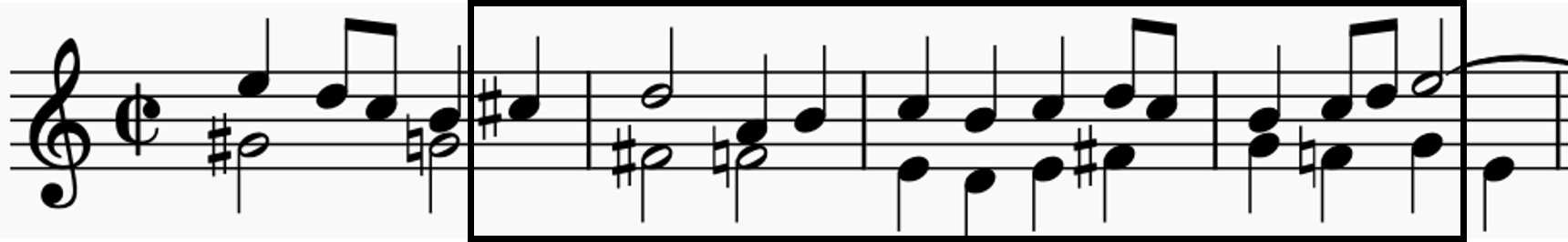}
    \end{tabularx}
    \caption{Examples of the alignments predicted by (a) the constant-tempo baseline model and (b) Deep Performer, our proposed model. (c) shows the input score.}
    \label{deepperformer:fig:results-alignment}
\end{figure}

\begin{figure}
    \small
    \centering
    \begin{tabularx}{\linewidth}{m{.02\linewidth} X}
        (a) &\includegraphics[width=\linewidth,height=1in]{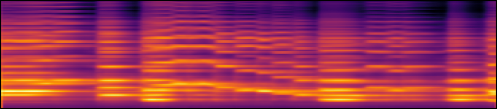}\\[-1ex]
        (b) &\includegraphics[width=\linewidth,height=.5in]{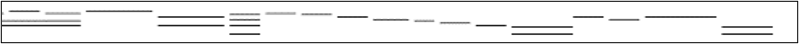}\\[2ex]
        (c) &\includegraphics[width=\linewidth,height=1in]{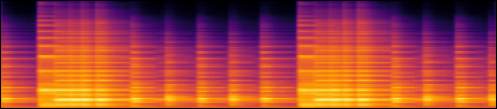}\\[-1ex]
        (d) &\includegraphics[width=\linewidth,height=.5in]{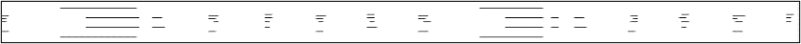}
    \end{tabularx}
    \caption{Examples of the mel spectrograms, in log scale, synthesized by our proposed model for (a) violin and (c) piano. (b) and (d) show the input scores for (a) and (c), respectively.}
    \label{deepperformer:fig:results-synth-poly}
\end{figure}

\subsection{Qualitative and quantitative results}
%------------------------------------------------
\label{deepperformer:sec:results}

\begin{figure}
    \centering
    \small
    \begin{tabularx}{\linewidth}{cX}
        \makecell[c]{(a)} &\makecell[c]{\includegraphics[width=\linewidth,height=1in]{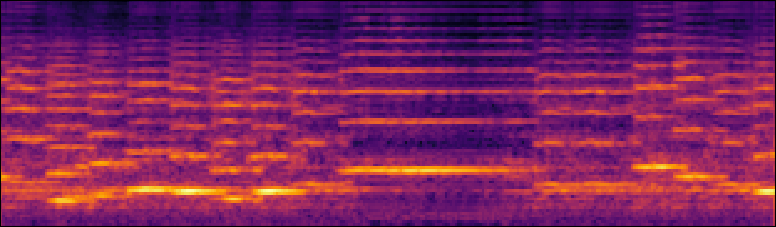}}\\[2ex]
        \makecell[c]{(b)} &\makecell[c]{\includegraphics[width=\linewidth,height=1in]{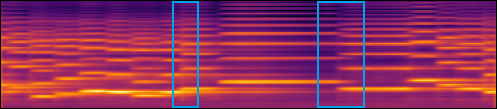}}\\[-1ex]
        \makecell[c]{(c)} &\makecell[c]{\includegraphics[width=\linewidth,height=.5in]{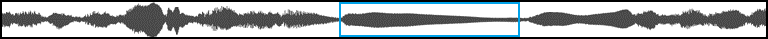}}\\[4ex]
        \makecell[c]{(d)} &\makecell[c]{\includegraphics[width=\linewidth,height=1in]{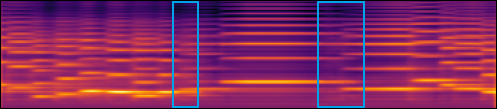}}\\[-1ex]
        \makecell[c]{(e)} &\makecell[c]{\includegraphics[width=\linewidth,height=.5in]{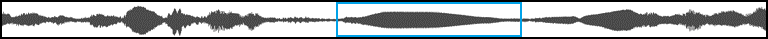}}\\[4ex]
        \makecell[c]{(f)} &\makecell[c]{\includegraphics[width=\linewidth,height=.5in]{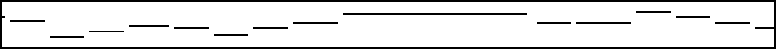}}
    \end{tabularx}
    \caption{Examples of the mel spectrograms, in log scale, synthesized by (a) the baseline model, (b) our proposed synthesis model, and (d) our proposed synthesis model without the note-wise positional encoding. (c) and (e) show the waveforms for (b) and (d), respectively. (f) shows the input score.}
    \label{deepperformer:fig:results-comparison}
\end{figure}

We show in \cref{deepperformer:fig:results-alignment} an example of the alignment predicted by our proposed alignment model alongside that generated by assuming a constant tempo. We can see that our proposed model is able to predict realistic timing and insert rests between notes. To showcase the effectiveness of the proposed polyphonic mixer, we present in \cref{deepperformer:fig:results-synth-poly} examples of the synthesized mel spectrograms for two polyphonic scores, where we can observe clear harmonic structures and polyphony.

Next, we compare our proposed synthesis model against a baseline model that uses a Hifi-GAN \citep{kong2020hifigan} to synthesize the waveform directly from an aligned piano roll. For a fair comparison, we condition this model with the performer IDs and provide a \textit{position roll} that encodes the note-wise position information. (A position roll is similar to a piano roll, but the values decrease linearly from $1$ to $0$, from the beginning of a note to its end.) As can be seen from \cref{deepperformer:fig:results-comparison}(a) and (b), our proposed model produces smoother contours and clearer harmonic structures, especially on the high frequency end, while the baseline model generates sharper yet noisier results. \cref{deepperformer:tab:results-loss} shows the final MSE between the synthesized mel spectrograms and the ground truths. We can see that our proposed model achieves a lower MSE than the baseline model on both datasets. Finally, due to the reduced temporal resolution of a mel spectrogram compared to that of a waveform, our proposed model is faster in training than the baseline model. Audio samples can be found on our project website.\cref{deepperformer:fn:website}

\begin{table}
    \centering
    \caption{Comparisons of the final MSE between the synthesized mel spectrograms and the ground truths, in log scales.}
    \label{deepperformer:tab:results-loss}
    \small
    \begin{tabular}{lll}
        \toprule
                                                          &Violin &Piano\\
        \midrule
        Hifi-GAN baseline                                 &0.892  &0.722\\
        Deep Performer (ours)                             &0.700  &0.436\\
        ~~-~without note-wise positional encoding       &0.700  &0.433\\
        ~~-~without performer embedding                 &1.030  &0.523\\
        ~~-~without encoder (using piano roll input)    &0.844  &0.621\\
        \bottomrule
    \end{tabular}
\end{table}

\begin{table*}
    \footnotesize
    \centering
    \caption[Results of the subjective listening test.]{Results of the subjective listening test. The mean opinion scores (MOS) and 95\% confidence intervals are reported.}
    \label{deepperformer:tab:results}
    \begin{tabular}{l@{~~~}l@{~~~}l@{~~~}l@{~~~}l@{~~~}l}
        \toprule
                                                          &Violin &&& &Piano\\
        \cmidrule(r){2-5} \cmidrule(r){6-6}
                                                          &Pitch accuracy  &Timbre          &Noise level     &Overall         &Overall\\
        \midrule
        Hifi-GAN baseline                                 &4.02 $\pm$ 0.31 &3.13 $\pm$ 0.26 &2.51 $\pm$ 0.29 &2.57 $\pm$ 0.22 &1.49 $\pm$ 0.17\\
        Deep Performer (ours)                             &4.22 $\pm$ 0.30 &3.26 $\pm$ 0.30 &2.67 $\pm$ 0.31 &2.58 $\pm$ 0.21 &2.17 $\pm$ 0.24\\
        ~~-~without note-wise positional encoding       &4.13 $\pm$ 0.29 &3.24 $\pm$ 0.27 &2.52 $\pm$ 0.29 &2.61 $\pm$ 0.23 &2.37 $\pm$ 0.23\\
        ~~-~without performer embedding                 &3.05 $\pm$ 0.52 &2.54 $\pm$ 0.42 &2.04 $\pm$ 0.31 &2.01 $\pm$ 0.25 &2.26 $\pm$ 0.25\\
        ~~-~without encoder (using piano roll input)    &4.30 $\pm$ 0.36 &2.91 $\pm$ 0.28 &2.39 $\pm$ 0.28 &2.22 $\pm$ 0.18 &1.43 $\pm$ 0.16\\
        \bottomrule
    \end{tabular}
\end{table*}

\subsection{Subjective listening test}
%-------------------------------------

To further evaluate our proposed system, we conduct a subjective listening test with 15 participants recruited from our social networks, where 14 of them plays a musical instrument. We randomly choose 5 musical scores from each dataset and synthesize them with different models. The participants are instructed to rate the synthesized audios in a 5-point Likert scale in terms of pitch accuracy, timbre and noise level as well as the overall quality. We report the results in \cref{deepperformer:tab:results}. We can see that our proposed model significantly outperforms the baseline model on the piano dataset and achieves comparable performance to the baseline on the violin dataset.

\subsection{Ablation study}
%--------------------------

To measure the contributions of different components of the proposed model, we consider three ablated versions of our model. The first removes the note-wise positional encoding. The second removes the performer embedding. The third removes the encoder and uses piano rolls and position rolls (see \cref{deepperformer:sec:results}) as the inputs to the decoder, while keeping the performer embedding. As we can see from \cref{deepperformer:fig:results-comparison}(b)--(e), note-wise positional encoding help the model produce clearer note transitions and a more realistic waveform envelope (see the highlighted regions). We also report in \cref{deepperformer:tab:results-loss,deepperformer:tab:results} the results for these ablated models. We can see that the performer embedding significantly improves the quality across all criteria. While we show above the effectiveness of the note-wise positional encoding, its impact does not reach statistical significance in our subjective listening test, possibly overshadowed by the artifacts produced by the models. Finally, including an encoder network improves the quality significantly, suggesting that the encoder can learn a more effective representation of the score as compared to the piano roll representation.

%===================
\section{Conclusion}
%===================
\label{deepperformer:sec:conclusion}

We presented a novel three-stage system for synthesizing natural music performance from unaligned musical scores. We proposed the polyphonic mixer for aligning the encoder and decoder with polyphonic inputs. In addition, we also proposed the note-wise positional encoding for providing a fined-grained conditioning to the synthesis model. Through the subjective listening test, we show that our proposed model significantly outperforms the baseline model on the piano dataset and achieves competitive quality against the baseline on the violin dataset. For future work, we plan to utilize the articulation marks and ornaments on scores to better model playing techniques \citep{yang2016violin,shih2017violin}, disentangle the timbre from room acoustics to enhance controllability \citep{engel2020ddsp}, and incorporate adversarial losses \citep{isola2017pix2pix,yang2021ganspeech} to improve the sharpness of the results.

\secbreak

\begin{adjustwidth}{.5in}{.5in}
    \itshape
    \hspace{\parindent}
    This chapter, in full, is a reprint of the material as it appears in ``Deep Performer: Score-to-Audio Music Performance Synthesis'' by Hao-Wen Dong, Cong Zhou, Taylor Berg-Kirkpatrick and Julian McAuley, which was published in the Proceedings of the IEEE International Conference on Acoustics, Speech and Signal Processing (ICASSP) in 2022. The dissertation author was the primary investigator and author of this paper.
\end{adjustwidth}

\clearpage

\begin{subappendices}

%==============================
\section{Preprocessing details}
%==============================

We downmix the recordings to mono and downsample them to 16 kHz using FFmpeg. We then convert them into mel spectrograms using librosa. For the mel spectrogram computation, we use a filter length of 1024, a hop length of 256 and a window size of 1024 in the short-time Fourier transform (STFT), and we use 80 mel bands in mel scale conversion. We summarize these parameters in \cref{tab:preprocessing}.

\begin{table}[h]
    \small
    \centering
    \caption{Preprocessing parameters}
    \begin{tabular}{ll}
        \toprule
        Parameter &Value\\
        \midrule
        Audio channels      &mono\\
        Sampling rate       &16 kHz\\
        STFT filter length  &1024\\
        STFT hop length     &256\\
        STFT window size    &1024\\
        Mel bands           &80\\
        \bottomrule
    \end{tabular}
    \label{tab:preprocessing}
\end{table}

%==============================
\section{Network architectures}
%==============================

%---------------------------
\subsection{Alignment model}
%---------------------------

\begin{figure}
    \small
    \centering
    \includegraphics[width=.8\linewidth]{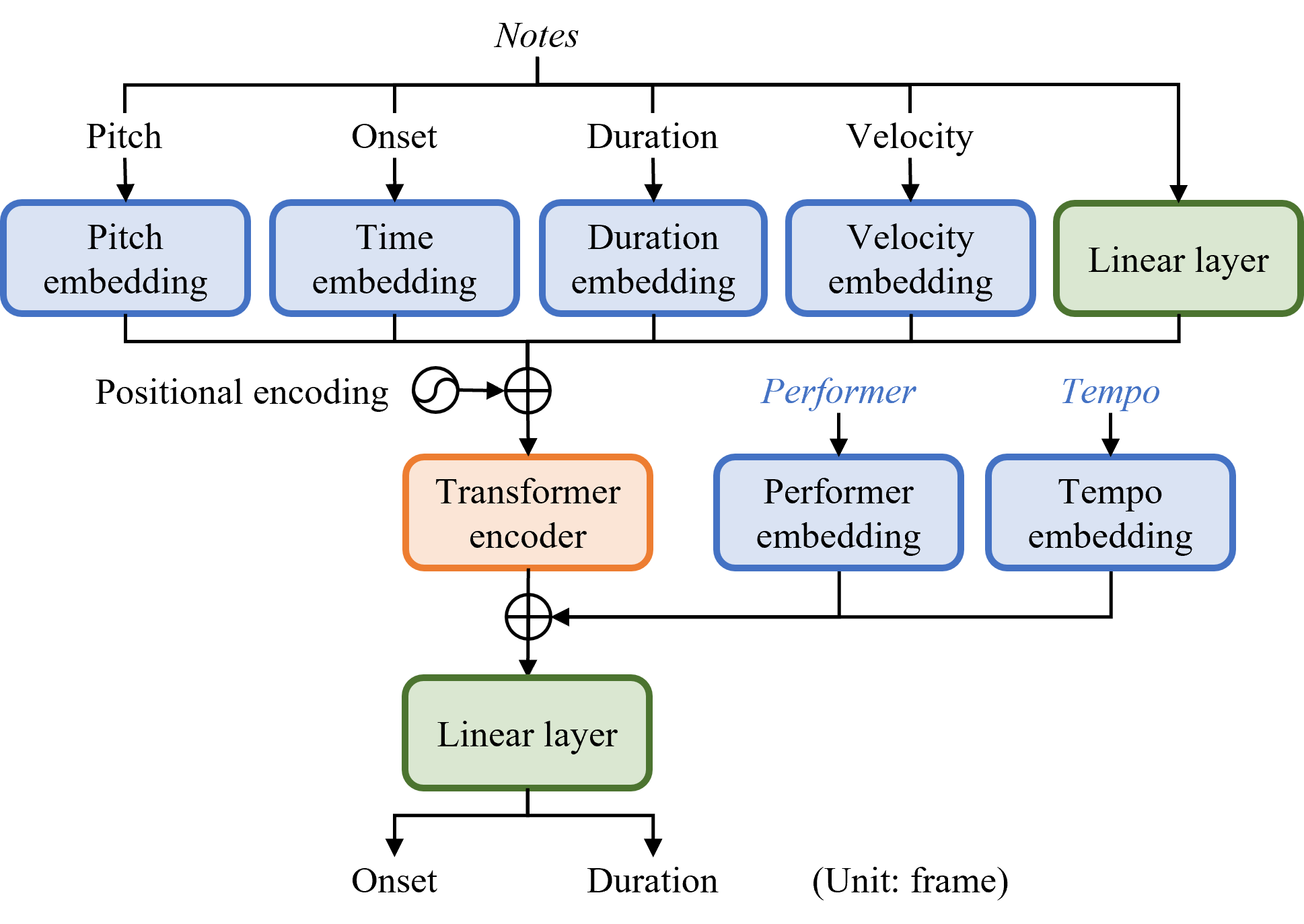}
    \caption{An illustration of the proposed alignment model.}
    \label{fig:alignment-model}
\end{figure}

We illustrate the proposed alignment model in \cref{fig:alignment-model}. We use 128 dimensions for all embeddings. For the transformer encoder, we use 3 transformer layers, each consisting of a multi-head attention (MHA) and a position-wise feed-forward network (FFN) sub-layer. We use 64 hidden neurons and 2 attention heads for each MHA layer. For each FFN layer, we use 256 hidden neurons with kernel sizes of 9 and 1 for the two convolutional layers. Further, we use a maximum sequence length of 1000 and clip the time and duration to 96. We summarize these hyperparameters in \cref{tab:alignment}.

\begin{table}
    \small
    \centering
    \caption{Alignment model architecture}
    \begin{tabular}{ll}
        \toprule
        Parameter &Value\\
        \midrule
        Encoder layers      &3\\
        MHA heads           &2\\
        MHA hidden neurons  &64\\
        FFN hidden neurons  &256\\
        FFN kernel sizes    &9, 1\\
        Max sequence length &1000\\
        Max time            &96\\
        Max duration        &96\\
        \bottomrule
    \end{tabular}
    \label{tab:alignment}
\end{table}

%---------------------------
\subsection{Synthesis model}
%---------------------------

For the synthesis model, we use 128 dimensions for all embeddings. For the transformer model, we use 3 and 6 transformer layers for the encoder and decoder, respectively. We use 128 hidden neurons and 2 attention heads for each MHA layer. For each FFN layer, we use 256 hidden neurons with kernel sizes of 9 and 1 for the two convolutional layers. In addition, we use a maximum sequence length of 1000. We also clip the time and duration to 96 and 100 for the violin and piano datasets, respectively. We summarize these hyperparameters in \cref{tab:synthesis}. We base our implementation on the source code kindly provided in \citep{chien2021tts}.\footnote{\url{https://github.com/ming024/FastSpeech2}}

\begin{table}[ht]
    \small
    \centering
    \caption{Synthesis model architecture}
    \begin{tabular}{ll}
        \toprule
        Parameter &Value\\
        \midrule
        Encoder layers      &3\\
        Decoder layers      &6\\
        MHA heads           &2\\
        MHA hidden neurons  &128\\
        FFN hidden neurons  &512\\
        FFN kernel sizes    &9, 1\\
        Max sequence length &1000\\
        Max time            &96$^*$\\
        Max duration        &96$^*$\\
        \bottomrule
        \footnotesize $^*$100 for the piano dataset
    \end{tabular}
    \label{tab:synthesis}
\end{table}

%---------------------------
\subsection{Inversion model}
%---------------------------

For the inversion model, we use the network architecture of the Hifi-GAN v2 model proposed in \citep{kong2020hifigan}. We base our implementation on the source code kindly provided in \citep{kong2020hifigan}.\footnote{\url{https://github.com/jik876/hifi-gan}}

%--------------------------
\subsection{Baseline model}
%--------------------------

We base the baseline model on the same Hifi-GAN v2 model \citep{kong2020hifigan}. In addition, we include an additional linear layer that maps the input piano roll to a hidden vector whose dimension matches the input dimension of the Hifi-GAN v2 model. Further, we include an additional embedding layer to condition the baseline model on the input performer IDs. The outputs of these two layers are summed up and fed as the input to the Hifi-GAN v2 model.

%=========================
\section{Training details}
%=========================

We use a batch size of 16 and apply a dropout rate of 0.2 after each sub-layer. We use the same optimizer settings as the original implementation of transformer \citep{vaswani2017transformer}. For the alignment model, we apply the learning rate annealing schedule used in \citep{chien2021tts}. We summarize these hyperparameters in \cref{tab:training}. Unlike \citep{ren2019fastspeech}, we train the alignment and synthesis models separately as we find that joint training hinders convergence. For the violin dataset, we train the alignment, synthesis and inversion models for 10K, 100K and 1M steps, respectively. For the piano dataset, we train the synthesis and inversion models for 250K and 1M steps, respectively. For each dataset, the inversion model is trained once and used with different synthesis models.

\begin{table}[h]
    \small
    \centering
    \caption{Training hyperparameters}
    \begin{tabular}{ll}
        \toprule
        Parameter &Value\\
        \midrule
        Batch size                          &16\\
        Dropout                             &0.2\\
        Adam optimizer $\beta_1$            &0.9\\
        Adam optimizer $\beta_2$            &0.98\\
        Adam optimizer $\epsilon$           &$10^{-9}$\\
        Gradient clipping threshold         &1.0\\
        Warm up steps (alignment model)     &1000\\
        Warm up steps (synthesis model)     &4000\\
        Learning rate annealing steps$^*$   &10K, 20K, 50K\\
        Learning rate annealing rate$^*$    &0.5\\
        \bottomrule
        \footnotesize $^*$Applied to the alignment model only
    \end{tabular}
    \label{tab:training}
\end{table}

\end{subappendices}

\graphicspath{{chapters/clipsep/figs/}}
\chapter{CLIPSep: Learning Text-queried Sound Separation with Noisy Unlabeled Videos}
\label{chap:clipsep}

\begin{abstract}
Recent years have seen progress beyond domain-specific sound separation for speech or music towards universal sound separation for arbitrary sounds. Prior work on universal sound separation has investigated separating a target sound out of an audio mixture given a text query. Such text-queried sound separation systems provide a natural and scalable interface for specifying arbitrary target sounds. However, supervised text-queried sound separation systems require costly labeled audio-text pairs for training. Moreover, the audio provided in existing datasets is often recorded in a controlled environment, causing a considerable generalization gap to noisy audio in the wild. In this work, we aim to approach text-queried universal sound separation by using only unlabeled data. We propose to leverage the visual modality as a bridge to learn the desired audio-textual correspondence. The proposed CLIPSep model first encodes the input query into a query vector using the contrastive language-image pretraining (CLIP) model, and the query vector is then used to condition an audio separation model to separate out the target sound. While the model is trained on image-audio pairs extracted from unlabeled videos, at test time we can instead query the model with text inputs in a zero-shot setting, thanks to the joint language-image embedding learned by the CLIP model. Further, videos in the wild often contain off-screen sounds and background noise that may hinder the model from learning the desired audio-textual correspondence. To address this problem, we further propose an approach called \textit{noise invariant training} for training a query-based sound separation model on noisy data. Experimental results show that the proposed models successfully learn text-queried universal sound separation using only noisy unlabeled videos, even achieving competitive performance against a supervised model in some settings.
\end{abstract}

%=====================
\section{Introduction}
%=====================

Humans can focus on to a specific sound in the environment and describe it using language. Such abilities are learned using multiple modalities---auditory for selective listening, vision for learning the concepts of sounding objects, and language for describing the objects or scenes for communication. In machine listening, selective listening is often cast as the problem of sound separation, which aims to separate sound sources from an audio mixture \citep{Cherry1953cocktailparty,Bach2005}. While text queries offer a natural interface for humans to specify the target sound to separate from a mixture \citep{Liu2022SeparateWY,Kilgour2022TextDrivenSO}, training a text-queried sound separation model in a supervised manner requires labeled audio-text paired data of single-source recordings of a vast number of sound types, which can be costly to acquire. Moreover, such isolated sounds are often recorded in controlled environments and have a considerable domain gap to recordings in the wild, which usually contain arbitrary noise and reverberations. In contrast, humans often leverage the visual modality to assist learning the sounds of various objects \citep{Baillarge2002}. For instance, by observing a dog barking, a human can associate the sound with the dog, and can separately learn that the animal is called a ``\textit{dog}.'' Further, such learning is possible even if the sound is observed in a noisy environment, e.g., when a car is passing by or someone is talking nearby, where humans can still associate the barking sound solely with the dog. Prior work in psychophysics also suggests the intertwined cognition of vision and hearing \citep{Sekuler1997,Shimojo2001,Rahne2007}. 

\begin{figure}
    \small
    \centering
    \includegraphics[width=.8\linewidth]{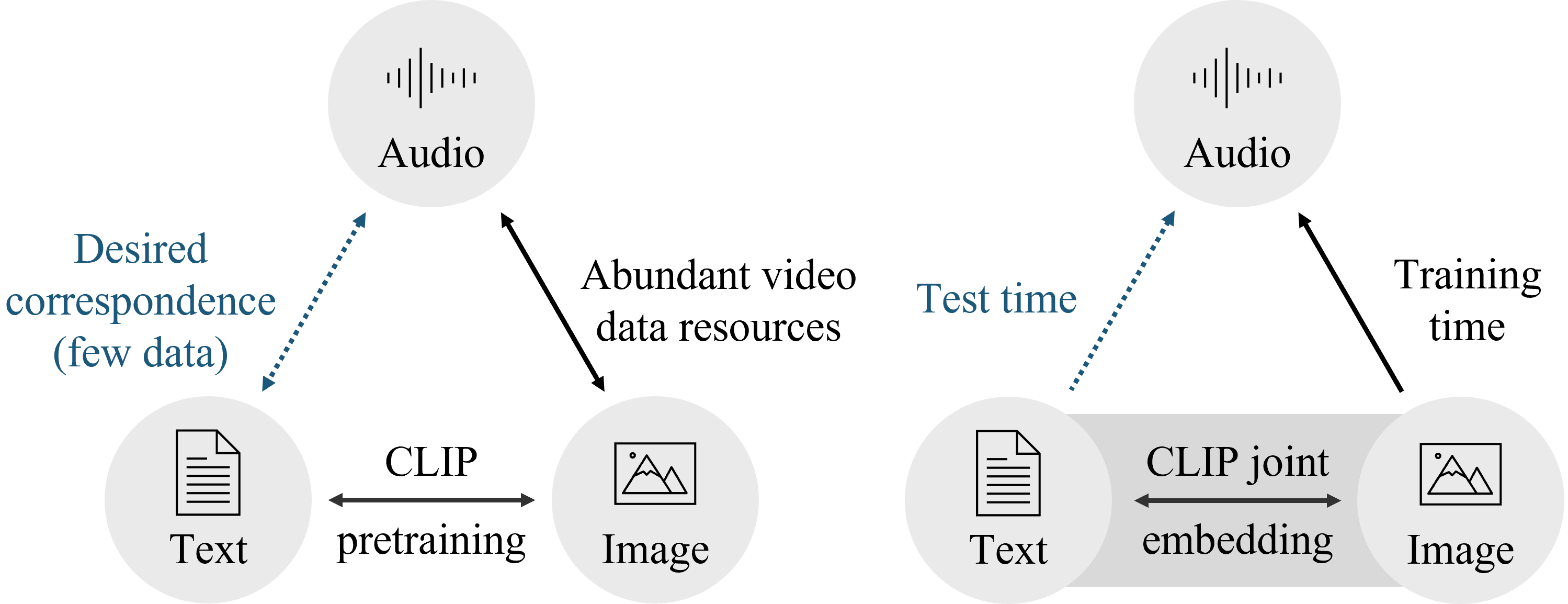}\\
    \caption{An illustration of modality transfer.}
    \label{clipsep:fig:modalities}
\end{figure}

Motivated by this observation, we aim to tackle text-queried sound separation using only unlabeled videos in the wild. We propose a text-queried sound separation model called CLIPSep that leverages abundant unlabeled video data resources by utilizing the contrastive image-language pretraining (CLIP) \citep{radford2021clip} model to bridge the audio and text modalities. As illustrated in \cref{clipsep:fig:modalities}, during training, the image feature extracted from a video frame by the CLIP-image encoder is used to condition a sound separation model, and the model is trained to separate the sound that corresponds to the image query in a self-supervised setting. Thanks to the properties of the CLIP model, which projects corresponding text and images to close embeddings, at test time we instead use the text feature obtained by the CLIP-text encoder from a text query in a zero-shot setting.

However, such zero-shot modality transfer can be challenging when we use videos in the wild for training as they often contain off-screen sounds and voice overs that can lead to undesired audio-visual associations. To address this problem, we propose the \emph{noise invariant training} (NIT), where query-based separation heads and permutation invariant separation heads jointly estimate the noisy target sounds. We validate in our experiments that the proposed noise invariant training reduces the zero-shot modality transfer gap when the model is trained on a noisy dataset, sometimes achieving competitive results against a fully supervised text-queried sound separation system. 

Our contributions can be summarized as follows: 1) We propose the first text-queried universal sound separation model that can be trained on unlabeled videos. 2) We propose a new approach called \textit{noise invariant training} for training a query-based sound separation model on noisy data in the wild. Audio samples can be found on an our demo website.\footnote{\url{https://sony.github.io/CLIPSep/}\label{clipsep:fn:demo}} For reproducibility, all source code, hyperparameters and pretrained models are available at: \url{https://github.com/sony/CLIPSep}.

%=====================
\section{Related Work}
%=====================

\paragraph{Universal sound separation.}
Much prior work on sound separation focuses on separating sounds for a specific domain such as speech \citep{wang2018speech} or music \citep{Takahashi21D3Net, mitsufuji2021music}. Recent advances in domain specific sound separation lead several attempts to generalize to arbitrary sound classes. \citet{Kavalerov19USS} reported successful results on separating arbitrary sounds with a fixed number of sources by adopting the \textit{permutation invariant training} (PIT) \citep{yu2017pit}, which was originally proposed for speech separation. While this approach does not require labeled data for training, a post-selection process is required as we cannot not tell what sounds are included in each separated result. Follow-up work \citep{Ochiai20USSLabel,Kong20USSLabel} addressed this issue by conditioning the separation model with a class label to specify the target sound in a supervised setting. However, these approaches still require labeled data for training, and the interface for selecting the target class becomes cumbersome when we need a large number of classes to handle open-domain data. \citet{Wisdom2020MixIT} later proposed an unsupervised method called mixture invariant training (MixIT) for learning sound separation on noisy data. MixIT is designed to separate all sources at a time and also requires a post-selection process such as using a pre-trained sound classifier \citep{Scott2021}, which requires labeled data for training, to identify the target sounds. We summarize and compare related work in \cref{clipsep:tab:related-work}.

\begin{table}
    \small
    \centering
    \caption[Comparisons of related work in sound separation.]{Comparisons of related work in sound separation. `($\checkmark$)' indicates that the problem of noisy training data is partially addressed. CLIPSep-NIT denotes the proposed CLIPSep model trained with the noise invariant training. To the best of our knowledge, no previous work has attempted the problem of label-free text-queried sound separation.}
    \label{clipsep:tab:related-work}
    \begin{tabular}{lccc}
        \toprule
        Method &Query type &Unlabeled data &Noisy data\\
        \midrule
        USS {\notesize\citep{Kavalerov19USS}} &$\times$ &$\checkmark$\\
        MixIT {\notesize\citep{Wisdom2020MixIT}} &$\times$ &$\checkmark$ &$\checkmark$\\
        \cmidrule(lr){1-4}
        Universal Sound Selector {\notesize\citep{Ochiai20USSLabel}} &Label\\
        USS-Label {\notesize\citep{Kong20USSLabel}} &Label & &($\checkmark$)\\
        \cmidrule(lr){1-4}
        PixelPlayer {\notesize\citep{zhao2018sop}} &Image &$\checkmark$ &($\checkmark$)\\
        AudioScope {\notesize\citep{Tzinis2021AudioScope}} &Image &$\checkmark$ &$\checkmark$\\
        SoundFilter {\notesize\citep{Gfeller2021OneShotCA}} &Audio &$\checkmark$\\
        Zero-shot audio separation {\notesize\citep{Chen22ZeroshotSS}} &Audio & &($\checkmark$)\\
        Text-Queried {\notesize\citep{Liu2022SeparateWY}} &Text\\
        Text/Audio-Queried {\notesize\citep{Kilgour2022TextDrivenSO}} &Text / Audio\\
        \cmidrule(lr){1-4}
        CLIPSep (ours) &Text / Image &$\checkmark$ & \\
        CLIPSep-NIT (ours) &Text / Image &$\checkmark$ &$\checkmark$ \\
        \bottomrule
    \end{tabular}
\end{table}

\paragraph{Query-based sound separation.}
Visual information has been used for selecting the target sound in speech  \citep{Ephrat2019L2L,Afouras2020SSLAV}, music \citep{zhao2018sop,zhao2019som,Tian21CoSep} and universal sounds \citep{Owens2018AVSA,Gao2018AVSNMF,Rouditchenko2019}. While many image-queried sound separation approaches require clean video data that contains isolated sources, \citet{Tzinis2021AudioScope} introduced an unsupervised method called AudioScope for separating on-screen sounds using noisy videos based on the MixIT model. While image queries can serve as a natural interface for specifying the target sound in certain use cases, images of target sounds become unavailable in low-light conditions and for sounds from out-of-screen objects.

Another line of research uses the audio modality to query acoustically similar sounds. \citet{Chen22ZeroshotSS} showed that such approach can generalize to unseen sounds. Later, \citet{Gfeller2021OneShotCA} cropped two disjoint segments from single recording and used them as a query-target pair to train a sound separation model, assuming both segments contain the same sound source. However, in many cases, it is impractical to prepare a reference audio sample for the desired sound as the query.

Most recently, text-queried sound separation has been studied as it provides a natural and scalable interface for specifying arbitrary target sounds as compared to systems that use a fixed set of class labels. \citet{Liu2022SeparateWY} employed a pretrained language model to encode the text query, and condition the model to separate the corresponding sounds. \citet{Kilgour2022TextDrivenSO} proposed a model that accepts audio or text queries in a hybrid manner. These approaches, however, require labeled text-audio paired data for training. Different from prior work, our goal is to learn text-queried sound separation for arbitrary sound \emph{without} labeled data, specifically using unlabeled noisy videos in the wild.

\paragraph{Contrastive language-image-audio pretraining.}
The CLIP model \citep{radford2021clip} has been used as a pretraining of joint embedding spaces among text, image and audio modalities for downstream tasks such as audio classification \citep{wu2022wav2clip, guzhov2022audioclip} and sound guided image manipulation \citep{Lee22SoundGuidedIM}. Pretraining is done either in a supervised manner using labels \citep{guzhov2022audioclip,Lee22SoundGuidedIM} or in a self-supervised manner by training an additional audio encoder to map input audio to the pretrained CLIP embedding space \citep{wu2022wav2clip}. In contrast, we explore the zero-shot modality transfer capability of the CLIP model by freezing the pre-trained CLIP model and directly optimizing the rest of the model for the target sound separation task.

%===============
\section{Method}
%===============

%---------------------------------------------------------------------------------
\subsection{CLIPSep---Learning text-queried sound separation without labeled data}
%---------------------------------------------------------------------------------

\begin{figure}
    \small
    \centering
    \includegraphics[width=\linewidth,clip]{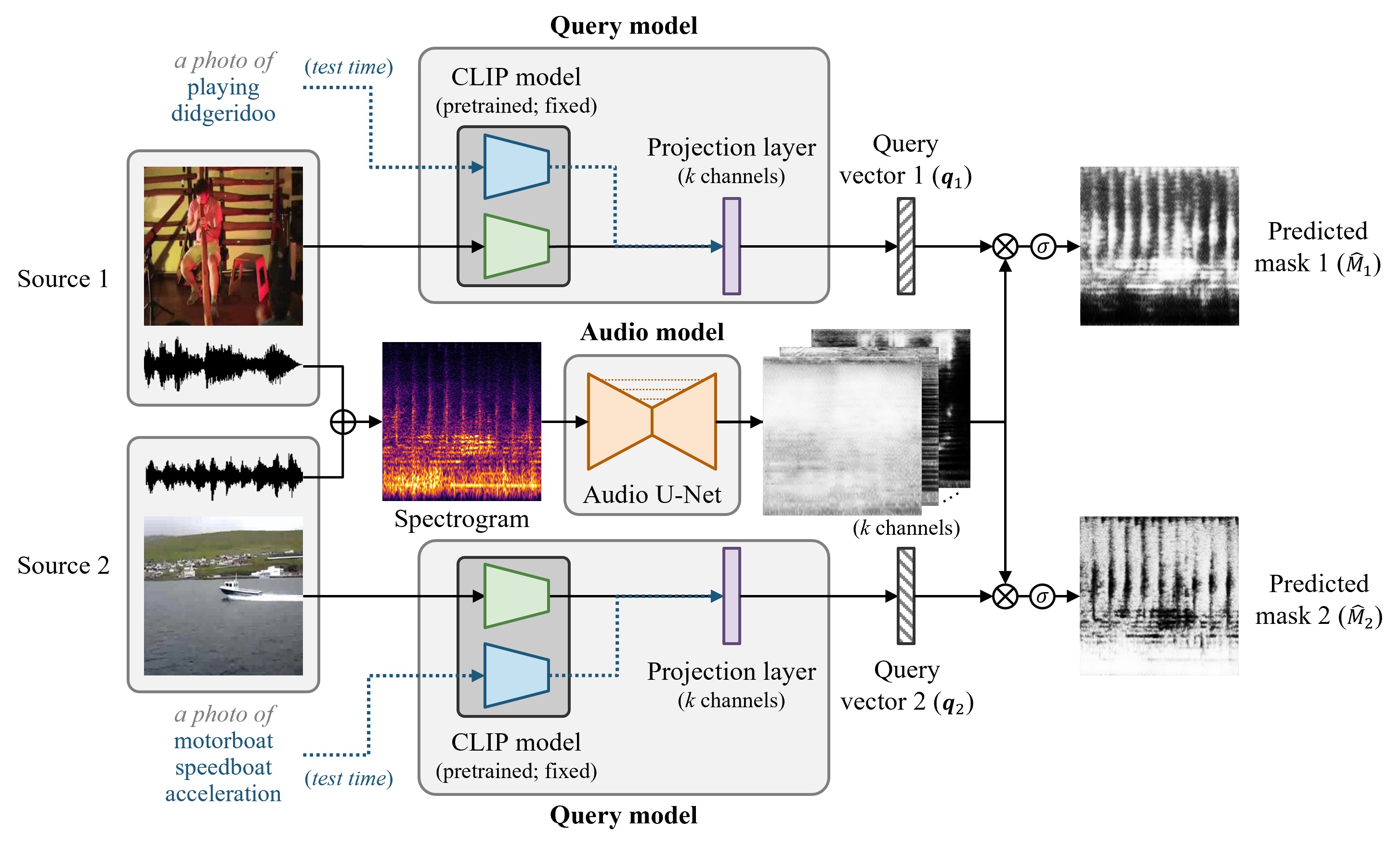}\\
    \caption[An illustration of the proposed CLIPSep model for $n = 2$.]{An illustration of the proposed CLIPSep model for $n = 2$. During training, we mix audio from two videos and train the model to separate each audio source given the corresponding video frame as the query. At test time, we instead use a text query in the form of ``\textit{a photo of [user input query]}'' to query the sound separation model. Thanks to the properties of the pretrained CLIP model, the query vectors we obtain for the image and text queries are expected to be close.}
    \label{clipsep:fig:clipsep}
\end{figure}

In this section, we propose the CLIPSep model for text-queried sound separation without using labeled data. We base the CLIPSep model on Sound-of-Pixels (SOP) \citep{zhao2018sop} and replace the video analysis network of the SOP model. As illustrated in \cref{clipsep:fig:clipsep}, during training, the model takes as inputs an audio mixture $\bm{x} = \sum_{i = 1}^n \bm{s}_i$, where $\bm{s}_1, \dots, \bm{s}_n$ are the $n$ audio tracks, along with their corresponding images $\bm{y}_1, \dots, \bm{y}_n$ extracted from the videos. We first transform the audio mixture $\bm{x}$ into a magnitude spectrogram $X$ and pass the spectrogram through an audio U-Net \citep{ronneberger2015unet,jansson2017unet} to produce $k$ ($\geq n$) intermediate masks $\tilde{M}_1, \dots, \tilde{M}_k$. On the other stream, each image is encoded by the pretrained CLIP model \citep{radford2021clip} into an embedding $\bm{e}_i \in \mathbb{R}^{512}$. The CLIP embedding $\bm{e}_i$ will further be projected to a \textit{query vector} $\bm{q}_i \in \mathbb{R}^k$ by a \textit{projection layer}, which is expected to extract only audio-relevant information from $\bm{e}_i$.\footnote{We extract three frames with 1-sec intervals and compute their mean CLIP embedding as the input to the projection layer to reduce the negative effects when the selected frame does not contain the objects of interest.} Finally, the query vector $\bm{q}_i$ will be used to mix the intermediate masks into the final predicted masks $\hat{M}_i = \sum_{j = 1}^k \sigma\big(w_{ij}q_{ij} \tilde{M}_j + b_i\big)$, where $\bm{w}_i \in \mathbb{R}^k$ is a learnable scale vector, $b_i \in \mathbb{R}$ a learnable bias, and $\sigma(\cdot)$ the sigmoid function. Now, suppose $M_i$ is the ground truth mask for source $\bm{s}_i$. The training objective of the model is the sum of the weighted binary cross entropy losses for each source:
\begin{equation}%\small
    \mathcal{L}_\mathit{CLIPSep} = \sum_{i = 1}^n \mathit{WBCE}(M_i, \hat{M_i}) = \sum_{i = 1}^n X \odot \bigg(-M_i \log \hat{M_i} - (1 - M_i) \log \left(1 - \hat{M}_i\right)\bigg)\,.
\end{equation}
At test time, thanks to the joint image-text embedding offered by the CLIP model, we feed a text query instead of an image to the query model to obtain the query vector and separate the target sounds accordingly (see \cref{clipsep:sec:inference} for an illustration). As suggested by \citet{radford2021clip}, we prefix the text query into the form of ``\textit{a photo of [user input query]}'' to reduce the generalization gap.\footnote{Similar to how we prepare the image queries, we create four queries from the input text query using four query templates (see \cref{clipsep:sec:queries}) and take their mean CLIP embedding as the input to the projection layer.}

%----------------------------------------------------------------------
\subsection{Noise invariant training---Handling noisy data in the wild}
%----------------------------------------------------------------------

\begin{figure}
    \small
    \centering
    \includegraphics[width=\linewidth]{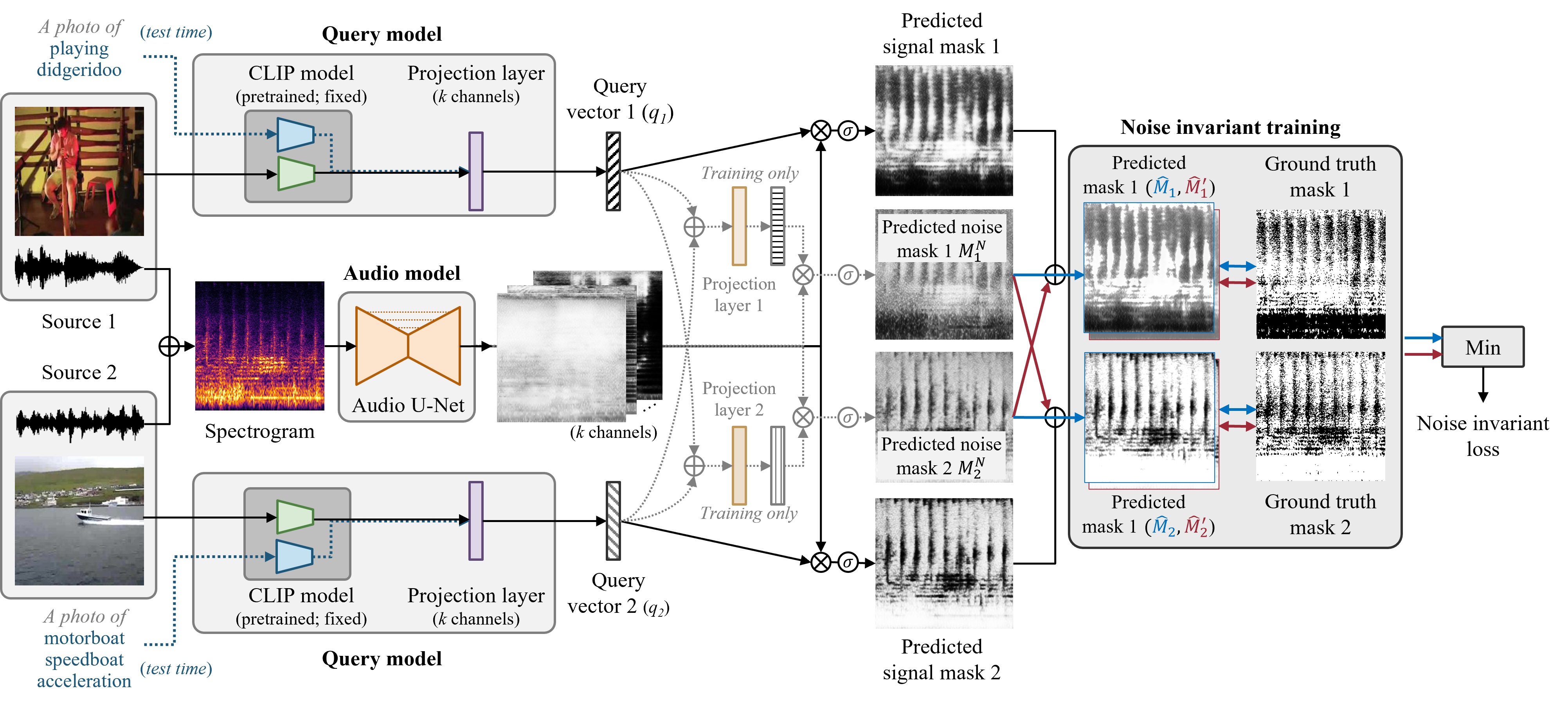}\\
    \caption[An illustration of the proposed CLIPSep-NIT model for $n = 2$.]{An illustration of the proposed CLIPSep-NIT model for $n = 2$. Similar to CLIPSep, we train the model to separate each audio source given the corresponding query image during training and switch to using a text query at test time. The two predicted noise masks are interchangeable for loss computation during training, and they are discarded at test time (grayed out paths).}
    \label{clipsep:fig:clipit}
\end{figure}

While the CLIPSep model can separate sounds given image or text queries, it assumes that the sources are clean and contain few query-irrelevant sounds. However, this assumption does not hold for videos in the wild as many of them contain out-of-screen sounds and various background noises. Inspired by the mixture invariant training (MixIT) proposed by \citet{Wisdom2020MixIT}, we further propose the \textit{noise invariant training} (NIT) to tackle the challenge of training with noisy data. As illustrated in \cref{clipsep:fig:clipit}, we introduce $n$ additional permutation invariant heads called \textit{noise heads} to the CLIPSep model, where the masks predicted by these heads are interchangeable during loss computation. Specifically, we introduce $n$ additional projection layers, and each of them takes as input the sum of all query vectors produced by the \textit{query heads} (i.e., $\sum_{i = 1}^n \bm{q}_i$) and produce a vector that is later used to mix the intermediate masks into the predicted \textit{noise mask}. In principle, the \textit{query masks} produced by the query vectors are expected to extract query-relevant sounds due to their stronger correlations to their corresponding queries, while the interchangeable noise masks should `soak up' other sounds. Mathematically, let $M^Q_1, \dots, M^Q_n$ be the predicted query masks and $M^N_1, \dots, M^N_n$ be the predicted noise masks. Then, the noise invariant loss is defined as:
\begin{equation}%\small
    \mathcal{L}_\mathit{NIT} = \min_{(j_1, \dots, j_n) \in \Sigma_n} \sum_{i = 1}^n \mathit{WBCE}\bigg(M_i, \min\left(1, \hat{M}^Q_i + \hat{M}^N_{j_i}\right)\bigg)\,,
\end{equation}
where $\Sigma_n$ denotes the set of all permutations of $\{1, \dots, n\}$.\footnote{We note that CLIPSep-NIT considers $2n$ sources in total as the model has $n$ queried heads and $n$ noise heads. While PIT \citep{yu2017pit} and MixIT \citep{Wisdom2020MixIT} respectively require $\mathcal{O}((2n)!)$ and $\mathcal{O}(2^{2n})$ search to consider $2n$ sources, the proposed NIT only requires $\mathcal{O}(n!)$ permutation in the loss computation.} Take $n = 2$ for example.\footnote{Since our goal is not to further separate the noise into individual sources but to separate the sounds that correspond to the query, $n$ may not need to be large. In practice, we find that the CLIPSep-NIT model with $n = 2$ already learns to handle the noise properly and can successfully transfer to the text-queried mode. Thus, we use $n = 2$ throughout this paper and leave the testing on larger $n$ as future work.} We consider the two possible ways for combining the query heads and the noise heads:
\begin{align}%\small
    \text{(Arrangement 1)} \qquad &\hat{M}_1 = \min\big(1, \hat{M}^Q_1 + \hat{M}^N_1\big), \quad \hat{M}_2 = \min\big(1, \hat{M}^Q_2 + \hat{M}^N_2\big)\,,\\
    %\small
    \text{(Arrangement 2)} \qquad &\hat{M}'_1 = \min\big(1, \hat{M}^Q_1 + \hat{M}^N_2\big), \quad \hat{M}'_2 = \min\big(1,\hat{M}^Q_2 + \hat{M}^N_1\big)\,.
\end{align}
Then, the noise invariant loss is defined as the smallest loss achievable:
\begin{equation}%\small
    \mathcal{L}_\mathit{NIT}^{(2)} = \min \Big(\mathit{WBCE}\big(M_1, \hat{M}_1\big) + \mathit{WBCE}\big(M_2, \hat{M}_2\big), \mathit{WBCE}\big(M_1, \hat{M}'_1\big) + \mathit{WBCE}\big(M_2, \hat{M}'_2\big)\Big)\,.
\end{equation}
Once the model is trained, we discard the noise heads and use only the query heads for inference (see \cref{clipsep:sec:inference} for an illustration). Unlike the MixIT model \citep{Wisdom2020MixIT}, our proposed noise invariant training still allows us to specify the target sound by an input query, and it does not require any post-selection process as we only use the query heads during inference.

In practice, we find that the model tends to assign part of the target sounds to the noise heads as these heads can freely enjoy the optimal permutation to minimize the loss. Hence, we further introduce a regularization term to penalize producing high activations on the noise masks:
\begin{equation}%\small
    \mathcal{L}_\mathit{REG} = \max \bigg(0, \sum_{i = 1}^n \mathrm{mean}\left(\hat{M}^N_i\right) - \gamma\bigg)\,,
    \label{clipsep:eq:regularization}
\end{equation}
where $\gamma \in [0, n]$ is a hyperparameter that we will refer to as the \textit{noise regularization level}. The proposed regularization has no effect when the sum of the means of all the noise masks is lower than a predefined threshold $\gamma$, while having a linearly growing penalty when the sum is higher than $\gamma$. Finally, the training objective of the CLIPSep-NIT model is a weighted sum of the noise invariant loss and regularization term: $\mathcal{L}_\mathit{CLIPSep\text{-}NIT} = \mathcal{L}_\mathit{NIT} + \lambda \mathcal{L}_\mathit{REG}$, where $\lambda \in \mathbb{R}$ is a weight hyperparameter. We set $\lambda = 0.1$ for all experiments, which we find work well across different settings.

%====================
\section{Experiments}
%====================

We base our implementations on the code provided by \citet{zhao2018sop} (\url{https://github.com/hangzhaomit/Sound-of-Pixels}). Implementation details can be found in \cref{clipsep:sec:implementation}.

\begin{table}
    \small
    \centering
    \caption[Results on the MUSIC dataset.]{Results on the MUSIC dataset. Standard errors are reported in the mean SDR column. Bold values indicate the largest SDR achieved per group.}
    \label{clipsep:tab:music}
    \begin{tabular}{lcccccc}
        \toprule
        \multirow{2}{*}[-3pt]{Model} &\multirow{2}{*}[-3pt]{\shortstack{Unlabeled\\data}} &\multirow{2}{*}[-1ex]{\shortstack{Post-proc.\\free}} &\multicolumn{2}{c}{Query type} &\multicolumn{2}{c}{SDR [dB]} \\
        \cmidrule(lr){4-5} \cmidrule(lr){6-7}
        &&&Training &Test &Mean &Median\\
        \midrule
        Mixture &- &- &- &- &0.00 $\pm$ 0.89 &0.00\\
        \midrule
        \textbf{Text-queried models}\\
        CLIPSep &$\checkmark$ &$\checkmark$ &Image &Text &\textbf{5.49 $\pm$ 0.72} &\textbf{4.97}\\
        \cmidrule(lr){1-7}
        CLIPSep-Text & &$\checkmark$ &Text &Text &7.91 $\pm$ 0.81 &7.46\\
        CLIPSep-Hybrid & &$\checkmark$ &Text + Image &Text &\textbf{8.36 $\pm$ 0.83} &\textbf{8.72}\\
        \midrule
        \textbf{Image-queried models}\\
        SOP {\notesize\citep{zhao2018sop}} &$\checkmark$ &$\checkmark$ &Image &Image &6.59 $\pm$ 0.85 &\textbf{6.22}\\
        CLIPSep &$\checkmark$ &$\checkmark$ &Image &Image &\textbf{7.03 $\pm$ 0.70} &5.85\\
        \cmidrule(lr){1-7}
        CLIPSep-Text & &$\checkmark$ &Text &Image &6.25 $\pm$ 0.72 &6.19\\
        CLIPSep-Hybrid & &$\checkmark$ &Text + Image &Image &\textbf{8.06 $\pm$ 0.79} &\textbf{8.01}\\
        \midrule
        \textbf{Nonqueried models}\\
        LabelSep & &$\checkmark$ &Label &Label &8.18 $\pm$ 0.80 &\textbf{7.82}\\
        PIT {\notesize\citep{yu2017pit}} &$\checkmark$ & &$\times$ &$\times$ &\textbf{8.68 $\pm$ 0.76} &7.67\\
        \bottomrule
    \end{tabular}
\end{table}

%-------------------------------------
\subsection{Experiments on clean data}
%-------------------------------------

We first evaluate the proposed CLIPSep model without the noise invariant training on musical instrument sound separation task using the MUSIC dataset, as done in \citep{zhao2018sop}. This experiment is designed to focus on evaluating the quality of the learned query vectors and the zero-shot modality transferability of the CLIPSep model on a small, clean dataset rather than showing its ability to separate arbitrary sounds. The MUSIC dataset is a collection of 536 video recordings of people playing a musical instrument out of 11 instrument classes. Since no existing work has trained a text-queried sound separation model using only unlabeled data to our knowledge, we compare the proposed CLIPSep model with two baselines that serve as upper bounds---the PIT model \citep[see \cref{clipsep:sec:pit} for an illustration]{yu2017pit} and a version of the CLIPSep model where the query model is replaced by learnable embeddings for the labels, which we will refer to as the LabelSep model. In addition, we also include the SOP model \citep{zhao2018sop} to investigate the quality of the query vectors as the CLIPSep and SOP models share the same network architecture except the query model.

\begin{figure}
    \small
    \centering
    \includegraphics[width=.6\linewidth]{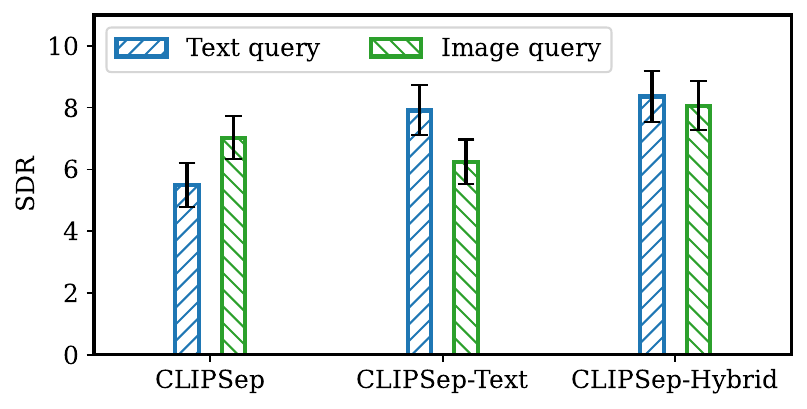}\\
    \caption{Mean SDR and standard errors of the models trained and tested on different modalities.}
    \label{clipsep:fig:clipsep-modalities}
\end{figure}

We report the results in \cref{clipsep:tab:music}. Our proposed CLIPSep model achieves a mean signal-to-distortion ratio (SDR) \citep{vincent2006performance} of 5.49 dB and a median SDR of 4.97 dB using text queries in a zero-shot modality transfer setting. When using image queries, the performance of the CLIPSep model is comparable to that of the SOP model. This indicates that the CLIP embeddings are as informative as those produced by the SOP model. The performance difference between the CLIPSep model using text and image queries at test time indicates the \textit{zero-shot modality transfer gap}. We observe 1.54 dB and 0.88 dB differences on the mean and median SDRs, respectively. Moreover, we also report in \cref{clipsep:tab:music} and \cref{clipsep:fig:clipsep-modalities} the performance of the CLIPSep models trained on different modalities to investigate their modality transferability in different settings. We notice that when we train the CLIPSep model using text queries, dubbed as CLIPSep-Text, the mean SDR using text queries increases to 7.91 dB. However, when we test this model using image queries, we observe a 1.66 dB difference on the mean SDR as compared to that using text queries, which is close to the mean SDR difference we observe for the model trained with image queries. Finally, we train a CLIPSep model using both text and image queries in alternation, dubbed as CLIPSep-Hybrid. We see that it leads to the best test performance for both text and image modalities, and there is only a mean SDR difference of 0.30 dB between using text and image queries. As a reference, the LabelSep model trained with labeled data performs worse than the CLIPSep-Hybrid model using text queries. Further, the PIT model achieves a mean SDR of 8.68 dB and a median SDR of 7.67 dB, but it requires post-processing to figure out the correct assignments.

\begin{table}
    \small
    \centering
    \caption[Results of the MUSIC$^+$ and VGGSound-Clean$^+$ evaluations (see \cref{clipsep:sec:vggsound-exp}).]{Results of the MUSIC$^+$ and VGGSound-Clean$^+$ evaluations (see \cref{clipsep:sec:vggsound-exp}). Standard errors are reported in the mean SDR [dB] columns. Bold values indicate the largest SDR achieved per group. We use $\gamma = 0.25$ for CLIPSep-NIT. Note that the LabelSep model does not work on the MUSIC dataset due to the different label taxonomies of the MUSIC and VGGSound datasets.}
    \label{clipsep:tab:vggsound-exp}
    \begin{tabular}{l@{~~~}c@{~~~}ccccc}
        \toprule
        &&&\multicolumn{2}{c}{MUSIC$^+$ } &\multicolumn{2}{c}{VGGSound-Clean$^+$ }\\
        \cmidrule(lr){4-5} \cmidrule(lr){6-7}
        \multirow{2}{*}{Model} &\multirow{2}{*}{\shortstack{Unlabeled\\data}} &\multirow{2}{*}{\shortstack{Post-proc.\\free}} &\multirow{2}{*}{Mean SDR} &Median &\multirow{2}{*}{Mean SDR} &Median\\
        &&&&SDR &&SDR\\
        \midrule
        Mixture &- &- &~~4.49 $\pm$ 1.41 &~~2.04 &-0.77 $\pm$ 1.31 &-0.84\\
        \midrule
        \textbf{Text-queried models}\\
        CLIPSep &$\checkmark$ &$\checkmark$ &~~9.71 $\pm$ 1.21 &~~8.73 &~2.76 $\pm$ 1.00 &~\textbf{3.95}\\
        CLIPSep-NIT &$\checkmark$ &$\checkmark$ &\textbf{10.27 $\pm$ 1.04} &\textbf{10.02} &~\textbf{3.05 $\pm$ 0.73} &~3.26\\
        \cmidrule(lr){1-7}
        BERTSep & &$\checkmark$ &~~4.67 $\pm$ 0.44 &~4.41 &~5.09 $\pm$ 0.80 &~5.49\\
        CLIPSep-Text & &$\checkmark$ &10.73 $\pm$ 0.99 &~9.93 &~5.49 $\pm$ 0.82 &~5.06\\
        \midrule
        \textbf{Image-queried models}\\
        SOP {\notesize\citep{zhao2018sop}} &$\checkmark$ &$\checkmark$ &11.44 $\pm$ 1.18 &11.18 &~2.99 $\pm$ 0.84 &~3.89\\
        CLIPSep &$\checkmark$ &$\checkmark$ &\textbf{12.20 $\pm$ 1.17} &\textbf{12.42} &~\textbf{5.46 $\pm$ 0.79} &~\textbf{5.35}\\
        CLIPSep-NIT &$\checkmark$ &$\checkmark$ &11.28 $\pm$ 1.08 &10.83 &~4.84 $\pm$ 0.66 &~3.57\\
        \cmidrule(lr){1-7}
        CLIPSep-Text & &$\checkmark$ &~~9.89 $\pm$ 1.04 &~~8.09 &~2.45 $\pm$ 0.70 &~1.74\\
        \midrule
        \textbf{Nonqueried models}\\
        PIT {\notesize\citep{yu2017pit}} &$\checkmark$ & &\textbf{12.24 $\pm$ 1.20} &\textbf{12.53} &~\textbf{5.73 $\pm$ 0.79} &~\textbf{4.97}\\
        LabelSep & &$\checkmark$ &- &- &~5.55 $\pm$ 0.81 &~5.29\\
        \bottomrule
    \end{tabular}
\end{table}

%-------------------------------------
\subsection{Experiments on noisy data}
%-------------------------------------
\label{clipsep:sec:vggsound-exp}

Next, we evaluate the proposed method on a large-scale dataset aiming at universal sound separation. We use the VGGSound dataset \citep{chen2020vggsound}, a large-scale audio-visual dataset containing more than 190,000 10-second videos in the wild out of more than 300 classes. We find that the audio in the VGGSound dataset is often noisy and contains off-screen sounds and background noise. Although we train the models on such noisy data, it is not suitable to use the noisy data as targets for evaluation because it fails to provide reliable results. For example, if the target sound labeled as ``dog barking'' also contains human speech, separating only the dog barking sound provides a lower SDR value than separating the mixture of dog barking sound and human speech even though the text query is ``dog barking''. (Note that we use the labels only for evaluation but not for training.) To avoid this issue, we consider the following two evaluation settings:
\begin{itemize}
    \item \textbf{MUSIC$^+$}: Samples in the MUSIC dataset are used as clean targets and mixed with a sample in the VGGSound dataset as an interference. The separation quality is evaluated on the clean target from the MUSIC dataset. As we do not use the MUSIC dataset for training, this can be considered as zero-shot transfer to a new data domain containing unseen sounds \citep{radford2019gpt2,brown2020gpt3}. To avoid the unexpected overlap of the target sound types in the MUSIC and VGGSound datasets caused by the label mismatch, we exclude all the musical instrument playing videos from  the VGGSound dataset in this setting.
    \item \textbf{VGGSound-Clean$^+$}: We manually collect 100 clean samples that contain distinct target sounds from the VGGSound test set, which we will refer to as VGGSound-Clean. We mix an audio sample in VGGSound-Clean with another in the test set of VGGSound. Similarly, we consider the VGGSound audio as an interference sound added to the relatively cleaner VGGSound-Clean audio and evaluate the separation quality on the VGGSound-Clean stem.
\end{itemize}

\begin{figure}
    \small
    \centering
    % \begin{tabular}{@{}c@{\hspace{3pt}}c@{\hspace{3pt}}c@{\hspace{3pt}}c@{\hspace{3pt}}c@{\hspace{3pt}}c@{\hspace{3pt}}c@{}}
    %     \includegraphics[width=.16\linewidth,height=.16\linewidth,trim={0 0 0 3cm},clip]{W3/mix.png} &\includegraphics[width=.16\linewidth,height=.16\linewidth,trim={0 0 0 3cm},clip]{W3/gtamp.png} &\includegraphics[width=.16\linewidth,height=.16\linewidth,trim={0 0 0 3cm},clip]{W3/intamp.png} &\includegraphics[width=.16\linewidth,height=.16\linewidth,trim={0 0 0 3cm},clip]{W3/predamp.png} &\includegraphics[width=.16\linewidth,height=.16\linewidth,trim={0 0 0 3cm},clip]{W3/pitmag1.png} &\includegraphics[width=.16\linewidth,height=.16\linewidth,trim={0 0 0 3cm},clip]{W3/pitmag2.png}\\
    %     (a) Input mixture &(b) Ground truth &(c) Interference &(d) Prediction &(e) Noise head 1 &(f) Noise head 2
    % \end{tabular}
    \begin{tabular}{ccc}
        \includegraphics[width=.3\linewidth,height=.3\linewidth,trim={0 0 0 3cm},clip]{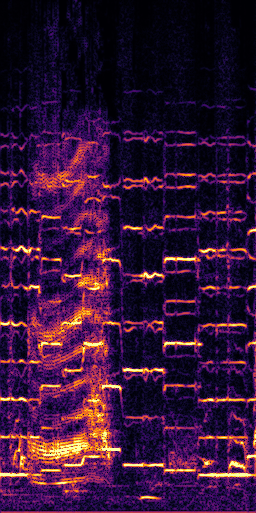} &\includegraphics[width=.3\linewidth,height=.3\linewidth,trim={0 0 0 3cm},clip]{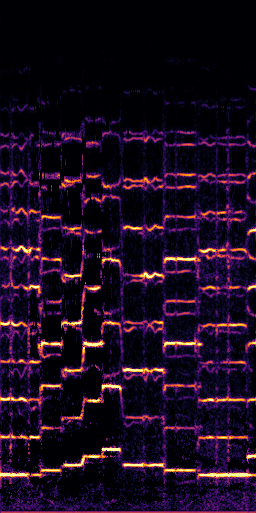} &\includegraphics[width=.3\linewidth,height=.3\linewidth,trim={0 0 0 3cm},clip]{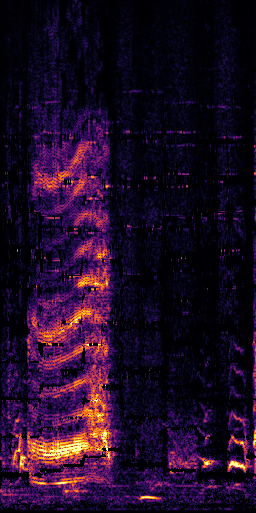}\\
        (a) Input mixture &(b) Ground truth &(c) Interference\\[2ex]
        \includegraphics[width=.3\linewidth,height=.3\linewidth,trim={0 0 0 3cm},clip]{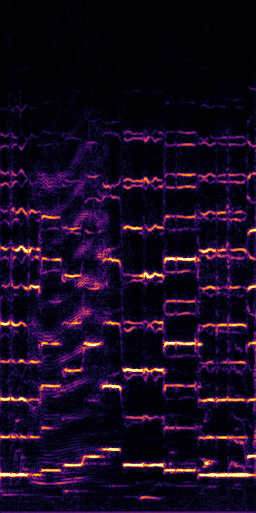} &\includegraphics[width=.3\linewidth,height=.3\linewidth,trim={0 0 0 3cm},clip]{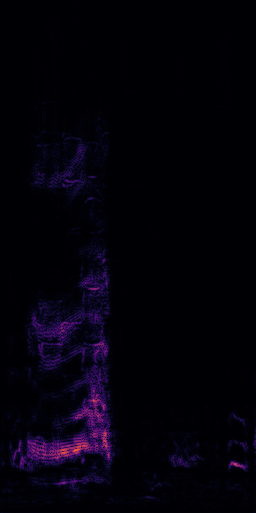} &\includegraphics[width=.3\linewidth,height=.3\linewidth,trim={0 0 0 3cm},clip]{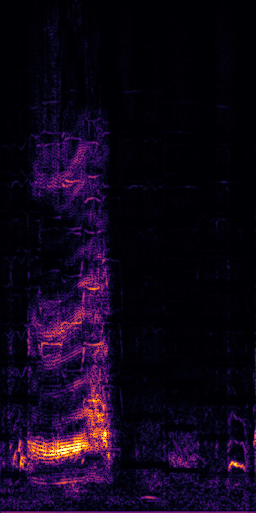}\\
        (d) Prediction &(e) Noise head 1 &(f) Noise head 2
    \end{tabular}
    \caption[Example results of the proposed CLIPSep-NIT model with $\gamma = 0.25$ on the MUSIC$^+$ dataset.]{Example results of the proposed CLIPSep-NIT model with $\gamma = 0.25$ on the MUSIC$^+$ dataset. We mix the an audio sample (``violin'' in this example) in the MUSIC dataset with an interference audio sample (``people sobbing'' in this example) in the VGGSound dataset to create an artificial mixture. (b) and (c) show the reconstructed signals using the ground truth ideal binary masks. The spectrograms are shown in the log frequency scale. We observe that the proposed model successfully separates the desired sounds (i.e., (d)) from query-irrelevant noise (i.e., (e) and (f)).}
    \label{clipsep:fig:example}
\end{figure}

\cref{clipsep:tab:vggsound-exp} shows the evaluation results. First, CLIPSep successfully learns text-queried sound separation even with noisy unlabeled data, achieving 5.22 dB and 3.53 dB SDR improvements over the mixture on MUSIC$^+$ and VGGSound-Clean$^+$, respectively. By comparing CLIPSep and CLIPSep-NIT, we observe that NIT improves the mean SDRs in both settings. Moreover, on MUSIC$^+$, CLIPSep-NIT's performance matches that of CLIPSep-Text, which utilizes labels for training, achieving only a 0.46 dB lower mean SDR and even a 0.05 dB higher median SDR. This result suggests that the proposed self-supervised text-queried sound separation method can learn separation capability competitive with the fully supervised model in some target sounds. In contrast, there is still a gap between them on VGGSound-Clean$^+$, possibly because the videos of non-music-instrument objects are more noisy in both audio and visual domains, thus resulting in a more challenging zero-shot modality transfer. This hypothesis is also supported by the higher zero-shot modality transfer gap (mean SDR difference of image- and text-queried mode) of 1.79 dB on VGGSound-Clean$^+$ than that of 1.01 dB on MUSIC$^+$ for CLIPSep-NIT. In addition, we consider another baseline model that replaces the CLIP model in CLIPSep with a BERT encoder \citep{devlin2019bert}, which we call BERTSep. Interestingly, although BERTSep performs similarly to CLIPSep-Text on VGGSound-Clean$^+$, the performance of BERTSep is significantly lower than that of CLIPSep-Text on MUISC$^+$, indicating that BERTSep fails to generalize to unseen text queries. We hypothesize that the CLIP text embedding captures the timbral similarity of musical instruments better than the BERT embedding do, because the CLIP model is aware of the visual similarity between musical instruments during training. Moreover, it is interesting to see that CLIPSep outperforms CLIPSep-NIT when an image query is used at test time (domain-matched condition), possibly because images contain richer context information such as objects nearby and backgrounds than labels, and the models can use such information to better separate the target sound. While CLIPSep has to fully utilize such information, CLIPSep-NIT can use the noise heads to model sounds that are less relevant to the image query. Since we remove the noise heads from CLIPSep-NIT during the evaluation, it can rely less on such information from the image, thus improving the zero-shot modality transferability. \cref{clipsep:fig:example} shows an example of the separation results on MUSIC$^+$ (see \cref{clipsep:fig:example1,clipsep:fig:example2,clipsep:fig:example3,clipsep:fig:example4} for more examples). We observe that the two noise heads contain mostly background noise. Audio samples can be found on our demo website.\cref{clipsep:fn:demo}

%------------------------------------------------------------------------------------------------
\subsection{Examining the effects of the noise regularization level \texorpdfstring{$\gamma$}{γ}}
%------------------------------------------------------------------------------------------------

\begin{figure}
    \small
    \centering
    % \begin{tabular}{@{}c@{\hspace{.9em}}c@{\hspace{.9em}}c@{}}
    %     \makebox[0pt][l]{(a)}\includegraphics[width=.32\linewidth,trim={0cm 0.25cm 0cm 0.2cm},clip]{sdr_gamma_music.pdf} &\makebox[0pt][l]{(b)}\includegraphics[width=.32\linewidth,trim={0cm 0.25cm 0cm 0.2cm},clip]{sdr_gamma.pdf} &\makebox[0pt][l]{(c)}\includegraphics[width=.32\linewidth,trim={0cm 0.25cm 0cm 0.2cm},clip]{activations-alt.pdf}\\
    % \end{tabular}
    \begin{tabular}{cc}
        \includegraphics[width=.45\linewidth,trim={0cm 0.25cm 0cm 0.2cm},clip]{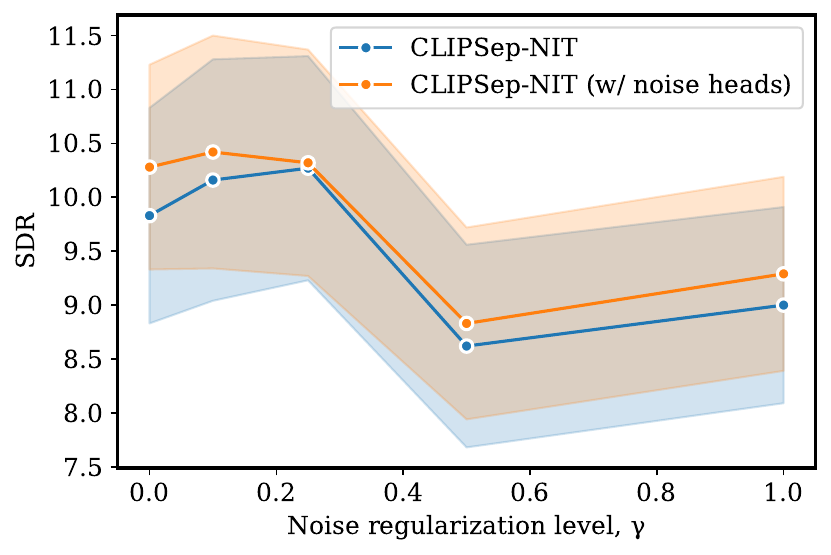} &\includegraphics[width=.45\linewidth,trim={0cm 0.25cm 0cm 0.2cm},clip]{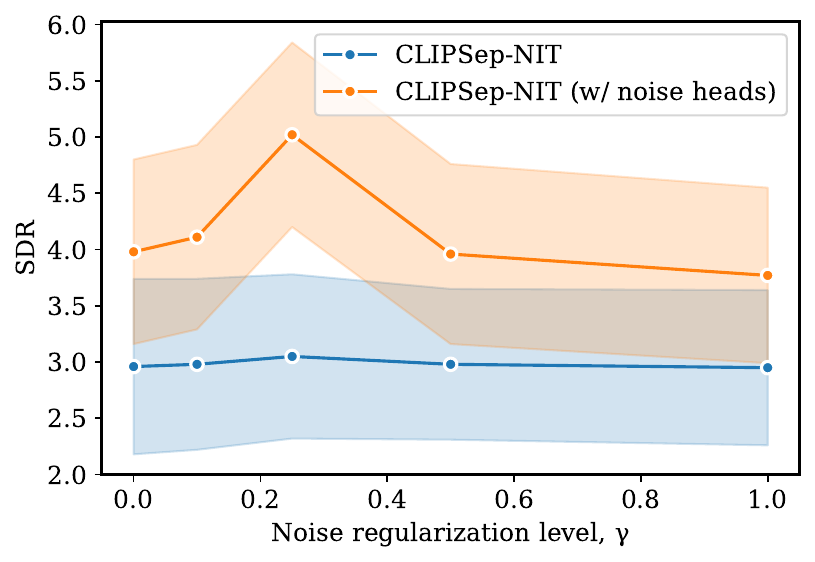}\\
        (a) &(b)
    \end{tabular}\\[2ex]
    \includegraphics[width=.45\linewidth,trim={0cm 0.25cm 0cm 0.2cm},clip]{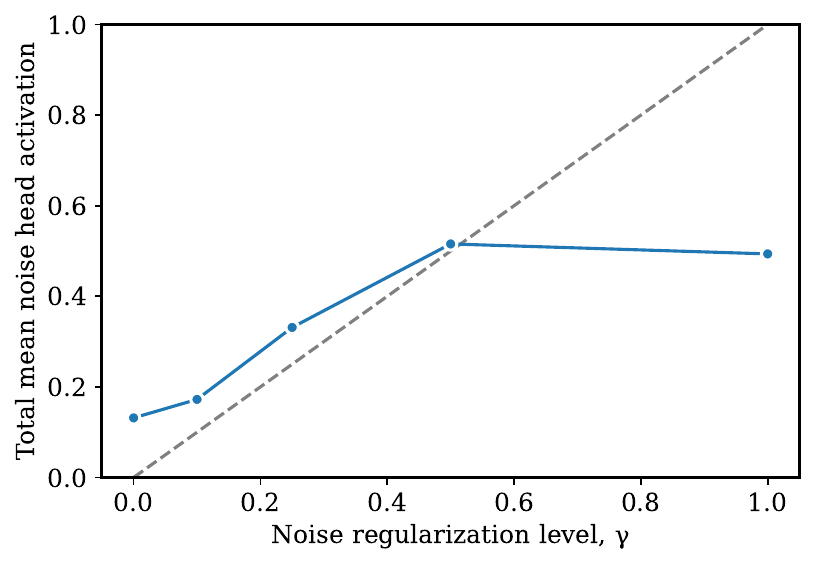}\\
    (c)
    \caption[Effects of the noise regularization level $\gamma$ for the proposed CLIPSep-NIT model---mean SDR for the (a) MUSIC$^+$ and (b) VGGSound-Clean$^+$ evaluations, and (c) the total mean noise head activation, $\sum_{i = 1}^n \mathrm{mean}(\hat{M}^N_i)$, on the validation set.]{Effects of the noise regularization level $\gamma$ for the proposed CLIPSep-NIT model---mean SDR for the (a) MUSIC$^+$ and (b) VGGSound-Clean$^+$ evaluations, and (c) the total mean noise head activation, $\sum_{i = 1}^n \mathrm{mean}(\hat{M}^N_i)$, on the validation set. The shaded areas show standard errors.}
    \label{clipsep:fig:gamma-exp}
\end{figure}

In this experiment, we examine the effects of the noise regularization level $\gamma$ in \cref{clipsep:eq:regularization} by changing the value from 0 to 1. As we can see from \cref{clipsep:fig:gamma-exp}~(a) and (b), CLIPSep-NIT with $\gamma = 0.25$ achieves the highest SDR on both evaluation settings. This suggests that the optimal  $\gamma$ value is not sensitive to the evaluation dataset. Further, we also report in \cref{clipsep:fig:gamma-exp}~(c) the total mean noise head activation, $\sum_{i = 1}^n \mathrm{mean}(\hat{M}^N_i)$, on the validation set. As $\hat{M}^N_i$ is the mask estimate for the noise, the total mean noise head activation value indicates to what extent signals are assigned to the noise head. We observe that the proposed regularizer successfully keeps the total mean noise head activation close to the desired level, $\gamma$, for $\gamma \leq0.5$. Interestingly, the total mean noise head activation is still around 0.5 when $\gamma = 1.0$, suggesting that the model inherently tries to use both the query-heads and the noise heads to predict the noisy target sounds. Moreover, while we discard the noise heads during evaluation in our experiments, keeping the noise heads can lead to a higher SDR as shown in \cref{clipsep:fig:gamma-exp}~(a) and (b), which can be helpful in certain use cases where a post-processing procedure similar to the PIT model \citep{yu2017pit} is acceptable.

%====================
\section{Discussions}
%====================

For the experiments presented in this paper, we work on labeled datasets so that we can evaluate the performance of the proposed models. However, our proposed models do not require any labeled data for training, and can thus be trained on larger unlabeled video collections in the wild. Moreover, we observe that the proposed model shows the capability of combing multiple queries, e.g., ``\textit{a photo of [query A] and [query B]},'' to extract multiple target sounds, and we report the results on the demo website. This offers a more natural user interface against having to separate each target sound and mix them via an additional post-processing step. We also show in \cref{clipsep:sec:queries-exp} that our proposed model is robust to different text queries and can extract the desired sounds.

In our experiments, we often observe a modality transfer gap greater than 1 dB difference of SDR. A future research direction is to explore different approaches to reduce the modality transfer gap. For example, the CLIP model is pretrained on a different dataset, and thus finetuning the CLIP model on the target dataset can help improve the underlying modality transferability within the CLIP model. Further, while the proposed noise invariant training is shown to improve the training on noisy data and reduce the modality transfer gap, it still requires a sufficient audio-visual correspondence for training video. In other words, if the audio and images are irrelevant in most videos, the model will struggle to learn the correspondence between the query and target sound. In practice, we find that the data in the VGGSound dataset often contains off-screen sounds and the labels sometimes correspond to only part of the video content. Hence, filtering on the training data to enhance its audio-visual correspondence can also help reduce the modality transfer gap. This can be achieved by self-supervised audio-visual correspondence prediction \citep{Arandjelovic2017L3,Arandjelovic2018} or temporal synchronization \citep{Korbar2018AVTS,Owens2018AVSA}. 

Another future direction is to explore the semi-supervised setting where a small subset of labeled data can be used to improve the modality transferability. We can also consider the proposed method as a pretraining on unlabeled data for other separation tasks in the low-resource regime. We include in \cref{clipsep:sec:ft-esc50} a preliminary experiment in this aspect using the ESC-50 dataset \citep{piczak2015esc}.

%===================
\section{Conclusion}
%===================

In this work, we have presented a novel text-queried universal sound separation model that can be trained on noisy unlabeled videos. In this end, we have proposed to use the contrastive image-language pretraining to bridge the audio and text modalities, and proposed the noise invariant training for training a query-based sound separation model on noisy data. We have shown that the proposed models can learn to separate an arbitrary sound specified by a text query out of a mixture, even achieving competitive performance against a fully supervised model in some settings. We believe our proposed approach closes the gap between the ways humans and machines learn to focus on a sound in a mixture, namely, the multi-modal self-supervised learning paradigm of humans against the supervised learning paradigm adopted by existing label-based machine learning approaches.

%==========================
\section{Acknowledgements}
%==========================

We would like to thank Stefan Uhlich, Giorgio Fabbro and Woosung Choi for their helpful comments during the preparation of this manuscript. We also thank Mayank Kumar Singh for supporting the setup of the subjective test in \cref{clipsep:sec:subjective}. Hao-Wen thank J. Yang and Family Foundation and Taiwan Ministry of Education for supporting his PhD study.

\secbreak

\begin{adjustwidth}{.5in}{.5in}
    \itshape
    \hspace{\parindent}
    This chapter, in full, is a reprint of the material as it appears in ``CLIPSep: Learning Text-queried Sound Separation with Noisy Unlabeled Videos'' by Hao-Wen Dong, Naoya Takahashi, Yuki Mitsufuji, Julian McAuley and Taylor Berg-Kirkpatrick, which was published in the Proceedings of the International Conference on Learning Representations (ICLR) in 2023. The dissertation author was the primary investigator and author of this paper.
\end{adjustwidth}

\clearpage

\begin{subappendices}

%======================================================
\section{Inference Pipeline of CLIPSep and CLIPSep-NIT}
%======================================================
\label{clipsep:sec:inference}

\begin{figure}
    \centering
    \includegraphics[width=\linewidth]{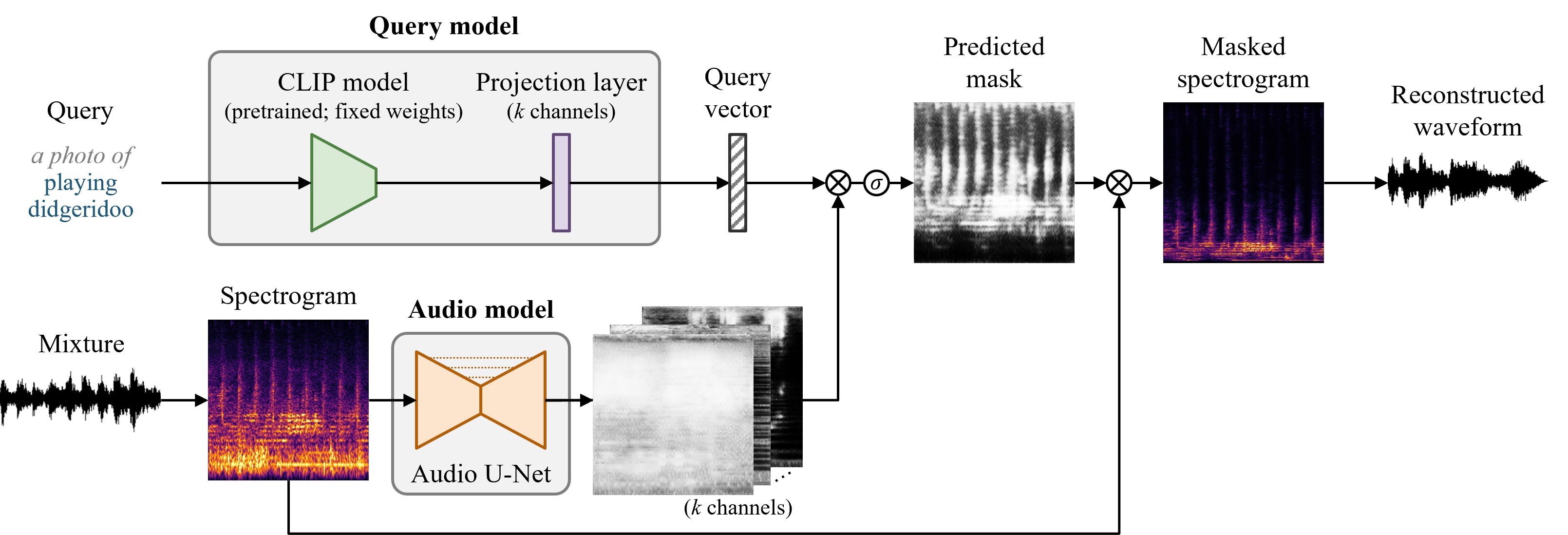}
    \caption{Inference pipeline of the proposed CLIPSep and CLIPSep-NIT models. Note that the mixture spectrogram and the masks are shown in log frequency scale, while the masked spectrogram is shown in linear frequency scale.}
    \label{clipsep:fig:clipsep-inference}
\end{figure}

\cref{clipsep:fig:clipsep-inference} illustrates the inference pipeline for the proposed CLIPSep and CLIPSep-NIT models. For the CLIPSep-NIT model, we discard the noise heads at test time. The masked spectrogram is combined with the input phase and converted to the waveform by the inverse short-time Fourier transform (STFT).

%==========================
\section{Query Ensembling}
%==========================
\label{clipsep:sec:queries}

\cite{radford2021clip} suggest that using a \textit{prompt template} in the form of ``\textit{a photo of [user input query]}'' helps bridge the distribution gap between text queries used for zero-shot image classification and text in the training dataset for the CLIP model. They further show that the ensemble of various prompt templates improve the generalizability. Motivated by this observation, we adopt a similar idea and use several query templates at test time (see \cref{clipsep:tab:query-templates}). These query templates are heuristically chosen to handle the noisy images extracted from videos.

%===============================
\section{Implementation Details}
%===============================
\label{clipsep:sec:implementation}

We implement the audio model as a 7-layer U-Net \citep{ronneberger2015unet}. We use $k = 32$. We use binary masks as the ground truth masks during training while using the raw, real-valued masks for evaluation. We train all the models for 200,000 steps with a batch size of 32. We use the Adam optimizer \citep{kingma2015adam} with $\beta_1 = 0.9$, $\beta_2 = 0.999$ and $\epsilon = 10^{-8}$. In addition, we clip the norm of the gradients to 1.0 \citep{zhang2020gradientclipping}. We adopt the following learning rate schedule with a warm-up---the learning rate starts from 0 and grows to $0.001$ after 5,000 steps, and then it linearly drops to $0.0001$ at 100,000 steps and keeps this value thereafter. We validate the model every 10,000 steps using image queries as we do not assume labeled data is available for the validation set. We use a sampling rate of 16,000 Hz and work on audio clips of length 65,535 samples ($\approx$ 4 seconds). During training, we randomly sample a center frame from a video and extract three frames (images) with 1-sec intervals and 4-sec audio around the center frame. During inference, for image-queried models, we extract three frames with 1-sec intervals around the center of the test clip. For the spectrogram computation, we use a filter length of 1024, a hop length of 256 and a window size of 1024 in the short-time Fourier transform (STFT). We resize images extracted from video to a size of 224-by-224 pixels. For the CLIPSep-Hybrid model, we alternatively train the model with text and image queries, i.e., one batch with all image queries and next with all text queries, and so on. We implement all the models using the PyTorch library \citep{paszke2019pytorch}. We compute the signal-to-distortion ratio (SDR) using museval \citep{stoter2018museval}.

\begin{table}
    \centering
    \small
    \caption{Query templates used in our experiments.}
    \label{clipsep:tab:query-templates}
    \begin{tabular}{ll}
        \toprule
        Query template &Example query for the label ``\textit{dog barking}''\\
        \midrule
        a photo of [\textit{user input query}] &a photo of dog barking\\
        a photo of the small [\textit{user input query}] &a photo of the small dog barking\\
        a low resolution photo of a [\textit{user input query}] &a low resolution photo of a dog barking\\
        a photo of many [\textit{user input query}] &a photo of many dog barking\\
        \bottomrule
    \end{tabular}
\end{table}

In our preliminary experiments, we also tried directly predicting the final mask by conditioning the audio model on the query vector. We applied this modification for both SOP and CLIPSep models, however, we observe that this architecture is prone to overfitting. We hypothesize that this is because the audio model is powerful enough to remember the subtle clues in the query vector, which hinder the generalization to a new sound and query. In contrast, the proposed architecture first predicts over-determined masks and then combines them on the basis of the query vector, which avoids the overfitting problem due to the simple fusion step.

%=======================================
\section{Permutation Invariant Training}
%=======================================
\label{clipsep:sec:pit}

\begin{figure}
    \centering
    \includegraphics[width=.9\linewidth]{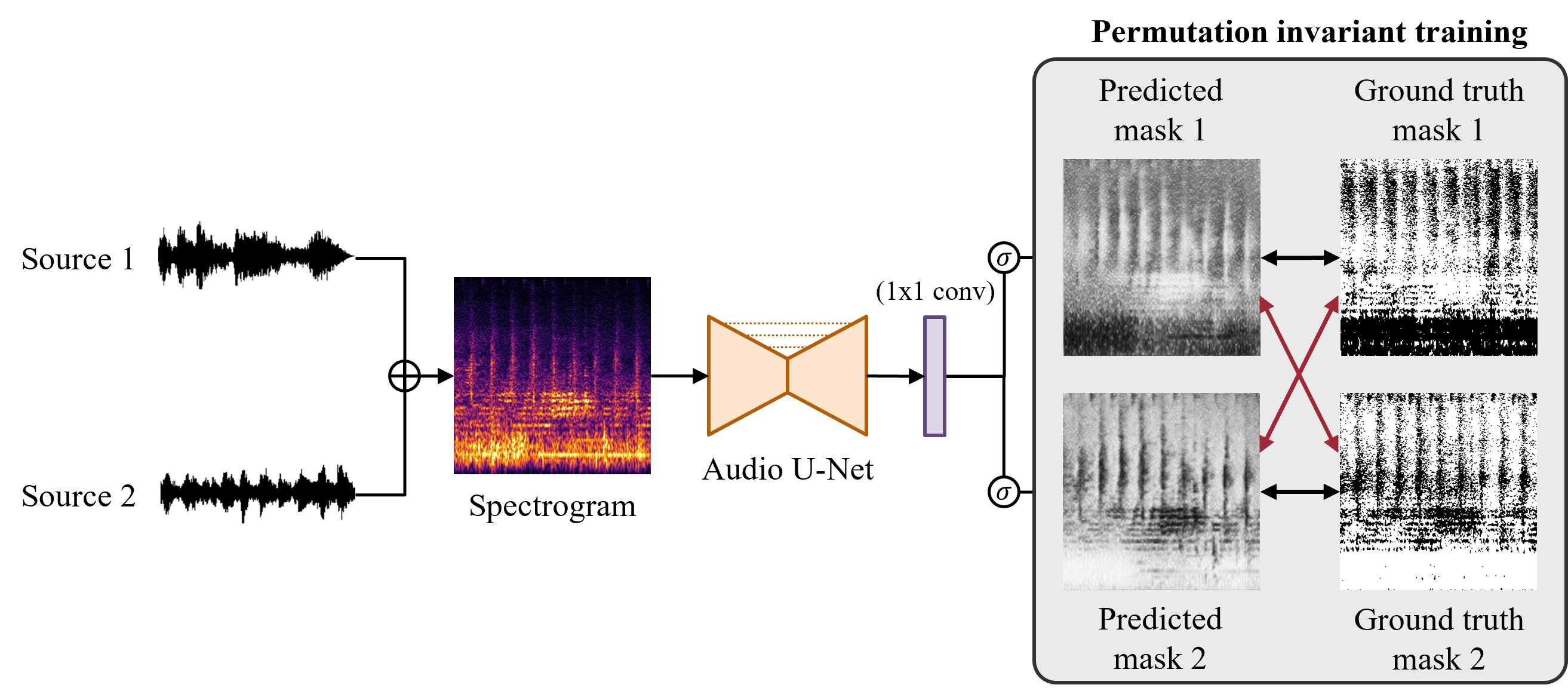}
    \caption{An illustration of the PIT model for $n = 2$. The two predicted masks are interchangeable during the loss computation. Since the two predicted masks are interchangeable, the PIT model requires an additional post-selection step to obtain the target sound.}
    \label{clipsep:fig:pit}
\end{figure}

\cref{clipsep:fig:pit} illustrates the permuatation invariant training (PIT) model \citep{yu2017pit}. The permutation invariant loss is defined as follows for $n = 2$.
\begin{equation}%\small
    \mathcal{L}_\mathit{PIT} = \min \big(\mathit{WBCE}(M_1, \hat{M}_1) + \mathit{WBCE}(M_2, \hat{M}_2), \mathit{WBCE}(M_1, \hat{M}_2) + \mathit{WBCE}(M_2, \hat{M}_1)\big)\,,
\end{equation}
where $\hat{M}_1$ and $\hat{M}_2$ are the predicted masks. Note that the PIT model requires an additional post-selection step to obtain the target sound.

%====================================
\section{Qualitative Example Results}
%====================================
\label{clipsep:sec:examples}

\begin{figure}
    \centering
    \small
    \setlength{\figwidth}{2in}
    \setlength{\figheight}{1in}
    \begin{tabular}{ccc}
        \multirow{1}{*}[1ex]{Mixture} &\multirow{1}{*}[1ex]{Ground truth} &\shortstack{Ground truth\\(Interference)}\\
        \includegraphics[width=\figwidth,height=\figheight]{W3/mix.png} &\includegraphics[width=\figwidth,height=\figheight]{W3/gtamp.png} &\includegraphics[width=\figwidth,height=\figheight]{W3/intamp.png}\\
        &\includegraphics[width=\figwidth,height=\figheight]{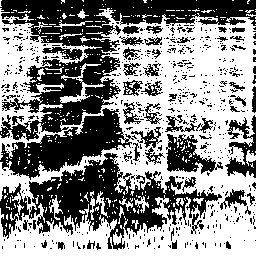} &\includegraphics[width=\figwidth,height=\figheight]{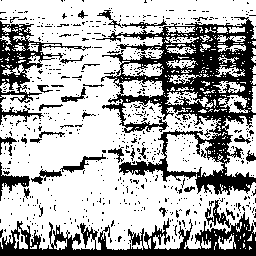}
    \end{tabular}\\[2ex]
    \begin{tabular}{ccc}
        \multirow{1}{*}[1ex]{CLIPSep} &\shortstack{CLIPSep-NIT\\($\gamma = 0.25$)} &\shortstack{PIT\\\notesize\citep{yu2017pit}}\\
        \includegraphics[width=\figwidth,height=\figheight]{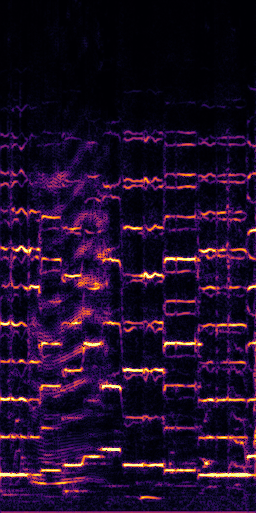} &\includegraphics[width=\figwidth,height=\figheight]{W3/predamp.png}
        &\includegraphics[width=\figwidth,height=\figheight]{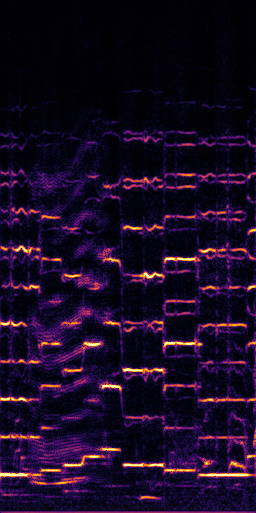}\\
        \includegraphics[width=\figwidth,height=\figheight]{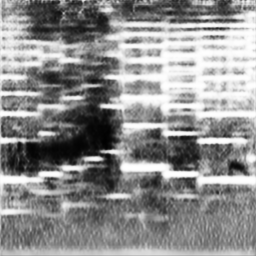} &\includegraphics[width=\figwidth,height=\figheight]{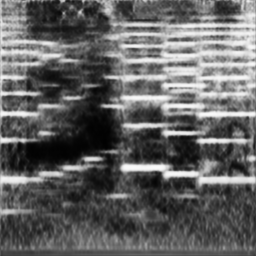} &\includegraphics[width=\figwidth,height=\figheight]{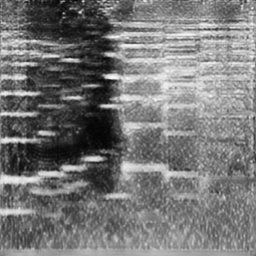}
    \end{tabular}\\[2ex]
    \begin{tabular}{cc}
        \shortstack{Noise head 1\\(CLIPSep-NIT)} &\shortstack{Noise head 2\\(CLIPSep-NIT)}\\
        \includegraphics[width=\figwidth,height=\figheight]{W3/pitmag1.png} &\includegraphics[width=\figwidth,height=\figheight]{W3/pitmag2.png}\\
        \includegraphics[width=\figwidth,height=\figheight]{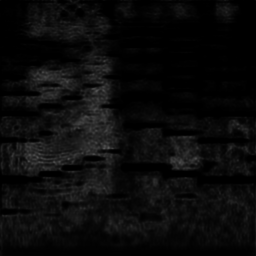} &\includegraphics[width=\figwidth,height=\figheight]{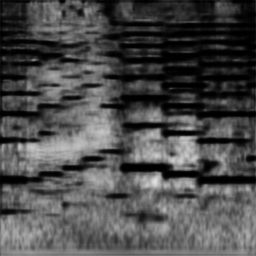}
    \end{tabular}
    \caption{Example results on the MUSIC$^+$ dataset. Target source---``violin''; interference---``people sobbing''; query---``\textit{violin}''. The spectrograms and masks are shown in the log and linear frequency scales, respectively.}
    \label{clipsep:fig:example1}
\end{figure}

\begin{figure}
    \centering
    \small
    \setlength{\figwidth}{2in}
    \setlength{\figheight}{1in}
    \begin{tabular}{ccc}
        \multirow{1}{*}[1ex]{Mixture} &\multirow{1}{*}[1ex]{Ground truth} &\shortstack{Ground truth\\(Interference)}\\
        \includegraphics[width=\figwidth,height=\figheight]{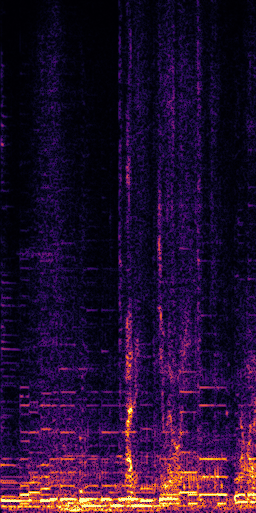} &\includegraphics[width=\figwidth,height=\figheight]{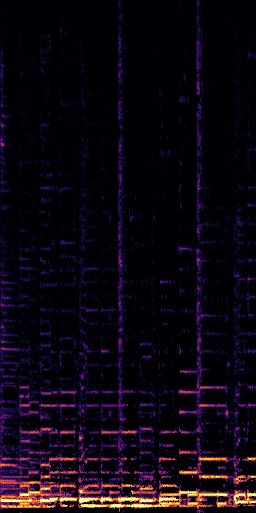} &\includegraphics[width=\figwidth,height=\figheight]{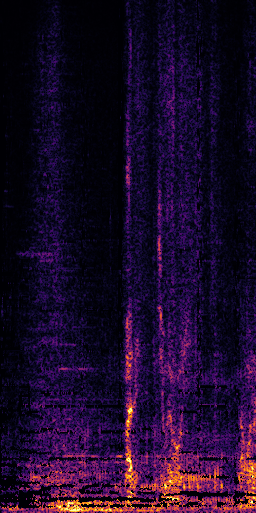}\\
        &\includegraphics[width=\figwidth,height=\figheight]{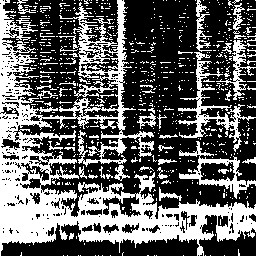} &\includegraphics[width=\figwidth,height=\figheight]{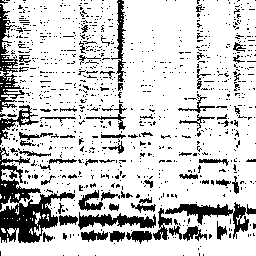}
    \end{tabular}\\[2ex]
    \begin{tabular}{ccc}
        \multirow{1}{*}[1ex]{CLIPSep} &\shortstack{CLIPSep-NIT\\($\gamma = 0.25$)} &\shortstack{PIT\\\notesize\citep{yu2017pit}}\\
        \includegraphics[width=\figwidth,height=\figheight]{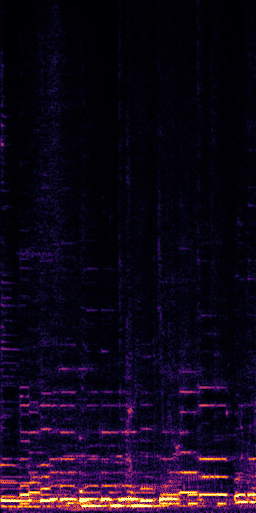} &\includegraphics[width=\figwidth,height=\figheight]{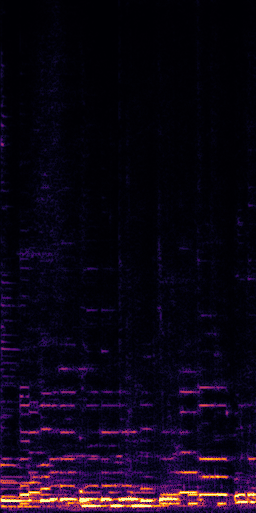}
        &\includegraphics[width=\figwidth,height=\figheight]{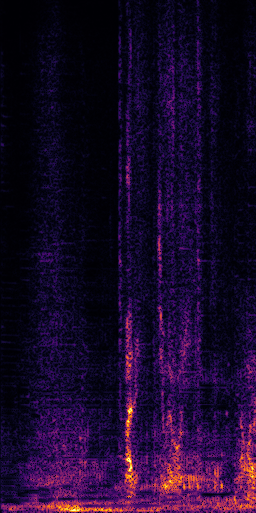}\\
        \includegraphics[width=\figwidth,height=\figheight]{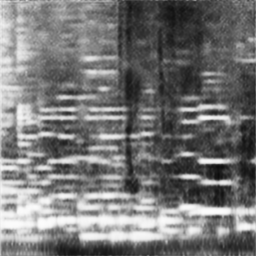} &\includegraphics[width=\figwidth,height=\figheight]{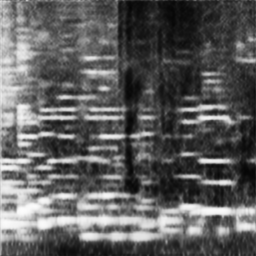} &\includegraphics[width=\figwidth,height=\figheight]{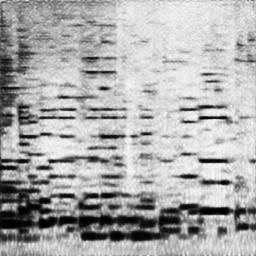}
    \end{tabular}\\[2ex]
    \begin{tabular}{cc}
        \shortstack{Noise head 1\\(CLIPSep-NIT)} &\shortstack{Noise head 2\\(CLIPSep-NIT)}\\
        \includegraphics[width=\figwidth,height=\figheight]{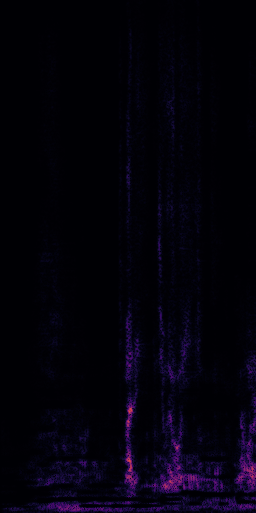} &\includegraphics[width=\figwidth,height=\figheight]{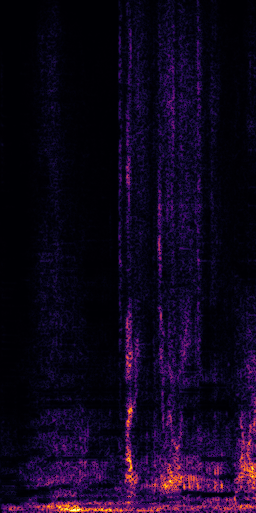}\\
        \includegraphics[width=\figwidth,height=\figheight]{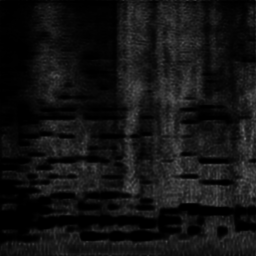} &\includegraphics[width=\figwidth,height=\figheight]{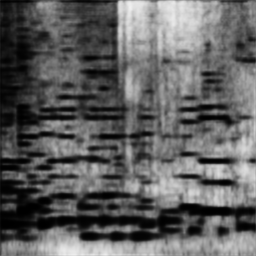}
    \end{tabular}
    \caption{Example results on the MUSIC$^+$ dataset. Target source---``acoustic guitar''; interference---``cheetah chirrup'', query---``\textit{acoustic guitar}''. The spectrograms and masks are shown in the log and linear frequency scales, respectively.}
    \label{clipsep:fig:example2}
\end{figure}

\begin{figure}
    \centering
    \small
    \setlength{\figwidth}{1.5in}
    \setlength{\figheight}{1in}
    \begin{tabular}{ccc}
        \multirow{1}{*}[1ex]{Mixture} &\multirow{1}{*}[1ex]{Ground truth} &\shortstack{Ground truth\\(Interference)}\\
        \includegraphics[width=\figwidth,height=\figheight]{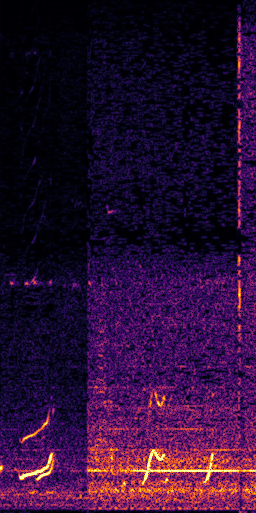} &\includegraphics[width=\figwidth,height=\figheight]{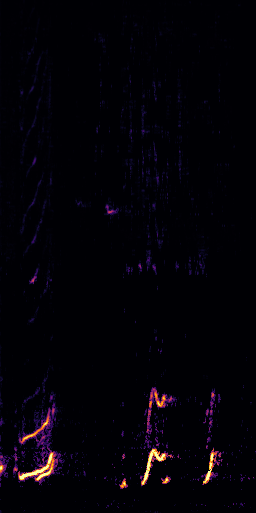} &\includegraphics[width=\figwidth,height=\figheight]{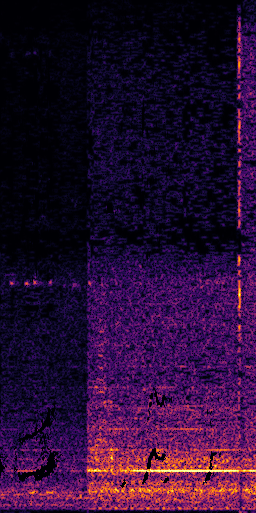}\\
        &\includegraphics[width=\figwidth,height=\figheight]{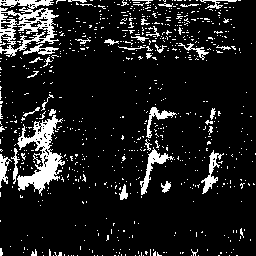} &\includegraphics[width=\figwidth,height=\figheight]{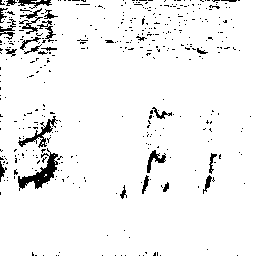}
    \end{tabular}\\[2ex]
    \begin{tabular}{@{}cccc@{}}
        \multirow{1}{*}[1ex]{CLIPSep} &\shortstack{CLIPSep-NIT\\($\gamma = 0.25$)} &\shortstack{PIT\\\notesize\citep{yu2017pit}} &\multirow{1}{*}[1ex]{LabelSep}\\
        \includegraphics[width=\figwidth,height=\figheight]{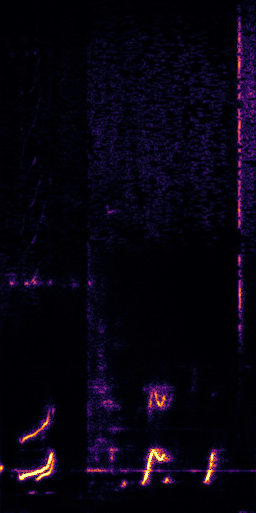} &\includegraphics[width=\figwidth,height=\figheight]{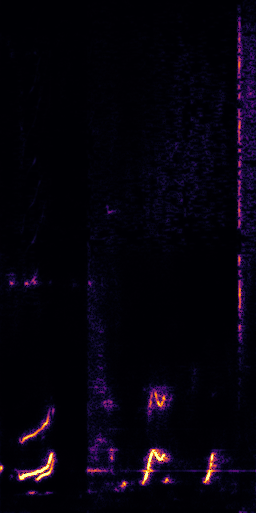}
        &\includegraphics[width=\figwidth,height=\figheight]{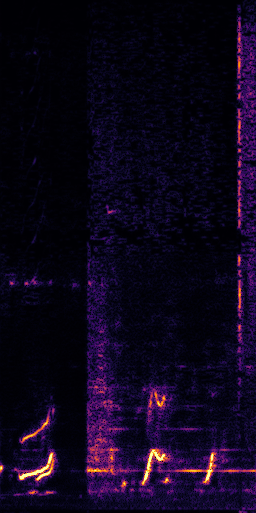} &\includegraphics[width=\figwidth,height=\figheight]{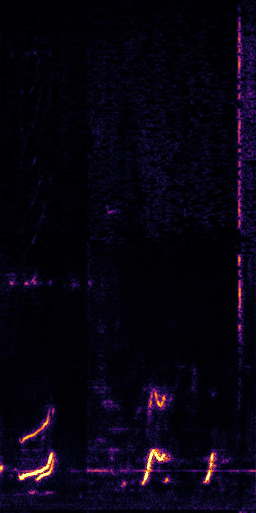}\\
        \includegraphics[width=\figwidth,height=\figheight]{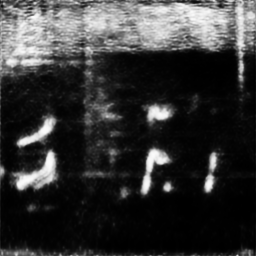} &\includegraphics[width=\figwidth,height=\figheight]{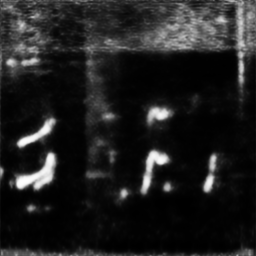} &\includegraphics[width=\figwidth,height=\figheight]{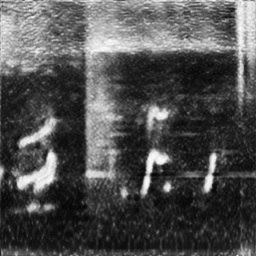} &\includegraphics[width=\figwidth,height=\figheight]{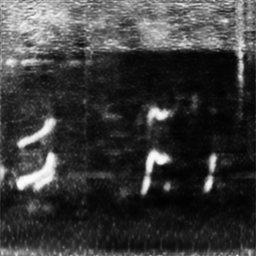}
    \end{tabular}\\[2ex]
    \begin{tabular}{cc}
        \shortstack{Noise head 1\\(CLIPSep-NIT)} &\shortstack{Noise head 2\\(CLIPSep-NIT)}\\
        \includegraphics[width=\figwidth,height=\figheight]{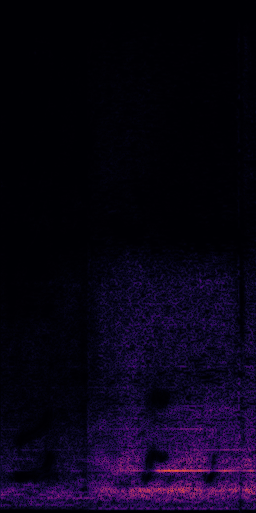} &\includegraphics[width=\figwidth,height=\figheight]{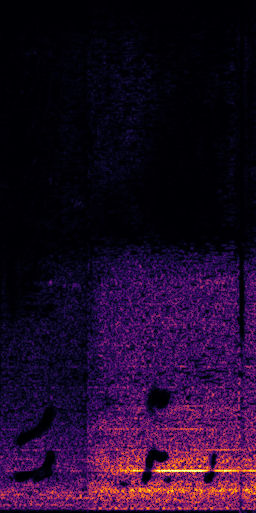}\\
        \includegraphics[width=\figwidth,height=\figheight]{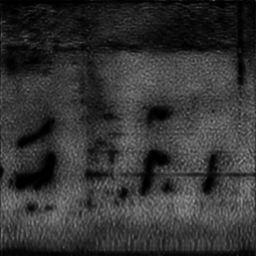} &\includegraphics[width=\figwidth,height=\figheight]{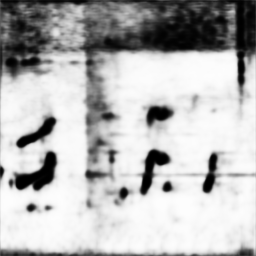}
    \end{tabular}
    \caption{Example results on the VGGSound-Clean$^+$ dataset. Target source---``cat growling''; interference---``railroad car train wagon''; query---``\textit{cat growling}''. The spectrograms and masks are shown in the log and linear frequency scales, respectively.}
    \label{clipsep:fig:example3}
\end{figure}

\begin{figure}
    \centering
    \small
    \setlength{\figwidth}{1.5in}
    \setlength{\figheight}{1in}
    \begin{tabular}{ccc}
        \multirow{1}{*}[1ex]{Mixture} &\multirow{1}{*}[1ex]{Ground truth} &\shortstack{Ground truth\\(Interference)}\\
        \includegraphics[width=\figwidth,height=\figheight]{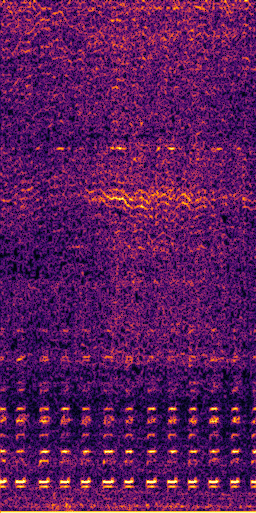} &\includegraphics[width=\figwidth,height=\figheight]{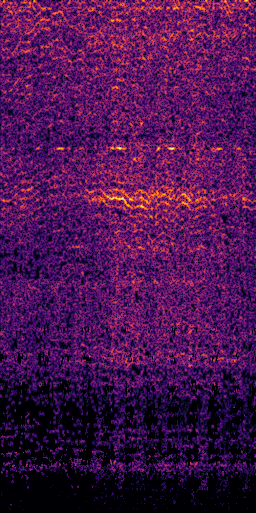} &\includegraphics[width=\figwidth,height=\figheight]{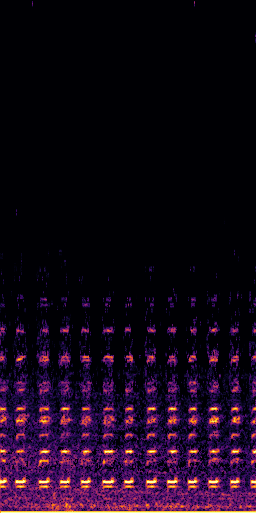}\\
        &\includegraphics[width=\figwidth,height=\figheight]{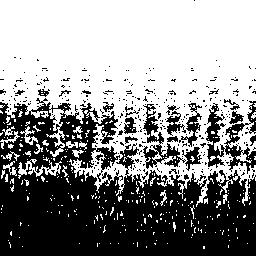} &\includegraphics[width=\figwidth,height=\figheight]{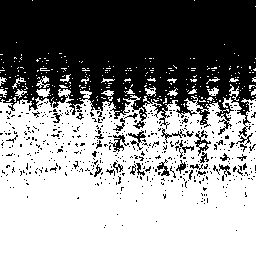}
    \end{tabular}\\[2ex]
    \begin{tabular}{@{}cccc@{}}
        \multirow{1}{*}[1ex]{CLIPSep} &\shortstack{CLIPSep-NIT\\($\gamma = 0.25$)} &\shortstack{PIT\\\notesize\citep{yu2017pit}} &\multirow{1}{*}[1ex]{LabelSep}\\
        \includegraphics[width=\figwidth,height=\figheight]{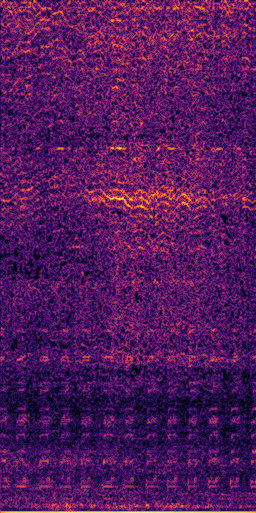} &\includegraphics[width=\figwidth,height=\figheight]{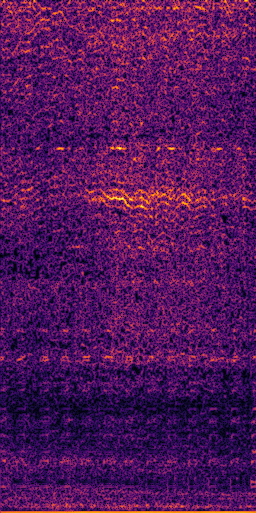}
        &\includegraphics[width=\figwidth,height=\figheight]{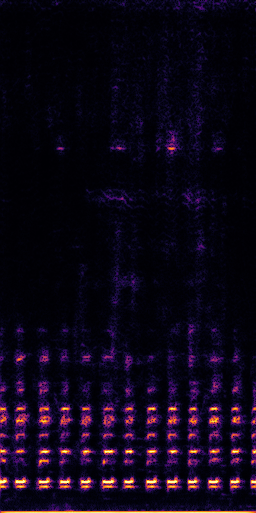} &\includegraphics[width=\figwidth,height=\figheight]{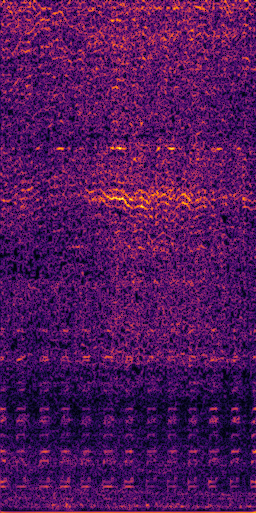}\\
        \includegraphics[width=\figwidth,height=\figheight]{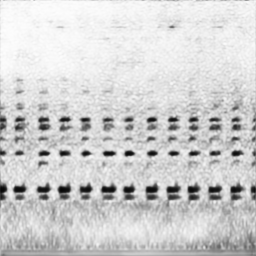} &\includegraphics[width=\figwidth,height=\figheight]{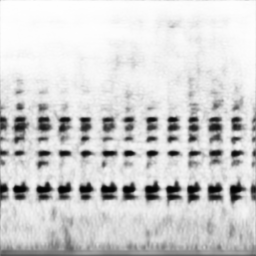} &\includegraphics[width=\figwidth,height=\figheight]{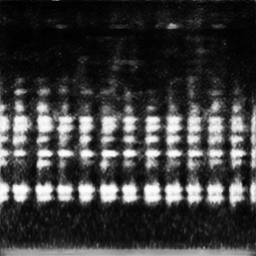} &\includegraphics[width=\figwidth,height=\figheight]{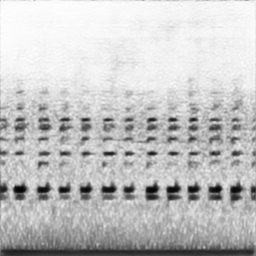}
    \end{tabular}\\[2ex]
    \begin{tabular}{cc}
        \shortstack{Noise head 1\\(CLIPSep-NIT)} &\shortstack{Noise head 2\\(CLIPSep-NIT)}\\
        \includegraphics[width=\figwidth,height=\figheight]{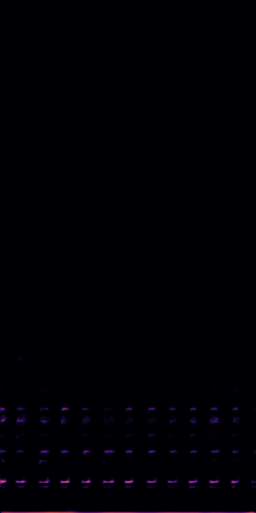} &\includegraphics[width=\figwidth,height=\figheight]{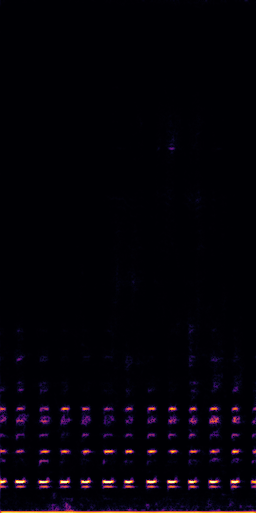}\\
        \includegraphics[width=\figwidth,height=\figheight]{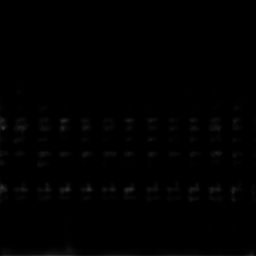} &\includegraphics[width=\figwidth,height=\figheight]{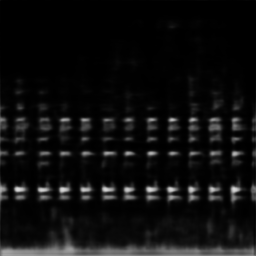}
    \end{tabular}
    \caption{Example results on the VGGSound-Clean$^+$ dataset. Target source---``electric grinder grinding''; interference---``vehicle horn car horn honking''; query---``\textit{electric grinder grinding}''. The spectrograms and masks are shown in the log and linear frequency scales, respectively. Note that the PIT model requires a post-selection step to get the correct source. Without the post-selection step, the PIT model return the right source in only a 50\% chance.}
    \label{clipsep:fig:example4}
\end{figure}

We show in \cref{clipsep:fig:example1,clipsep:fig:example2,clipsep:fig:example3,clipsep:fig:example4} some example results. More results and audio samples can be found at \url{https://sony.github.io/CLIPSep/}.

%==============================
\section{Subjective Evaluation}
%==============================
\label{clipsep:sec:subjective}

We conduct a subjective test to evaluate whether the SDR results aligned with perceptual quality. As done in the Sound of Pixel \citep{zhao2018sop}, separated audio samples are randomly presented to evaluators, and the following question is asked: “Which sound do you hear? 1. A, 2. B, 3. Both, or 4. None of them”. Here A and B are replaced by labels of their mixture sources, e.g. A=accordion, B=engine accelerating. Ten samples (including naturally occurring mixture) are evaluated for each model and 16 evaluators have participated in the evaluation. \cref{clipsep:tab:subj} shows the percentages
of samples which are correctly identified the target sound class (Correct), which are incorrectly identified the target sound sources (Wrong), which are selected as both sounds are audible (Both), and which are  selected as neither of the sounds are audible (None).
The results indicate that the evaluators more often choose the correct sound source for CLIPSep-NIT (83.8\%) than CLIPSep (66.3\%) with text queries. Notably, CLIPSep-NIT with text-query obtained a higher correct score than that with image-query, which matches the training mode. This is probably because image queries often contain information about backgrounds and environments, hence, some noise and off-screen sounds are also suggested by the image-queries and leak to the query head. In contrast, text-queries purely contain the information of target sounds, thus, the query head more aggressively extract the target sounds.

\begin{table}
    \centering
    \small
    \caption{Subjective test results.}
    \label{clipsep:tab:subj}
    \begin{tabular}{lccccc}
        \toprule
        Model &Query type &Correct [\%] & Wrong [\%] &Both [\%]    &None [\%]\\
        \midrule
        Mixture &-  &17.5	&10.0	&72.5	&0.0\\
        \cmidrule(lr){1-6}
        \multirow{2}{*}{CLIPSep} &Image &70.6	&0.0	&29.4	&0.0\\
         &Text &66.3	&3.8	&30.0	&0.0\\
        \cmidrule(lr){1-6}
        \multirow{2}{*}{CLIPSep-NIT} &Image  &68.8	&1.9	&28.1	&1.3\\
         &Text   &83.8	&0.6	&15.0	&0.6\\
        \bottomrule
    \end{tabular}
\end{table}

%========================================
\section{Robustness to different queries}
%========================================
\label{clipsep:sec:queries-exp}

\begin{figure}
    \centering
    \small
    \setlength{\figwidth}{2in}
    \setlength{\figheight}{1.5in}
    \begin{tabular}{ccc}
        \multirow{1}{*}[1ex]{Mixture} &\multirow{1}{*}[1ex]{Ground truth} &\shortstack{Ground truth\\(Interference)}\\
        \includegraphics[width=\figwidth,height=\figheight]{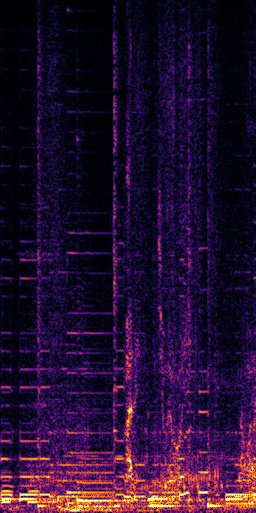} &\includegraphics[width=\figwidth,height=\figheight]{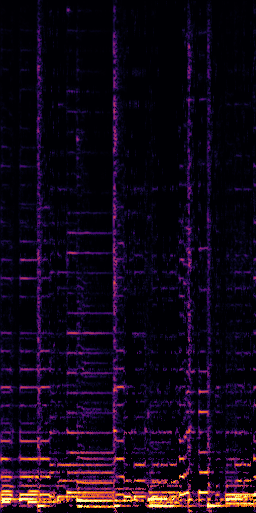} &\includegraphics[width=\figwidth,height=\figheight]{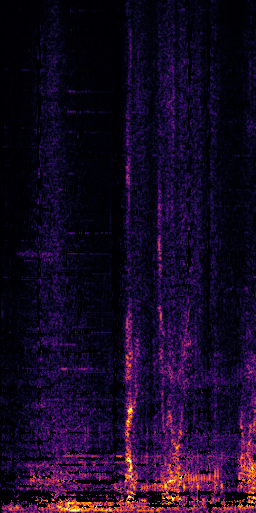}
    \end{tabular}\\[4ex]
    \begin{tabular}{ccc}
        \shortstack{Prediction\\\notesize(Query: "\textit{acoustic guitar}")\\[.5ex]} &\shortstack{Prediction\\\notesize(Query: "\textit{guitar}")\\[.5ex]} &\shortstack{Prediction\\\notesize(Query: "\textit{a man is playing}\\\textit{acoustic guitar}")}\\
        \includegraphics[width=\figwidth,height=\figheight]{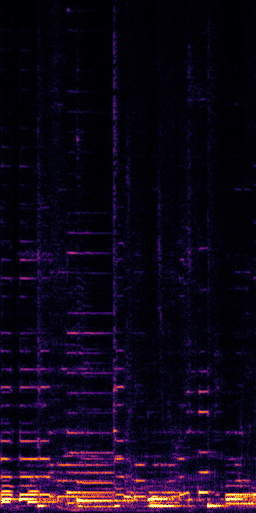} &\includegraphics[width=\figwidth,height=\figheight]{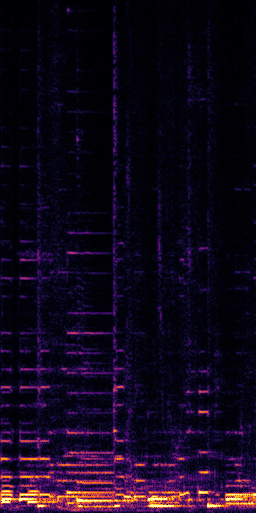}
        &\includegraphics[width=\figwidth,height=\figheight]{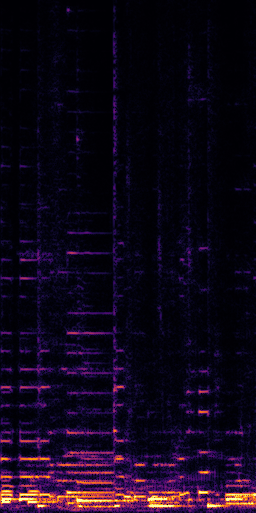}
    \end{tabular}\\[4ex]
    \begin{tabular}{cc}
        \shortstack{Prediction\\\notesize(Query: "\textit{a man is playing}\\\textit{acoustic guitar in a room}")} &\shortstack{Prediction\\\notesize(Query: "\textit{car engine}")\\[.5ex]}\\
        \includegraphics[width=\figwidth,height=\figheight]{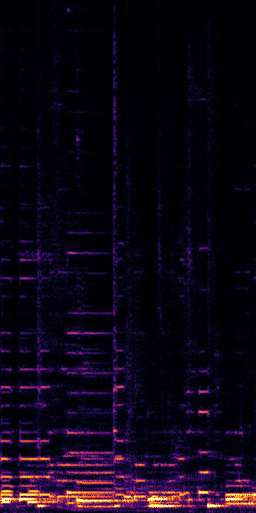} &\includegraphics[width=\figwidth,height=\figheight]{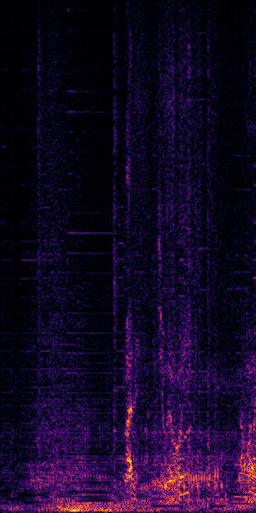}
    \end{tabular}
    \caption{Query robustness experiment on the MUSIC$^+$ dataset. Target source---``acoustic guitar''; interference---``cheetah chirrup''. The spectrograms are shown in the log frequency scale.}
    \label{clipsep:fig:queries}
\end{figure}

To examine the model’s robustness to different queries, we take the same input mixture and query the model with different text queries. We use the CLIPSep-NIT model on the MUSIC$^+$ dataset and report in \cref{clipsep:fig:queries} the results. We see that the model is robust to different text queries and can extract the desired sounds. Audio samples can be found at \url{https://sony.github.io/CLIPSep/}.

%=====================================================
\section{Finetuning Experiments on the ESC-50 Dataset}
%=====================================================
\label{clipsep:sec:ft-esc50}

In this experiment, we aim to examine the possibilities of having a clean dataset for further finetuning. We consider the ESC-50 dataset \citep{piczak2015esc}, a collection of 2,000 high-quality environmental audio recordings, as the clean dataset here.\footnote{\url{https://github.com/karolpiczak/ESC-50}} We report the experimental results in \cref{clipsep:tab:esc50-finetuning}. We can see that the model pretrained on VGGSound does not generalize well to the ESC-50 dataset as the ESC-50 contains much cleaner sounds, i.e., without query-irrelevant sounds and background noise. Further, if we train the CLIPSep model from scratch on the ESC-50 dataset, it can only achieve a mean SDR of 5.18 dB and a median SDR of 5.09 dB. However, if we take the model pretrained on the VGGSound dataset and finetune it on the ESC-50 dataset, it can achieve a mean SDR of 6.73 dB and a median SDR of 4.89 dB, resulting in an improvement of 1.55 dB on the mean SDR.

\begin{table}
    \centering
    \small
    \caption{Results on the ESC-50 dataset. Standard errors are reported in the mean SDR column.}
    \label{clipsep:tab:esc50-finetuning}
    \begin{tabular}{lccccc}
        \toprule
        \multirow{2}{*}{Model} &\multirow{2}{*}[-1ex]{\shortstack{Post-processing\\free}} &\multicolumn{2}{c}{Dataset} &\multicolumn{2}{c}{SDR}\\
        \cmidrule(lr){3-4} \cmidrule(lr){5-6}
        &&Training &Finetuning &Mean &Median\\
        \midrule
        Mixture &- &- &- &0.00 $\pm$ 0.44 &0.00\\
        \cmidrule(lr){1-6}
        PIT &$\times$ &VGGSound &- &4.90 $\pm$ 0.26 &2.44\\
        \cmidrule(lr){1-6}
        \multirow{3}{*}{CLIPSep} &$\checkmark$ &VGGSound &- &1.07 $\pm$ 0.28 &2.34\\
        &$\checkmark$ &ESC-50 &- &5.18 $\pm$ 0.26 &5.09\\
        &$\checkmark$ &VGGSound &ESC-50 &6.73 $\pm$ 0.26 &5.89\\
        \bottomrule
    \end{tabular}
\end{table}

%===========================
\section{Training Behaviors}
%===========================
\label{clipsep:sec:training}

\begin{figure}
    \centering
    \begin{tabular}{@{}c@{}c@{}c@{}}
        \includegraphics[width=.33\linewidth]{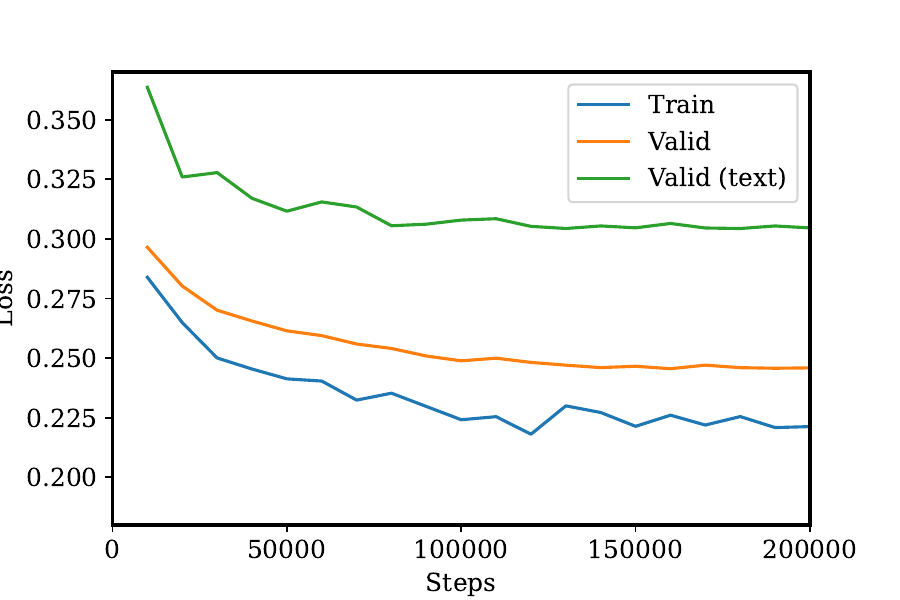} &\includegraphics[width=.33\linewidth]{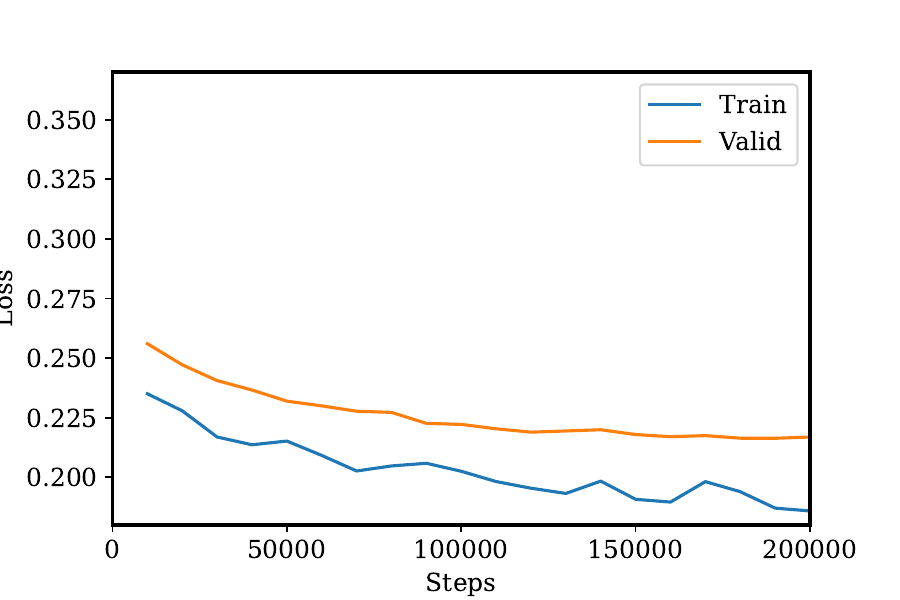} &\includegraphics[width=.33\linewidth]{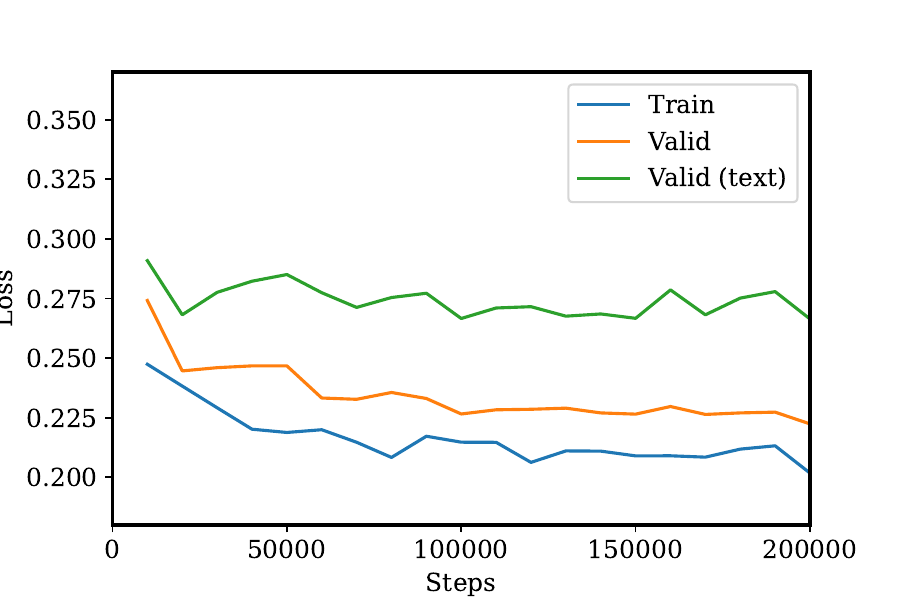} \\
        (a) CLIPSep &(b) PIT &(c) CLIPSep-NIT {\small(unregularized)}\\[1ex]
        \includegraphics[width=.33\linewidth]{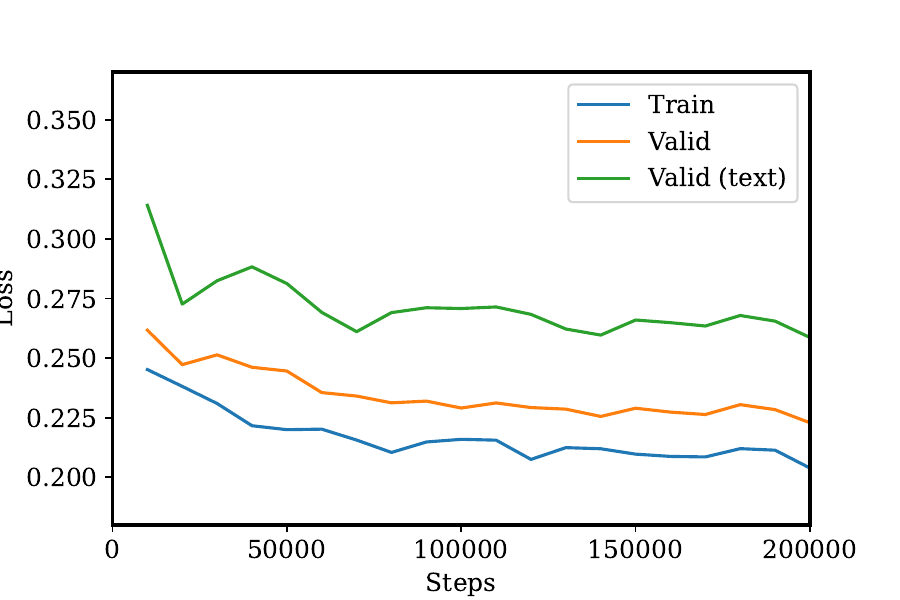} &\includegraphics[width=.33\linewidth]{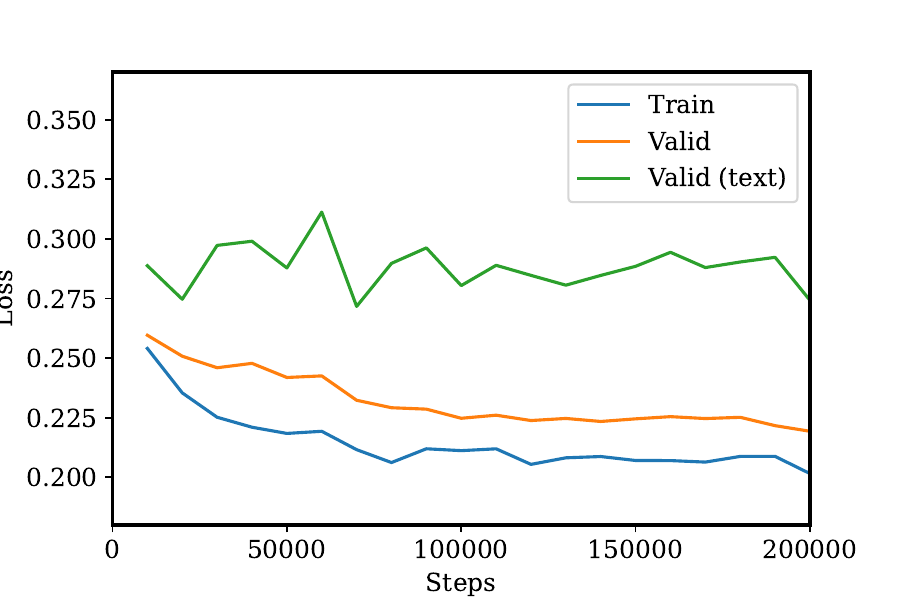} &\includegraphics[width=.33\linewidth]{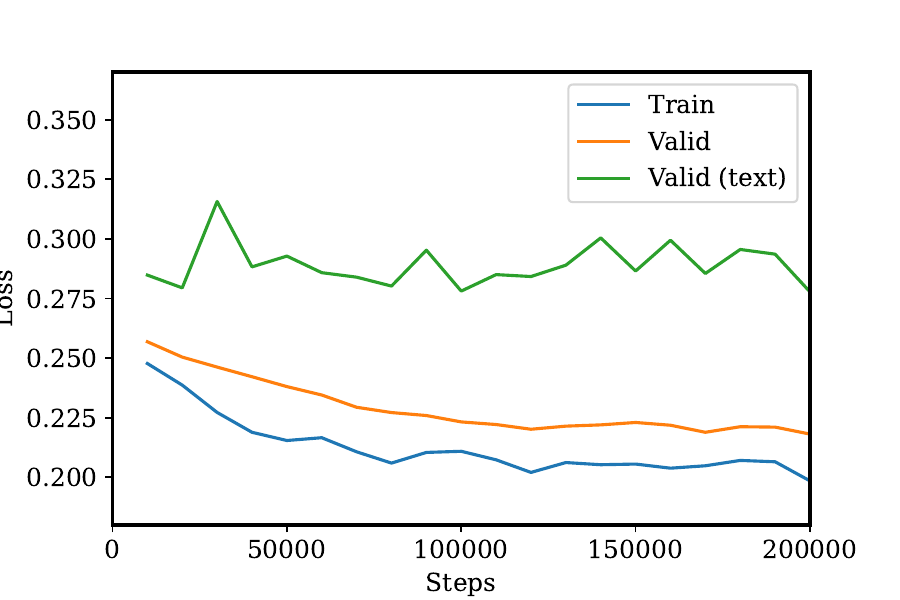}\\
        (d) CLIPSep-NIT ($\gamma = 0.5$) &(e) CLIPSep-NIT ($\gamma = 0.25$) &(f) CLIPSep-NIT ($\gamma = 0.1$)
    \end{tabular}
    \caption{Training and validation losses along the training progress on the VGGSound dataset. We also include the losses computed using text queries instead of image queries. The y-axes are intentionally set to the same range for easy comparison. Note that we do not use the validation results obtained with text queries for choosing the best model.}
    \label{clipsep:fig:losses}
\end{figure}

\begin{figure}
    \centering
    \begin{tabular}{@{}c@{}c@{}c@{}}
        \includegraphics[width=.33\linewidth]{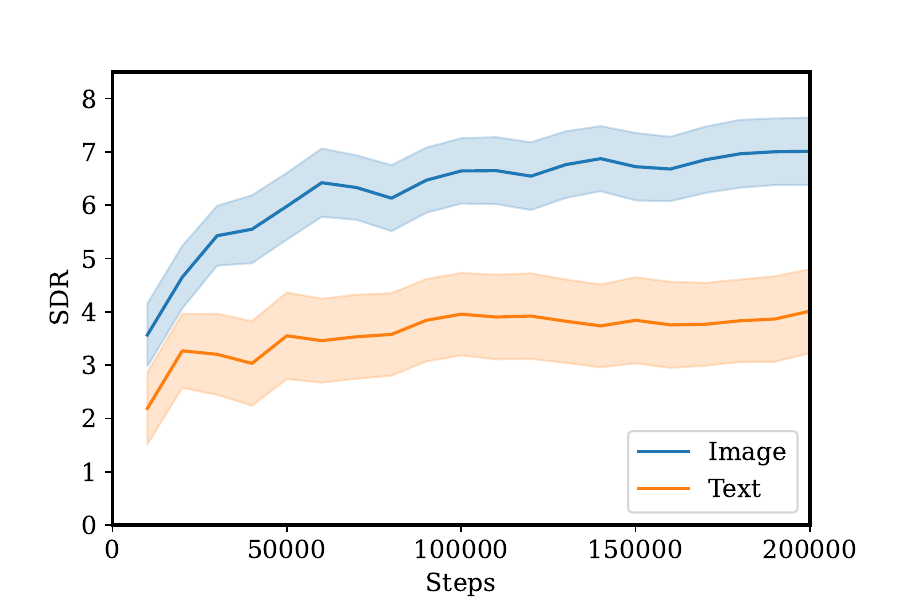} &\includegraphics[width=.33\linewidth]{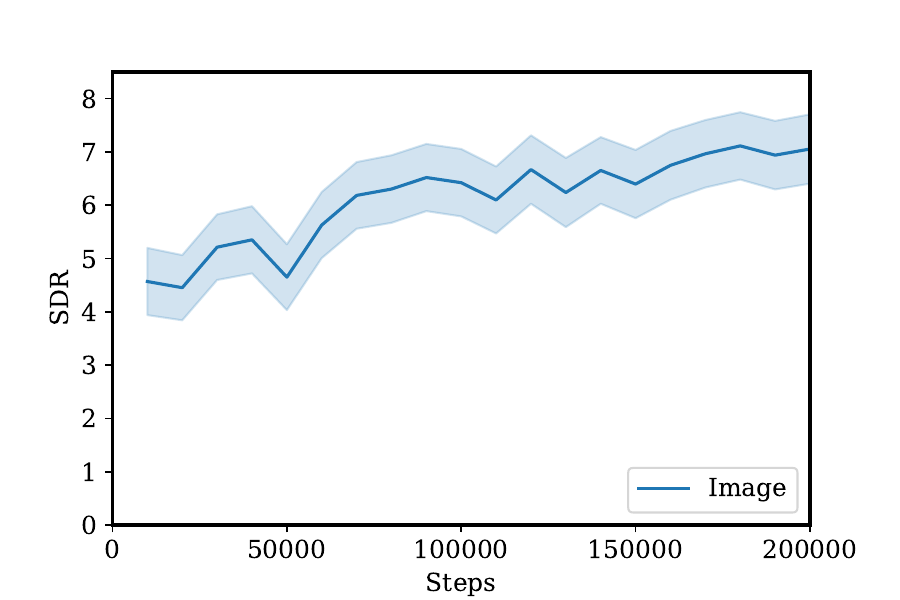} &\includegraphics[width=.33\linewidth]{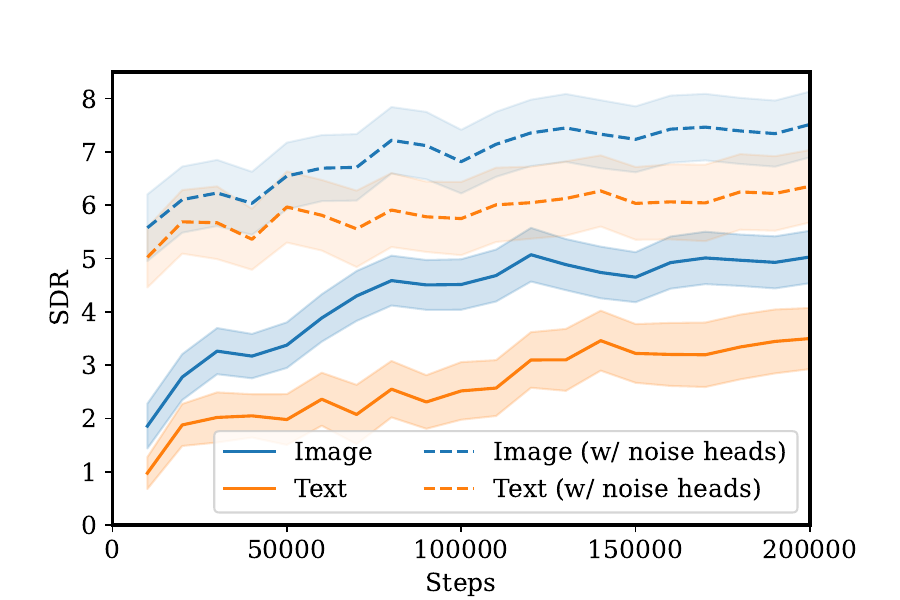}\\
        (a) CLIPSep &(b) PIT &(c) CLIPSep-NIT {\small(unregularized)}\\[1ex]
        \includegraphics[width=.33\linewidth]{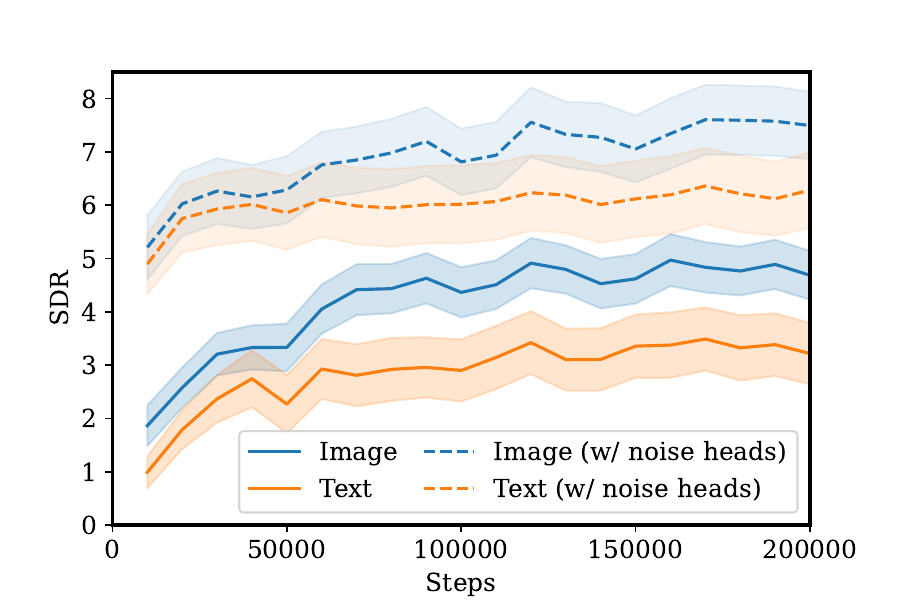} &\includegraphics[width=.33\linewidth]{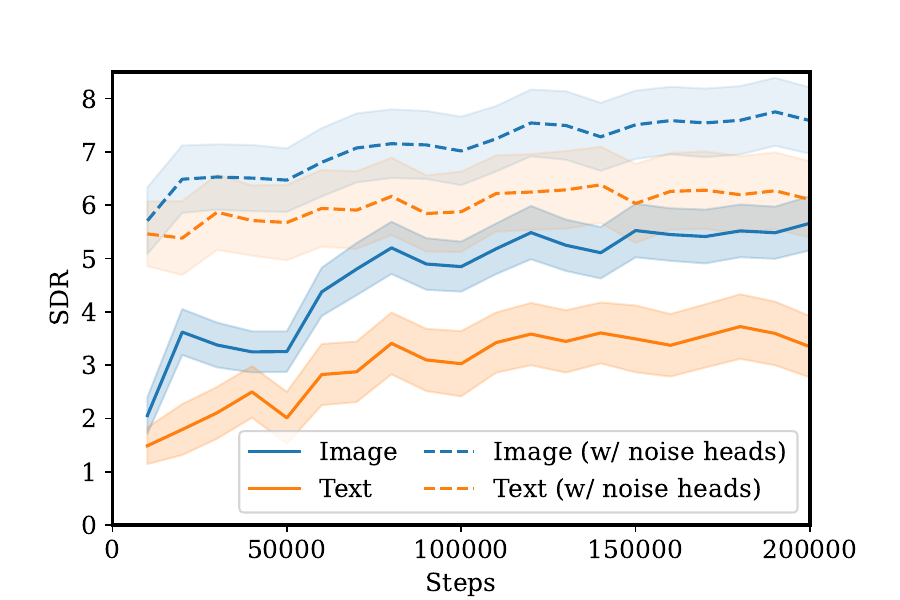} &\includegraphics[width=.33\linewidth]{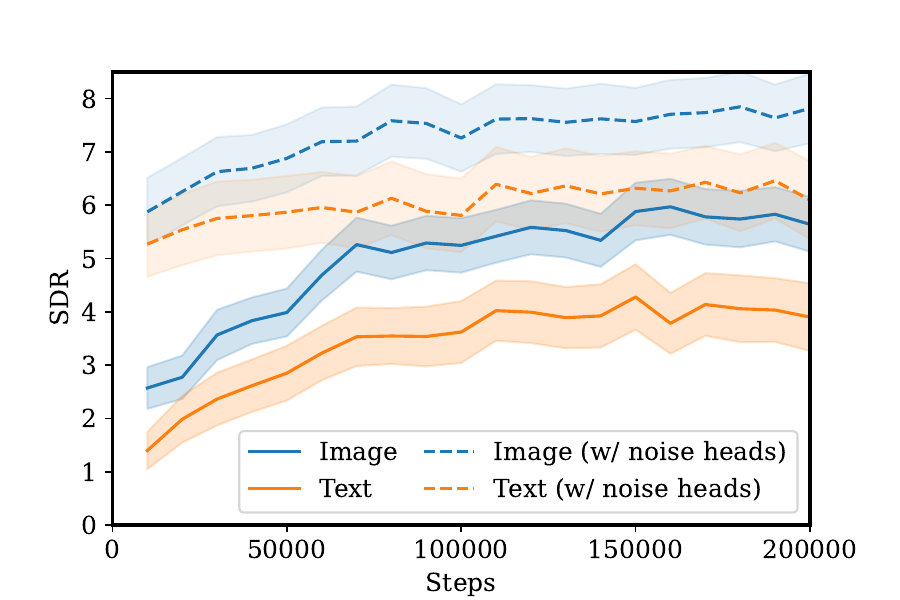}\\
        (d) CLIPSep-NIT ($\gamma = 0.5$) &(e) CLIPSep-NIT ($\gamma = 0.25$) &(f) CLIPSep-NIT ($\gamma = 0.1$)
    \end{tabular}
    \caption{Test SDR along the training progress on the VGGSound-Clean dataset. The y-axes are intentionally set to the same range for easy comparison.}
    \label{clipsep:fig:sdr}
\end{figure}

\begin{figure}
    \centering
    \includegraphics[width=.6\linewidth]{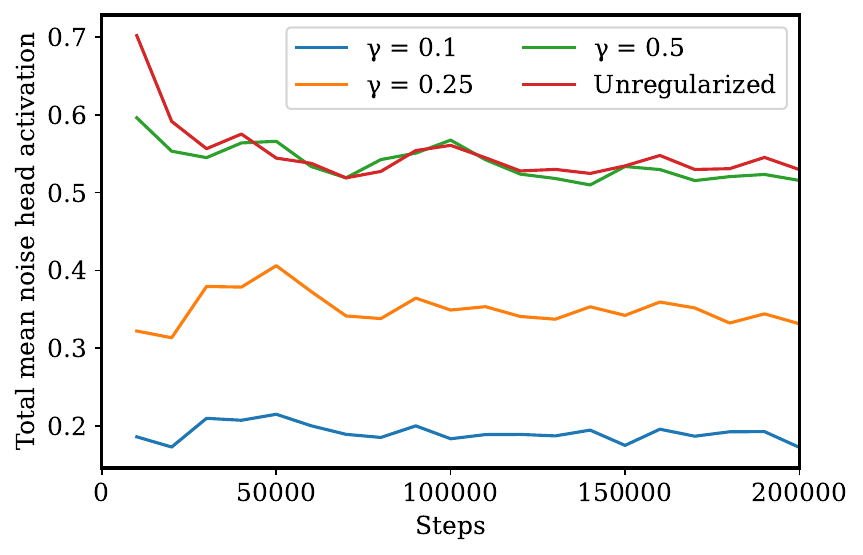}
    \caption{Total mean noise head activation, $\sum_{i = 1}^n \mathrm{mean}(\hat{M}^N_i)$, on the validation set for the CLIPSep-NIT models along the training progress.}
    \label{clipsep:fig:activations-history}
\end{figure}

We present in \cref{clipsep:fig:losses} the training and validation losses along the training progress. Please note that we only show the results obtained using text queries for reference but do not use them for choosing the best model. We also evaluate the intermediate checkpoints every 10,000 steps and present in \cref{clipsep:fig:sdr} the test SDR along the training progress. In addition, for the CLIPSep-NIT model, we visualize in \cref{clipsep:fig:activations-history} the total mean noise head activation, $\sum_{i = 1}^n \mathrm{mean}(\hat{M}^N_i)$, along the training progress. We can see that the total mean noise head activation stays around the desired level for $\gamma = 0.1, 0.25$. For $\gamma = 0.5$ and the unregularized version, the total mean noise head activation converges to a similar value around 0.55.

\end{subappendices}

\graphicspath{{chapters/clipsonic/figs/}}
\chapter{CLIPSonic: Text-to-Audio Synthesis with Unlabeled Videos and Pretrained Language-Vision Models}
\label{chap:clipsonic}

\begin{abstract}
Recent work has studied text-to-audio synthesis using large amounts of paired text-audio data. However, audio recordings with high-quality text annotations can be difficult to acquire. In this work, we approach text-to-audio synthesis using unlabeled videos and pretrained language-vision models. We propose to learn the desired text-audio correspondence by leveraging the visual modality as a bridge. We train a conditional diffusion model to generate the audio track of a video, given a video frame encoded by a pretrained contrastive language-image pretraining (CLIP) model. At test time, we first explore performing a zero-shot modality transfer and condition the diffusion model with a CLIP-encoded text query. However, we observe a noticeable performance drop with respect to image queries. To close this gap, we further adopt a pretrained diffusion prior model to generate a CLIP image embedding given a CLIP text embedding. Our results show the effectiveness of the proposed method, and that the pretrained diffusion prior can reduce the modality transfer gap. While we focus on text-to-audio synthesis, the proposed model can also generate audio from image queries, and it shows competitive performance against a state-of-the-art image-to-audio synthesis model in a subjective listening test. This study offers a new direction of approaching text-to-audio synthesis that leverages the naturally-occurring audio-visual correspondence in videos and the power of pretrained language-vision models.
\end{abstract}

%=====================
\section{Introduction}
%=====================
\label{clipsonic:sec:intro}

\begin{figure}
    \centering
    \includegraphics[width=0.8\linewidth]{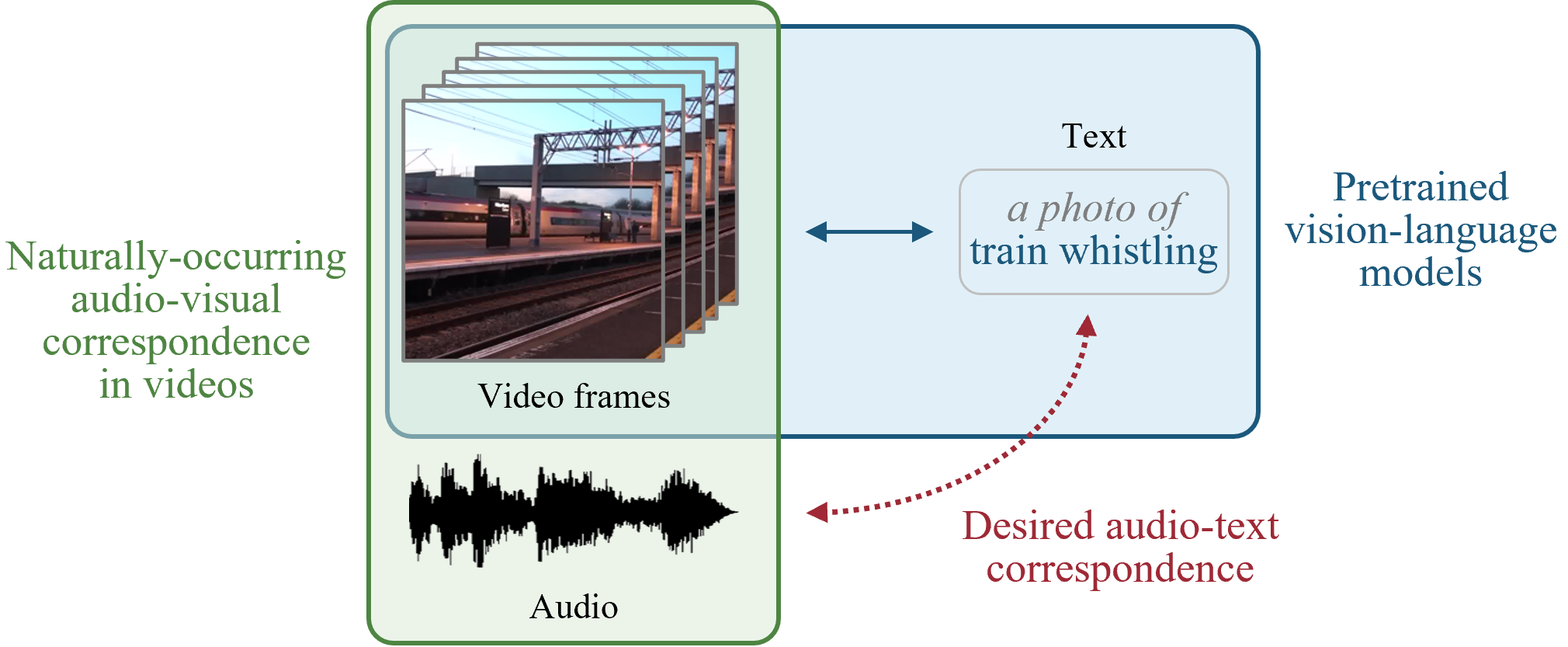}
    \caption{We learn the text-audio correspondence by leveraging the audio-visual correspondences in videos and the multimodal representation learned by pretrained language-vision models.}
    \label{clipsonic:fig:bridge}
\end{figure}

With the advance of generative modeling \citep{radford2019gpt2,ho2020ddpm,rombach2022ldm} and language-audio contrastive learning \citep{wu2022clap,huang2022mulan,guzhov2022audioclip}, various deep learning-based text-to-audio synthesis systems have recently emerged \citep{yang2022diffsound,kreuk2022audiogen,liu2023audioldm,huang2023makeanaudio,huang2023noise2music,agostinelli2023musiclm}. However, these systems typically require a large amount of paired text-audio data for training. Despite extensive human annotation efforts, the current largest public text-audio dataset contains around 630\,k text-audio pairs \citep{wu2022clap}. Given the relative scarcity of text-audio data on the web as compared to text-image data, it remains unclear whether we can scale up text-audio datasets to a size comparable with large scale text-image datasets, e.g., the LAION-5B dataset \citep{schuhmann2022laion5b}, which contains 5.85 billion text-image pairs. In this work, we approach text-to-audio synthesis without text-audio pairs through leveraging the naturally-occurring audio-visual correspondence in videos and the multimodal representation learned by pretrained language-vision models (see \cref{clipsonic:fig:bridge}).

The proposed CLIPSonic model is based on a conditional diffusion model \citep{nichol2021ddpm}, a constrastive language-image pretraining (CLIP) model \citep{radford2021clip}, and a pretrained diffusion prior model \citep{ramesh2022dalle2}, as illustrated in \cref{clipsonic:fig:model}. Given a video, CLIPSonic is trained to synthesize the mel spectrogram of the audio given a CLIP-encoded frame, randomly selected from the video. Since CLIP embeds images and texts into a cross-modal semantic space, CLIPSonic learns to map the CLIP embedding space to audio. At test time, we first explore performing a zero-shot modality transfer and conditioning the diffusion model directly with a CLIP-encoded text query. However, we observe in practice a noticeable performance drop with respect to image queries. To close this gap, we adopt a pretrained diffusion prior model to generate a CLIP image embedding given a CLIP text embedding. We note that our proposed system requires only 1) unlabeled videos, for training the conditional diffusion model, and 2) image-text pairs, for pre-training the language-vision models. Through a subjective listening test and an objective evaluation, our experimental results demonstrate the effectiveness of the proposed method. Audio samples are available on our demo website.\footnote{\url{https://salu133445.github.io/clipsonic/}\label{clipsonic:fn:demo}}

Our study differs from prior work in several ways. Existing text-to-audio models rely on large amounts of text-audio training pairs \citep{yang2022diffsound,kreuk2022audiogen,liu2023audioldm,huang2023makeanaudio,huang2023noise2music,agostinelli2023musiclm}, 
%To the best of our knowledge, 
whereas CLIPSonic
%represents the first system that 
learns text-queried audio synthesis without text-audio pairs. Prior work studied image-to-audio synthesis \citep{owens2016vis,iashin2021specvqgan,sheffer2022im2wav}, but they do not examine the zero-shot modality transfer between texts and images. CLIPSep \citep{dong2023clipsep} and CLIPSynth \citep{dong2023clipsynth} propose to learn text-queried source separation and audio synthesis from unlabeled videos, respectively, but they do not address the issue of the zero-shot modality transfer gap. DALL-E 2 \citep{ramesh2022dalle2} proposes the diffusion prior model to address the zero-shot modality transfer gap in CLIP-based text-to-image synthesis, and we explore leveraging a pretrained diffusion prior model to transfer the knowledge learned from videos for text-to-audio synthesis. Other related works are AudioLDM \citep{liu2023audioldm} and MusicLM \citep{agostinelli2023musiclm}, which rely on language-audio models \citep{wu2022clap,huang2022mulan} to perform a zero-shot audio-to-text modality transfer, but such language-audio models are trained on audio-text pairs.

%==================
\section{CLIPSonic}
%==================
\label{clipsonic:sec:method}

\begin{figure*}
    \centering
    \includegraphics[width=.9\linewidth]{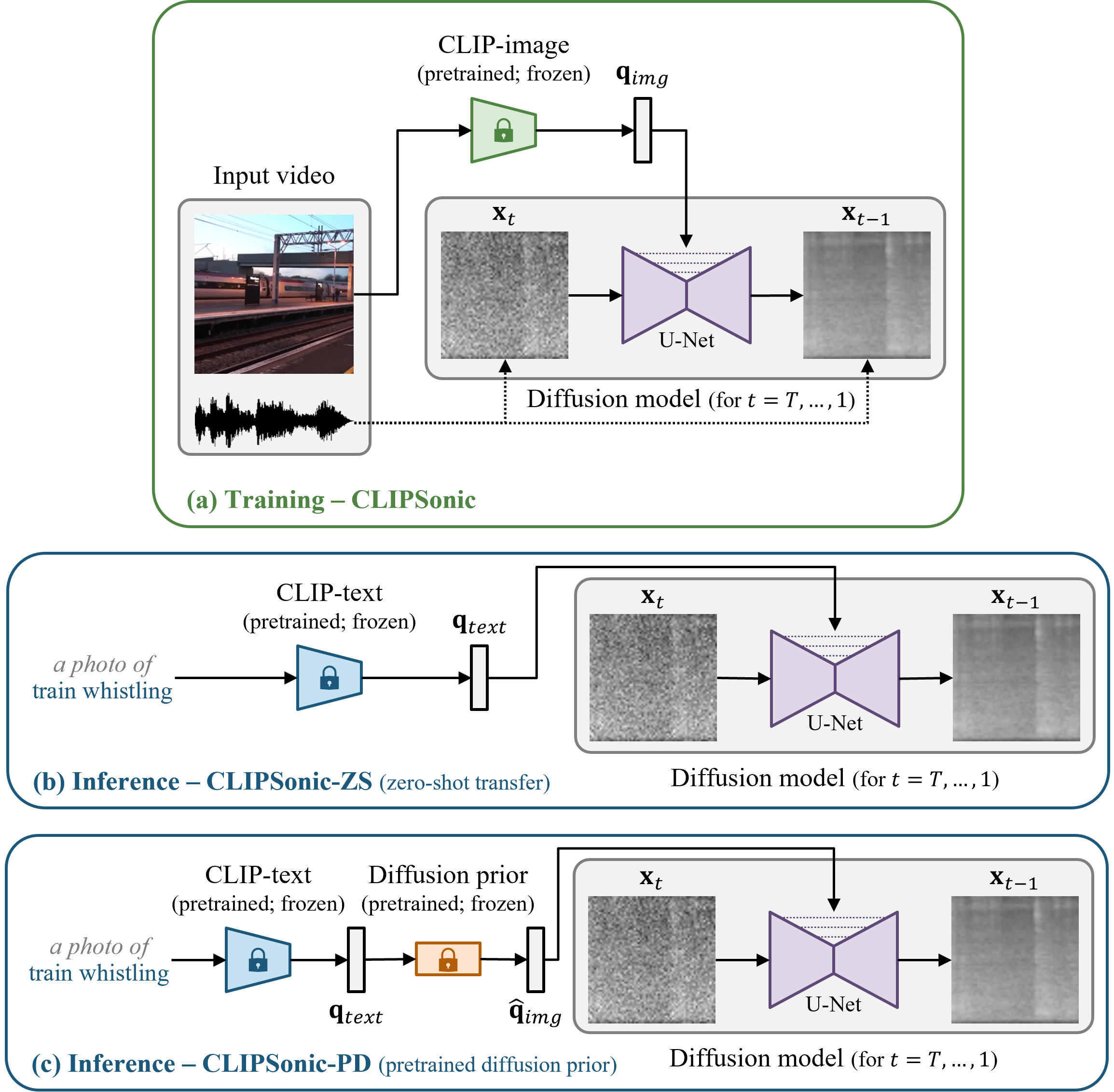}
    \caption[Proposed CLIPSonic model.]{Proposed CLIPSonic model. During training, CLIPSonic learns to synthesize the audio track of a video given the image in a video frame. At inference time, we feed a text query in the form of ``a photo of [label]'' to approach text-to-audio synthesis or use a pretrained diffusion prior model to close the gap between  the text queries (used for inference) and the image queries (used for training). $\textbf{x}_t$ represents a noisy spectrogram at diffusion step $t$. The generated mel spectrogram $\hat{\textbf{x}}_0$ is inverted back to waveform by a pretrained BigVGAN model \citep{lee2023bigvgan}.}
    \label{clipsonic:fig:model}
\end{figure*}

In this section, we introduce the proposed CLIPSonic model for learning text-to-audio synthesis from unlabeled videos. As illustrated in \cref{clipsonic:fig:model}(a), CLIPSonic uses a mel spectrogram-based diffusion model for audio synthesis. We adopt the diffusion framework for its strong performance in audio synthesis \citep{kong2021diffwave,pascual2022diffusion,liu2023audioldm}. Given a video, CLIPSonic is trained to synthesize the mel spectrogram of the audio from the image in a randomly extracted video frame. Specifically, we first use a pretrained CLIP image encoder to encode the image into a query vector $\mathbf{q}_\mathit{img}$. Then, this query vector is used as a conditional signal to guide the diffusion model to generate a mel spectrogram $\hat{\mathbf{x}}_0$. We adopt a denoising diffusion probabilistic model \citep{nichol2021ddpm} and classifier-free guidance \citep{ho2021classifierfree}, which allows us to control the degree of conditioning signal through the guidance level variable $w$ during inference.\footnote{We use the formulation: $\nabla_\mathbf{x} \log p_w (\mathbf{x}\,|\,\mathbf{q}) = (1 - w) \nabla_\mathbf{x} \log p(\mathbf{x}) + w \nabla_\mathbf{x} \log p(\mathbf{x}\,|\,\mathbf{q})$. A larger $w$ leads to a stronger conditioning signal, and $w = 1$ corresponds to a conditional model without classifier-free guidance.}
%  where $w = 0$ corresponds to an unconditional model
The generated mel spectrograms are inverted back to waveforms using a separately-trained BigVGAN \citep{lee2023bigvgan}. We choose to perform diffusion on the mel spectrogram domain for its lower dimensionality than waveforms, and because BigVGAN shows good quality when synthesizing general audio from mel spectrograms.

\paragraph{CLIPSonic-ZS (zero-shot modality transfer).}
At inference time, we aim to leverage the language-vision embedding space learned by CLIP to achieve text-to-audio synthesis. CLIPSonic-ZS explores swapping the CLIP image embeddings for the CLIP text embeddings, as a way to use text queries in a zero-shot modality transfer setting. As illustrated in \cref{clipsonic:fig:model}(b), we use the CLIP text encoder to encode the input text query into a query vector $\mathbf{q}_\mathit{text}$, which is fed as a condition to the diffusion model. We refer to this model as CLIPSonic-ZS, where ``ZS'' stands for \underline{z}ero-\underline{s}hot modality transfer.

\paragraph{CLIPSonic-PD (pretrained diffusion prior).}
As to be shown in \cref{clipsonic:sec:results}, we observe a modality gap between CLIP's text and image embedding spaces. Following DALL-E 2 \citep{ramesh2022dalle2}, we explore relying on a diffusion prior model to bridge this gap. As illustrated in \cref{clipsonic:fig:model}(c), we first encode the input text query into a CLIP text embedding vector $\mathbf{q}_\mathit{text}$ and then generate a CLIP image embedding vector $\hat{\mathbf{q}}_\mathit{img}$ from $\mathbf{q}_\mathit{text}$ using the pretrained diffusion prior model. The generated query vector $\hat{\mathbf{q}}_\mathit{img}$ is then passed as the conditioning signal to the diffusion model. We refer to this model as \mbox{CLIPSonic-PD} (\underline{p}retrained \underline{d}iffusion prior). Note that both CLIPSonic-ZS and CLIPSonic-PD require no text-audio pairs for training. Further, both the CLIP and diffusion prior models can be pretrained using only text-image pairs, hence suppressing the need for paired audio-text data.

\paragraph{CLIPSonic-IQ and CLIPSonic-SD.}
While here we focus on text-to-audio, CLIPSonic can also be used as an image-to-audio synthesis model by using ${\mathbf{q}}_\mathit{img}$ queries. We will refer to this variant as CLIPSonic-IQ (\underline{i}mage-\underline{q}ueried). Moreover, we find that it is possible to train the diffusion prior model from scratch on domain-specific datasets, and hence we also consider a variant called CLIPSonic-SD (\underline{s}upervised \underline{d}iffusion prior), where we train the diffusion prior model from scratch using text-image pairs in our datasets. As will be specified in \cref{clipsonic:sec:setup}, since the text data used to train the diffusion prior in CLIPSonic-SD comes from audio labels in this work, CLIPSonic-SD serves as an oracle model against CLIPSonic-PD. By comparing CLIPSonic-PD to CLIPSonic-SD, we intend to study the effectiveness of using a diffusion prior model pretrained on a massive amount of data against one trained on the target dataset.

%===========================
\section{Experimental setup}
%===========================
\label{clipsonic:sec:setup}

\paragraph{Data.}
We consider two datasets: VGGSound \citep{chen2020vggsound} and MUSIC \citep{zhao2018sop}. The VGGSound dataset consists of 171,899 10-sec YouTube videos, covering 310 classes of sounds in the wild, and we follow the train-test split provided with the dataset. The MUSIC dataset consists of 1,055 full-length YouTube videos of people playing a musical instrument, with 21 instrument types in total. We randomly split the dataset into a 9:1 train-test split.
VGGSound represents a large, diverse dataset captured from unstructured sources in the wild, whereas MUSIC represents a small, curated dataset of a specific domain of interest. As both datasets come with only class labels, we convert such labels \mbox{into pseudo text in the form of ``a photo of [label]''.}

\paragraph{Baseline models.}
We compare CLIPSonic models against the following text-to-audio (TTA) and reconstruction models. 
\begin{itemize}
    \item \textbf{CLIP-TTA} is the supervised version of CLIPSonic where we use text-audio pairs for training. The pretrained CLIP-text embedding is used as conditioning.
    \item \textbf{CLAP-TTA} is the same as CLIP-TTA but uses pretrained CLAP-text embeddings \citep{wu2022clap}, where we use a prompt in the form of ``the sound of [label]''. Unlike CLIP-text embeddings, CLAP-text embeddings are expected to encode audio-grounded features rather than visually-grounded features.
    \item \textbf{BigVGAN mel spectrogram reconstruction} are waveforms reconstructed from the ground-truth mel spectrograms by the BigVGAN model. This serves as an upper bound of spectrogram-based synthesis systems that use BigVGAN as the inversion model.
\end{itemize}

\paragraph{Implementation details.}
For mel spectrogram computation, we use a sampling rate of 16\,kHz, a hop size of 512, an FFT filter size of 2048, and 64~mel bands. During training, we use mel spectrograms of size 64$\times$64, which corresponds to two seconds of audio. For the diffusion model, we follow the network architecture proposed in \citep{nichol2021ddpm} and use the open-source code in \citep{improved_diff_github}. We use a cosine noise schedule with 4000 diffusion steps during training and 1000 steps at inference time. We use AdamW with a learning rate of 0.0001, a batch size of 32, and a dropout rate of 0.1 in classifier-free guidance. All diffusion models are trained for 200\,k steps on MUSIC and 500\,k steps on VGGSound using two NVIDIA RTX 2080 Ti GPUs, which takes a day on MUSIC and two days on VGGSound. For the pretrained CLIP model, we use the ``ViT-L/14'' version trained on 400~million image-text pairs \citep{clip_github}. We use a pretrained transformer-based diffusion prior model trained on 2~billion image-text pairs using the same backbone CLIP model \citep{dalle2_huggingface}. For training the diffusion prior model CLIPSonic-SD from scratch, we follow the same architecture as in CLIPSonic-PD and use the code in \citep{dalle2_github}. We use AdamW with a learning rate of 0.0001 and a batch size of 32. The diffusion prior models are trained on MUSIC and VGGSound, respectively, until convergence at around 200\,k steps, which takes a day on a NVIDIA RTX 2080 Ti GPU. For the CLAP model, we use the ``630k-audioset-fusion'' version released in \citep{clap_github}. For the BigVGAN model, we pretrain it on VGGSound for 500\,k steps using the code in \citep{bigvgan_github} and use this pretrained version in all of our experiments.

\paragraph{Evaluation metrics.}
To compare the performance of our method against the baselines, we sample 512 audio samples from each model and compute the Fréchet audio distance (FAD) \citep{kilgour2019fad} and the CLAP score \citep{huang2023makeanaudio, wu2022clap}. The FAD measures how close the generated audio samples are to the reference audio in terms of quality and diversity.\footnote{Following \citep{yang2022diffsound,liu2023audioldm} we also computed the Fréchet inception distance (FID) of the generated spectrograms, and found that the trend of FID aligned well with that of FAD. For brevity, we only report and discuss the FAD results.} We adopt the open-source implementation provided in \citep{fad_github} and use the VGGish \citep{hershey2017cnn} as the backbone model for FAD. The CLAP score measures the relevance between the generated audio and the input query text, and it is formally defined as the cosine similarity between the CLAP embedding of the audio and that of the input text query.

\paragraph{Subjective test.}
We conduct a listening test to study the fidelity of the generated audio and their relevance to textual (text-to-audio) and visual (image-to-audio) prompts. We ask 21 expert listeners to rate the generated audio samples on a 1--5 scale in terms of fidelity and relevance. Fidelity experiments study the quality of the generated audio (without evaluating its semantic grounding) while relevance experiments study the semantic correspondence with respect to the prompt (without evaluating its audio quality). The audio samples used for this test are available on our demo website.\cref{clipsonic:fn:demo}

%================
\section{Results}
%================
\label{clipsonic:sec:results}

\begin{figure*}[t]
    \centering
    \includegraphics[width=\linewidth]{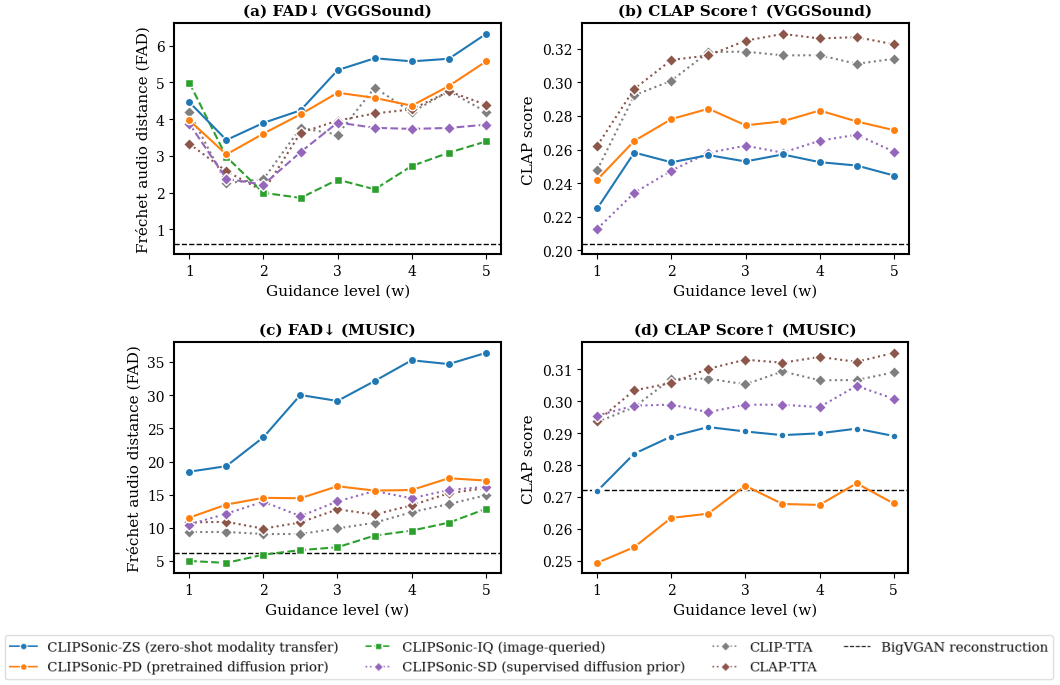}
    \caption{Objective evaluation results on VGGSound and MUSIC.}
    \label{clipsonic:fig:results}
\end{figure*}

%----------------------------------------
\subsection{Objective Evaluation Results}
%----------------------------------------
\label{clipsonic:sec:obj}

\paragraph{Guidance.}
\cref{clipsonic:fig:results} shows the results of the studied models as a function of the classifier-free guidance scale $w$. As noted in \citep{ho2021classifierfree} using conceptually-similar measures, the different curves between FAD and CLAP scores imply a trade-off between quality/diversity (represented by FAD) and query-sample relevance (represented by CLAP score). Noticeably, in terms of CLAP score, all models (except CLIPSonic-PD on MUSIC) outperform the BigVGAN reconstruction on both datasets (see \cref{clipsonic:fig:results}(b) and (d)). We attribute the higher CLAP scores to the classifier-free guidance as it is shown to improve adherence to the conditioning \citep{ho2021classifierfree} but at the cost of diversity---note the increasing FAD in \cref{clipsonic:fig:results}(a) and (c) as $w$ increases. As such, practitioners can choose $w$ based on their specific requirements. We use $w = 1.5$ as it offers a good balance between quality/diversity and relevance, and we report the results in \cref{clipsonic:tab:results}.

\begin{table*}[t]
    \centering
    \caption{Evaluation results on VGGSound and MUSIC datasets, evaluated at $w = 1.5$.}
    \footnotesize
    \begin{tabular}{l@{~~}cc@{~~}cc@{~~}cc@{~~}c}
        \toprule
        \multirow{2}{*}{Model} &\multirow{2}{*}[-3pt]{\shortstack{Without\\text-audio\\pairs}} &\multicolumn{2}{c}{Query modality} &\multicolumn{2}{c}{VGGSound} &\multicolumn{2}{c}{MUSIC}\\
        \cmidrule(r){3-4} \cmidrule(r){5-6} \cmidrule(r){7-8}
        & & Training & Inference &FAD$\downarrow$ &\makecell[c]{CLAP\\score$\uparrow$} &FAD$\downarrow$  &\makecell[c]{CLAP\\score$\uparrow$}\\
        \midrule
        CLIPSonic-IQ {\scriptsize(image-queried)}               &-      &Image &Image &2.97 &-     &4.71  &-\\
        CLIPSonic-ZS {\scriptsize(zero-shot modality transfer)} &\cmark &Image &Text  &3.43 &0.258 &19.30 &0.284 \\
        CLIPSonic-PD {\scriptsize(pretrained diffusion prior)}  &\cmark &Image &Text  &3.04 &0.265 &13.51 &0.254 \\
        \midrule
        CLIPSonic-SD {\scriptsize(supervised diffusion prior)}  &\xmark &Image &Text  &2.37 &0.234 &12.13 &0.299\\
        CLIP-TTA                                                  &\xmark &Text  &Text  &2.26 &0.292 &9.39  &0.298\\
        CLAP-TTA                                                  &\xmark &Text  &Text  &2.58 &0.296 &10.92 &0.303\\
        \midrule
        BigVGAN mel spectrogram reconstruction                    &-      &-     &-     &0.60 &0.204 &6.21  &0.272\\
        \bottomrule
    \end{tabular}
    \label{clipsonic:tab:results}
\end{table*}

\paragraph{Models without text-audio pairs.}
First, we discuss CLIPSonic models in \cref{clipsonic:tab:results} that do not use text-audio pairs during training. CLIPSonic-IQ (image-queried) achieves a strong performance on both datasets. Yet, when we switch to using text queries in a zero-shot setting with CLIPSonic-ZS, we observe a performance drop in terms of FAD on both datasets. This performance drop suggests a modality gap between CLIP's image (used during training) and text (used during inference) embedding spaces. In contrast, with the pretrained diffusion prior model, CLIPSonic-PD achieves a lower FAD than CLIPSonic-ZS across different $w$ values (see also \cref{clipsonic:fig:results}). To further investigate this, we report in \cref{clipsonic:tab:cosine_sim} the average cosine similarity between the query embedding ($\mathbf{q}_\mathit{text}$ or $\hat{\mathbf{q}}_\mathit{img}$) and the ground truth CLIP-image embedding $\mathbf{q}_\mathit{img}$. We note that CLIPSonic-ZS leads to a low cosine similarity, which supports our hypothesis that there is a modality gap in CLIP's embedding space. In contrast, CLIPSonic-PD achieves a significantly higher cosine similarity, showing that the pretrained diffusion prior model can effectively bridge the modality gap. Moreover, while we observe a lower CLAP score for CLIPSonic-PD on MUSIC, we observe little difference in the relevance criterion in the listening test to be discussed in \cref{clipsonic:sec:subjective} (see \cref{clipsonic:tab:subjective}), suggesting that all these models have passed a reasonable level of audio-text relevance.

\begin{table}
    \small
    \centering
    \caption{Cosine similarities between various query embeddings.}
    \begin{tabular}{lccc}
        \toprule
        Model        &Similarity computed &VGGSound &MUSIC \\ 
        \midrule
        CLIPSonic-ZS &$\mathrm{sim}(\mathbf{q}_\mathit{text}, \mathbf{q}_\mathit{img})$ &0.205 &0.245\\
        CLIPSonic-PD &$\mathrm{sim}(\hat{\mathbf{q}}_\mathit{img}, \mathbf{q}_\mathit{img})$ &0.647 &0.720\\
        CLIPSonic-SD &$\mathrm{sim}(\hat{\mathbf{q}}_\mathit{img}, \mathbf{q}_\mathit{img})$ &0.711 &0.824\\
        \bottomrule
    \end{tabular}
    \label{clipsonic:tab:cosine_sim}
\end{table}

\paragraph{Models using text-audio pairs.}
We now compare the baseline models that do use text-audio pairs for training against the previous CLIPSonic variants. 
First, we see that CLIPSonic-SD, with a diffusion prior trained directly on the target dataset, achieves a lower FAD than CLIPSonic-PD, which uses the pretrained diffusion prior. This is possibly due to the distribution mismatch between the target datasets and the LAION-2B dataset used to train the pretrained prior.\footnote{We note that there is also a mismatch in the semantics of the textual queries, where the target datasets contain audio-specific labels while LAION-2B contains visually-grounded labels. However, the similar performance of CLIP-TTA and CLAP-TTA suggests that this is a minor effect.} From \cref{clipsonic:tab:cosine_sim}, we can also see that CLIPSonic-SD can generate a CLIP-image embedding closer to the ground truth embedding on the target datasets than CLIPSonic-PD. Yet, in our subjective evaluation below we will see that CLIPSonic-PD still exhibits a favorable degree of generalization to downstream datasets since it consistently outperforms CLIPSonic-ZS. Moreover, we observe a gap between the performance of CLIPSonic-PD and that of CLIP-TTA and CLAP-TTA. However, we note that this is an unfair comparison as CLIP-TTA and CLAP-TTA are trained on audio-text pairs, while CLIPSonic-PD does not use audio-text pairs in training.

%---------------------------------------------
\subsection{Subjective Listening Test Results}
%---------------------------------------------
\label{clipsonic:sec:subjective}

\begin{table}
    \small
    \centering
    \caption{Listening test results for text-to-audio synthesis (MOS).}
    \begin{tabular}{lcccc}
        \toprule
        \multirow{2}{*}[-.5ex]{Model} &\multicolumn{2}{c}{VGGSound} &\multicolumn{2}{c}{MUSIC}\\
        \cmidrule(lr){2-3} \cmidrule(lr){4-5}
        &Fidelity &Relevance &Fidelity &Relevance\\
        \midrule
        CLIPSonic-ZS &{2.55 $\pm$ 0.22} &{2.01 $\pm$ 0.27} &{2.98 $\pm$ 0.23} &{3.87 $\pm$ 0.24}\\
        CLIPSonic-PD &\textbf{3.04 $\pm$ 0.20} &{2.86 $\pm$ 0.25} &\textbf{3.67 $\pm$ 0.18} &{3.91 $\pm$ 0.24}\\
        CLIPSonic-SD &{2.96 $\pm$ 0.21} &\textbf{3.49 $\pm$ 0.28} &{3.36 $\pm$ 0.20} &\textbf{4.07 $\pm$ 0.22}\\
        \cmidrule(lr){1-5}
        Ground truth &{3.78 $\pm$ 0.19} &{3.54 $\pm$ 0.29} &{3.90 $\pm$ 0.17} &{4.34 $\pm$ 0.18}\\
        \bottomrule
    \end{tabular}
    \label{clipsonic:tab:subjective}
\end{table}

\begin{table}
    \centering
    \small
    \caption{Listening test results for image-to-audio synthesis (MOS).}
    \begin{tabular}{lcc}
        \toprule
        Model &Fidelity &Relevance\\
        \midrule
        CLIPSonic-IQ (image-queried)         &\textbf{3.29 $\pm$ 0.16} &3.80 $\pm$ 0.19\\
        SpecVQGAN \citep{iashin2021specvqgan} &2.15 $\pm$ 0.17 & 2.54 $\pm$ 0.23\\
        \textsc{Im2Wav} \citep{sheffer2022im2wav}      &2.19 $\pm$ 0.15 & \textbf{3.90 $\pm$ 0.22}\\
        \bottomrule
    \end{tabular}
    \label{clipsonic:tab:subjective_image}
\end{table}

\paragraph{Text-to-audio synthesis.}
We conduct an ablation study to compare CLIPSonic-ZS, -PD and -SD variants on MUSIC and VGGSound. As shown in \cref{clipsonic:tab:subjective}, CLIPSonic-ZS consistently underperforms, arguably because of the aforementioned mismatch between text and image embeddings. The two contributed variants, i.e., CLIPSonic-PD and -SD, consistently achieve higher MOS than CLIPSonic-ZS, both in terms of relevance and fidelity. Notably, the ground truth scores are relatively low (an MOS between 3 to 4), especially noticeable for VGGSound as it is noisier than the MUSIC dataset.

\paragraph{Image-to-audio synthesis.}
While our focus is to study text-to-audio synthesis, CLIPSonic-IQ can also generate audio from image queries. We compare it against SpecVQGAN \citep{iashin2021specvqgan}, a representative image-to-audio model, and \textsc{Im2Wav} \citep{sheffer2022im2wav}, a state-of-the-art model for image-to-audio synthesis. 
All three models are trained on VGGSound and tested on out-of-distribution samples from \textsc{ImageHear} \citep{sheffer2022im2wav}. The selected samples conform a challenging benchmark for us because they are 1) selected from \textsc{Im2Wav}'s demo website and 2) out-of-distribution. As shown in \cref{clipsonic:tab:subjective_image}, CLIPSonic-IQ outperforms the state-of-the-art in fidelity, while remaining competitive in terms of relevance. The improved fidelity can possibly be attributed to the fact that we use a continuous representation (mel spectrogram) with a state-of-the-art inversion model (BigVGAN), as compared to the discrete VQ-VAE representation used in \textsc{Im2Wav}.

%===================
\section{Conclusion}
%===================
\label{clipsonic:sec:conclusion}

We explored approaching text-to-audio synthesis without text-audio pairs by using unlabeled videos and pretrained language-vision models. Through both objective and subjective evaluations, we showed that the proposed models can effectively learn text-to-audio synthesis without text-audio pairs, and the pretrained diffusion prior can reduce the modality transfer gap caused by the mismatch between CLIP’s image (used for training) and text (used for inference) embedding spaces. Moreover, in a subjective listening test, the image-to-audio synthesis model that we base our modality transfer upon achieves competitive performance against a state-of-the-art image-to-audio synthesis model. Finally, we argue that images provide rich conditioning signals for audio synthesis, and leveraging such rich signals to improve text-to-audio synthesis is a promising research direction. Along this direction, CLIPSonic represents an example using videos and pretrained language-vision models. For future work, we intend to scale up the proposed method to a larger amount of videos, and explore using tri-modal audio-vision-language models \citep{guzhov2022audioclip,wu2022wav2clip,rouditchenko2021avlnet}.

\secbreak

\begin{adjustwidth}{.5in}{.5in}
    \itshape
    \hspace{\parindent}
    This chapter, in full, is a reprint of the material as it appears in ``CLIPSonic: Text-to-Audio Synthesis with Unlabeled Videos and Pretrained Language-Vision Models'' by Hao-Wen Dong, Xiaoyu Liu, Jordi Pons, Gautam Bhattacharya, Santiago Pascual, Joan Serrà, Taylor Berg-Kirkpatrick and Julian McAuley, which was published in the Proceedings of the IEEE Workshop on Applications of Signal Processing to Audio and Acoustics (WASPAA) in 2023. The dissertation author was the primary investigator and author of this paper.
\end{adjustwidth}

\clearpage

\begin{subappendices}

%=============================================================
\section{Implementation Details of the Diffusion Prior Models}
%=============================================================
\label{clipsonic:sec:detail_prior}

The diffusion prior models used in this paper are based on the open-source implementation of DALL-E 2 in  \citep{dalle2_github}. Specifically, the input to the models is a sequence formed in the order of the encoded CLIP text tokens, the CLIP text embedding, the diffusion step embedding, the noised CLIP image embedding, and a learnable final input embedding. This sequence is fed to a 12-layer transformer consisting of causal multi-head self-attention and feed-forward networks. The last layer's final output vector corresponding to the final input embedding serves as the prediction of the target CLIP image embedding.

For the diffusion prior model used in CLIPSonic-SD, we use a cosine noise schedule with 1000 diffusion steps during training, and 64 steps at inference time. At each diffusion step during training, we minimize the mean squared error between the predicted and the target CLIP image embeddings. Based on DALL-E 2 \citep{ramesh2022dalle2}, we also explore the classifier-free guidance for training the diffusion prior models by randomly replacing the encoded text tokens and the CLIP text embedding with learnable placeholders 10\% of the time. However, at inference time, we empirically find that using no guidance yields the best results. At inference time, for each CLIP text embedding, we generate two CLIP image embeddings from the diffusion prior model, and select the one with a higher cosine similarity to the CLIP text embedding. To train the model, we use the AdamW optimizer with a learning rate of 0.0001, a batch size of 32, a weight decay of 0.06, and we apply an
exponential moving average on the model parameters with a decay factor of 0.9999. The diffusion prior models in CLIPSonic-SD are trained on MUSIC and VGGSound independently
until convergence at around 200\,k steps.

%================================================
\section{CLAP Scores for BigVGAN Reconstructions}
%================================================
\label{clipsonic:sec:bigvgan_clap}

In \cref{clipsonic:fig:results}, we observe that the CLAP scores of the BigVGAN reconstruction using the ground truth mel spectrogram, in many cases, are lower than those of the proposed systems, which indicates lower relevance between the ground truth audio and the text query. In order to adhere to the length of the test data, the BigVGAN CLAP scores are obtained based on the entire 10-sec audio samples. However, empirical listening finds that some segments within the 10-sec samples correspond poorly to the text queries. To further investigate the correspondence, We also compute the BigVGAN CLAP scores using a 4-sec sliding window (consistent with the synthesized sample length) with a hop size of 0.5 sec, and report the maximum, mean, and the minimum scores over all 4-sec segments within a 10-sec sample as the overall score of that sample.

As shown in \cref{clipsonic:tab:slide_clap}, the maximum scores on both datasets are higher than the rest, which supports our observation by listening. On VGGSound, the maximum CLAP score also exceeds those of CLIPSonic-ZS, CLIPSonic-PD, and CLIPSonic-SD (see \cref{clipsonic:tab:results}). On MUSIC, there is a smaller gap between the maximum CLAP score and that obtained using the entire 10-sec audio, indicating a more uniform relevance level within a sample. However, the studied models trained on MUSIC still outperform the BigVGAN reconstruction in terms of the maximum CLAP score (except for CLIPSonic-PD, and CLIPSonic-ZS without using the classifier free guidance, see \cref{clipsonic:fig:results}). In addition to the contribution of the classifier free guidance (\cref{clipsonic:sec:obj}), the remaining reason requires further investigation. Possible directions include manually inspecting and removing samples with poor audio-text correspondence, and also finetuning CLAP on MUSIC.

\begin{table}
    \centering
    \caption{CLAP scores computed on BigVGAN reconstructions using a sliding window.}
    \label{clipsonic:tab:slide_clap}
    \begin{tabular}{ccc@{~~~}c}
        \toprule
        Window size &Mode &VGGSound &MUSIC \\ 
        \midrule
        4 sec       &Max  &0.273    &0.280\\
        4 sec       &Mean &0.195    &0.234\\
        4 sec       &Min  &0.111    &0.185\\
        \cmidrule(lr){1-4}
        10 sec      &-    &0.204    &0.272\\
        \bottomrule
    \end{tabular}
\end{table}

%====================
\section{Limitations}
%====================
\label{clipsonic:sec:limitations}

We observe some limitations of the proposed method. First, as CLIPSonic is conditioned on the CLIP embedding of a single video frame, it is not readily applicable to handle more complex text queries that involve sequences of events or dynamic interactions between objects. A more powerful language-vision model that can understand videos is required to apply our proposed method to leverage the rich temporal information in videos. Second, since the conditioning signals are extracted from videos, CLIPSonic cannot learn audio concepts that have little meaning in the visual domain, such as pitch, prosody, genre, and tempo. This represents one of the fundamental limitations of approaches that use the visual domain as a bridge to learn the text-audio correspondence. Finally, CLIPSonic offers limited controllability in generating semantically complex audio, such as speech or music given specific words or scores, respectively. However, the proposed method may serve as a pretraining approach for training language-audio models, where we can first pretrain a language-audio model on a large dataset with only unlabeled videos and later finetune the model on a small dataset with audio-text pairs.

\end{subappendices}

\graphicspath{{chapters/conclusion/figs/}}
\chapter{Conclusion}
\label{chap:conclusion}

\begin{center}
    \itshape
    Works of art make rules;\\
    rules do not make works of art.\\
    \normalfont
    ---Claude Debussy
\end{center}

\bigskip

In this dissertation, I have presented my research in three directions: \textit{multitrack music generation} (\cref{chap:muspy,chap:mmt}), \textit{assistive music creation tools} (\cref{chap:arranger,chap:deepperformer}) and \textit{multimodal learning for audio and music} (\cref{chap:clipsep,chap:clipsonic}).

\paragraph{Multitrack music generation.}
%---------------------------------------

In \cref{chap:muspy}, I presented MusPy, a new Python library for symbolic music generation \citep{dong2020muspy}. With MusPy, I conducted the first large-scale experiment that measures the cross-dataset generalizability of music generation models, a process which is made easier by MusPy’s dataset management system. In \cref{chap:mmt}, I proposed a new deep learning model for generating multi-instrument music that achieves comparable performance against state-of-the-art systems \citep{dong2023mmt}. Moreover, I also presented the first systematic analysis of musical self-attention and showed that our proposed model learns relative self-attention in certain aspects of music such as beats, positions and pitches.

\paragraph{Assistive music creation tools.}
%------------------------------------------

In \cref{chap:arranger}, I developed the first deep learning model for automatic instrumentation that can find applications in assistive composing tools and intelligent keyboards \citep{dong2021arranger}. I also demonstrated the potential for our proposed models to produce alternative convincing instrumentations for an existing arrangement. In \cref{chap:deepperformer}, I presented the first deep neural network-based polyphonic synthesizer that can synthesize a score into a natural, expressive performance, achieving competitive quality against the baseline model, a conditional generative audio model, in terms of pitch accuracy, timbre and noise level \citep{dong2021deepperformer}.

\paragraph{Multimodal learning for audio and music.}
%---------------------------------------------------

In \cref{chap:clipsep}, I developed the first text-queried sound separation model that can be trained without any text-audio pairs, achieving competitive performance against a supervised model in some settings \citep{dong2023clipsep}. In \cref{chap:clipsonic}, I presented the first text-to-audio synthesis model that requires no text-audio pairs during training \citep{dong2023clipsonic}. Further, while we focus on text-to-audio synthesis, the proposed model can also generate audio from image queries, and it achieves competitive performance against a state-of-the-art image-to-audio synthesis model in the subjective listening test.

%==========================
\section{Future Directions}
%==========================

\paragraph{Multimodal generative AI with music and audio.}
%---------------------------------------------------------

Multimodal content generation has quickly become the next frontier of generative AI. Many recently-released large pretrained multimodal contrastive models lay the foundations for exciting creative applications to film, video and audiobook generation. I am particularly interested in working on multimodal generative models for background music and sound effect generation for videos, audiobooks and games. I would also like to explore fusing multiple controlling signals from different modalities (e.g., text, image, video, audio, emotion measurements, etc.) for controllable music and audio generation. My long-term goal along this direction is to \textit{develop next-generation interfaces for music and audio editing equipped with intuitive multimodal controls}. I will seek collaborations with other faculty members in comp
uter vision and natural language processing to pursue research along this direction.

\begin{figure}
    \centering
    \includegraphics[width=\linewidth]{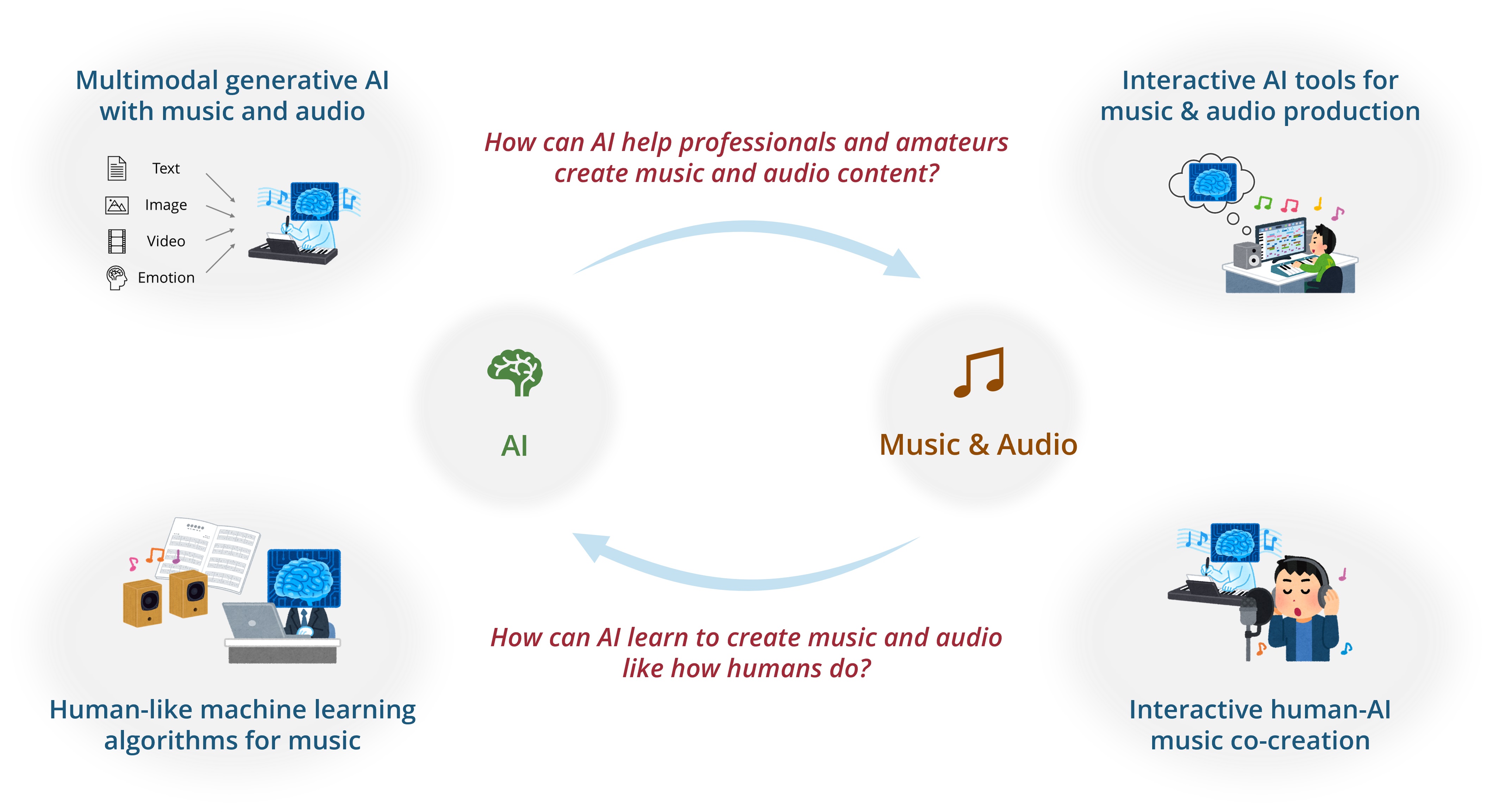}
    \caption{An overview of my future research directions.}
    \label{conclusion:fig:future_work}
\end{figure}

\paragraph{Interactive AI tools for music and audio production.}
%---------------------------------------------------------------

While recent deep learning-based music and audio generation systems can create short, plausible music excerpts, they offer limited usability and controllability for humans to step in. Instead of building a fully-automated generation system, I want to \textit{develop interactable music and audio production tools equipped with intermediate controls that humans can interact with}. For example, recently we have seen preliminary results on leveraging large language models (LLMs) for building a compositional, human-usable audio generation system \citep{liu2023wavjourney}. Moreover, subject-driven personalization \citep{ruiz2023dreambooth} and instruction-based editing \citep{brooks2023instructpix2pix} has lately been attracting attentions in the image generation community, and I would like to explore opportunities in these directions with an eye to integrate these tools into professional creative workflows in music and audio production software. I will seek collaborations with other faculty members in human-computer interaction to explore new creative interfaces for music and audio production.

\paragraph{Human-like machine leaning algorithms for music.}
%-----------------------------------------------------------

Richard Feynman once said: ``What I cannot create, I do not understand.'' This echos my motivations of pursuing music generation research---generation represents the highest-level of understanding. A long-term direction of my research is to \textit{develop human-like machine learning algorithms that can learn to create music in a way similar to how humans learn music}. For example, existing data-driven approaches for music generation usually rely on \textit{reading} a large collection of musical scores. Unlike machines, however, humans learn music mostly through listening and practicing music rather than reading scores over and over again. I am thus interested in exploring novel machine learning models that can learn symbolic music composition through listening to a large collection of musical audio data. Some recent work \citep{castellon2021codified} has shown preliminary results towards this direction. In my view, music possesses a unique complexity that might lead to new breakthroughs in AI and contribute towards the long-lasting pursuit of artificial general intelligence.

%========================
\section{Broader Impacts}
%========================

I envision my research to be integrated into the audio content creation workflow for professional artists and amateurs. Through providing new tools and interfaces to make music, my research could lower the barrier for music composition and empower novices to create their own music. Moreover, it could provide content creators (e.g., TikTokers, YouTubers and Twitch streamers) with royalty-free materials to avoid unintended copyright infringement. My research could also find applications in music education and therapy, where creating personalized courses can be costly. Finally, we could gain insights into the future of human-AI music co-creation though the interactions between human and automatic music composition systems. I envision this to foster the discussions in human-AI relationships in other fields.

% \secbreak

\vfill

\begin{center}
    \itshape
    Without deviation from the norm,\\
    progress is not possible.\\
    \normalfont
    ---Frank Zappa
\end{center}

\vfill

% “Without deviation from the norm, progress is not possible.”
% ― Frank Zappa

% “Music expresses that which cannot be put into words and that which cannot remain silent”
% ― Victor Hugo

% “I am enough of an artist to draw freely upon my imagination. Imagination is more important than knowledge. Knowledge is limited. Imagination encircles the world.”
% ― Albert Einstein

% \appendix

% Stuff at the end of the dissertation goes in the back matter
\backmatter
% \bibliographystyle{plain} % Or whatever style you want like plainnat
% \bibliography{ref}
\singlespacing
\printbibliography

@inproceedings{dong2020muspy,
  author    = {Hao-Wen Dong and Ke Chen and Julian McAuley and Taylor Berg-Kirkpatrick},
  title     = {{M}us{P}y: A Toolkit for Symbolic Music Generation},
  booktitle = {Proceedings of the 21st International Society for Music Information Retrieval Conference (ISMIR)},
  year      = 2020,
}

@inproceedings{raffel2014prettymidi,
  author    = {Colin Raffel and Daniel P. W. Ellis},
  title     = {Intuitive Analysis, Creation and Manipulation of {MIDI} Data with pretty\_midi},
  booktitle = {Late-Breaking Demos of the 15th International Society for Music Information Retrieval Conference (ISMIR)},
  year      = 2014,
}

@inproceedings{dong2018pypianoroll,
  author    = {Hao-Wen Dong and Wen-Yi Hsiao and Yi-Hsuan Yang},
  title     = {Pypianoroll: Open Source {Python} Package for Handling Multitrack Pianorolls},
  booktitle = {Late-Breaking Demos of the 19th International Society for Music Information Retrieval Conference (ISMIR)},
  year      = 2018,
}

@inproceedings{cuthbert2010music21,
  author    = {Cuthbert, Michael Scott and Ariza, Christopher},
  title     = {Music21: A Toolkit for Computer-Aided Musicology and Symbolic Music Data},
  booktitle = {Proceedings of the 11th International Society for Music Information Retrieval Conference (ISMIR)},
  year      = 2010,
}

@inproceedings{mckay2006jsymbolic,
  author    = {Mckay, Cory and Fujinaga, Ichiro},
  title     = {{JS}ymbolic: A feature extractor for {MIDI} files},
  booktitle = {Proceedings of the 2006 International Computer Music Conference (ICMC)},
  year      = 2006,
}

@inproceedings{bittner2019mirdata,
  author    = {Bittner, Rachel M and Fuentes, Magdalena and Rubinstein, David and Jansson, Andreas and Choi, Keunwoo and Kell, Thor},
  title     = {mirdata: Software for Reproducible Usage of Datasets},
  booktitle = {Proceedings of the 20th International Society for Music Information Retrieval Conference (ISMIR)},
  year      = 2019,
}

@inproceedings{raffel2014mireval,
  author    = {Colin Raffel and Brian McFee and Eric J. Humphrey and Justin Salamon and Oriol Nieto and Dawen Liang and Daniel P. W. Ellis},
  title     = {mir\_eval: A Transparent Implementation of Common {MIR} Metrics},
  booktitle = {Proceedings of the 15th International Society for Music Information Retrieval Conference (ISMIR)},
  year      = 2014,
}

@incollection{hewlett1997musedata, 
  author    = {Hewlett, Walter B.},
  title     = {{MuseData}: Multipurpose Representation},
  editor    = {Selfridge-Field, Eleanor},
  booktitle = {Beyond {MIDI}: The Handbook of Musical Codes},
  chapter   = 27,
  pages     = {402--447},
  year      = 1997,
}

@incollection{huron1997humdrum, 
  author    = {David Huron},
  title     = {Humdrum and {K}ern: Selective Feature Encoding},
  editor    = {Selfridge-Field, Eleanor},
  booktitle = {Beyond {MIDI}: The Handbook of Musical Codes},
  chapter   = 27,
  pages     = {375--401},
  year      = 1997,
}

@inproceedings{abadi2016tensorflow,
  author    = {Martin Abadi and Paul Barham and Jianmin Chen and Zhifeng Chen and Andy Davis and Jeffrey Dean and Matthieu Devin and Sanjay Ghemawat and Geoffrey Irving and Michael Isard and Manjunath Kudlur and Josh Levenberg and Rajat Monga and Sherry Moore and Derek G. Murray and Benoit Steiner and Paul Tucker and Vijay Vasudevan and Pete Warden and Martin Wicke and Yuan Yu and Xiaoqiang Zheng},
  title     = {{T}ensor{F}low: A system for large-scale machine learning},
  booktitle = {Proceedings of the 12th USENIX Symp. on Operating Systems Design and Implementation (OSDI)},
  year      = 2016,
}

@inproceedings{paszke2019pytorch,
  author    = {Paszke, Adam and Gross, Sam and Massa, Francisco and Lerer, Adam and Bradbury, James and Chanan, Gregory and Killeen, Trevor and Lin, Zeming and Gimelshein, Natalia and Antiga, Luca and Desmaison, Alban and Kopf, Andreas and Yang, Edward and DeVito, Zachary and Raison, Martin and Tejani, Alykhan and Chilamkurthy, Sasank and Steiner, Benoit and Fang, Lu and Bai, Junjie and Chintala, Soumith},
  title     = {{P}y{T}orch: An Imperative Style, High-Performance Deep Learning Library},
  booktitle = {Advances in Neural Information Processing Systems 32 (NeurIPS)},
  pages     = {8024--8035},
  year      = 2019,
}

@article{briot2017survey,
  author    = {Jean-Pierre Briot and Gaëtan Hadjeres and François Pachet},
  title     = {Deep Learning Techniques for Music Generation: A Survey},
  journal   = {arXiv preprint arXiv:1709.01620},
  year      = 2017,
}

@inproceedings{engel2017nsynth,
  author    = {Jesse Engel and Cinjon Resnick and Adam Roberts and Sander Dieleman and Mohammad Norouzi and Douglas Eck and Karen Simonyan},
  title     = {Neural Audio Synthesis of Musical Notes with {W}ave{Net} Autoencoders},
  booktitle = {Proceedings of the 34th International Conference on Machine Learning (ICML)},
  year      = 2017,
}

@inproceedings{wang2019performancenet,
  author    = {Wang, Bryan and Yang, Yi-Hsuan},
  title     = {{P}erformance{N}et: Score-to-audio music generation with multi-band convolutional residual network},
  booktitle = {Proceedings of the AAAI Conference on Artificial Intelligence (AAAI)},
  pages     = {1174--1181},
  year      = 2019,
}

@article{dannenberg1993survey,
  author    = {Dannenberg, Roger B},
  title     = {A brief survey of music representation issues, techniques, and systems},
  journal   = {Computer Music Journal},
  volume    = {17},
  number    = {3},
  pages     = {20--30},
  year      = 1993,
}

@inproceedings{hankinson2011mei,
  author    = {Hankinson, Andrew and Roland, Perry and Fujinaga, Ichiro},
  title     = {The Music Encoding Initiative as a Document-Encoding Framework},
  booktitle = {Proceedings of the 12th International Society for Music Information Retrieval Conference (ISMIR)},
  year      = 2011,
}

@inproceedings{torchvision,
  author    = {Marcel, Sébastien and Rodriguez, Yann},
  title     = {Torchvision the Machine-Vision Package of {T}orch},
  booktitle = {Proceedings of the 18th ACM International Conference on Multimedia},
  year      = 2010,
}

@incollection{good2001musicxml,
  author    = {Michael Good},
  title     = {MusicXML for Notation and Analysis},
  booktitle = {The {V}irtual {S}core: Representation, Retrieval, Restoration},
  editor    = {Walter B. Hewlett and Eleanor Selfridge-Field},
  address   = {Cambridge, Massachusetts},
  chapter   = {8},
  pages     = {113--124},
  year      = 2001,
}

@inproceedings{donahue2018nesmdb,
  author    = {Chris Donahue and Huanru Henry Mao and Julian McAuley},
  title     = {The {NES} Music Database: A multi-instrumental dataset with expressive performance attributes},
  booktitle = {Proceedings of the 19th International Society for Music Information Retrieval Conference (ISMIR)},
  year      = 2018,
}

@inproceedings{bl2012jsb,
  author    = {Nicolas Boulanger-Lewandowski and Yoshua Bengio and Pascal Vincent},
  title     = {Modeling Temporal Dependencies in High-Dimensional Sequences: Application to Polyphonic Music Generation and Transcription},
  booktitle = {Proceedings of the 29th International Conference on Machine Learning (ICML)},
  year      = 2012,
}

@inproceedings{dong2018musegan,
  author    = {Hao-Wen Dong and Wen-Yi Hsiao and Li-Chia Yang and Yi-Hsuan Yang},
  title     = {{M}use{GAN}: Multi-track Sequential Generative Adversarial Networks for Symbolic Music Generation and Accompaniment},
  booktitle = {Proceedings of the 32nd AAAI Conference on Artificial Intelligence (AAAI)},
  year      = 2018,
}

@inproceedings{mogren2016crnngan,
  author    = {Olof Mogren},
  title     = {{C-RNN-GAN}: Continuous Recurrent neural networks with adversarial training},
  booktitle = {NeuIPS Worshop on Constructive Machine Learning},
  year      = 2016,
}

@article{yang2018evaluation,
  author    = {Yang, Li-Chia and Lerch, Alexander},
  title     = {On the evaluation of generative models in music},
  journal   = {Neural Computing and Applications},
  volume    = {32},
  pages     = {4773--4784},
  year      = 2018,
}

@inproceedings{kingma2015adam,
  author    = {Diederik P. Kingma and Jimmy Ba},
  title     = {Adam: A Method for Stochastic Optimization},
  booktitle = {Proceedings of the 3rd International Conference for Learning Representations (ICLR)},
  year      = 2015,
}

@article{hochreiter1997lstm,
  author    = {Hochreiter, Sepp and Schmidhuber, Jürgen},
  title     = {Long short-term memory},
  journal   = {Neural computation},
  volume    = {9},
  number    = {8},
  pages     = {1735--1780},
  year      = 1997,
}

@article{schuster1997bilstm,
  author = {Schuster, Mike and Paliwal, Kuldip},
  title = {Bidirectional Recurrent Neural Networks},
  journal = {IEEE Transactions on Signal Processing},
  volume = {45},
  number = {11},
  pages = {2673--2681},
  year = {1997},
}

@inproceedings{cho2014gru,
  author    = {Kyunghyun Cho and {van Merrienboer}, B and Caglar Gulcehre and F Bougares and H Schwenk and Yoshua Bengio},
  title     = {Learning phrase representations using {RNN} encoder-decoder for statistical machine translation},
  booktitle = {Proceedings of the Conference on Empirical Methods in Natural Language Processing (EMNLP)},
  year      = 2014,
}

@inproceedings{vaswani2017transformer,
  author    = {Vaswani, Ashish and Shazeer, Noam and Parmar, Niki and Uszkoreit, Jakob and Jones, Llion and Gomez, Aidan N and Kaiser, {\L}ukasz and Polosukhin, Illia},
  title     = {Attention is all you need},
  booktitle = {Advances in Neural Information Processing Systems 30 (NeurIPS)},
  year      = 2017,
}

@inproceedings{roberts2018musicvae,
  author    = {Roberts, Adam and Engel, Jesse and Raffel, Colin and Hawthorne, Curtis and Eck, Douglas},
  title     = {A hierarchical latent vector model for learning long-term structure in music},
  booktitle = {Proceedings of the 35th International Conference on Machine Learning (ICML)},
  year      = 2018,
}

@inproceedings{donahue2019lakhnes,
  author    = {Donahue, Chris and Mao, Huanru Henry and Li, Yiting Ethan and Cottrell, Garrison W. and McAuley, Julian},
  title     = {{L}akh{NES}: Improving multi-instrumental music generation with cross-domain pre-training},
  booktitle = {Proceedings of the 20th International Society for Music Information Retrieval Conference (ISMIR)},
  year      = 2019,
}

@inproceedings{eck2002lstm,
  author    = {Eck, Douglas and Schmidhuber, Juergen},
  title     = {Finding temporal structure in music: Blues improvisation with {LSTM} recurrent networks},
  booktitle = {Proceedings of the IEEE Workshop on Neural Networks for Signal Processing},
  pages     = {747--756},
  year      = 2002,
}

@article{mozer1994neural,
  author    = {Mozer, Michael},
  title     = {Neural Network Music Composition by Prediction: Exploring the Benefits of Psychoacoustic Constraints and Multi-scale Processing},
  journal   = {Connection Science},
  volume    = {6},
  pages     = {247-280},
  year      = 1994,
}

@inproceedings{huang2019musictransformer,
  author    = {Cheng-Zhi Anna Huang and Ashish Vaswani and Jakob Uszkoreit and Ian Simon and Curtis Hawthorne and Noam
Shazeer and Andrew M. Dai and Matthew D. Hoffman and Monica Dinculescu and Douglas Eck},
  title     = {Music {T}ransformer: Generating music with long-term structure},
  booktitle = {Proceedings of the 7th International Conference for Learning Representations (ICLR)},
  year      = 2019,
}

@inproceedings{yang2018midinet,
  author    = {Li-Chia Yang and Szu-Yu Chou and Yi-Hsuan Yang},
  title     = {{M}idi{N}et: A Convolutional Generative Adversarial Network for Symbolic-domain Music Generation},
  booktitle = {Proceedings of the 18th International Society for Music Information Retrieval Conference (ISMIR)},
  year      = 2017,
}

@misc{mido,
  title = {{M}ido: {MIDI} Objects for {P}ython},
  url = {https://github.com/mido/mido},
}

@misc{lilypond,
  title = {{L}ily{P}ond},
  url = {https://lilypond.org/},
}

@misc{wikifonia,
  title = {Wikifonia},
  url = {http://www.wikifonia.org/},
}

@misc{essen,
  title = {Essen Folk Song Database},
  url = {https://ifdo.ca/~seymour/runabc/esac/esacdatabase.html},
}

@misc{hymnal,
  title = {Hymnal},
  url = {https://www.hymnal.net/},
}

@misc{nmd,
  title = {Nottingham Database},
  url = {https://ifdo.ca/~seymour/nottingham/nottingham.html},
}

@misc{tfds,
  title = {{T}ensor{F}low Datasets},
  url = {https://www.tensorflow.org/datasets},
}

@misc{magenta,
  title = {Magenta},
  url = {https://magenta.tensorflow.org/},
}

@inproceedings{wu2020jazz,
  author = {Shih-Lun Wu and Yi-Hsuan Yang},
  title = {The {J}azz {T}ransformer on the Front Line: Exploring the Shortcomings of {AI}-composed Music through Quantitative Measures},
  booktitle = {Proceedings of the 21st International Society for Music Information Retrieval Conference (ISMIR)},
  year = 2020,
}

@inproceedings{thickstun2017musicnet,
  author = {John Thickstun and Zaid Harchaoui and Sham M. Kakade},
  title = {Learning Features of Music from Scratch},
  booktitle = {Proceedings of the 5th International Conference on Learning Representations (ICLR)},
  year = 2017,
}

@inproceedings{hadjakos2019hands,
  author = {Hadjakos, Aristotelis AND Waloschek, Simon AND Leemhuis, Alexander},
  title = {Detecting Hands from Piano {MIDI} Data},
  booktitle = {Mensch und Computer 2019 - Workshopband},
  year = 2019,
}

@inproceedings{guiomardkagan2015voice,
  author = {Nicolas Guiomard-Kagan and Mathieu Giraud and Richard Groult and Florence Levé},
  title = {Comparing Voice and Stream Segmentation Algorithms},
  booktitle = {Proceedings of the 16th International Society for Music Information Retrieval Conference (ISMIR)},
  year = 2015,
}

@inproceedings{kilian2002voiceseparation,
  author = {Jürgen Kilian and Holger H. Hoos},
  title = {Voice Separation --- A Local Optimisation Approach},
  booktitle = {Proceedings of the 3rd International Conference on Music Information Retrieval (ISMIR)},
  year = 2002,
}

@inproceedings{karydis2007visa,
  author = {Ioannis Karydis and Alexandros Nanopoulos and Apostolos Papadopoulos and Emilios Cambouropoulos},
  title = {{VISA}: The Voice Integration/Segregation Algorithm},
  booktitle = {Proceedings of the 8th International Conference on Music Information Retrieval (ISMIR)},
  year = 2007,
}

@inproceedings{cambouropoulos2006voice,
  author = {Emilios Cambouropoulos},
  title = {`{V}oice' separation: theoretical, perceptual and computational perspectives},
  booktitle = {Proceedings of the 9th International Conference on Music Perception \& Cognition (ICMPC)},
  year = 2006,
}

@article{cambouropoulos2008voice,
  author = {Emilios Cambouropoulos},
  title = {Voice and Stream: Perceptual and Computational Modeling of Voice Separation},
  journal = {Music Perception},
  volume = 26,
  number = 1,
  page = {75--94},
  year = 2008,
}

@inproceedings{devalk2019deepseparation,
  author = {de Valk, Reinier and Tillman Weyde},
  title = {Deep neural networks with voice entry estimation heuristics for voice separation in symbolic music representations},
  booktitle = {Proceedings of the 19th International Society for Music Information Retrieval Conference (ISMIR)},
  year = 2018,
}

@article{gray2020voiceseparation,
  author = {Patrick Gray and Razvan Bunescu},
  title = {From Note-Level to Chord-Level Neural Network Models for Voice Separation in Symbolic Music},
  journal = {arXiv preprint arXiv:2011.03028},
  year = 2020,
}

@inproceedings{gray2016voiceseparation,
  author = {Patrick Gray and Razvan Bunescu},
  title = {A Neural Greedy Model for Voice Separation in Symbolic Music},
  booktitle = {Proceedings of the 17th International Society for Music Information Retrieval Conference (ISMIR)},
  year = 2016,
}

@inproceedings{ba2016layernorm,
  author = {Jimmy Lei Ba and Jamie Ryan Kiros and Geoffrey E. Hinton},
  title = {Layer Normalization},
  booktitle = {NeurIPS 2016 Deep Learning Symposium},
  year = 2016,
}

@article{srivastava2014dropout,
  author  = {Nitish Srivastava and Geoffrey Hinton and Alex Krizhevsky and Ilya Sutskever and Ruslan Salakhutdinov},
  title   = {{D}ropout: A Simple Way to Prevent Neural Networks from Overfitting},
  journal = {Journal of Machine Learning Research (JMLR)},
  volume  = {15},
  number  = {56},
  pages   = {1929-1958},
  year    = {2014},
}

@inproceedings{hung2019improving,
  title={Improving automatic jazz melody generation by transfer learning techniques},
  author={Hung, Hsiao-Tzu and Wang, Chung-Yang and Yang, Yi-Hsuan and Wang, Hsin-Min},
  booktitle={2019 Asia-Pacific Signal and Information Processing Association Annual Summit and Conference (APSIPA ASC)},
  year={2019},
}

@misc{musenet,
  author = {Payne, Christine},
  title = {{M}use{N}et},
  date = {2019-04-25},
  howpublished = {{O}pen{AI}},
  url = {https://openai.com/blog/musenet/},
  year = {2019},
}

@misc{simon2017performance,
    author = {Ian Simon and Sageev Oore},
    title = {{P}erformance {RNN}: Generating Music with Expressive Timing and Dynamics},
    howpublished = {{M}agenta Blog},
    url = {https://magenta.tensorflow.org/performance-rnn},
    year = {2017},
}

@article{ens2020mmm,
  author = {Jeff Ens and Philippe Pasquier},
  title = {{MMM}: Exploring Conditional Multi-Track Music Generation with the {T}ransformer},
  journal = {arXiv preprint arXiv:2008.06048},
  year = {2020},
}

@article{hsiao2021transformer,
  author = {Wen-Yi Hsiao and Jen-Yu Liu and Yin-Cheng Yeh and Yi-Hsuan Yang},
  title = {Compound Word {T}ransformer: Learning to Compose Full-Song Music over Dynamic Directed Hypergraphs},
  journal = {Proceedings of the 35th AAAI Conference on Artificial Intelligence (AAAI)},
  year = {2021},
}

@inproceedings{simon2018multitrackmusicvae,
  author = {Ian Simon and Adam Roberts and Colin Raffel and Jesse Engel and Curtis Hawthorne and Douglas Eck},
  title	= {Learning a Latent Space of Multitrack Measures},
  booktitle = {NeurIPS Workshop on Machine Learning for Creativity and Design},
  year = {2018},
}

@inproceedings{muhamed2021transformergan,
  author = {Aashiq Muhamed and Liang Li and Xingjian Shi and Suri Yaddanapudi and Wayne Chi and Dylan Jackson and Rahul Suresh and Zachary C. Lipton and Alex J. Smola},
  title = {Symbolic Music Generation with {T}ransformer-{GAN}s},
  booktitle = {Proceedings of the 35th AAAI Conference on Artificial Intelligence (AAAI)},
  year = {2021},
}

@inproceedings{chiu2009automatic,
  title={Automatic system for the arrangement of piano reductions},
  author={Chiu, Shih-Chuan and Shan, Man-Kwan and Huang, Jiun-Long},
  booktitle={Proceedings of the 2009 11th IEEE International Symposium on Multimedia (ISM)},
  year={2009},
}

@inproceedings{onuma2010piano,
  title={Piano Arrangement System Based On Composers' Arrangement Processes},
  author={Onuma, Sho and Hamanaka, Masatoshi},
  booktitle={Proceedings of the 2010 International Computer Music Conference (ICMC)},
  year={2010}
}

@article{huang2012towards,
  title={Towards an automatic music arrangement framework using score reduction},
  author={Huang, Jiun-Long and Chiu, Shih-Chuan and Shan, Man-Kwan},
  journal={ACM Transactions on Multimedia Computing, Communications, and Applications (TOMM)},
  volume={8},
  number={1},
  pages={1--23},
  year={2012},
  publisher={ACM New York, NY, USA}
}

@inproceedings{nakamura2015automatic,
  title={Automatic piano reduction from ensemble scores based on merged-output hidden {M}arkov model},
  author={Nakamura, Eita and Sagayama, Shigeki},
  booktitle={Proceedings of the 2005 International Computer Music Conference (ICMC)},
  year={2015}
}

@inproceedings{takamori2017automatic,
  title={Automatic arranging musical score for piano using important musical elements},
  author={Takamori, Hirofumi and Sato, Haruki and Nakatsuka, Takayuki and Morishima, Shigeo},
  booktitle={Proceedings of the 14th Sound and Music Computing Conference, Aalto, Finland},
  pages={35--41},
  year={2017}
}

@article{nakamura2018statistical,
  title={Statistical piano reduction controlling performance difficulty},
  author={Nakamura, Eita and Yoshii, Kazuyoshi},
  journal={APSIPA Transactions on Signal and Information Processing},
  volume={7},
  year={2018},
  publisher={Cambridge University Press}
}

@inproceedings{tuohy2005genetic,
  title={A genetic algorithm for the automatic generation of playable guitar tablature},
  author={Tuohy, Daniel R and Potter, Walter D},
  booktitle={Proceedings of the 2005 International Computer Music Conference (ICMC)},
  year={2005}
}

@inproceedings{hori2012automatic,
  title={Automatic arrangement for guitars using hidden {M}arkov model},
  author={Hori, G and Yoshinaga, Y and Fukayama, S and Kameoka, H and Sagayama, S},
  booktitle={Proceedings of the 9th Sound and Music Computing Conference (SMC)},
  pages={450--456},
  year={2012}
}

@article{hori2013input,
  title={Input-output {HMM} applied to automatic arrangement for guitars},
  author={Hori, Gen and Kameoka, Hirokazu and Sagayama, Shigeki},
  journal={Information and Media Technologies},
  volume={8},
  number={2},
  pages={477--484},
  year={2013},
  publisher={Information and Media Technologies Editorial Board}
}

@article{abe2012automatic,
  title={Automatic arrangement for the bass guitar in popular music using principle component analysis},
  author={Abe, Yuzo and Murakami, Yuki and Miura, Masanobu},
  journal={Acoustical Science and Technology},
  volume={33},
  number={4},
  pages={229--238},
  year={2012},
  publisher={Acoustical Society of Japan}
}

@article{crestel2016live,
  title={Live orchestral piano, a system for real-time orchestral music generation},
  author={Crestel, L{\'e}opold and Esling, Philippe},
  journal={arXiv preprint arXiv:1609.01203},
  year={2016}
}

@inproceedings{chen2020vggsound,
    author = {Honglie Chen and Weidi Xie and Andrea Vedaldi and Andrew Zisserman},
    title = {{VGGS}ound: A Large-scale Audio-Visual Dataset},
    booktitle = {Proc. ICASSP}, 
    year = 2020,
}

@inproceedings{piczak2015esc,
    author = {Piczak, Karol J.},
    title = {{ESC}: Dataset for Environmental Sound Classification},
    booktitle = {Proc. MM},
    year = 2015,
}

@inproceedings{jansson2017unet,
    author = {Andreas Jansson and Eric Humphrey and Nicola Montecchio and Rachel Bittner and Aparna Kumar and Tillman Weyde},
    title = {Singing Voice Separation with Deep {U}-{N}et Convolutional Networks},
    booktitle = {Proc. ISMIR},
    year = 2017,
}

@inproceedings{ronneberger2015unet,
    author = {Olaf Ronneberger and Philipp Fischer and Thomas Brox},
    title = {{U}-{N}et: Convolutional Networks for Biomedical Image Segmentation},
    booktitle = {Proc. MICCAI},
    year = 2015,
}

@inproceedings{yu2017pit,
    author = {Dong Yu and Morten Kolbæk and Zheng-Hua Tan and Jesper Jensen},
    title = {Permutation invariant training of deep models for speaker-independent multi-talker speech separation},
    booktitle = {Proc. ICASSP},
    year = 2017,
}

@inproceedings{zhang2020gradientclipping,
    author = {Jingzhao Zhang and Tianxing He and Suvrit Sra and Ali Jadbabaie},
    title = {Why Gradient Clipping Accelerates Training: A Theoretical Justification for Adaptivity},
    booktitle = {Proc. ICLR},
    year = 2020,
}

@inproceedings{stoter2018museval,
    author = {Stöter , Fabian-Robert and Liutkus, Antoine and Ito, Nobutaka},
    title = {The 2018 Signal Separation Evaluation Campaign},
    booktitle = {Proc. LVA/ICA},
    year = 2018,
}

@inproceedings{radford2021clip,
    author = {Radford, Alec and Kim, Jong Wook and Hallacy, Chris and Ramesh, Aditya and Goh, Gabriel and Agarwal, Sandhini and Sastry, Girish and Askell, Amanda and Mishkin, Pamela and Clark, Jack and Krueger, Gretchen and Sutskever, Ilya},
    title = {Learning Transferable Visual Models From Natural Language Supervision},
    booktitle = {Proc. ICML},
    year = 2021,
}

@article{radford2019gpt2,
    author = {Radford, Alec and Wu, Jeff and Child, Rewon and Luan, David and Amodei, Dario and Sutskever, Ilya},
    title = {Language Models are Unsupervised Multitask Learners},
    journal = {Technical Report of OpenAI},
    year = 2019,
}

@inproceedings{brown2020gpt3,
    author = {Brown, Tom and Mann, Benjamin and Ryder, Nick and Subbiah, Melanie and Kaplan, Jared D and Dhariwal, Prafulla and Neelakantan, Arvind and Shyam, Pranav and Sastry, Girish and Askell, Amanda and Agarwal, Sandhini and Herbert-Voss, Ariel and Krueger, Gretchen and Henighan, Tom and Child, Rewon and Ramesh, Aditya and Ziegler, Daniel and Wu, Jeffrey and Winter, Clemens and Hesse, Chris and Chen, Mark and Sigler, Eric and Litwin, Mateusz and Gray, Scott and Chess, Benjamin and Clark, Jack and Berner, Christopher and McCandlish, Sam and Radford, Alec and Sutskever, Ilya and Amodei, Dario},
    title = {Language Models are Few-Shot Learners},
    booktitle = {Proc. NeurIPS},
    year = 2020,
}

@inproceedings{Liu2022SeparateWY,
    author = {Xubo Liu and Haohe Liu and Qiuqiang Kong and Xinhao Mei and Jinzheng Zhao and Qiushi Huang and MarkD . Plumbley and Wenwu Wang},
    title = {Separate What You Describe: Language-Queried Audio Source Separation},
    booktitle = {Proc. INTERSPEECH},
    year = 2022,
}

@inproceedings{Kilgour2022TextDrivenSO,
    author = {Kevin Kilgour and Beat Gfeller and Qingqing Huang and Aren Jansen and Scott Wisdom and Marco Tagliasacchi},
    title = {Text-Driven Separation of Arbitrary Sounds},
    booktitle = {Proc. INTERSPEECH},
    year = 2022,
}

@inproceedings{Tzinis2021AudioScope,
    title = {Into the Wild with {A}udio{S}cope: Unsupervised Audio-Visual Separation of On-Screen Sounds},
    author = {Efthymios Tzinis and Scott Wisdom and Aren Jansen and Shawn Hershey and Tal Remez and Daniel P. W. Ellis and John R. Hershey},
    booktitle = {Proc. ICLR},
    year = 2021,
}

@inproceedings{Wisdom2020MixIT,
    author = {Scott Wisdom and Efthymios Tzinis  and Hakan Erdogan and Ron J. Weiss and Kevin Wilson and John R. Hershey},
    title = {Unsupervised Sound Separation Using Mixture Invariant Training},
    booktitle = {Proc. NeurIPS},
    year = 2020,
}

@inproceedings{Kavalerov19USS,
    author = {Ilya Kavalerov and Scott Wisdom and Hakan Erdogan and Brian Patton and Kevin Wilson and Jonathan Le Roux and John R. Hershey},
    title = {Universal Sound Separation},
    booktitle = {Proc. WASPAA},
    year = 2019,
}

@inproceedings{Gfeller2021OneShotCA,
    author = {Beat Gfeller and Dominik Roblek and Marco Tagliasacchi},
    title = {One-Shot Conditional Audio Filtering of Arbitrary Sounds},
    booktitle={Proc. ICASSP},
    year = 2021,
}

@inproceedings{Chen22ZeroshotSS,
    author = {Ke Chen and Xingjian Du and Bilei Zhu and Zejun Ma and Taylor Berg-Kirkpatrick and Shlomo Dubnov},
    title = {Zero-shot Audio Source Separation through Query-based Learning from Weakly-labeled Data},
    booktitle = {Proc. AAAI},
    year = 2022,
}

@inproceedings{Kong20USSLabel,
    author = {Qiuqiang Kong and Yuxuan Wang and Xuchen Song and Yin Cao and Wenwu Wang and Mark D. Plumbley},
    title = {Source separation with weakly labelled data: An approach to computational auditory scene analysis},
    booktitle = {Proc. ICASSP},
    year = 2020,
}

@inproceedings{Ochiai20USSLabel,
    author = {Tsubasa Ochiai and Marc Delcroix and Yuma Koizumi and Hiroaki Ito and Keisuke Kinoshita and Shoko Araki},
    title = {Listen to What You Want: Neural Network-based Universal Sound Selector},
    booktitle = {Proc. INTERSPEECH},
    year = 2020,
}

@inproceedings{zhao2018sop,
    author = {Zhao, Hang and Gan, Chuang and Rouditchenko, Andrew and Vondrick, Carl and McDermott, Josh and Torralba, Antonio},
    title = {The Sound of Pixels},
    booktitle = {Proc. ECCV},
    year = 2018,
}

@inproceedings{Gao2018AVSNMF,
    author = {Ruohan Gao and Rogerio Feris and Kristen Grauman},
    title = {Learning to Separate Object Sounds by Watching Unlabeled Video},
    booktitle = {Proc. ECCV},
    year = 2018,
}

@inproceedings{Rouditchenko2019,
    author = {Andrew Rouditchenko and Hang Zhao and Chuang Gan and Josh McDermott and Antonio Torralba},
    title = {Self-Supervised Audio-Visual Co-Segmentation},
    booktitle = {Proc. ICASSP},
    year = 2019,
}

@article{Ephrat2019L2L,
    author = {Ariel Ephrat and Inbar Mosseri and Oran Lang and Tali Dekel and Kevin Wilson and Avinatan Hassidim and William T Freeman and Michael Rubinstein},
    title = {Looking to Listen at the Cocktail Party: A Speaker-Independent Audio-Visual Model for Speech Separation},
    journal = {ACM Transactions on Graphics},
    year = 2019,
    volume = 37,
    issue = 4,
    page = {1--11},
}

@inproceedings{Afouras2020SSLAV,
    author = {Triantafyllos Afouras and Andrew Owens and Joon Son Chung and Andrew Zisserman},
    title = {Self-Supervised Learning of Audio-Visual Objects from Video},
    booktitle = {Proc. ECCV},
    year = 2020,
}

@inproceedings{Owens2018AVSA,
    author = {Andrew Owens and Alexei A Efros},
    title = {Audio-Visual Scene Analysis with Self-Supervised Multisensory Features},
    booktitle = {Proc. ECCV},
    year = 2018,
}

@inproceedings{Tian21CoSep,
    author = {Yapeng Tian and Di Hu and Chenliang Xu},
    title = {Cyclic Co-Learning of Sounding Object Visual Grounding and Sound Separation},
    booktitle = {Proc. CVPR},
    year = 2021,
}

@inproceedings{zhao2019som,
    author = {Hang Zhao and Chuang Gan and Wei-Chiu Ma and Antonio Torralba},
    title = {The Sound of Motions},
    booktitle = {Proc. ICCV},
    year = 2019,
}

@inproceedings{Lee22SoundGuidedIM,
    title = {Sound-Guided Semantic Image Manipulation},
    author = {Seung Hyun Lee and Wonseok Roh and Wonmin Byeon and Sang Ho Yoon and Chan Young Kim and Jinkyu Kim and Sangpil Kim},
    booktitle = {Proc. CVPR},
    year = 2022,
}

@inproceedings{devlin2019bert,
    author = {Jacob Devlin and Ming-Wei Chang and Kenton Lee and Kristina Toutanova},
    title = {{BERT}: Pre-training of Deep Bidirectional Transformers for Language Understanding},
    booktitle = {Proc. NAACL},
    year = 2019,
}

@inproceedings{Arandjelovic2017L3,
  title = {Look, listen and learn},
  author = {Relja Arandjelović and Andrew Zisserman},
  booktitle = {Proc. ICCV},
  year = 2017,
}

@inproceedings{Arandjelovic2018,
  title = {Objects that Sound},
  author = {Relja Arandjelovi\'{c} and Andrew Zisserman},
  booktitle = {Proc. ECCV},
  year = 2017,
}

@inproceedings{Korbar2018AVTS,
  title = {Cooperative learning of audio and video models from self-supervised synchronization},
  author = {Bruno Korbar and Du Tran and Lorenzo Torresani},
  booktitle = {Proc. NeurIPS},
  year = 2018,
}

@inproceedings{Bach2005,
  title = {Blind one-microphone speech separation: A spectral learning approach},
  author = {Francis R. Bach and Michael I. Jordan},
  booktitle = {Proc. NIPS},
  year = 2005,
}

@article{Cherry1953cocktailparty,
    author = {E Colin Cherry},
    title = {Some experiments on the recognition of speech, with one and
with two ears},
    journal = {The Journal of the acoustical society of America},
    year = 1953,
    volume = 25,
    issue = 5,
    page = {975-–979},
}

@inproceedings{Scott2021,
    author = {Wisdom Scott and Aren Jansen and Ron J. Weiss and Hakan Erdogan and John R. Hershey},
    title = {Sparse, efficient, and semantic mixture invariant training: Taming in-the-wild unsupervised sound separation},
    booktitle = {Proc. WASPAA},
    year = 2021,
}

@inproceedings{Takahashi21D3Net,
    author = {Naoya Takahashi and Yuki Mitsufuji},
    booktitle = {Proc. CVPR},
    title = {Densely connected multidilated convolutional networks for dense prediction tasks},
    year = {2021}
}

@article{Shimojo2001,
    author = {Shinsuke Shimojo and Ladan Shams},
    title = {Sensory modalities are not separate modalities: plasticity and interactions},
    journal = {Current Opinion in Neurobiology},
    volume = 11,
    issue = 4,
    year = 2001,
    page = {505--509}
}

@article{Baillarge2002,
    author = {Renée Baillargeon},
    title = {The acquisition of physical knowledge in infancy: A summary in eight lessons},
    journal = {Blackwell handbook of childhood cognitive development},
    page = {46–83},
    year = 2002,
}

@article{Rahne2007,
    author = {Torsten Rahne and Martin Böckmann and Hellmut von Specht and Elyse S Sussman},
    title = {Visual cues can modulate integration and segregation of objects in auditory scene analysis},
    journal = {Brain research},
    year = 2007,
}

@article{Sekuler1997,
    author = {Robert Sekuler and Allison B. Sekuler and Renee Lau},
    title = {Sound alters visual motion perception},
    journal = {Nature},
    year = 1997,
    volume = 385,
    issue = 308 ,
}

@article{vincent2006performance,
  author = {Vincent, Emmanuel and Gribonval, Rémi and Févotte, Cédric},
  title = {Performance measurement in blind audio source separation},
  journal = {IEEE Transactions on Audio, Speech, and Language Processing},
  volume = 14,
  number = 4,
  pages = {1462--1469},
  year = 2006
}

@misc{mitsufuji2021music,
  title = {Music Demixing Challenge 2021},
  author = {Yuki Mitsufuji and Giorgio Fabbro and Stefan Uhlich and Fabian-Robert Stöter and Alexandre Défossez and Minseok Kim and Woosung Choi and Chin-Yun Yu and Kin-Wai Cheuk},
  year = 2021,
  journal = {arxiv preprint arxiv:2108.13559},
}

@article{wang2018speech,
  author = {Wang, DeLiang and Chen, Jitong},
  journal = {IEEE/ACM Transactions on Audio, Speech, and Language Processing}, 
  title = {Supervised Speech Separation Based on Deep Learning: An Overview}, 
  year = 2018,
  volume = 26,
  number = 10,
  pages = {1702--1726},
}

@inproceedings{dong2023clipsep,
    author = {Hao-Wen Dong and Naoya Takahashi and Yuki Mitsufuji and Julian McAuley and Taylor Berg-Kirkpatrick},
    title = {{{CLIPS}ep: Learning Text-queried Sound Separation with Noisy Unlabeled Videos}},
    booktitle = {Proc. ICLR},
    year = 2023,
}

@article{yang2022diffsound,
    author={Yang, Dongchao and Yu, Jianwei and Wang, Helin and Wang, Wen and Weng, Chao and Zou, Yuexian and Yu, Dong},
    title = {{Diffsound: Discrete Diffusion Model for Text-to-sound Generation}},
    journal = {arXiv preprint arXiv:2207.09983},
    year = 2022,
}

@inproceedings{kreuk2022audiogen,
  title = {{{A}udio{G}en: Textually Guided Audio Generation}},
  author = {Felix Kreuk and Gabriel Synnaeve and Adam Polyak and Uriel Singer and Alexandre Défossez and Jade Copet and Devi Parikh and Yaniv Taigman and Yossi Adi},
  booktitle = {Proc. ICLR},
  year = {2023},
}

@inproceedings{hershey2017cnn,
  title={{CNN Architectures for Large-scale Audio Classification}},
  author={Hershey, Shawn and Chaudhuri, Sourish and Ellis, Daniel PW and Gemmeke, Jort F and Jansen, Aren and Moore, R Channing and Plakal, Manoj and Platt, Devin and Saurous, Rif A and Seybold, Bryan and others},
  booktitle={Proc. ICASSP},
  pages={131--135},
  year={2017},
}

@inproceedings{kilgour2019fad,
    author = {Kevin Kilgour and Mauricio Zuluaga and Dominik Roblek and Matthew Sharifi},
    title = {{Fréchet Audio Distance: A Metric for Evaluating Music Enhancement Algorithms}},
    booktitle = {Proc. INTERSPEECH},
    pages = {2350-2354},
    year = 2019,
}

@inproceedings{nichol2021ddpm,
    author = {Alex Nichol and Prafulla Dhariwal},
    title = {{Improved Denoising Diffusion Probabilistic Models}},
    booktitle = {Proc. ICML},
    year = 2019,
}

@inproceedings{ho2020ddpm,
    author = {Jonathan Ho and Ajay Jain and Pieter Abbeel},
    title = {{Denoising Diffusion Probabilistic Models}},
    booktitle = {Proc. NeurIPS},
    year = 2020,
}

@inproceedings{kong2020hifigan,
    author = {Jungil Kong and Jaehyeon Kim and Jaekyoung Bae},
    title = {{{H}i{F}i-{GAN}: Generative Adversarial Networks for Efficient and High Fidelity Speech Synthesis}},
    booktitle = {Proc. NeurIPS},
    year = 2020,
}

@inproceedings{engel2019gansynth,
  author = {Jesse Engel and Kumar Krishna Agrawal and Shuo Chen and Ishaan Gulrajani and Chris Donahue and Adam Roberts},
  title = {{{GANS}ynth: Adversarial Neural Audio Synthesis}},
  booktitle = {Proc. ICLR},
  year = 2019,
}

@inproceedings{kumar2019melgan,
    author = {Kundan Kumar and Rithesh Kumar and de Boissiere, Thibault and Lucas Gestin and Wei Zhen Teoh and Jose Sotelo and de Brebisson, Alexandre and Yoshua Bengio and Aaron Courville},
    title = {Mel{GAN}: Generative Adversarial Networks for Conditional Waveform Synthesis},
    booktitle = {Proc. NeurIPS},
    year = 2019,
}

@inproceedings{prenger2019waveglow,
    author = {Ryan Prenger and Rafael Valle and Bryan Catanzaro},
    title = {Wave{G}low: A Flow-based Generative Network for Speech Synthesis},
    booktitle = {Proc. ICASSP},
    year = 2019,
}

@inproceedings{hayes2021waveshaping,
    author = {Ben Hayes and Charalampos Saitis and George Fazekas},
    title = {Neural Waveshaping Synthesis},
    booktitle = {Proc. ISMIR},
    year = 2021,
}

@inproceedings{kong2021diffwave,
    author = {Zhifeng Kong and Wei Ping and Jiaji Huang and Kexin Zhao and Bryan Catanzaro},
    title = {{Diff{W}ave: A Versatile Diffusion Model for Audio Synthesis}},
    booktitle = {Proc. ICLR},
    year = 2021,
}

@inproceedings{iashin2021specvqgan,
    title = {{Taming Visually Guided Sound Generation}},
    author = {Iashin, Vladimir and Rahtu, Esa},
    booktitle = {Proc. BMVC},
    year = 2021,
}

@inproceedings{wu2022clap,
    author = {Yusong Wu and Ke Chen and Tianyu Zhang and Yuchen Hui and Taylor Berg-Kirkpatrick and Shlomo Dubnov},
    title = {{Large-scale Contrastive Language-Audio Pretraining with Feature Fusion and Keyword-to-Caption Augmentation}},
    booktitle = {Proc. ICASSP},
    year = 2023,
}

@inproceedings{huang2022mulan,
    title = {{{M}u{L}an: A Joint Embedding of Music Audio and Natural Language}},
    author = {Qingqing Huang and Aren Jansen and Joonseok Lee and Ravi Ganti and Judith Yue Li and Daniel P. W. Ellis},
    booktitle = {Proc. ISMIR},
    year = 2022
}

@inproceedings{wu2022wav2clip,
    author = {Ho-Hsiang Wu and Prem Seetharaman and Kundan Kumar and Juan Pablo Bello},
    title = {{{W}av2{CLIP}: Learning Robust Audio Representations From {CLIP}}},
    booktitle = {Proc. ICASSP},
    pages = {4563--4567},
    year = 2022,
}

@inproceedings{guzhov2022audioclip,
    author = {Andrey Guzhov and Federico Raue and Jörn Hees and Andreas Dengel},
    title = {{Audio{CLIP}: Extending {CLIP} to Image, Text and Audio}},
    booktitle = {Proc. ICASSP},
    pages = {976--980},
    year = 2022,
}

@inproceedings{owens2016vis,
    author = {Andrew Owens and Phillip Isola and Josh McDermott and Antonio Torralba and Edward H. Adelson and William T. Freeman},
    title = {{Visually Indicated Sounds}},
    booktitle = {Proc. CVPR},
    pages={2405--2413},
    year = 2016,
}

@inproceedings{lee2023bigvgan,
    author = {Sanggil Lee and Wei Ping and Boris Ginsburg and Bryan Catanzaro and Sungroh Yoon},
    title = {{Big{VGAN}: A Universal Neural Vocoder with Large-Scale Training}},
    booktitle = {Proc. ICLR},
    year = 2023,
}

@inproceedings{sheffer2022im2wav,
    title = {{I Hear Your True Colors: Image Guided Audio Generation}},
    author = {Roy Sheffer and Yossi Adi},
    booktitle = {Proc. ICASSP},
    year = 2023,
}

@article{liu2023audioldm,
    author = {Haohe Liu and Zehua Chen and Yi Yuan and Xinhao Mei and Xubo Liu and Danilo Mandic and Wenwu Wang and Mark D. Plumbley},
    title = {{{A}udio{LDM}: Text-to-Audio Generation with Latent Diffusion Models}},
    journal = {Proc. ICML},
    year = 2023,
}

@inproceedings{huang2023makeanaudio,
    author = {Rongjie Huang and Jiawei Huang and Dongchao Yang and Yi Ren and Luping Liu and Mingze Li and Zhenhui Ye and Jinglin Liu and Xiang Yin and Zhou Zhao},
    title = {{{M}ake-{A}n-{A}udio: Text-To-Audio Generation with Prompt-Enhanced Diffusion Models}},
    booktitle = {Proc. ICML},
    year = 2023,
}

@article{huang2023noise2music,
    author = {Qingqing Huang and Daniel S. Park and Tao Wang and Timo I. Denk and Andy Ly and Nanxin Chen and Zhengdong Zhang and Zhishuai Zhang and Jiahui Yu and Christian Frank and Jesse Engel and Quoc V. Le and William Chan and Zhifeng Chen and Wei Han},
    title = {{{N}oise2{M}usic: Text-conditioned Music Generation with Diffusion Models}},
    journal = {arXiv preprint arXiv:2302.03917},
    year = 2023,
}

@article{agostinelli2023musiclm,
    author = {Andrea Agostinelli and Timo I. Denk and Zalán Borsos and Jesse Engel and Mauro Verzetti and Antoine Caillon and Qingqing Huang and Aren Jansen and Adam Roberts and Marco Tagliasacchi and Matt Sharifi and Neil Zeghidour and Christian Frank},
    title = {{{M}usic{LM}: Generating Music From Text}},
    journal = {arXiv preprint arXiv:2302.03917},
    year = 2023,
}

@inproceedings{ho2021classifierfree,
    author = {Jonathan Ho, Tim Salimans},
    title = {{Classifier-Free Diffusion Guidance}},
    booktitle = {NeurIPS Workshop on Deep Generative Models and Downstream Applications},
    year = 2021,
}

@article{ramesh2022dalle2,
    author = {Aditya Ramesh and Prafulla Dhariwal and Alex Nichol and Casey Chu and Mark Chen},
    title = {{Hierarchical Text-Conditional Image Generation with CLIP Latents}},
    journal = {arXiv preprint arXiv:2204.06125},
    year = 2022,
}

@inproceedings{rombach2022ldm,
  title = {{High-resolution image synthesis with latent diffusion models}},
  author = {Robin Rombach and Andreas Blattmann and Dominik Lorenz and Patrick Esser and Björn Ommer},
  booktitle = {Proc. CVPR},
  pages = {10684--10695},
  year = 2022,
}

@inproceedings{pascual2022diffusion,
  title = {{Full-band General Audio Synthesis with Score-based Diffusion}},
  author = {Santiago Pascual and Gautam Bhattacharya and Chunghsin Yeh and Jordi Pons and Joan Serrà},
  booktitle = {Proc. ICASSP},
  year = 2023,
}

@inproceedings{schuhmann2022laion5b,
    title = {{LAION-5B: An open large-scale dataset for training next generation image-text models}},
    author = {Christoph Schuhmann and Romain Beaumont and Richard Vencu and Cade Gordon and Ross Wightman and Mehdi Cherti and Theo Coombes and Aarush Katta and Clayton Mullis and Mitchell Wortsman and Patrick Schramowski and Srivatsa Kundurthy and Katherine Crowson and Ludwig Schmidt and Robert Kaczmarczyk and Jenia Jitsev},
    booktitle = {NeurIPS 2022 Datasets and Benchmarks},
    year = 2022,
}

@inproceedings{rouditchenko2021avlnet,
    author = {Andrew Rouditchenko and Angie Boggust and David Harwath and Brian Chen and Dhiraj Joshi and Samuel Thomas and Kartik Audhkhasi and Hilde Kuehne and Rameswar Panda and Rogerio Feris and Brian Kingsbury and Michael Picheny and Antonio Torralba and James Glass},
    title = {{{AVL}net: Learning Audio-Visual Language Representations from Instructional Videos}},
    booktitle = {Proc. INTERSPEECH},
    pages={1584--1588},
    year = 2021,
}

@misc{clip_github,
    howpublished = {\url{https://github.com/openai/CLIP}},
}

@misc{improved_diff_github,
    howpublished = {\url{https://github.com/openai/improved-diffusion}},
}

@misc{dalle2_huggingface,
    howpublished = {\url{https://huggingface.co/laion/DALLE2-PyTorch}},
}

@misc{dalle2_github,
    howpublished = {\url{https://github.com/lucidrains/DALLE2-pytorch}},
}

@misc{bigvgan_github,
    howpublished = {\url{https://github.com/NVIDIA/BigVGAN}},
}

@misc{clap_github,
    howpublished = {\url{https://github.com/LAION-AI/CLAP}},
}

@misc{fad_github,
    howpublished = {\url{https://github.com/gudgud96/frechet-audio-distance}},
}

@inproceedings{castellon2021codified,
  author = {Rodrigo Castellon and Chris Donahue and Percy Liang},
  title = {Codified audio language modeling learns useful representations for music information retrieval},
  booktitle = {International Society for Music Information Retrieval Conference (ISMIR)},
  year = 2021,
}

@inproceedings{hawthorne2019maestro,
  author = {Curtis Hawthorne and Andriy Stasyuk and Adam Roberts and Ian Simon and Cheng-Zhi Anna Huang and Sander Dieleman and Erich Elsen and Jesse Engel and Douglas Eck},
  title = {Enabling Factorized Piano Music Modeling and Generation with the {MAESTRO} Dataset},
  booktitle = {ICLR},
  year = 2019,
}

@inproceedings{ren2019fastspeech,
  author = {Yi Ren and Yangjun Ruan and Xu Tan and Tao Qin and Sheng Zhao and Zhou Zhao and Tie-Yan Liu},
  title = {Fast{S}peech: Fast, Robust and Controllable Text to Speech},
  booktitle = {NeurIPS},
  year = 2019,
}

@inproceedings{yu2020durian,
  author = {Chengzhu Yu and Heng Lu and Na Hu and Meng Yu and Chao Weng and Kun Xu and Peng Liu and Deyi Tuo and Shiyin Kang and Guangzhi Lei and Dan Su and Dong Yu},
  title = {Dur{IAN}: Duration Informed Attention Network For Multimodal Synthesis},
  booktitle = {INTERSPEECH},
  year = 2020,
}

@inproceedings{kim2019mel2mel,
  author = {Jong Wook Kim and Rachel Bittner and Aparna Kumar and Juan Pablo Bello},
  title = {Neural Music Synthesis for Flexible Timbre Control},
  booktitle = {ICASSP},
  year = 2019,
}

@inproceedings{engel2020ddsp,
  author = {Jesse Engel and Lamtharn Hantrakul and Chenjie Gu and Adam Roberts},
  title = {{DDSP}: Differentiable Digital Signal Processing},
  booktitle = {ICLR},
  year = 2020,
}

@inproceedings{chien2021tts,
  author = {Chung-Ming Chien and Jheng-Hao Lin and Chien-Yu Huang and Po-Chun Hsu and Hung-Yi Lee},
  title = {Investigating on Incorporating Pretrained and Learnable Speaker Representations for Multi-Speaker Multi-Style Text-to-Speech},
  booktitle = {ICASSP},
  year = 2021,
}

@article{oore2020performancernn,
  author = {Sageev Oore and Ian Simon and Sander Dieleman and Douglas Eck and Karen Simonyan},
  title = {This Time with Feeling: Learning Expressive Musical Performance},
  journal = {Neural Computing and Applications},
  volume = 32,
  pages = {955--967},
  year = 2020,
}

@inproceedings{shen2018tacotron2,
  author = {Jonathan Shen and Ruoming Pang and Ron J. Weiss and Mike Schuster and Navdeep Jaitly and Zongheng Yang and Zhifeng Chen and Yu Zhang and Yuxuan Wang and RJ Skerry-Ryan and Rif A. Saurous and Yannis Agiomyrgiannakis and Yonghui Wu},
  title = {Natural {TTS} Synthesis by Conditioning {W}ave{N}et on Mel Spectrogram Predictions},
  booktitle = {ICASSP},
  year = 2018,
}

@inproceedings{ren2020deepsinger,
  author = {Yi Ren and Xu Tan and Tao Qin and Jian Luan and Zhou Zhao and Tie-Yan Liu},
  title = {Deep{S}inger: Singing Voice Synthesis with Data Mined From the Web},
  booktitle = {KDD},
  year = 2020,
}

@inproceedings{defossez2018sing,
  author = {Alexandre Défossez and Neil Zeghidour and Nicolas Usunier and Léon Bottou and Francis Bach},
  title = {{SING}: Symbol-to-Instrument Neural Generator},
  booktitle = {NeurIPS},
  year = 2018,
}

@inproceedings{manzelli2018wavenet,
  author = {Rachel Manzelli and Vijay Thakkar and Ali Siahkamari and Brian Kulis},
  title = {Conditioning Deep Generative Raw Audio Models for Structured Automatic Music},
  booktitle = {ISMIR},
  year = 2018,
}

@inproceedings{schimbinschi2019synthnet,
  author = {Florin Schimbinschi and Christian Walder and Sarah M. Erfani and James Bailey},
  title = {Synth{N}et: Learning to Synthesize Music End-to-End},
  booktitle = {IJCAI},
  year = 2019,
}

@article{tan2021ttssurvey,
  author = {Xu Tan and Tao Qin and Frank Soong and Tie-Yan Liu},
  title = {A Survey on Neural Speech Synthesis},
  journal = {arXiv preprint arXiv:2106.15561},
  year = 2021,
}

@inproceedings{yang2016violin,
  author = {Chih-Hong Yang and Pei-Ching Li and Alvin W. Y. Su and Li Su and Yi-Hsuan Yang},
  title = {Automatic Violin Synthesis Using Expressive Musical Term Features},
  booktitle = {DAFx},
  year = 2016,
}

@inproceedings{shih2017violin,
  author = {Chi-Ching Shih and Pei-Ching Li and Yi-Ju Lin and Yu-Lin Wang and Alvin W. Y. Su and Li Su and Yi-Hsuan Yang},
  title = {Analysis and Synthesis of the Violin Playing Style of {H}eifetz and {O}istrakh},
  booktitle = {DAFx},
  year = 2017,
}

@inproceedings{isola2017pix2pix,
  author = {Phillip Isola and Jun-Yan Zhu and Tinghui Zhou and Alexei A. Efros},
  title = {Image-to-Image Translation with Conditional Adversarial Networks},
  booktitle = {CVPR},
  year = 2017,
}

@inproceedings{yang2021ganspeech,
  author = {Jinhyeok Yang and Jae-Sung Bae and Taejun Bak and Youngik Kim and Hoon-Young Cho},
  title = {{GANS}peech: Adversarial Training for High-Fidelity Multi-Speaker Speech Synthesis},
  booktitle = {INTERSPEECH},
  year = 2021,
}

@inproceedings{wang2017tacotron,
  author = {Yuxuan Wang and RJ Skerry-Ryan and Daisy Stanton and Yonghui Wu and Ron J. Weiss and Navdeep Jaitly and Zongheng Yang and Ying Xiao and Zhifeng Chen and Samy Bengio and Quoc Le and Yannis Agiomyrgiannakis and Rob Clark and Rif A. Saurous},
  title = {Tacotron: Towards End-to-End Speech Synthesis},
  booktitle = {INTERSPEECH},
  year = 2017,
}

@inproceedings{anonymous2022mididdsp,
  author = {Yusong Wu and Ethan Manilow and Yi Deng and Rigel Swavely and Kyle Kastner and Tim Cooijmans and Aaron Courville and Cheng-Zhi Anna Huang and Jesse Engel},
  title = {{MIDI}-{DDSP}: Detailed Control of Musical Performance via Hierarchical Modeling},
  booktitle = {ICLR},
  year = 2022,
}

@misc{piano-e-competition,
  title = {International {P}iano-e-Competition},
  howpublished = {\url{https://www.piano-e-competition.com/}},
}

@misc{fluidsynth,
  title = {Fluid{S}ynth},
  howpublished = {\url{https://www.fluidsynth.org/}},
}

@inproceedings{huang2020aisongcontest,
  author = {Cheng-Zhi Anna Huang and Hendrik Vincent Koops and Ed NewtonRex and Monica Dinculescu and Carrie J. Cai},
  title = {AI Song Contest: Human-AI Co-Creation in Songwriting},
  booktitle = {International Society for Music Information Retrieval Conference (ISMIR)},
  year = 2020,
}

@book{deman2019production,
  author = {De Man, Brecht and Ryan Stables and Joshua D. Reiss},
  title = {Intelligent Music Production},
  publisher = {Routledge},
  year = 2019,
}

@article{liu2023wavjourney,
  author = {Xubo Liu and Zhongkai Zhu and Haohe Liu and Yi Yuan and Meng Cui and Qiushi Huang and Jinhua Liang and Yin Cao and Qiuqiang Kong and Mark D. Plumbley and Wenwu Wang},
  title	= {Wav{J}ourney: Compositional Audio Creation with Large Language Models},
  journal = {arXiv preprint arXiv:2307.14335},
  year = 2023,
}

@inproceedings{ruiz2023dreambooth,
  author = {Nataniel Ruiz and Yuanzhen Li and Varun Jampani and Yael Pritch and Michael Rubinstein and Kfir Aberman},
  title	= {Dream{B}ooth: Fine Tuning Text-to-Image Diffusion Models for Subject-Driven Generation},
  booktitle = {IEEE / CVF Computer Vision and Pattern Recognition Conference (CVPR)},
  year = 2023,
}

@inproceedings{brooks2023instructpix2pix,
  author = {Tim Brooks and Aleksander Holynski and Alexei A. Efros},
  title	= {Instruct{P}ix2{P}ix: Learning to Follow Image Editing Instructions},
  booktitle = {IEEE / CVF Computer Vision and Pattern Recognition Conference (CVPR)},
  year = 2023,
}

@inproceedings{dong2023clipsonic,
  author    = {Hao-Wen Dong and Xiaoyu Liu and Jordi Pons and Gautam Bhattacharya and Santiago Pascual and Joan Serrà and Taylor Berg-Kirkpatrick and Julian McAuley},
  title     = {CLIPSonic: Text-to-Audio Synthesis with Unlabeled Videos and Pretrained Language-Vision Models},
  booktitle = {IEEE Workshop on Applications of Signal Processing to Audio and Acoustics (WASPAA)},
  year      = 2023,
}

@inproceedings{dong2023mmt,
  author    = {Hao-Wen Dong and Ke Chen and Shlomo Dubnov and Julian McAuley and Taylor Berg-Kirkpatrick},
  title     = {Multitrack Music Transformer},
  booktitle = {IEEE International Conference on Acoustics, Speech and Signal Processing (ICASSP)},
  year      = 2023,
}

@inproceedings{dong2021deepperformer,
  author    = {Hao-Wen Dong and Cong Zhou and Taylor Berg-Kirkpatrick and Julian McAuley},
  title     = {{D}eep {P}erformer: Score-to-Audio Music Performance Synthesis},
  booktitle = {IEEE International Conference on Acoustics, Speech and Signal Processing (ICASSP)},
  year      = 2022,
}

@inproceedings{dong2021arranger,
  author    = {Hao-Wen Dong and Chris Donahue and Taylor Berg-Kirkpatrick and Julian McAuley},
  title     = {Towards Automatic Instrumentation by Learning to Separate Parts in Symbolic Multitrack Music},
  booktitle = {International Society for Music Information Retrieval Conference (ISMIR)},
  year      = 2021,
}

@inproceedings{dong2018binarymusegan,
  author    = {Hao-Wen Dong and Yi-Hsuan Yang},
  title     = {Convolutional Generative Adversarial Networks with Binary Neurons for Polyphonic Music Generation},
  booktitle = {International Society for Music Information Retrieval Conference (ISMIR)},
  year      = 2018,
}

@article{yeh2020harmonization,
  author = {Yin-Cheng Yeh and Wen-Yi Hsiao and Satoru Fukayama and Tetsuro Kitahara and Benjamin Genchel and Hao-Min Liu and Hao-Wen Dong and Yian Chen and Terence Leong and Yi-Hsuan Yang},
  title = {Automatic Melody Harmonization with Triad Chords: A Comparative Study},
  journal = {Journal of New Music Research (JNMR)},
  volume = 50,
  number = 1,
  pages = {37--51},
  year = 2021,
}

@inproceedings{dong2017musegandemo,
  author    = {Hao-Wen Dong and Wen-Yi Hsiao and Li-Chia Yang and Yi-Hsuan Yang},
  title     = {{M}use{GAN}: Demonstration of a Convolutional {GAN} Based Model for Generating Multi-track Piano-rolls},
  booktitle = {ISMIR Late-Breaking Demos},
  year      = 2017,
  author+an = {1=first; 2=first},
}

@inproceedings{xu2023mmt2,
  author    = {Weihan Xu and Julian McAuley and Shlomo Dubnov and Hao-Wen Dong},
  title     = {Equipping Pretrained Unconditional Music Transformers with Instrument and Genre Controls},
  booktitle = {IEEE Big Data Workshop on AI Music Generation (AIMG)},
  year      = 2023,
}

@inproceedings{dong2023clipsynth,
  author    = {Hao-Wen Dong and Gunnar A. Sigurdsson and Chenyang Tao and Jiun-Yu Kao and Yu-Hsiang Lin and Anjali Narayan-Chen and Arpit Gupta and Tagyoung Chung and Jing Huang and Nanyun Peng and Wenbo Zhao},
  title     = {CLIPSynth: Learning Text-to-audio Synthesis from Videos using CLIP and Diffusion Models},
  booktitle = {CVPR Workshop on Sight and Sound},
  year      = 2023,
}

@inproceedings{liu2018musegangtc,
  author    = {Hao-Min Liu and Hao-Wen Dong and Wen-Yi Hsiao and Yi-Hsuan Yang},
  title     = {Lead sheet and Multi-track Piano-roll generation using {M}use{GAN}},
  booktitle = {GPU Technology Conference (GTC) Taiwan},
  year      = 2018,
}

@phdthesis{raffel16lmd,
  author    = {Colin Raffel},
  title     = {Learning-Based Methods for Comparing Sequences, with Applications to Audio-to-{MIDI} Alignment and Matching},
  school    = {Columbia University},
  year      = 2016,
  url       = {https://colinraffel.com/projects/lmd/},
}

@inproceedings{huang2020remi,
  author = {Yu-Siang Huang and Yi-Hsuan Yang},
  title = {Pop Music Transformer: Generating Music with Rhythm and Harmony},
  booktitle = {Proc. MM},
  year = 2020,
}

@inproceedings{crestel2017lop,
  author = {Léopold Crestel and Philippe Esling and Lena Heng and Stephen McAdams},
  title = {A database linking piano and orchestral {MIDI} scores with application to automatic projective orchestration},
  booktitle = {Proc. ISMIR},
  year = 2017,
  url = {https://qsdfo.github.io/LOP/database.html},
}

@inproceedings{liu2018transformerdecoder,
  author = {Peter J. Liu and Mohammad Saleh and Etienne Pot and Ben Goodrich and Ryan Sepassi and Lukasz Kaiser and Noam Shazeer},
  title = {Generating Wikipedia by Summarizing Long Sequences},
  booktitle = {Proc. ICLR},
  year = 2018,
}

@inproceedings{chang2021infiliing,
  author = {Chin-Jui Chang and Chun-Yi Lee and Yi-Hsuan Yang},
  title = {Variable-Length Music Score Infilling via {XLN}et and Musically Specialized Positional Encoding},
  booktitle = {Proc. ISMIR},
  year = 2021,
}

@article{shih2022themetransformer,
  author = {Yi-Jen Shih and Shih-Lun Wu and Frank Zalkow and Meinard Müller and Yi-Hsuan Yang},
  title = {Theme Transformer: Symbolic Music Generation with Theme-Conditioned Transformer},
  journal = {IEEE Transactions on Multimedia},
  year = 2022,
}

@inproceedings{huang2018visualizing,
  author = {Anna Huang and Monica Dinculescu and Ashish Vaswani and Douglas Eck},
  title = {Visualizing Music Self-Attention},
  booktitle = {Proc. NeurIPS Workshop on Interpretability and Robustness in Audio, Speech, and Language},
  year = 2018,
}

@article{vonrutte2022figaro,
  author = {Dimitri von Rütte and Luca Biggio and Yannic Kilcher and Thomas Hofmann},
  title = {{FIGARO}: Generating Symbolic Music with Fine-Grained Artistic Control},
  journal = {arXiv preprint arXiv:2201.10936},
  year = 2022,
}

@article{ji2020survey,
  author = {Shulei Ji and Jing Luo and Xinyu Yang},
  title = {A Comprehensive Survey on Deep Music Generation: Multi-level Representations, Algorithms, Evaluations, and Future Directions},
  journal = {arXiv preprint arXiv:2011.06801},
  year = 2020,
}

@inproceedings{chen2020guitar,
  title = {Automatic Composition of Guitar Tabs by Transformers and Groove Modeling},
  author = {Yu-Hua Chen and Yu-Siang Huang and Wen-Yi Hsiao and Yi-Hsuan Yang},
  booktitle = {Proc. ISMIR},
  year = 2020
}

@article{chen2021transformer,
  author = {Tsung-Ping Chen and Li Su},
  title = {Attend to Chords: Improving Harmonic Analysis of Symbolic Music Using Transformer-Based Models},
  journal = {Transactions of ISMIR},
  volume = 4,
  number = 1,
  page = {1--13},
  year = 2021,
}

@inproceedings{ziyu2021musebert,
  author = {Ziyu Wang and Gus Xia},
  title = {{M}use{BERT}: Pre-training of Music Representation for Music Understanding and Controllable Generation},
  booktitle = {Proc. ISMIR},
  year = 2021,
}

@article{chou2021midibert,
  author = {Yi-Hui Chou and I-Chun Chen and Chin-Jui Chang and Joann Ching and Yi-Hsuan Yang},
  title = {{M}idi{BERT}-Piano: Large-scale Pre-training for Symbolic Music Understanding},
  journal = {arXiv preprint arXiv:2107.05223},
  year = 2021,
}

@inproceedings{zeng2021musicbert,
  author = {Mingliang Zeng and Xu Tan and Rui Wang and Zeqian Ju and Tao Qin and Tie-Yan Liu},
  title = {{M}usic{BERT}: Symbolic Music Understanding with Large-Scale Pre-Training},
  booktitle = {Proc. Findings of ACL},
  year = 2021,
}

@misc{musescoresoundfont,
  title = {Muse{S}core General SoundFont},
  url = {https://musescore.org/en/handbook/3/soundfonts-and-sfz-files},
}

\end{document}